# Fetal Brain Tissue Annotation and Segmentation Challenge Results

Kelly Payette[a,b*], Hongwei Li[c,d], Priscille de Dumast[e,f], Roxane Licandro[g,h], Hui Ji[a,b], Md Mahfuzur Rahman Siddiquee[i,j], Daguang Xu[j], Andriy Myronenko[j], Hao Liu[k], Yuchen Pei[k], Lisheng Wang[k], Ying Peng[l], Juanying Xie[l], Huiquan Zhang[l], Guiming Dong[jj], Hao Fu[jj], Guotai Wang[jj], ZunHyan Rieu[m], Donghyeon Kim[m], Hyun Gi Kim[n], Davood Karimi[o], Ali Gholipour[o], Helena R. Torres[p,q,r,s], Bruno Oliveira[p,q,r,s], João L. Vilaça[p], Yang Lin[kk], Netanell Avisdris[t,u], Ori Ben-Zvi[u,v], Dafna Ben Bashat[u,v,w], Lucas Fidon[x], Michael Aertsen[y], Tom Vercauteren[x], Daniel Sobotka[z], Georg Langs[z], Mireia Alenyà[aa], Maria Inmaculada Villanueva[bb,cc], Oscar Camara[aa], Bella Specktor Fadida[t], Leo Joskowicz[t], Liao Weibin[ll], Lv Yi[ll], Li Xuesong[ll], Moona Mazher[dd], Abdul Qayyum[ee], Domenec Puig[dd], Hamza Kebiri[e,f], Zelin Zhang[ff], Xinyi Xu[ff], Dan Wu[ff], KuanLun Liao[gg], YiXuan Wu[gg], JinTai Chen[gg], Yunzhi Xu[ff], Li Zhao[ff], Lana Vasung[hh,ii], Bjoern Menze[mm], Meritxell Bach Cuadra[f,e], Andras Jakab[a,b]

a. Center for MR Research, University Children's Hospital Zurich, University of Zurich, Zurich, Switzerland ; b. Neuroscience Center Zurich, University of Zurich, Zurich, Switzerland; c. Department of Quantitative Biomedicine, University of Zurich, Zurich, Switzerland; d. Department of Informatics, Technical University of Munich, Munich, Germany; e. Medical Image Analysis Laboratory, Department of Diagnostic and Interventional Radiology, Lausanne University Hospital and University of Lausanne, Lausanne, Switzerland; f. CIBM, Center for Biomedical Imaging, Lausanne, Switzerland; g. Laboratory for Computational Neuroimaging, Athinoula A. Martinos Center for Biomedical Imaging, Massachusetts General Hospital/Harvard Medical School, Charlestown, MA, USA; h. Department of Biomedical Imaging and Image-guided Therapy, Computational Imaging Research Lab (CIR), Medical University of Vienna, Vienna, Austria; i. Arizona State University; j. NVIDIA; k. Shanghai Jiaotong University; l. School of Computer Science, Shaanxi Normal University, Xi'an 710119, PR China; m. Research Institute, NEUROPHET Inc., Seoul 06247, Korea; n. Department of Radiology, The Catholic University of Korea, Eunpyeong St. Mary's Hospital, Seoul 06247, Korea; o.Boston Children's Hospital and Harvard Medical School, Boston, MA, USA; p.  2Ai – School of Technology, IPCA, Barcelos, Portugal; q. Algoritmi Center, School of Engineering, University of Minho, Guimarães, Portugal; r.Life and Health Sciences Research Institute (ICVS), School of Medicine, University of Minho, Braga, Portugal; s. ICVS/3B's - PT Government Associate Laboratory, Braga/Guimarães, Portugal; t. School of Computer Science and Engineering, The Hebrew University of Jerusalem, Israel; u. Sagol Brain Institute, Tel Aviv Sourasky Medical Center, Israel; v. Sagol School of Neuroscience, Tel Aviv University, Israel; w. Sackler Faculty of Medicine, Tel Aviv University, Israel; x. School of Biomedical Engineering & Imaging Sciences, King's College London, London, SE1 7EU, UK; y. Department of Radiology, University Hospitals Leuven, 3000 Leuven, Belgium; z. Computational Imaging Research Lab, Department of Biomedical Imaging and Image-guided Therapy, Medical University of Vienna, Vienna, Austria; aa. BCN-MedTech, Department of Information and Communications Technologies, Universitat Pompeu Fabra, Barcelona, Spain; bb.  Department of Information and Communications Technologies, Universitat Pompeu Fabra, Barcelona, Spain; cc. Institut d'Investigacions Biomèdiques August Pi i Sunyer, Barcelona, Spain; dd. Department of Computer Engineering and Mathematics, University Rovira i Virgili,Spain; ee. Université de Bourgogne, France; ff. Key Laboratory for Biomedical Engineering of Ministry of Education, Department of Biomedical Engineering, College of Biomedical Engineering & Instrument Science, Zhejiang University, Yuquan Campus, Hangzhou, China; gg. College of Computer Science, Zhejiang University, Hangzhou, China; hh. Division of Newborn Medicine, Department of Pediatrics, Boston Children's Hospital; ii. Department of Pediatrics, Harvard Medical School; jj. School of Mechanical and Electrical Engineering, University of Electronic Science and Technology of China, Chengdu, China; kk. Department of Computer Science, Hong Kong University of Science and Technology; ll. School of Computer Science, Beijing Institute of Technology; mm. Department of Quantitative Biomedicine, University of Zurich, Zurich, Switzerland

Corresponding author: Kelly Payette, Email: kelly.payette@kispi.uzh.ch

## Highlights

- benchmark for future automatic multi-tissue fetal brain segmentation algorithms
- used the largest publicly available fetal brain dataset with manual annotations
- U-Net is the dominant method for automatic fetal brain segmentation
- results using the U-Net have reached a plateau
- challenge results analyzed from both technical and clinical perspectives

## Abstract

In-utero fetal MRI is emerging as an important tool in the diagnosis and analysis of the developing human brain. Automatic segmentation of the developing fetal brain is a vital step in the quantitative analysis of prenatal neurodevelopment both in the research and clinical context. However, manual segmentation of cerebral structures is time-consuming and prone to error and inter-observer variability. Therefore, we organized the Fetal Tissue Annotation (FeTA) Challenge in 2021 in order to encourage the development of automatic segmentation algorithms on an international level. The challenge utilized FeTA Dataset, an open dataset of fetal brain MRI reconstructions segmented into seven different tissues (external cerebrospinal fluid, grey matter, white matter, ventricles, cerebellum, brainstem, deep grey matter). 20 international teams participated in this challenge, submitting a total of 21 algorithms for evaluation. In this paper, we provide a detailed analysis of the results from both a technical and clinical perspective. All participants relied on deep learning methods, mainly U-Nets, with some variability present in the network architecture, optimization, and image pre- and post-processing. The majority of teams used existing medical imaging deep learning frameworks. The main differences between the submissions were the fine tuning done during training, and the specific pre- and post-processing steps performed. The challenge results showed that almost all submissions performed similarly. Four of the top five teams used ensemble learning methods. However, one team's algorithm performed significantly superior to the other submissions, and consisted of an asymmetrical U-Net network architecture. This paper provides a first of its kind benchmark for future automatic multi-tissue segmentation algorithms for the developing human brain in utero.

## Graphical Abstract

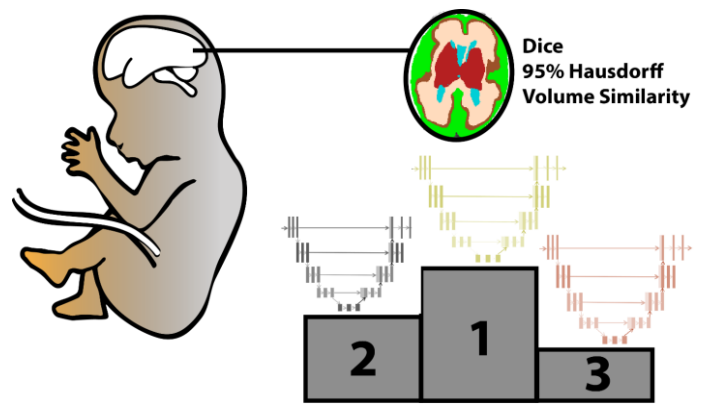

## Introduction

Fetal in-utero magnetic resonance imaging (MRI) is a powerful tool to investigate the developing human brain in fetuses with and without pathological features (De Asis-Cruz et al., 2021; Hosny and Elghawabi, 2010). It can be used to portray the complex neurodevelopmental events during human gestation, which remain to be completely characterized (Vasung et al., 2019). Clinically, it is becoming an important adjunct to ultrasound in the detection and diagnosis of congenital disorders (Hart et al., 2020), and can be used to aid during prenatal care (Gholipour et al., 2014).

Automated segmentation and quantification of the highly complex and rapidly changing brain morphology using MRI prior to birth has great potential to improve the diagnostic process, as manual segmentation is both time consuming and subject to human error and inter-rater variability. It is clinically relevant to analyze the morphometry of the developing brain, where measures such as the volume or the shape can be objectively compared with population-based references of normative development. Many congenital and acquired disorders manifest in reduced brain volume or altered anatomical structure of cerebral tissue compartments, for example, slower cortical growth (Clouchoux et al., 2013; Egaña-Ugrinovic et al., 2013) or reduced white matter volume (Rollins et al., 2021). Existing MRI based data of brain growth is mainly based on normally developing brains (Jarvis et al., 2019; Kyriakopoulou et al., 2017; Prayer et al., 2006), leaving brain growth in various numerous pathologies and congenital disorders largely unexplored.

From a technical standpoint, there are many challenges that an automatic segmentation method of the fetal brain would need to overcome. The cerebral structures are constantly growing and developing in complexity throughout gestation, which results in a gradually changing appearance in shape, size, and image intensity on MRI. In addition, the quality of the images can be poor due to fetal and maternal movement and imaging artefacts (Glenn, 2010). The boundary between tissues is often unclear on MR images due to partial volume effects (Bach Cuadra et al., 2009). Furthermore, fetal brains with abnormal features can have radically different morphology than those in a non-pathological brain. This can make it challenging for an automatic method to correctly identify these structures.

Fetal MRI requires no special MRI equipment, is noninvasive, safe (Gowland, 2011; Zvi et al., 2020), and its value in the diagnosis of certain central nervous system or somatic disorders is being increasingly recognized (Griffiths et al., 2019; Nagaraj et al., 2022). The development of ultra-fast MRI sequences such as the single shot T2-weighted sequence have also led to the increasing popularity of fetal MRI as these images have excellent soft tissue contrast and reduced motion artefact (Gholipour et al., 2014). As a result, fetal MRI is more frequently performed at diagnostic and surgical centers worldwide. There is also an increase in the number of studies focused on developing computational tools to quantitatively analyze the fetal brain. Some studies have focused on segmenting a specific tissue for analysis, such as the cortical plate (Benkarim et al., 2018; de Dumast et al., 2020; Fetit et al., 2020; Hong et al., 2020). Other studies have developed multi-tissue segmentation algorithms using a limited in-house dataset (e.g., clinically acquired anisotropic coronal images of normal fetuses (Khalili et al., 2019), or images of a specific pathology (Sanroma et al., 2018), or with atlas based frameworks (Dittrich et al., 2011; Gholipour et al., 2017; Licandro et al., 2016)). However, the field of developing automated tools for fetal MRI has been understudied due to both challenges in imaging and the lack of public, curated, and annotated ground truth data. Such shared datasets are currently the backbone for developing computer-aided diagnostic support systems.

In this paper we describe the Fetal Brain Tissue Annotation and Segmentation Challenge (FeTA) and outline the challenge organization, the submitted segmentation frameworks, and a detailed evaluation of the challenge results, with reporting based on the BIAS method (Maier-Hein et al., 2020). The aim of the FeTA Challenge was to develop reliable, valid, and reproducible methods of analyzing high resolution reconstructed MR images of the developing fetal brain from gestational week 20-35. The FeTA Challenge used an expanded version of the original FeTA Dataset to develop automatic fetal brain tissue segmentation methods (Payette et al., 2021a). Our evaluation compares and analyzes the algorithms on a test dataset hidden to the participants. The submitted algorithms are also tested on various subsets of the testing dataset in order to determine whether they perform better or worse under various circumstances such as image quality or reconstruction method. Finally, we investigated two real life applications outside the scope of the FeTA Challenge evaluation: First, the performance of the submitted algorithms to estimate intracranial volume was evaluated, an application relevant to the characterization of developmental delay in many conditions, such as intrauterine growth restriction or congenital heart defects (Polat et al., 2017; Sadhwani et al., 2022; Skotting et al., 2021). Second, we looked at the ability of the algorithms to segment younger (<29 weeks) versus older (≥29 weeks) fetal brains). The algorithms developed as part of the FeTA Challenge will have the potential to help better understand the underlying causes of congenital disorders and ultimately to guide the development of perinatal guidelines and clinical risk stratification tools for early interventions, treatments, and care management decisions.

## Materials and Methods

**Challenge Organization:** The FeTA Challenge was held as part of the international Medical Image Computing and Computer Assisted Intervention (MICCAI) 2021 Conference (https://feta.grand-challenge.org/). Participants were to create a fully automatic multi-class segmentation algorithm of the fetal brain (with optional inputs of gestational age and whether the brain was pathological or not). The training dataset was made available to the participants on May 3rd, 2021 on Synapse (Payette and Jakab, 2021) to train their own methods. Participants were able to use other publicly available datasets for training if they wished to, as long as it was documented in their algorithm description. Participants created a Docker container which stored the algorithm, and submitted this container to the organizers by July 30, 2021. Organizers were allowed to submit containers, but were not eligible for prizes. This container was run by the challenge organizers locally on the hidden testing dataset in order to compare the algorithms. Re-submission of the Docker container was only allowed in cases of technical difficulties or bugs identified during evaluation. The top teams received their results on September 1, 2021 in order to prepare presentations. The complete results and awards to the top three teams were presented on Oct 1, 2021 at the MICCAI Conference FeTA Challenge Session. Dockers of teams who provided permission are available on Dockerhub (https://hub.docker.com/u/fetachallenge). For the complete overview of the challenge, see the final challenge proposal (Payette et al., 2021b).

**Mission of the Challenge:** The mission of the FeTA Challenge is to boost the development of accurate and automatic multi-class segmentation algorithms for the developing human brain with fetal MRI, and to create a benchmark for future algorithms. There were a total of eight classes: external cerebrospinal fluid (eCSF), grey matter (GM), white matter (WM), ventricles (including cavum), cerebellum, deep grey matter (deep GM), brainstem, and background. The target cohort for the FeTA Challenge were pregnant mothers who, after an initial ultrasound examination, were clinically referred for a fetal MRI. The acquired fetal MRI images were then reconstructed into a 3-dimensional volume using a super-resolution method (for details see Section 0). The task of the challenge was to segment these super-resolution volumes into different brain tissues. The challenge cohort was made up of two subgroups: fetuses with normal and abnormal development of the nervous system, and covers a gestational age (GA) range of 20-35 weeks. The accuracy of the automatically generated fetal brain segmentations was evaluated in the challenge cohort in order to determine the optimal segmentation method for fetal brain MRI.

**Challenge Dataset:** For the challenge, a clinically acquired dataset from a single institution was used for both the training and testing data. 120 fetal MRI brain scans were acquired. Recorded gestational age was modified by a random value within the range of ±3 days to further anonymize the data. Several T2-weighted single shot Fast Spin Echo (ssFSE) images were acquired for each subject in all three planes with a reconstructed resolution of 0.5mm x 0.5mm x 3 to 5mm. The images were acquired on either a 1.5T or 3T clinical GE whole-body MRI scanners (Signa Discovery MR450 and MR750) using an 8-channel cardiac or body coil with the following sequence parameters: TR: 2000–3500 ms, TE: 120 ms (minimum), flip angle: 90°, sampling percentage 55%. Field of view (200–240 mm) and image matrix (1.5T: 256×224; 3T: 320×224) were adjusted depending on the gestational age and size of the fetus. The data was acquired at the University Children's Hospital Zurich in Zurich, Switzerland by trained radiographers using clinically defined protocols.

For each subject, the acquired images were reviewed, and images of good quality, at least one image in each of the axial, sagittal, coronal planes with respect to the fetal brain, were chosen. A high-resolution fetal brain reconstruction was performed with the chosen scans using a super-resolution (SR) method (60 cases reconstructed with the mialSR method (Pierre Deman et al., 2020; Tourbier et al., 2019, 2015) and 60 cases reconstructed with the Simple IRTK method (Kuklisova-Murgasova et al., 2012)). Fetal brain masks were created where necessitated by the SR algorithm, either manually or with a custom MeVisLab module (Pierre Deman et al., 2020; Tourbier et al., 2015). Cases reconstructed with mialSR were reoriented prior to reconstruction through the MeVisLab module. Cases reconstructed with the Simple IRTK method were registered to an atlas after reconstruction (Serag et al., 2012). After reconstruction, each fetal brain volume had an isotropic resolution of approximately 0.5mmx0.5mmx0.5mm, with some deviation in exact dimensions between the SR methods. Each reconstructed image was then histogram-matched using Slicer (Kikinis et al., 2014), and zero-padded to be 256x256x256 voxels. For each reconstruction method, 40 cases were included in the training dataset available to the challenge participants (for a total of 80 cases), and 20 cases were included in testing dataset not available to the participants (for a total of 40 cases). Note that maternal tissue was excluded from the super-resolution reconstruction, only the fetal brain was reconstructed. Examples of non-pathological fetal brains across the range of gestational ages included in the dataset and their corresponding label maps can be seen in Figure 1.

The training and testing datasets consisted of fetuses with both typical and atypical features. In the group with atypical features, a variety of cerebral pathologies of varying severities were included (such as Chiari-II malformation or ventricular dysmorphology seen in ventriculomegaly). There were slightly more pathological than neurotypical cases, as in the clinic where the scans were performed it is more common to see pathologic brains (Figure 2). Fetuses with a gestational age range of 20 to 35 gestational weeks were included (mean gestational age: 27.0±3.60 weeks), with the distribution of ages and pathologies equal between the training and testing datasets (see Figure 3). The gestational age and the label of "neurotypical/pathological" was made available to the participants. Each case's label map was manually segmented by individuals with experience in segmenting medical images using the method described in (Payette et al., 2021a). Each case consists of a 3D super-resolution reconstruction of a fetal brain (256x256x256 voxels) and the associated manually segmented label map. There is no overlap of subjects between the training and testing dataset, each dataset is unique. The dataset and affiliated custom license is publicly available on Synapse (Payette and Jakab, 2021).

Mothers of the healthy fetuses participating in the BrainDNIU study were prospectively informed about the inclusion in the FeTA Dataset by members of the research team and gave written consent for their participation. Mothers of all other fetuses included in the current work were scanned as part of their routine clinical care and gave informed written consent for the re-use of their data for research purposes. The ethical committee of the Canton of Zurich, Switzerland approved the prospective and retrospective studies that collected and analyzed the MRI data (Decision numbers: 2017-00885, 2016-01019, 2017-00167), and a waiver for an ethical approval was acquired for the release of a fully anonymous dataset for research purposes.

Participants were free to choose if they wanted to work with the data in a 2D or 3D format. A validation dataset was not provided to the participants, it was up to the team's discretion to decide how to train their data and what to use for validation. The following section outlines the evaluation metrics used to determine the ranking of participants in the challenge.

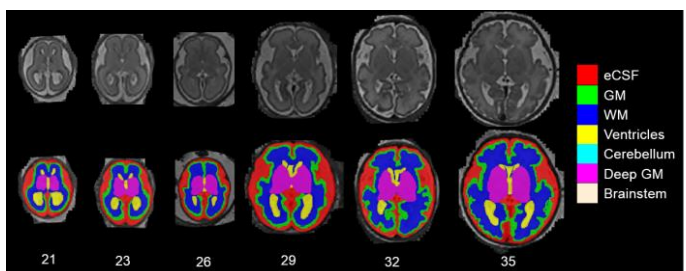

*Figure 1: Fetal Brain Segmentations by gestational age*

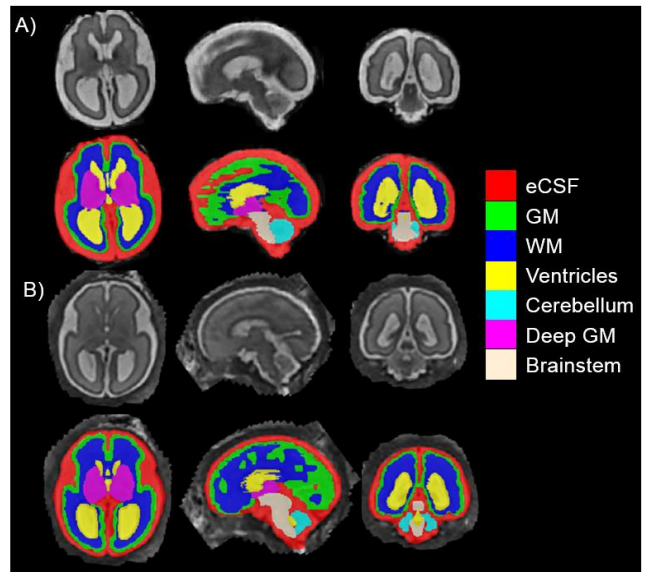

*Figure 2: Pathological fetal brain viewed in axial, sagittal, and coronal directions A): mialSR reconstruction, 27.3 GA; B) Simple IRTK SR Reconstruction, 26.9 GA*

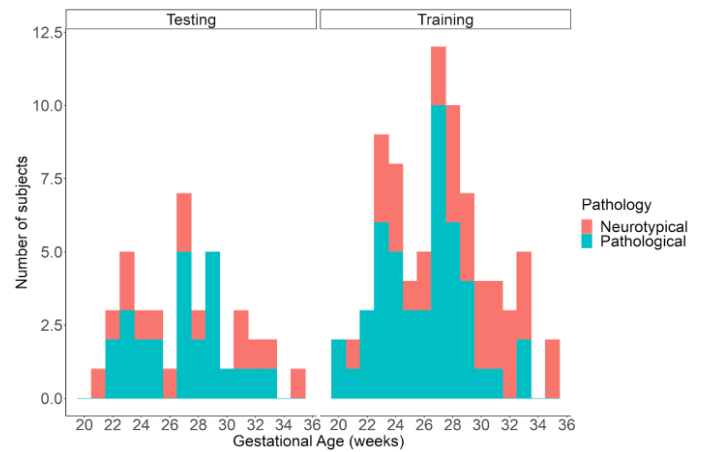

*Figure 3: Dataset Age Range: Histogram of the gestational age range of neurotypical and pathological cases within the testing and training dataset for the FeTA Challenge.*

**Assessment Method - Evaluation Metrics:** Three different metrics were chosen to compute the rankings of the FeTA Challenge: the Dice Similarity Coefficient (DSC), The Volume Similarity (VS), and the Hausdorff distance (HD). The DSC was chosen, as it is the most popular segmentation overlap metric for segmentation evaluation (Dice, 1945). However, we were also interested in assessing volume, as relevant biomarker for fetal development, and surface-based error. Therefore the HD (surface, (Hausdorff, 1991)), and VS (volume) metrics were chosen as well (Taha and Hanbury, 2015), and the final ranking will take all three metrics into account.

The DSC measures the amount of overlap between the manual segmentations (MS) label and the new segmentation (NS) generated by the participant's algorithm, and is defined as

$$DSC = \frac{2 \, |MS \cap NS|}{|MS| + |NS|}$$

The VS is a volumetric metric that measures the similarity between the volume of the GT and NS label map and is defined as

$$VS = 1 - \frac{|MS_{vol} - NS_{vol}|}{MS_{vol} + NS_{vol}}$$

The HD is a distance metric that evaluates the distance between two finite point sets A and B.

The Hausdorff distance (HD) is a spatial metric helpful in evaluating the contours of segmentations as well as the spatial positions of the voxels. The HD between two finite point sets A and B is defined as

$$HD(A,B) = max(h(A,B), h(B,A))$$

$$h(A,B) = ||a - b||$$

Note: The original challenge design had stated that the 95th percentile HD of maximum distances would be used to exclude possible outliers. However, after the challenge it was discovered that there was an error in the implementation of the 95th percentile, and the values reported were close to the maximum HD, and therefore these are the scores reported in this paper. This makes the HD values reported within this report slightly more susceptible to outliers. However, as we take three different metrics into account for the final ranking, the overall impact of outliers is reduced. For each metric, the implementation described in (Taha and Hanbury, 2015) was used (EvaluateSegmentation Tool, v2017.04.25).

**Assessment Method – Ranking:** Each of the participating teams was ranked based on each evaluation metric, and then the final rankings combined the rankings from all of the metrics (DSC, HD95, VS). The DSC, HD95, and VS were calculated for each label within each of the corresponding predicted label maps of the fetal brain volumes in the testing set. The mean and standard deviation of each label for all test cases was calculated, and the participating algorithms were ranked from low to high (HD95), where the lowest score received the highest scoring rank (best), and from high to low (DSC, VS), where the highest value received highest scoring rank (best) based on the calculated mean across all labels and test cases. If there were missing results, the worst possible value is used. For example, if a label does not exist in the NS label map but is present in the GT label map, it will receive a DSC and VS score of 0, and the HD95 score will be double the max value of the other algorithms submitted. This ranking procedure was developed in order to take three different metric types equally into account.

**Assessment Method - Further Analysis:** In addition to the ranking above, several other analyses were performed on the submitted algorithms. Per-label rankings of the entire dataset were analyzed. In addition, the algorithms were evaluated in the categories 'Non Pathological cases' and 'Pathological cases', SR reconstruction method (mialSR and Simple IRTK) as well as 'Excellent Quality', 'Good Quality' and 'Poor Quality', with the identical ranking methodology for each category. The pathology of each fetal brain was determined by an experienced radiologist. The quality of the fetal brain SR reconstructions were determined based on ratings (Excellent, Good, Poor) from three independent raters, and the correlation of the reviewers was calculated using the Gwet AC coefficient using R (v4.0.2, (Gwet, 2019)). As the ratings are ordinal data, the median of the ratings were considered to be the final rating of the SR volume. The participating algorithms were also evaluated on the different SR reconstruction methods.

Intracranial volume was calculated and compared to the manual segmentation's intracranial volume as well, but not used in the rankings. Intracranial volume was calculated by adding all labels except the background together.

An analysis of the performance of the algorithms based on gestational age was also performed, as the structure of the fetal brain changes greatly throughout development, especially in the cortex where there is increased cortical complexity, disappearance of transient subplate zone related to cortical maturation (blurring of white matter and grey matter border) and partial volumes (blurring of white matter/grey matter border in gyral crest, blurring of CSF/grey matter border because of narrow sulci). Because of this, the random error in segmentation of the grey matter between 29-35 GW might be increased. Therefore, in order to determine if gestational age impacts the success of a segmentation algorithm, our testing dataset was split into two age groups (21-28 weeks, and 29-35 weeks), and the differences between these two groups was analyzed by looking for differences in the evaluation metrics between each label for each of the submitted algorithms.

## Results

**Training and Testing Data:** A Kolmogorov–Smirnov test was performed in R v4.0.0 (R Core Team, 2020) in order to compare the distribution of GA and non-pathological/pathological fetal brains between the training and testing data. No significant differences were found between the training and testing datasets (GA: p = 0.88; pathology: p = 1).

**Challenge Submission:** In total, 21 teams submitted algorithms to the FeTA Challenge. One team's Docker container was not able to be fixed prior to the deadline and was thereby excluded. One team (Ichi-love) submitted two algorithms, meaning the total number of participating teams was 20, and the total number of valid submissions was 21. Each team submitted a written description of their algorithm, which can be found in the Appendix. Each algorithm is summarized in Table 1, and the pre-processing and data augmentation used by each team is outlined in

*Table 2*. All teams submitted a deep learning-based method, most of which were variants based on the U-Net architecture (Çiçek et al., 2016; Ronneberger et al., 2015). The top five teams used similar loss functions (mainly the combination of Dice loss and cross-entropy loss), and four of the five (excluding Neurophet) used an ensemble learning method. Every method used a 3x3 (or 3x3x3) convolutional kernel except for one team (Physense-UPF) who used a 2x2x2 kernel. Most submissions (14 of 21) used a random initialization of network parameters. The networks were of varying depths, between 3 and 6 layers on each of the ascending and descending layers of the networks. Thirteen of the submitted networks were 3D networks, the remainder were 2D or 2.5D. Seven teams used cross-validation. A variety of different data augmentation strategies were used, and only two team did not employ data augmentation at all. Only four teams used external datasets, either during the training step or used pre-trained network backbones trained on publicly available datasets.

**Metric Values and Rankings:** Statistical analysis of the metrics of the challenge and images displayed in this section were created using the ChallengeR tool (Wiesenfarth et al., 2021). The individual metrics for each team (all labels combined) can be found in Figures 4-6. The final ranking of all teams and their average evaluation metrics can be found in Table 3. The full reports (DSC, HD95, VS of all labels combined) created by the ChallengeR Tool, including details on the statistical tests performed can be found in Sections 2-4 of the Appendix. In the significance maps displayed, the testing was done using a one-sided Wilcoxon signed rank test at a 5% significance level, with adjustments for multiple comparisons. In all cases, the x-axis in the boxplots are ranked according to the mean values of the respective evaluation metric, and the black bar indicates the median value.

The top three teams according to the DSC were NVAUTO, SJTU_EIEE_2-426Lab, and Neurophet. The top three teams according to the HD95 were NVAUTO, Hilab, and 2Ai. The top three teams according to the VS were ichilove-ax, NVAUTO, and SJTU_EIEE_2-426Lab. With a few exceptions, there was no statistically significant differences between the top 10-12 teams in all three metrics, suggesting that a plateau has been reached. The highest and lowest average DSC were: 0.786 (team NVAUTO) and 0.534 (team A3). The lowest and highest average HD95 were: 14.012 (team NVAUTO) and 39.608 (team A3). The highest and lowest average VS were: 0.888 (team ichilove-ax) and 0.791 (team A3). However, when the bootstrapping and significance maps are investigated, it is clear that NVAUTO is the top team for the DSC metric, placing first in 100% of the bootstrap sampling, and is statistically significant to all but one of the algorithms (Team pengyy). There is no difference between teams in places 2 to 4 for the DSC metric in both the bootstrapping and statistical significance testing. The same trend appears when looking at the HD95 metric, with NVAUTO being the clear winner, and the teams in places 2 to 4 performing equivalently. Some differences exist in the VS metrics, with ichilove-ax as the first place, but with not as clear of a lead with no statistical difference from any of the top teams, and a less clear winner when looking at the bootstrapping.

Table 1: Overview of algorithms submitted to the FeTA Challenge ordered from best to worst

| Team Name | Network | Loss Function | 2D/3D | Patch Size | Post-Processing | Convolution Kernel Size | Optimizer | Initialization | Learning Rate | Cross-Validation | Epochs | GPU Used | # of Layers | # of Trainable Parameters |
|---|---|---|---|---|---|---|---|---|---|---|---|---|---|---|
| NVAUTO | MONAI (SegResNet), OCR modules | Dice | 3D | 224x 224x 144 | Ensemble learning | 3x3x3 | AdamW | Random | 0.0002, decrease to 0 at final epoch with cosine annealing scheduler | 5-fold | 300 | 4 x Nvidia V100 32GB | 5 desc / 5 asc | 75 819 624 |
| SJTU_EIEE_2-426Lab | two steps: coarse to fine, 1. nnU-net and 3D UNet with residual architecture; 2. 5 3D Res-Unets | 1. Cross-entropy and Dice; 2. Hausdorff and Dice | 3D | 128x 128x 128 - nn-UNet. only | Ensemble learning | 3x3x3 | Adam | Random | 1. 1E-3; 2. 1E-4 | No | 1. 500; 2. 1000 | Nvidia RTX 3090 | 6 desc / 6 asc | 1. 2 235 680 (UNet); 31 199 584 (nnUNet) 2. 214 58 929 (first UNet); 85 823 969 (other 4 UNets) |
| pengyy | nnU-Net | Cross-entropy and Dice | 3D | 128x 128x 128 | Ensemble learning | 3x3x3 | Stochastic Gradient Descent | Random | 0.01 with reduction | 10-fold | 1000 | Nvidia GeForce RTX 3090 | 6 desc / 6 asc | 72 142 688 |
| Hilab | nnU-Net | Cross-entropy and Dice | 3D | 128x 128x 128 | Ensemble learning | 3x3x3 | Stochastic Gradient Descent | Random | 0.01 with decay | 5-fold | 400 | Nvidia GeForce RTX 2080 Ti | 6 desc / 6 asc | 30 847 564 |
| Neurophet | U-Net | sum of Cross-entropy and Dice | 3D | 64x64x64 | Isolated segmented voxels removed | 3x3x3 | AdamW | Random | 1.00E-05 | No | 500 | 3 x Tesla V100 | 5 desc / 5 asc | 314 999 688 |
| davoodkari-mi | U-Net with additional short and long skip connections | Novel loss function derived from mean absolute error | 3D | 128x 128x 128 | Label Fusion | 3x3x3 | Adam | He | 1E-4 with reduction | No | 400 | Nvidia GeForce GTX 1080 | 5 desc / 5 asc | 18 500 000 |
| 2Ai | U-Net/nnU-Net | Dice | 3D | 128x 128x 128 | Isolated segmented voxels removed | 3x3x3 | Adam | Xavier | 1E-3 with decay | No | 800 | 1 x GTX1070 | 6 desc / 6 asc | 29 971 032 |
| xlab | U-Net/nnUnet | Cross-entropy and Dice | 2D | No | None | 3x3 | Adam | Random | 3.00E-04 | 5-fold | 1000 | Nvidia RTX 3090 | 5 desc / 5 asc | - |
| Ichilove-ax | Two step networks. Dynamic U-Net with pre-trained ResNET34 network blocks (desc) and pixelShuffle ICNR blocks (asc) | Lovasz-Softmax loss | 2D | No | None | 3x3 | OneCycle | ResNet34 - encoder, ICNR - Decoder | 1.00E-03 | No | 60 | 1 x GTX1080Ti | 4 desc / 4 asc | 41 221 768 |
| TRABIT | DynU-Net from MONAI; 10 networks | Label-set Loss function: Leaf-Dice and marginalized cross entropy | 3D | 128x 160x 128 | Ensemble learning | 3x3x3 | Stochastic Gradient Descent | He | 0.01 with decay | No | 2200 | 1 x Tesla V100-SXM2-32GB | 6 desc / 6 asc | 31 195 784 |
| Ichilove-Combi | Two step networks. One for ROI, 3 for each axis (Coronal, Axial, Sagittal). Dynamic U-Net with pre-trained ResNET34 network blocks (desc) and pixelShuffle ICNR blocks (asc) | Lovasz-Softmax loss | 2D | No | Label Fusion | 3x3 | OneCycle | ResNet34 - encoder, ICNR - Decoder | 1.00E-03 | No | 60 | 1 x GTX1080Ti | 4 desc / 4 asc | 103 054 420 |
| muw_dsobotka | multi-task U-Net with two decoders (segmentation and reconstruction) | Homoscedastic uncertainty, cross-entropy, mean squared error | 3D | 128x 96x96 | None | 3x3x3 | Adam | Random | 0.001 | No | 100 | Nvidia GeForce RTX 2080 Ti | 3 desc / 3 asc | 6 491 385 |
| Physense-UPF Team | nnU-Net | Cross-entropy and Generalized Dice | 3D | 128x128x128 | None | 2x2x2 | Stochastic Gradient Descent | Random | 0.01 | 5-fold | 100 | 1x Nvidia GEFORCE GTX 1080 Ti | 6 desc / 6 asc | 31 199 584 |
| SingleNets | U-Net | Soft Dice and Contour Dice | 3D | 96x96x96 | Majority Voting, clipping using "skull" (background-foreground) network response with threshold 0.5 | 3x3x3 | Adam | Fine tuning from previously trained networks on smaller training set | 0.005 with reduction | No | 100 | Tesla M60 | 5 desc / 5 asc | 4 727 841 |
| BIT_LILAB | CNN-Transformer Hybrid (Trans-U-Net) | Cross-entropy and Dice | 2D | 16 x 16 | None | 3x3 | Stochastic Gradient Descent | Pre-trained ResNet-50 and ViT | 1E-2 with decay | No | 150 | 4 x Nvidia GTX 1080Ti GPU | 5 desc / 5 asc | 54 000 000 |
| Moona Mazher | DenseNet | Binary Cross-entropy | 2D | No | Label Fusion | 3x3 | Adam | Random | 0.0003 | 5-fold | 1000 | 4 x Nvidia V100 | 5 desc / 5 asc | 49 510 728 |
| MIAL | U-Net | hybrid loss (Dice and Cross-entropy) | 2D | 64x64 | Majority Voting | 3x3 | Adam | Random | 0.001 with decay | 5-fold | 100 | NVIDIA RTX 2070 | 5 desc / 5 asc | - |
| ZJUWULAB | U-Net with conv downsampling instead of max pooling downsampling | L1 Regularization and feature-matching with a pre-trained VGG19 Network | 2D | No | None | 3x3 | Adam | Random | 0.002 | No | 100 | 4 x RTX 3080Ti | 5 desc / 5 asc | 7 765 442 |
| FeVer | Res-Unet | Dice | 3D | 48x224x224 | Ensemble learning | 3x3x3 | QHAdam | Random | 0.005-0.0005 | No | 300 | 1 x RTX 3090 | 5 desc / 5 asc | 2 369 496 |
| Anonymous | U-Net | Focal Loss | 2D | No | None | 3x3 | Adam | Random | 0.002 | No | 30, backbone frozen for 15 | - | - | - |
| A3 | V-Net with PReLU activation | Binary Cross-entropy | 3D | Crop-ped & pad-ded to 192x 192x 192; down-sampled to 128 x 128 x 128 | None | 3x3x3 | Adam | Random | 1E-4 with reduction | No | 200 | 2 x NVIDIA P100, | 3 desc / 3 asc | 283 886 304 |

Table 2: Overview of the data augmentation, and pre-processing used in each submission.

| Team Name | Data Augmentation | External Dataset used | Pre-processing |
|---|---|---|---|
| NVAUTO | Rotation, Flipping, Zoom, contrast adjustment, Gaussian noise, Gaussian smoothing | No | Normalize images to zero mean |
| SJTU_EIEE_2-426Lab | Rotation, Scaling, Flipping | No | Normalize images to zero mean, cropping in 2nd stage |
| pengyy | Rotation, scaling, elastic deformation, mirroring, Gaussian noise, Gamma Correction | No | resample dimensions to .5x.5x.5mm; z-score normalization |
| Hilab | Pathological Cases copied 3 times in training data, rotation, scaling, Gaussian noise, Gaussian blur, brightness, contrast, simulation of low resolution, gamma augmentation, mirroring | No | Cropping and normalization |
| Neurophet | Affine, Blur | No | Intensity Normalization, classification of images into poor and good quality |
| davoodkarimi | Flipping, rotation, elastic deformation, label perturbation and smoothing | No | Intensity Normalization |
| 2Ai | Flipping, rotation, scaling, grid distortion, optical distortion, elastic transformations, noise, brightness, contrast, gamma transformations | No | Image normalization (mean value zero) |
| xlab | Mirroring, rotation, scaling, gamma correction, random elastic transformation | No | nnUNet standard preprocessing |
| Ichilove-ax | Intensity, contrast, scaling, normalization, rotation, intensity inhomogeneity | No | No |
| TRABIT | Flipping, zooming, rotation, Gaussian noise, spatial smoothing, gamma augmentation | Uses external fetal brain atlases and neonatal MRI's segmented with dHCP | generated brain mask (from atlas + niftyReg), registered brain to atlas, resampled to 0.8mm isotropic; skull stripping; thresholding of intensity percentiles; z-score normalization |
| Ichilove-Combi | Intensity, contrast, scaling, normalization, rotation, intensity inhomogeneity | No | No |
| muw_dsobotka | Elastic deformation, flipping, rotation, contrast, Gaussian noise, Poisson noise | No | z-score normalized patches |
| Physense-UPF Team | Rotation, elastic deformation, scaling, Gaussian noise, Gaussian blur, gamma transform, mirror, brightness, contrast, low resolution simulation, zoom | No | Cropping, resampling, normalization, classification of quality of images, registration to Gholipour atlas |
| SingleNets | Flipping, rotation, translation, scaling, Poisson noise, contrast, intensity | No | Thresholding, cropping, windowing, normalization, downscaling by 0.5 in all axes |
| BIT_LILAB | Rotation and flipping | Yes, Synapse multi-organ segmentation dataset (for pre-training) | None |
| Moona Mazher | Cropping, Flipping, Brightness and Contrast, Random Gamma | No | None |
| MIAL | Flipping, Rotation | No | Intensity standardization |
| ZJUWULAB | No | yes, pre-trained VGG19network | Normalize, colour map labels |
| FeVer | Flipping, Mixup | No | Intensity-based image filtering, resampling voxels to have equal spacing, remove slices with only background label |
| Anonymous | No | ResNet backbone pre-trained on Kinect400 | Intensity re-scaling; ResNet backbone pre-trained on ImageNet |
| A3 | Shifting, rotation, flipping | No | Image normalization (mean value zero) |

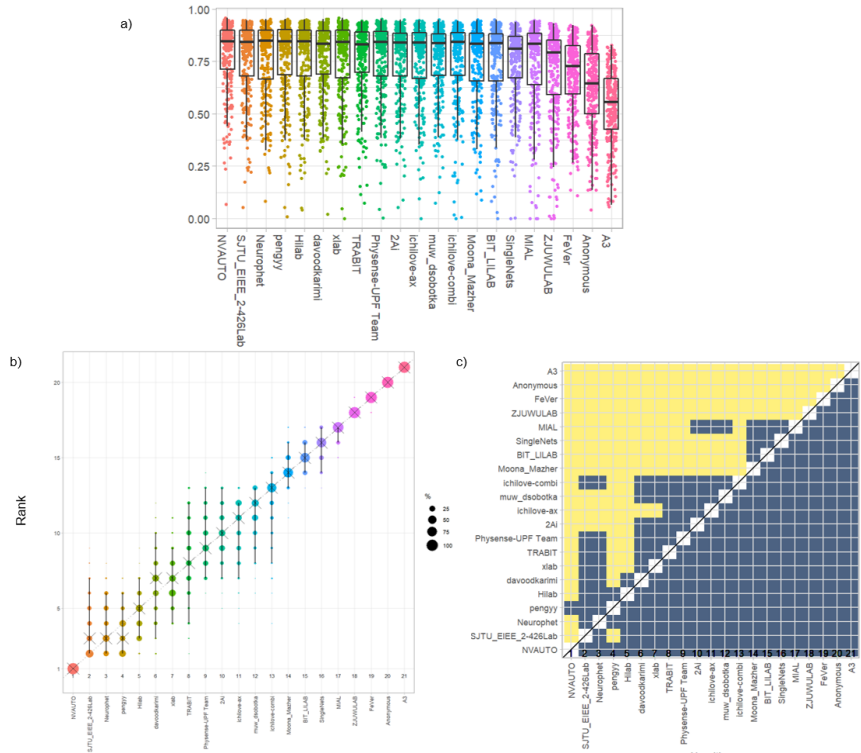

*Figure 4: DSC values of FeTA Challenge participants a) Dot and box plot; b) Blob plot for visualizing ranking stability based on bootstrap sampling, black cross indicated the median rank for each algorithm and 95% bootstrap intervals across samples are indicated by black lines; c) Significance maps for visualizing ranking stability based on statistical significance (Yellow: metrics from the algorithm on the x-axis were significantly superior to the algorithm on the y-axis, blue color indicates no significant difference). Figures were created using the ChallengeR Tool (Wiesenfarth et al., 2021).*

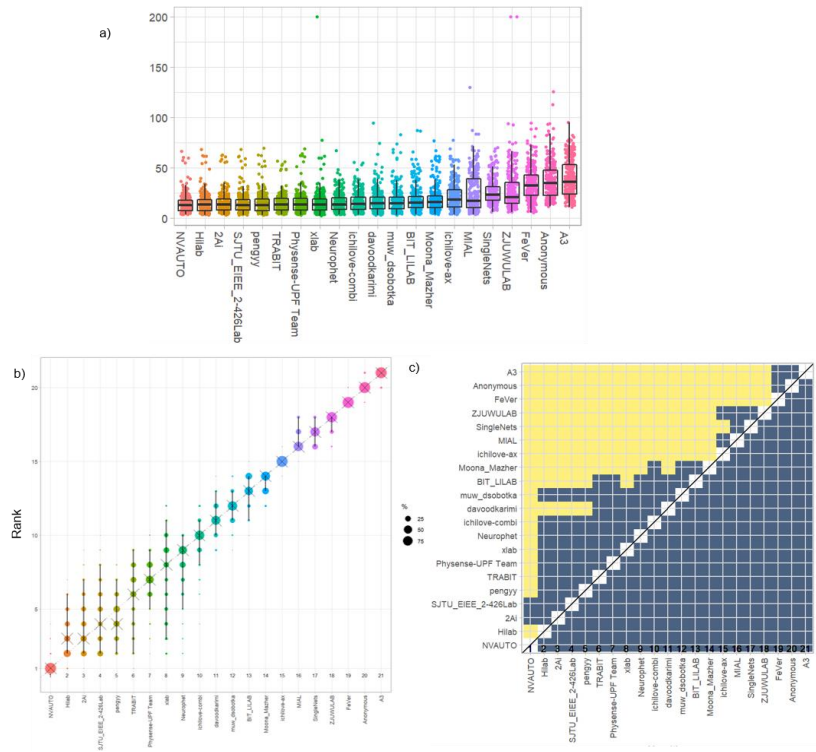

*Figure 5: HD95 values of FeTA Challenge participants a) Dot and box plot; b) Blob plot for visualizing ranking stability based on bootstrap sampling, black cross indicated the median rank for each algorithm and 95% bootstrap intervals across samples are indicated by black lines; c) Significance maps for visualizing ranking stability based on statistical significance (Yellow: metrics from the algorithm on the x-axis were significantly superior to the algorithm on the y-axis, blue color indicates no significant difference.)*

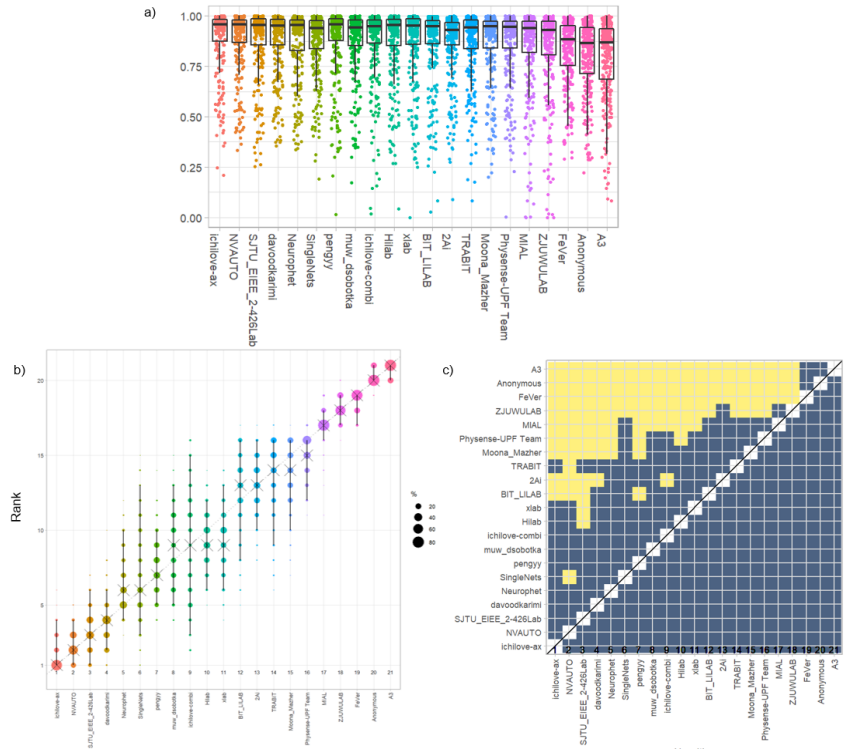

*Figure 6: VS values of FeTA Challenge participants a) Dot and box plot; b) Blob plot for visualizing ranking stability based on bootstrap sampling, black cross indicated the median rank for each algorithm and 95% bootstrap intervals across samples are indicated by black lines; c) Significance maps for visualizing ranking stability based on statistical significance (Yellow: metrics from the algorithm on the x-axis were significantly superior to the algorithm on the y-axis, blue color indicates no significant difference.)*

*Table 3: Final FeTA Ranking; * indicates a tie*

| Ranking | Team Name | Average DSC | Average HD95 (voxels) | Average VS |
|---|---|---|---|---|
| 1 | NVAUTO | 0.786 ± 0.161 | 14.012 ± 9.285 | 0.885 ± 0.156 |
| 2 | SJTU_EIEE_2-426Lab | 0.775 ± 0.173 | 14.671 ± 9.917 | 0.883 ± 0.166 |
| 3 | Pengyy | 0.774 ± 0.182 | 14.699 ± 10.049 | 0.875 ± 0.182 |
| 4 | Hilab* | 0.774 ± 0.181 | 14.569 ± 9.954 | 0.873 ± 0.180 |
| 4 | Neurophet* | 0.775 ± 0.171 | 15.375 ± 9.277 | 0.877 ± 0.165 |
| 6 | davoodkarimi | 0.771 ± 0.171 | 16.755 ± 11.443 | 0.882 ± 0.156 |
| 7 | 2Ai* | 0.767 ± 0.170 | 14.625 ± 9.892 | 0.867 ± 0.166 |
| 7 | xlab* | 0.771 ± 0.183 | 15.262 ± 14.769 | 0.873 ± 0.182 |
| 9 | ichilove-ax | 0.766 ± 0.176 | 21.329 ± 13.241 | 0.888 ± 0.158 |
| 10 | TRABIT | 0.769 ± 0.174 | 14.901 ± 9.049 | 0.866 ± 0.173 |
| 11 | ichilove-combi* | 0.762 ± 0.188 | 16.039 ± 9.395 | 0.873 ± 0.183 |
| 11 | muw_dsobotka* | 0.765 ± 0.171 | 17.159 ± 11.905 | 0.874 ± 0.168 |
| 11 | Physense-UPF Team* | 0.767 ± 0.182 | 15.018 ± 10.145 | 0.863 ± 0.180 |
| 14 | SingleNets | 0.748 ± 0.172 | 26.121 ± 12.072 | 0.876 ± 0.154 |
| 15 | BIT_LILAB | 0.752 ± 0.190 | 18.162 ± 12.644 | 0.868 ± 0.183 |
| 16 | Moona Mazher | 0.755 ± 0.183 | 18.548 ± 12.739 | 0.866 ± 0.179 |
| 17 | MIAL | 0.740 ± 0.211 | 25.107 ± 19.425 | 0.845 ± 0.213 |
| 18 | ZJUWULAB | 0.703 ± 0.217 | 27.948 ± 22.400 | 0.835 ± 0.218 |
| 19 | FeVer | 0.683 ± 0.180 | 34.419 ± 15.990 | 0.828 ± 0.164 |
| 20 | Anonymous | 0.621 ± 0.192 | 37.385 ± 18.249 | 0.801 ± 0.181 |
| 21 | A3 | 0.534 ± 0.178 | 39.608 ± 18.249 | 0.791 ± 0.199 |

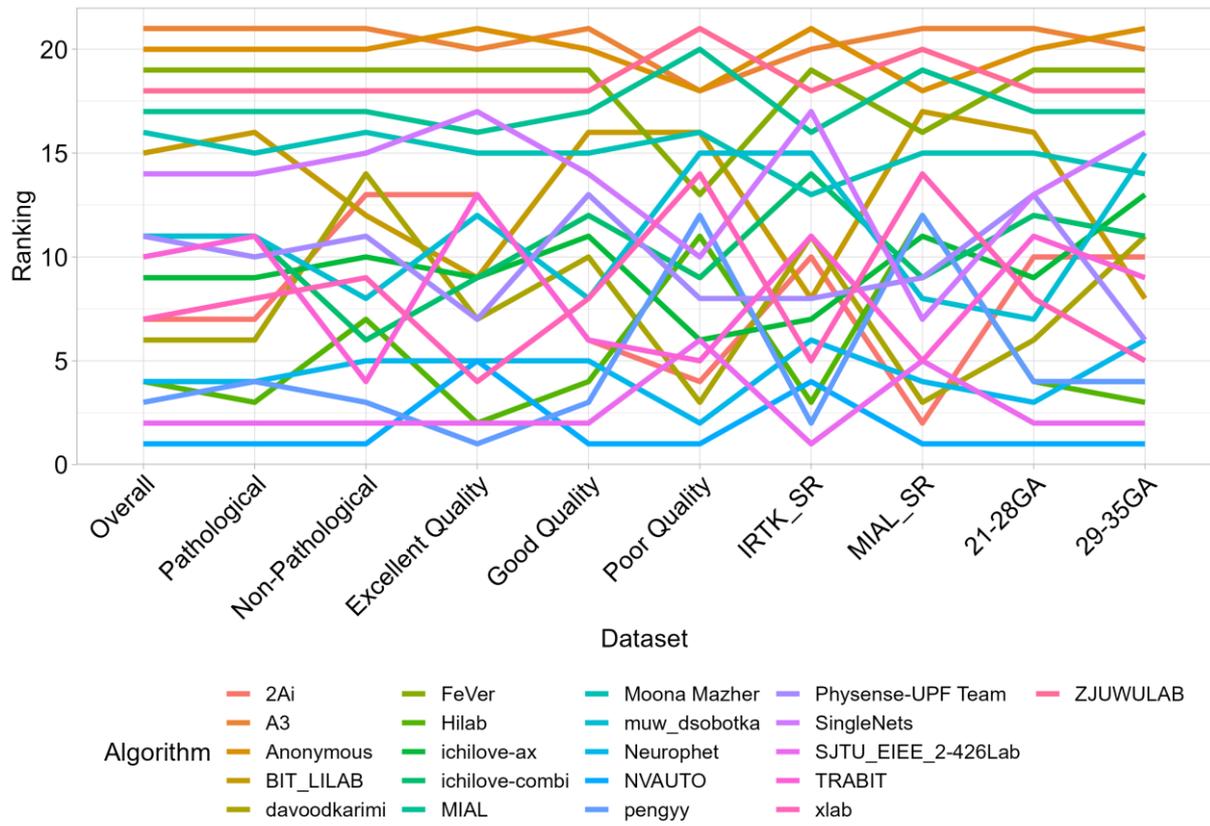

*Figure 7: Ranking of each algorithm for each subset of data in Section 3.3.*

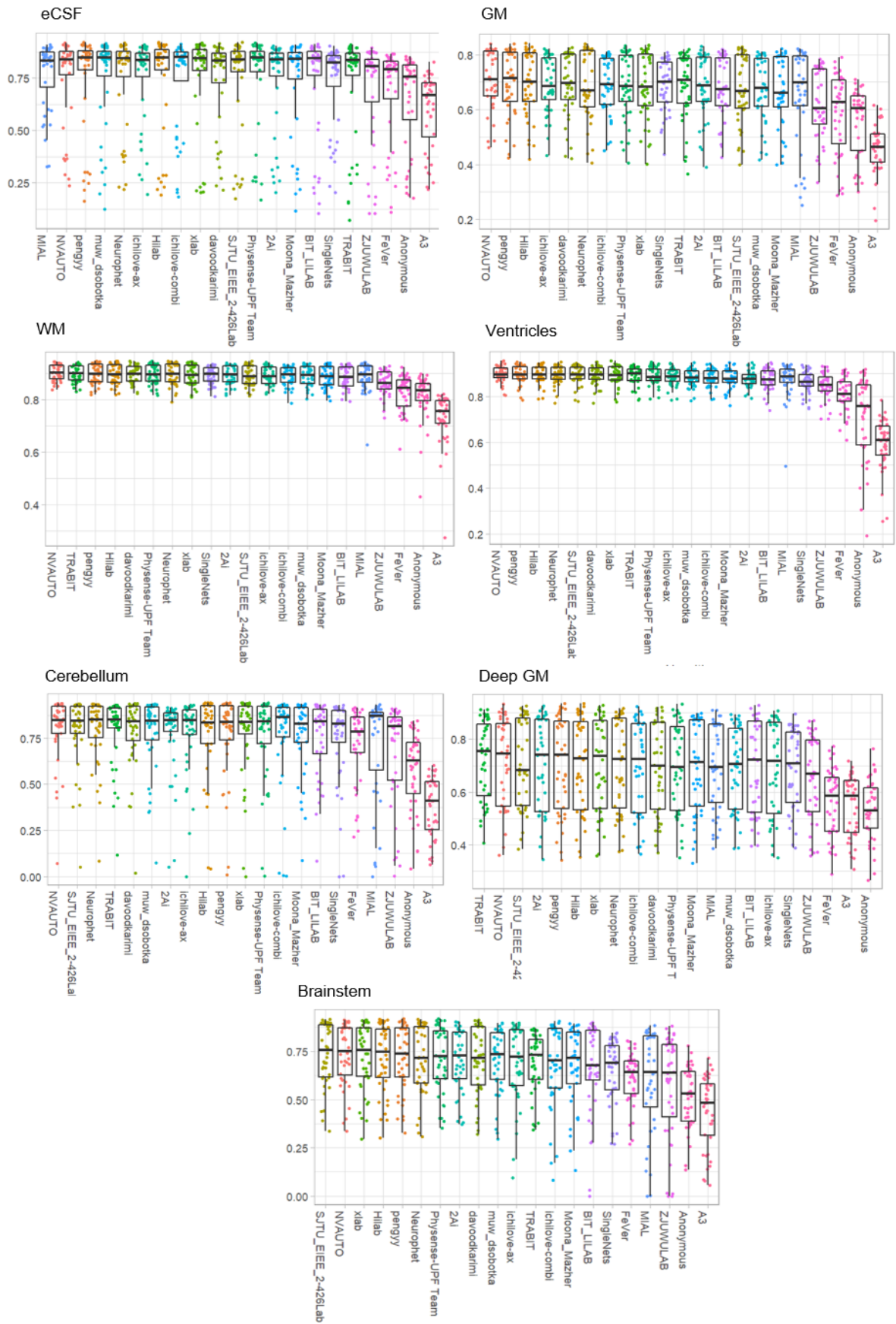

*Figure 8: DSC per label for each team (teams ranked from best to worst are visualized from left to right on the x-axis of each graph)*

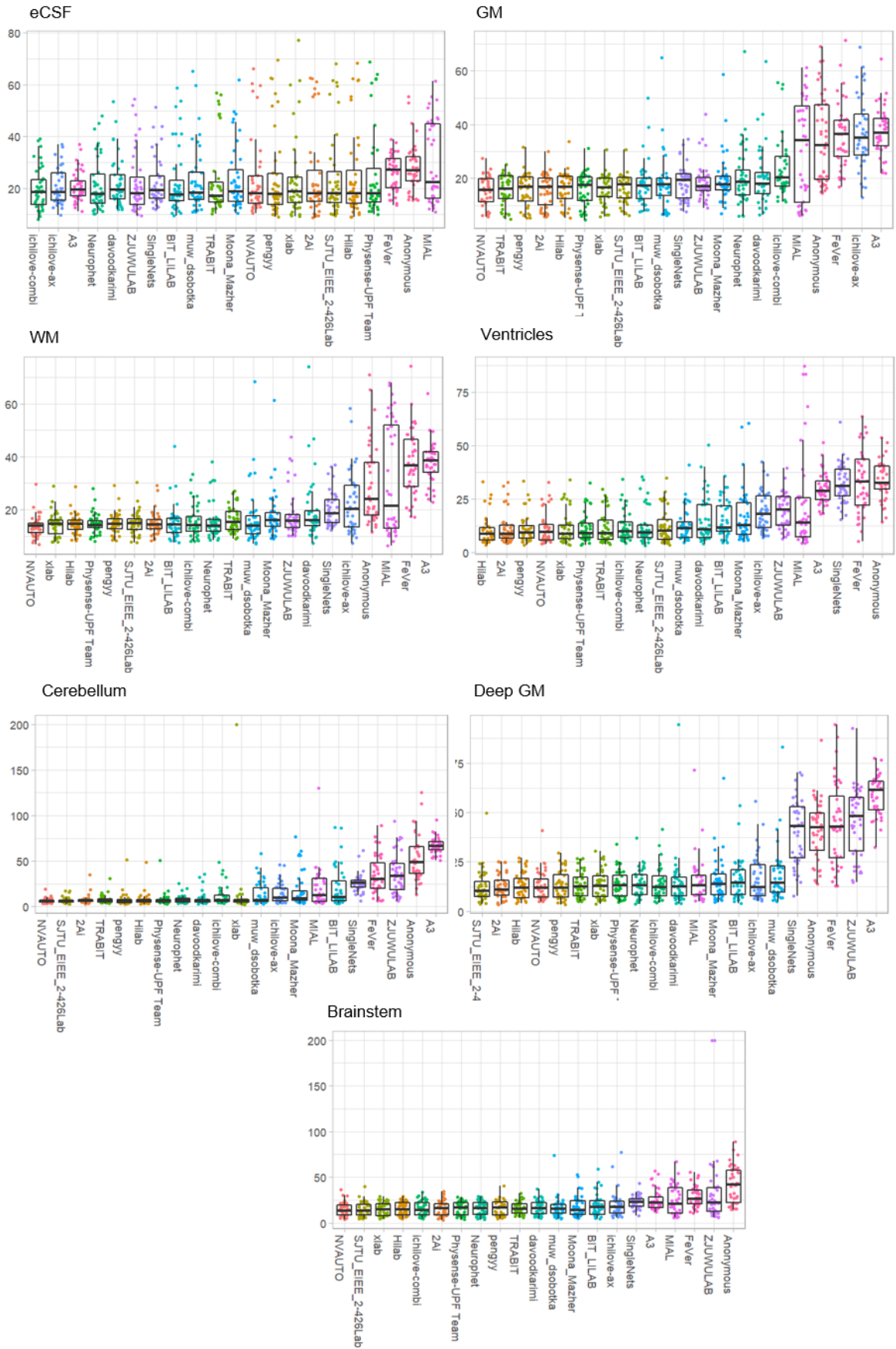

*Figure 9: HD95 per label for each team (teams ranked from best to worst are visualized from left to right on the x-axis of each graph)*

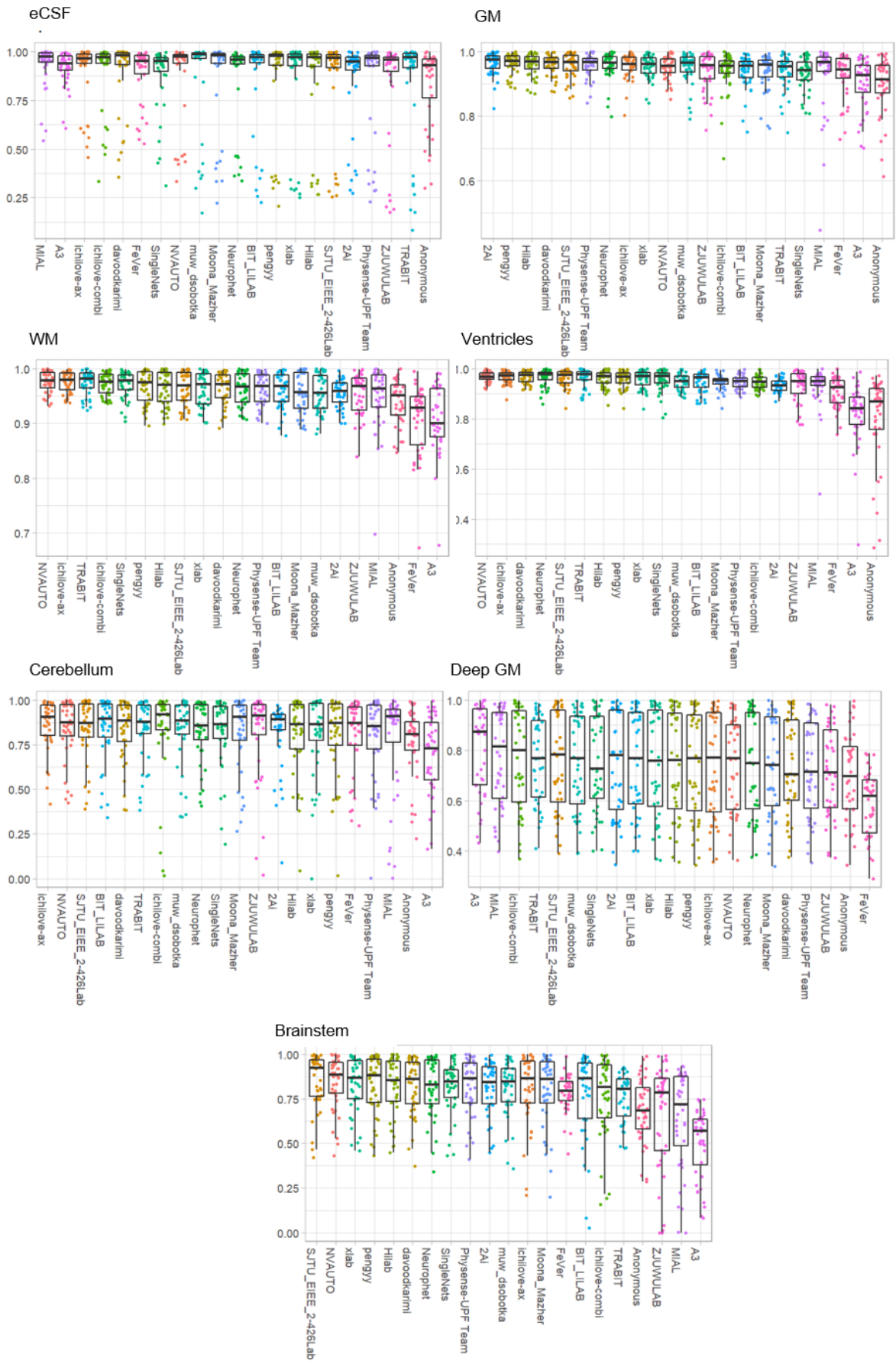

*Figure 10: VS per label for each team (teams ranked from best to worst are visualized from left to right on the x-axis of each graph)*

Table 4: Complete Rankings (DSC, HD95, VS combined) for each label; * indicates a tie, with the rank in brackets following the team name

| Rank | eCSF | GM | WM | Ventricle | Cerebel-lum | Deep GM | Brainstem |
|---|---|---|---|---|---|---|---|
| 1 | ichilove-ax | pengyy | NVAUTO | NVAUTO | NVAUTO | SJTU_EIEE_2-426Lab | SJTU_EIEE_2-426Lab |
| 2 | ichilove-combi | Hilab | Hilab* (2) | Hilab | SJTU_EIEE_2-426Lab | TRABIT | NVAUTO |
| 3 | Neurophet* (3) | NVAUTO | pengyy* (2) | pengyy | TRABIT | 2Ai | xlab |
| 4 | Davoodkarimi* (3) | 2Ai | TRABIT | Neurophet | davoodkarimi | NVAUTO* (4) | Hilab |
| 5 | NVAUTO* (5) | Physense-UPF Team | xlab | SJTU_EIEE_2-426Lab | Neurophet | Hilab* (4) | pengyy |
| 6 | muw_dsobotka* (5) | davoodkarimi | Physense-UPF Team | Davoodkarimi* (6) | ichilove-ax | ichilove-combi* (6) | Neurophet |
| 7 | MIAL | xlab | SJTU_EIEE_2-426Lab | TRABIT* (6) | 2Ai | pengyy* (6) | Physense-UPF Team |
| 8 | A3 | SJTU_EIEE_2-426Lab* (8) | ichilove-combi | xlab* (6) | muw_dsobotka | xlab | 2Ai |
| 9 | pengyy | Neurophet* (8) | Neurophet | ichilove-ax | Hilab | MIAL | davoodkarimi |
| 10 | SingleNets | TRABIT | SingleNets* (10) | Physense-UPF Team | ichilove-combi | Neurophet | muw_dsobotka |
| 11 | Moona_Mazher* (11) | ichilove-ax | Davoodkarimi* (10) | 2Ai | pengyy | muw_dsobotka | ichilove-combi |
| 12 | BIT_LILAB* (11) | ichilove-combi* (12) | ichilove-ax | muw_dsobotka | BIT_LILAB | Physense-UPF Team | ichilove-ax |
| 13 | xlab | muw_dsobotka* (12) | 2Ai | ichilove-combi | Physense-UPF Team* (13) | Davoodkarimi* (13) | TRABIT |
| 14 | Hilab | BIT_LILAB* (12) | BIT_LILAB | Moona_Mazher* (14) | xlab* (13) | BIT_LILAB* (13) | SingleNets* (14) |
| 15 | SJTU_EIEE_2-426Lab* (15) | SingleNets | muw_dsobotka | BIT_LILAB* (14) | Moona_Mazher | SingleNets* (15) | Moona_Mazher* (14) |
| 16 | ZJUWULAB* (15) | ZJUWULAB | Moona_Mazher | SingleNets | SingleNets | Moona_Mazher* (15) | BIT_LILAB |
| 17 | FeVer | Moona_Mazher | ZJUWULAB | MIAL* (17) | ZJUWULAB | A3 | FeVer |
| 18 | 2Ai | MIAL | MIAL | ZJUWULAB* (17) | FeVer* (18) | ichilove-ax | MIAL |
| 19 | TRABIT | FeVer | Anonymous | FeVer | MIAL* (18) | ZJUWULAB | ZJUWULAB |
| 20 | Physense-UPF Team | Anonymous | FeVer | A3 | Anonymous | Anonymous * (20) | Anonymous * (20) |
| 21 | Anonymous | A3 | A3 | Anonymous | A3 | FeVer* (20) | A3* (20) |

**Further Analysis:** A variety of subsets of the data were created in order to determine if the algorithms perform better or worse based on various criteria such as image quality, SR method used, and normal vs pathological brains. The rankings of the teams based on the different subsets can be seen in Figure 7. A large amount of variability in the rankings is present depending on the subset of data being investigated. However, NVAUTO remains in the number 1 ranking spot in all subsets except two (Excellent Quality and IRTK_SR).

**Per-Label Metric Values and Ranking:** Each team's algorithm was analyzed separately per tissue label. The average DSC, HD95, and VS scores for each team and label can be found in Figures 8 - 10. The order of the teams on the x-axis in each graph is ordered from best to worst, left to right. When looking at the DSC, team NVAUTO placed first in all labels except eCSF (MIAL), deepGM (TRABIT), and brainstem (SJTU_EIEE_2-426Lab). When looking at the HD95, team NVAUTO placed first in all labels except eCSF (ichilove-combi), Ventricles (Hilab), and deepGM (SJTU_EIEE_2-426Lab). In the VS metric, almost every label had a different top team: eCSF (MIAL), GM (2Ai), WM (NVAUTO), Ventricles (NVAUTO), Cerebellum (ichilove-ax), deepGM (A3), and brainstem (SJTU_EIEE_2-426Lab).

**Image Quality:** The dataset was split into three subsets based on the quality of the SR reconstructions as determined by experienced raters (excellent quality SR: n=11 (mialSR/IRTK: 1/10); good quality SR: n=25 (mialSR/IRTK: 15/10); poor quality SR: n=4 (mialSR/IRTK: 4/0)). Each team's algorithm was analyzed with the average metrics across all labels. The average DSC, HD95, and VS scores for each team and label can be found in Figure 11. The order of the teams on the x-axis in each graph is ordered from best to worst, left to right. Team pengyy performed the best (according to the DSC) when the fetal brain reconstructions were of excellent quality, while NVAUTO performed the best for good and poor quality reconstructions. Complete ranking information taking all three metrics into account based on SR reconstruction quality can be found in Figure 7.

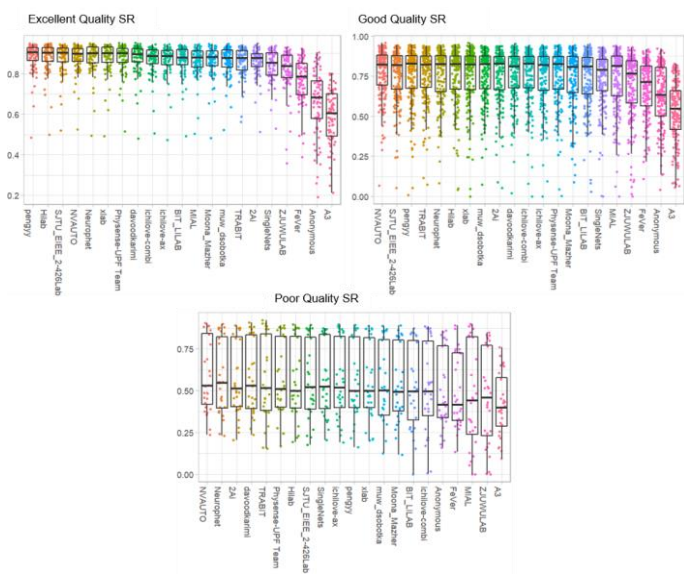

*Figure 11:* DSC across all labels for each team, based on the quality of the SR reconstruction (teams ranked from best to worst are visualized from left to right on the x-axis of each graph)

**SR Reconstruction:** The dataset was split into two subsets based on SR reconstruction method used. Each team's algorithm was analyzed with the average metrics across all labels. The average DSC, HD95, and VS scores for each team and label can be found in Figure 12. The order of the teams on the x-axis in each graph is ordered from best to worst, left to right. Team NVAUTO performed the best (according to the DSC) with the mialSRSR reconstruction, and Team SJTU_EIEE_2-426Lab performed the best (according to the DSC) with the IRTK SR reconstruction. Complete ranking information taking all three metrics into account for each SR reconstruction can be found in Figure 7.

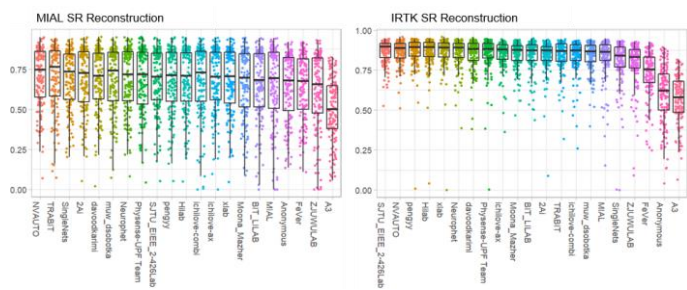

*Figure 12:* DSC across all labels for each team, based on the SR reconstruction used (teams ranked from best to worst are visualized from left to right on the x-axis of each graph)

**Pathology:** The dataset was split into two subsets based on whether the fetal brain contained a pathology (n=25 (mialSR/IRTK: 14/11)) or not (neurologically normal, n=15 (mialSR/IRTK: 6/9)). Each team's algorithm was analyzed with the average metrics across all labels. The average DSC, HD95, and VS scores for each team and label can be found in Figure 13. The order of the teams on the x-axis in each graph is ordered from best to worst, left to right. Team NVAUTO performed best for both pathological and non-pathological brains. No details of the specific pathologies were available to the challenge participants. Complete ranking information taking all three metrics into account for the pathological and non-pathological datasets can be found in Figure 7.

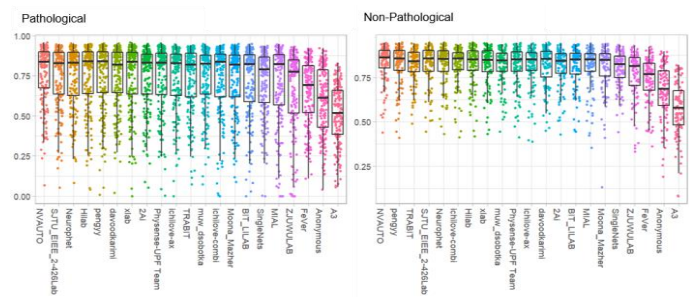

*Figure 13:* DSC across all labels for each team, based on whether the fetal brain was pathological or non-pathological (teams ranked from best to worst are visualized from left to right on the x-axis of each graph)

**Intracranial Volume:** The intracranial volume of each case in the test set was calculated using all labels (excluding the background) and compared to the intracranial volumes determined by each participant in the challenge. While most methods had some outliers, all teams except for five had a median percent difference from GT within ±1% (Figure 14).

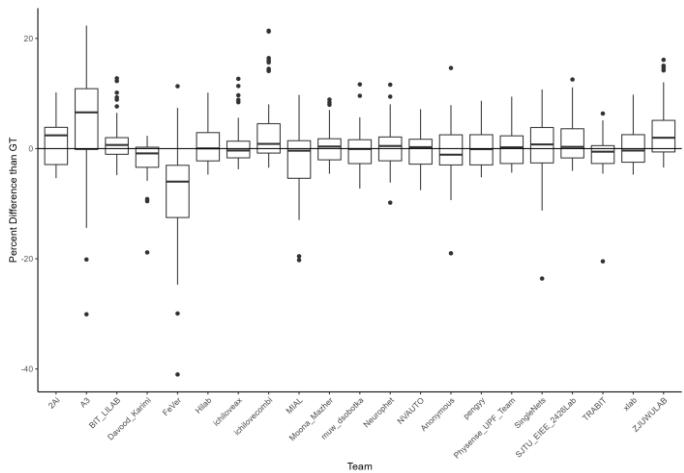

*Figure 14: Percent difference in intracranial volume between the submitted algorithms and the reference label map.*

**Gestational Age Comparison in the GM:** The evaluation metrics for the cortex (grey matter label) were calculated based on age of the fetus. The top-scoring teams for the younger fetuses (GA 21-28; n=28) were pengyy, NVAUTO, and Hilab.

The top scoring teams for the older fetuses (GA 29-35; n=12) were xlab, pengyy, and Hilab. When all ages were combined together, the top teams for the GM label were pengyy, Hilab, and NVAUTO (see Table 4), showing that gestational age does play a small role in the success of the algorithms in segmenting the cortex. There are fewer cases included in the older group mainly due to the smaller gestational age range, and the fact that the majority of fetal scans at the center used to collect the data happen by the 32nd gestational week. The evaluation metrics of the GM from both the older and younger fetuses can be seen in Figure 15 and 16.

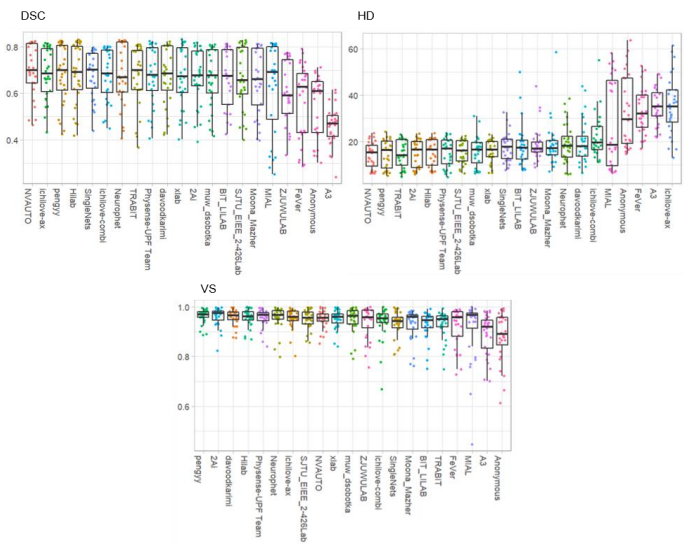

*Figure 15: Evaluation metrics (DSC, HD95, VS) of the GM label from younger GA fetuses (21-28GA)*

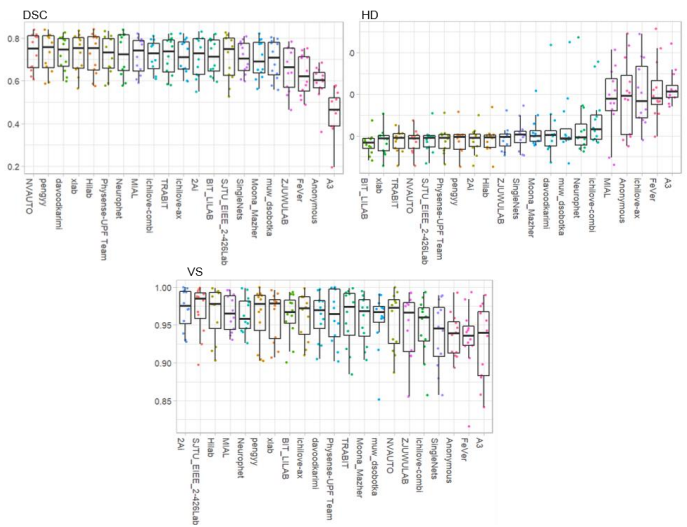

*Figure 16: Evaluation metrics (DSC, HD95, VS) of the GM label from older GA fetuses (29-35GA)*

## Discussion and Conclusion

In this paper we present the results of the first FeTA Challenge held at the MICCAI 2021 conference. All submissions to the FeTA Challenge were deep-learning based submissions. Other machine-learning methods or purely atlas-based approaches were not submitted. This demonstrates that deep learning is currently the leading method for fetal brain medical image segmentation, and confirms its dominance in medical image segmentation more broadly. Indeed, the top three teams all had very similar network architectures. The majority of participating teams obtained very similar evaluation metrics; however, one team performed significantly better than all other teams on the complete testing dataset.

**Top Methods:** When all labels are combined together and the entire testing dataset is used, Team NVAUTO submitted the top algorithm of the challenge. They ranked first in two out of the three evaluation metrics (DCS and HD95), and came second in the third (VS). In addition, the bootstrapping and significance testing showed that NVAUTO was the clear winner in the DSC and HD95 coefficients. The VS metric was more ambiguous across all participants, with no statistically significant difference among the first 9 teams. This suggests that while VS is a relevant biomarker, it is potentially not an ideal evaluation metric for this challenge as it is unable to show differences in the performance.

There were many methodological similarities among the top five ranking teams. All were 3D U-Nets, all used either a Dice or cross-entropy/Dice combination loss function, none used an external dataset, and all used standard data augmentation techniques such as rotation, flipping, scaling, addition of Gaussian noise, Gaussian smoothing, gamma correction, affine transformations, and contrast adjustment. Four of the teams used Pytorch, while the fifth used MONAI, which is Pytorch-based (MONAI Consortium, 2020). All used the same convolution kernel size (3x3x3) with random initialization. Four out of five used an ensemble learning strategy, three teams out of the top five used cross-validation. The main differences in networks appeared to be in the training procedures, such as the number of epochs, and in how the learning rate was manipulated throughout training.

When looking at the changes in rankings based on different subsets of the data the interpretation of the results become challenging. As shown in Figure 7, the rankings change considerably depending on the data subset tested. This is relevant, as different centers have different data, different age ranges at which fetal MRI is acquired and one algorithm may not work well across all sites. As the submitted algorithms were all similar deep learning-based methods, this suggests that fine tuning the networks plays a key role in any potential practical application of these algorithms, depending on the specific clinical or research usage.

**Performance of Submitted Algorithms:** As mentioned already, all submissions were deep learning-based submissions. To go one step further, it was not just that all submissions used deep learning, but 19 out of 21 submissions used some form of U- Net, consisting of a contracting and an expanding path forming a U-shaped network. During the former, higher resolution information is sacrificed for more context. However, U-Net has the capability, using skip connections, of combining this information with the corresponding output from the expanding path. There were many differences within each U-Net, but the overall shape and structure of the network remained consistent, including the depth of the network. Eight teams used the pre-existing medical imaging neural network frameworks nnU-Net (Isensee et al., 2021) or MONAI (MONAI Consortium, 2020). The main differences across the submissions were in how the training was performed (such as the use of cross-validation or changes in the learning rate decay), or in the pre-processing (patch size, how the data was normalized) and post-processing (such as ensemble learning, removal of external label 'blobs'). The plateauing of the top team entries is interesting as well, potentially suggesting that U-Nets have a performance limit in multi-class segmentation tasks with limited data.

The most likely labels to fail to be segmented (that is, where the algorithm was unable to detect any voxels with the specific tissue) were the brainstem and the cerebellum, in particular in the pathological cases. This could potentially be explained by unclear demarcations of the brainstem and cerebellum in pathological groups which contained some cases of the Chiari-II malformation. Overall, the most challenging labels to segment were cortical GM, deep GM, and the brainstem. This can be seen in Figure 8 - Figure 10, where these three tissue labels have worse performances than the other tissues, along with a larger distribution of evaluation metrics in each team. The potential reasons for this are multifold. The lateral and ventral borders of the deep GM and ventral portion of the brainstem are not well defined and are challenging for experienced radiologists to delineate. In the GM, the contrast between WM and GM changes throughout gestation due to neuronal migration and axonal outgrowth, while the surface pattern of the cortical GM becomes increasingly complex.

In general, the pathological brains were more challenging to segment than the non-pathological brains due to the larger variations in neuroanatomy. Selective data augmentation on these pathological cases could be a potential solution to this. The results of the image quality and SR reconstruction methods are related to each other, as the majority of the low quality images were done with the mialSR method, and the excellent quality brain volumes included were reconstructed with the Simple IRTK method. We would like to emphasize this is not a comment on the SR methods themselves, only a reflection of what cases were chosen for each reconstruction method. As expected, the low quality images, and therefore also the mialSR reconstructions were more challenging to accurately segment than the high quality and IRTK SR reconstructions, with lower DSC scores and a wider range of variability as can be seen in Figure 12.

**Clinical Applications:** Potential applications of the fully automatic and highly accurate fetal brain MRI segmentation algorithms are broad and span from neuroscience (characterizing spatio-temporal lateralization of the cortex (Kasprian et al., 2011; Vasung et al., 2020), virtopsies (identification and analysis of the details of demise (Rüegger et al., 2014)), surgery (clinical guidelines for early fetal surgery (Clewell et al., 1982; Meuli et al., 1997; Meuli and Moehrlen, 2013)), medicine (identification of biomarkers of outcome needed for stratification tools and development of early interventions (Rollins et al., 2021)), volumetric studies (Polat et al., 2017; Sadhwani et al., 2022) and development of new public health policies (prenatal programs focused on reduction of stress during pregnancy (van den Heuvel et al., 2021; Wu et al., 2020)).

Fetal MRI offers a unique possibility to study the human specific aspects of neurodevelopment. It remains the only non-invasive in vivo imaging modality to study connectivity, function, and structural anatomy of the fetal brain in a single session (Jakab et al., 2021, 2015). From the perspective of neuroscience, it is critically important to study the relationship between brain structure and function. However, this requires parcellation of the brain and cortex into regions or areas (e.g. (Amunts et al., 2020; Desikan et al., 2006; Klein and Tourville, 2012)). A first crucial step toward this is to perform reliable segmentation of the developing cerebral cortex, which was the objective of this FeTA Challenge.

Furthermore, normative charts showing age-related changes in volume of different brain structures throughout the lifespan, similar to head circumference in the pediatric population, have just started to emerge (Bethlehem et al., 2021). Nonetheless, in addition to obvious challenges of fetal MRI acquisition, the harmonization of MRI acquisition protocols across sites and the development of robust and automatic algorithms for accurate and precise segmentation of fetal brain remain prerequisites for any future clinical application.

**Limitations and Future Considerations:** Some limitations of the challenge include the fact that all images included were acquired from a single center, and therefore algorithms developed with this dataset are unlikely to be generalizable to other centers. In addition, while the total number of cases included is relatively large for the type of dataset, it is relatively small when compared to other datasets used for training neural networks (Bakas et al., 2019; Menze et al., 2015). The manual segmentations included in both the training and testing dataset were not perfect, and therefore there are mislabeled voxels. Annotations were made mainly in the axial plane, leading to some noisy labels and discontinuity in the annotations in the coronal and sagittal planes. The manual annotations were especially challenging in the mialSR reconstructions, as it was the low resolution scans that underwent reorientation rather than the final reconstructed volume, resulting in a final reconstruction that was not exactly 'in plane' according to standard fetal atlases. This led to the phenomenon of participants' algorithms performing quite well visually but receiving mid-range evaluation metrics. One team even performed their own revisions on the manual segmentations using their own in house experts, and then used them in their training dataset (Fidon et al., 2021). While organizing the challenge we were aware of these errors, and therefore included three different metrics in order to reduce the reliance on any one metric. Future work includes improving the manual segmentations included in the FeTA Dataset. Further research into inter-rater variability in fetal brain segmentations is also required to understand what values of evaluation metrics are considered 'good enough'. Preliminary research has been done using a very small sample set (Payette et al., 2021a), but a more extensive study should be performed.

In the future, we aim to expand the FeTA Dataset to include data from multiple centers in order to increase the generalizability of algorithms trained using this dataset. We also hope to extend the number of different pathologies included, and to increase the number of cases at the outer range of the gestational ages, especially at older gestational ages.

**Conclusion:** The algorithms developed as part of the FeTA Challenge provide a benchmark for future segmentation algorithms and can already be used to research fetal neurodevelopment. Our study found that most groups working on segmentation methods are using U-Nets, and that 3D U-Nets seem to be superior to 2D based on the evaluation metrics. In a dataset with large variation, such as the FeTA Dataset, the variation plays a role in the success of the algorithm. There was not one algorithm that was the 'best' when specific subsets of the data was analyzed, although there was a 'best' algorithm when the testing dataset was assessed as a whole. There are still many opportunities for improvement in developing multi-class segmentation techniques for the fetal brain throughout gestation, and therefore this challenge is the starting point for further development of such algorithms.

# CRediT Author Statements

Kelly Payette: Conceptualization, Writing - Original Draft, Data Curation, Visualization, Software, Validation, Investigation

Hongwei Li: Writing - Review & Editing, Software, Validation, Conceptualization

Priscille de Dumast: Writing - Review & Editing, Conceptualization

Roxane Licandro: Writing - Review & Editing, Conceptualization

Hui Ji: Data Curation

Lana Vasung: Writing - Review & Editing, Conceptualization

Bjoern Menze: Supervision

Meritxell Bach Cuadra: Conceptualization, Writing - Review & Editing, Supervision, Investigation

## Funding Sources and Acknowledgements

The authors would like to acknowledge funding from the following funding sources: the OPO Foundation, the Prof. Dr. Max Cloetta Foundation, the Anna Müller Grocholski Foundation, the Foundation for Research in Science and the Humanities at the UZH, the EMDO Foundation, the Hasler Foundation, the FZK Grant, the Swiss National Science Foundation (project 205321-182602), the Forschungskredit (Grant NO. FK-21-125) from University of Zurich, the ZNZ PhD Grant, the EU H2020 Marie Sklodowska-Curie [765148], Austrian Science Fund FWF [P 35189], Vienna Science and Technology Fund WWTF [LS20-065], and the Austrian Research Fund Grant I3925-B27 in collaboration with the French National Research Agency (ANR). We acknowledge access to the expertise of the CIBM Center for Biomedical Imaging, a Swiss research center of excellence founded and supported by Lausanne University Hospital (CHUV), University of Lausanne (UNIL), Ecole polytechnique fédérale de Lausanne (EPFL), University of Geneva (UNIGE) and Geneva University Hospitals (HUG). We would also like to acknowledge funding from the European Union's Horizon 2020 research and innovation program under the Marie Sklodowska-Curie grant agreement TRABIT No 765148, as well as from core and project funding from the Wellcome [203148/Z/16/Z; 203145Z/16/Z; WT101957], and EP-SRC [NS/A000049/1; NS/A000050/1; NS/A000027/1]. TV is supported by a Medtronic / RAEng Research Chair [RCSRF1819\7\34]. The authors would also like to thank NVIDIA for providing access to computing resources.

**This manuscript is the author's original version.**

## References

Amunts, K., Mohlberg, H., Bludau, S., Zilles, K., 2020. Julich-Brain: A 3D probabilistic atlas of the human brain's cytoarchitecture. Science 369, 988–992. https://doi.org/10.1126/science.abb4588

Bach Cuadra, M., Schaer, M., Andre, A., Guibaud, L., Eliez, S., Thiran, J.-P. (Eds.), 2009. Brain tissue segmentation of fetal MR images, in: Workshop on Image Analysis for Developing Brain, in 12th International Conference on Medical Image Computing and Computer Assisted Intervention. Presented at the Workshop on Image Analysis for Developing Brain, in 12th International Conference on Medical Image Computing and Computer Assisted Intervention, London, UK.

Bakas, S., Reyes, M., Jakab, A., Bauer, S., Rempfler, M., Crimi, A., Shinohara, R.T., Berger, C., Ha, S.M., Rozycki, M., Prastawa, M., Alberts, E., Lipkova, J., Freymann, J., Kirby, J., Bilello, M., Fathallah-Shaykh, H., Wiest, R., Kirschke, J., Wiestler, B., Colen, R., Kotrotsou, A., Lamontagne, P., Marcus, D., Milchenko, M., Nazeri, A., Weber, M.-A., Mahajan, A., Baid, U., Gerstner, E., Kwon, D., Acharya, G., Agarwal, M., Alam, M., Albiol, Alberto, Albiol, Antonio, Albiol, F.J., Alex, V., Allinson, N., Amorim, P.H.A., Amrutkar, A., Anand, G., Andermatt, S., Arbel, T., Arbelaez, P., Avery, A., Azmat, M., B., P., Bai, W., Banerjee, S., Barth, B., Batchelder, T., Batmanghelich, K., Battistella, E., Beers, A., Belyaev, M., Bendszus, M., Benson, E., Bernal, J., Bharath, H.N., Biros, G., Bisdas, S., Brown, J., Cabezas, M., Cao, S., Cardoso, J.M., Carver, E.N., Casamitjana, A., Castillo, L.S., Catà, M., Cattin, P., Cerigues, A., Chagas, V.S., Chandra, S., Chang, Y.-J., Chang, S., Chang, K., Chazalon, J., Chen, S., Chen, W., Chen, J.W., Chen, Z., Cheng, K., Choudhury, A.R., Chylla, R., Clérigues, A., Colleman, S., Colmeiro, R.G.R., Combalia, M., Costa, A., Cui, X., Dai, Z., Dai, L., Daza, L.A., Deutsch, E., Ding, C., Dong, C., Dong, S., Dudzik, W., Eaton-Rosen, Z., Egan, G., Escudero, G., Estienne, T., Everson, R., Fabrizio, J., Fan, Y., Fang, L., Feng, X., Ferrante, E., Fidon, L., Fischer, M., French, A.P., Fridman, N., Fu, H., Fuentes, D., Gao, Y., Gates, E., Gering, D., Gholami, A., Gierke, W., Glocker, B., Gong, M., González-Villá, S., Grosges, T., Guan, Y., Guo, S., Gupta, S., Han, W.-S., Han, I.S., Harmuth, K., He, H., Hernández-Sabaté, A., Herrmann, E., Himthani, N., Hsu, W., Hsu, C., Hu, Xiaojun, Hu, Xiaobin, Hu, Yan, Hu, Yifan, Hua, R., Huang, T.-Y., Huang, W., Van Huffel, S., Huo, Q., HV, V., Iftekharuddin, K.M., Isensee, F., Islam, M., Jackson, A.S., Jambawalikar, S.R., Jesson, A., Jian, W., Jin, P., Jose, V.J.M., Jungo, A., Kainz, B., Kamnitsas, K., Kao, P.-Y., Karnawat, A., Kellermeier, T., Kermi, A., Keutzer, K., Khadir, M.T., Khened, M., Kickingereder, P., Kim, G., King, N., Knapp, H., Knecht, U., Kohli, L., Kong, D., Kong, X., Koppers, S., Kori, A., Krishnamurthi, G., Krivov, E., Kumar, P., Kushibar, K., Lachinov, D., Lambrou, T., Lee, J., Lee, C., Lee, Y., Lee, M., Lefkovits, S., Lefkovits, L., Levitt, J., Li, T., Li, Hongwei, Li, W., Li, Hongyang, Li, Xiaochuan, Li, Y., Li, Heng, Li, Zhenye, Li, Xiaoyu, Li, Zeju, Li, XiaoGang, Li, W., Lin, Z.-S., Lin, F., Lio, P., Liu, C., Liu, Z., Liu, X., Liu, M., Liu, J., Liu, L., Llado, X., Lopez, M.M., Lorenzo, P.R., Lu, Z., Luo, L., Luo, Z., Ma, J., Ma, K., Mackie, T., Madabushi, A., Mahmoudi, I., Maier-Hein, K.H., Maji, P., Mammen, C.P., Mang, A., Manjunath, B.S., Marcinkiewicz, M., McDonagh, S., McKenna, S., McKinley, R., Mehl, M., Mehta, S., Mehta, R., Meier, R., Meinel, C., Merhof, D., Meyer, C., Miller, R., Mitra, S., Moiyadi, A., Molina-Garcia, D., Monteiro, M.A.B., Mrukwa, G., Myronenko, A., Nalepa, J., Ngo, T., Nie, D., Ning, H., Niu, C., Nuechterlein, N.K., Oermann, E., Oliveira, A., Oliveira, D.D.C., Oliver, A., Osman, A.F.I., Ou, Y.-N., Ourselin, S., Paragios, N., Park, M.S., Paschke, B., Pauloski, J.G., Pawar, K., Pawlowski, N., Pei, L., Peng, S., Pereira, S.M., Perez-Beteta, J., Perez-Garcia, V.M., Pezold, S., Pham, B., Phophalia, A., Piella, G., Pillai, G.N., Piraud, M., Pisov, M., Popli, A., Pound, M.P., Pourreza, R., Prasanna, P., Prkovska, V., Pridmore, T.P., Puch, S., Puybareau, É., Qian, B., Qiao, X., Rajchl, M., Rane, S., Rebsamen, M., Ren, H., Ren, X., Revanuru, K., Rezaei, M., Rippel, O., Rivera, L.C., Robert, C., Rosen, B., Rueckert, D., Safwan, M., Salem, M., Salvi, J., Sanchez, I., Sánchez, I., Santos, H.M., Sartor, E., Schellingerhout, D., Scheufele, K., Scott, M.R., Scussel, A.A., Sedlar, S., Serrano-Rubio, J.P., Shah, N.J., Shah, N., Shaikh, M., Shankar, B.U., Shboul, Z., Shen, Haipeng, Shen, D., Shen, L., Shen, Haocheng, Shenoy, V., Shi, F., Shin, H.E., Shu, H., Sima, D., Sinclair, M., Smedby, O., Snyder, J.M., Soltaninejad, M., Song, G., Soni, M., Stawiaski, J., Subramanian, S., Sun, L., Sun, R., Sun, J., Sun, K., Sun, Y., Sun, G., Sun, S., Suter, Y.R., Szilagyi, L., Talbar, S., Tao, D., Tao, D., Teng, Z., Thakur, S., Thakur, M.H., Tharakan, S., Tiwari, P., Tochon, G., Tran, T., Tsai, Y.M., Tseng, K.-L., Tuan, T.A., Turlapov, V., Tustison, N., Vakalopoulou, M., Valverde, S., Vanguri, R., Vasiliev, E., Ventura, J., Vera, L., Vercauteren, T., Verrastro, C.A., Vidyaratne, L., Vilaplana, V., Vivekanandan, A., Wang, G., Wang, Q., Wang, C.J., Wang, W., Wang, D., Wang, R., Wang, Y., Wang, C., Wang, G., Wen, N., Wen, X., Weninger, L., Wick, W., Wu, S., Wu, Q., Wu, Y., Xia, Y., Xu, Y., Xu, X., Xu, P., Yang, T.-L., Yang, X., Yang, H.-Y., Yang, J., Yang, H., Yang, G., Yao, H., Ye, X., Yin, C., Young-Moxon, B., Yu, J., Yue, X., Zhang, S., Zhang, A., Zhang, K., Zhang, Xuejie, Zhang, Lichi, Zhang, Xiaoyue, Zhang, Y., Zhang, Lei, Zhang, J., Zhang, Xiang, Zhang, T., Zhao, S., Zhao, Y., Zhao, X., Zhao, L., Zheng, Y., Zhong, L., Zhou, C., Zhou, X., Zhou, F., Zhu, H., Zhu, J., Zhuge, Y., Zong, W., Kalpathy-Cramer, J., Farahani, K., Davatzikos, C., van Leemput, K., Menze, B., 2019. Identifying the Best Machine Learning Algorithms for Brain Tumor Segmentation, Progression Assessment, and Overall Survival Prediction in the BRATS Challenge. arXiv:1811.02629 [cs, stat].

Benkarim, O.M., Hahner, N., Piella, G., Gratacos, E., González Ballester, M.A., Eixarch, E., Sanroma, G., 2018. Cortical folding alterations in fetuses with isolated non-severe ventriculomegaly. Neuroimage Clin 18, 103–114. https://doi.org/10.1016/j.nicl.2018.01.006

Bethlehem, R. a. I., Seidlitz, J., White, S.R., Vogel, J.W., Anderson, K.M., Adamson, C., Adler, S., Alexopoulos, G.S., Anagnostou, E., Areces-Gonzalez, A., Astle, D.E., Auyeung, B., Ayub, M., Ball, G., Baron-Cohen, S., Beare, R., Bedford, S.A., Benegal, V., Beyer, F., Bae, J.B., Blangero, J., Cábez, M.B., Boardman, J.P., Borzage, M., Bosch-Bayard, J.F., Bourke, N., Calhoun, V.D., Chakravarty, M.M., Chen, C., Chertavian, C., Chetelat, G., Chong, Y.S., Cole, J.H., Corvin, A., Courchesne, E., Crivello, F., Cropley, V.L., Crosbie, J., Crossley, N., Delarue, M., Desrivieres, S., Devenyi, G., Biase, M.A.D., Dolan, R., Donald, K.A., Donohoe, G., Dunlop, K., Edwards, A.D., Elison, J.T., Ellis, C.T., Elman, J.A., Eyler, L., Fair, D.A., Fletcher, P.C., Fonagy, P., Franz, C.E., Galan-Garcia, L., Gholipour, A., Giedd, J., Gilmore, J.H., Glahn, D.C., Goodyer, I., Grant, P.E., Groenewold, N.A., Gunning, F.M., Gur, R.E., Gur, R.C., Hammill, C.F., Hansson, O., Hedden, T., Heinz, A., Henson, R., Heuer, K., Hoare, J., Holla, B., Holmes, A.J., Holt, R., Huang, H., Im, K., Ipser, J., Jack, C.R., Jackowski, A.P., Jia, T., Johnson, K.A., Jones, P.B., Jones, D.T., Kahn, R., Karlsson, H., Karlsson, L.,

Kawashima, R., Kelley, E.A., Kern, S., Kim, K., Kitzbichler, M.G., Kremen, W.S., Lalonde, F., Landeau, B., Lee, S., Lerch, J., Lewis, J.D., Li, J., Liao, W., Linares, D.P., Liston, C., Lombardo, M.V., Lv, J., Lynch, C., Mallard, T.T., Marcelis, M., Markello, R.D., Mazoyer, B., McGuire, P., Meaney, M.J., Mechelli, A., Medic, N., Misic, B., Morgan, S.E., Mothersill, D., Nigg, J., Ong, M.Q.W., Ortinau, C., Ossenkoppele, R., Ouyang, M., Palaniyappan, L., Paly, L., Pan, P.M., Pantelis, C., Park, M.M., Paus, T., Pausova, Z., Binette, A.P., Pierce, K., Qian, X., Qiu, J., Qiu, A., Raznahan, A., Rittman, T., Rollins, C.K., Romero-Garcia, R., Ronan, L., Rosenberg, M.D., Rowitch, D.H., Salum, G.A., Satterthwaite, T.D., Schaare, H.L., Schachar, R.J., Schultz, A.P., Schumann, G., Schöll, M., Sharp, D., Shinohara, R.T., Skoog, I., Smyser, C.D., Sperling, R.A., Stein, D.J., Stolicyn, A., Suckling, J., Sullivan, G., Taki, Y., Thyreau, B., Toro, R., Tsvetanov, K.A., Turk-Browne, N.B., Tuulari, J.J., Tzourio, C., Vachon-Presseau, É., Valdes-Sosa, M.J., Valdes-Sosa, P.A., Valk, S.L., Amelsvoort, T. van, Vandekar, S.N., Vasung, L., Victoria, L.W., Villeneuve, S., Villringer, A., Vértes, P.E., Wagstyl, K., Wang, Y.S., Warfield, S.K., Warrier, V., Westman, E., Westwater, M.L., Whalley, H.C., Witte, A.V., Yang, N., Yeo, B.T.T., Yun, H.J., Zalesky, A., Zar, H.J., Zettergren, A., Zhou, J.H., Ziauddeen, H., Zugman, A., Zuo, X.N., Aibl, Initiative, A.D.N., Investigators, A.D.R.W.B., Asrb, Team, C., Cam-CAN, Ccnp, 3r-Brain, Cobre, Group, E.D.B.A. working, FinnBrain, Study, H.A.B., Imagen, K., Nspn, Oasis-3, Project, O., Pond, The PREVENT-AD Research Group, V., Alexander-Bloch, A.F., 2021. Brain charts for the human lifespan. https://doi.org/10.1101/2021.06.08.447489

Çiçek, Ö., Abdulkadir, A., Lienkamp, S.S., Brox, T., Ronneberger, O., 2016. 3D U-Net: learning dense volumetric segmentation from sparse annotation, in: International Conference on Medical Image Computing and Computer-Assisted Intervention. Springer, pp. 424–432.

Clewell, W.H., Johnson, M.L., Meier, P.R., Newkirk, J.B., Zide, S.L., Hendee, R.W., Bowes, W.A., Hecht, F., O'Keeffe, D., Henry, G.P., Shikes, R.H., 1982. A Surgical Approach to the Treatment of Fetal Hydrocephalus. New England Journal of Medicine 306, 1320–1325. https://doi.org/10.1056/NEJM198206033062202

Clouchoux, C., du Plessis, A.J., Bouyssi-Kobar, M., Tworetzky, W., McElhinney, D.B., Brown, D.W., Gholipour, A., Kudelski, D., Warfield, S.K., McCarter, R.J., Robertson, R.L., Evans, A.C., Newburger, J.W., Limperopoulos, C., 2013. Delayed Cortical Development in Fetuses with Complex Congenital Heart Disease. Cereb Cortex 23, 2932–2943. https://doi.org/10.1093/cercor/bhs281

De Asis-Cruz, J., Andescavage, N., Limperopoulos, C., 2021. Adverse Prenatal Exposures and Fetal Brain Development: Insights From Advanced Fetal Magnetic Resonance Imaging. Biological Psychiatry: Cognitive Neuroscience and Neuroimaging. https://doi.org/10.1016/j.bpsc.2021.11.009

de Dumast, P., Kebiri, H., Atat, C., Dunet, V., Koob, M., Cuadra, M.B., 2020. Segmentation of the cortical plate in fetal brain MRI with a topological loss. arXiv:2010.12391 [cs, eess].

Desikan, R.S., Ségonne, F., Fischl, B., Quinn, B.T., Dickerson, B.C., Blacker, D., Buckner, R.L., Dale, A.M., Maguire, R.P., Hyman, B.T., Albert, M.S., Killiany, R.J., 2006. An automated labeling system for subdividing the human cerebral cortex on MRI scans into gyral based regions of interest. Neuroimage 31, 968–980. https://doi.org/10.1016/j.neuroimage.2006.01.021

Dice, L.R., 1945. Measures of the Amount of Ecologic Association Between Species. Ecology 26, 297–302. https://doi.org/10.2307/1932409

Dittrich, E., Kasprian, G.J., Prayer, D., Langs, G., 2011. Atlas Learning in Fetal Brain Development. Topics in Magnetic Resonance Imaging 22, 107–111.

Egaña-Ugrinovic, G., Sanz-Cortes, M., Figueras, F., Bargalló, N., Gratacós, E., 2013. Differences in cortical development assessed by fetal MRI in late-onset intrauterine growth restriction. American Journal of Obstetrics and Gynecology 209, 126.e1-126.e8. https://doi.org/10.1016/j.ajog.2013.04.008

Fetit, A.E., Alansary, A., Cordero-Grande, L., Cupitt, J., Davidson, A.B., Edwards, A.D., Hajnal, J.V., Hughes, E., Kamnitsas, K., Kyriakopoulou, V., Makropoulos, A., Patkee, P.A., Price, A.N., Rutherford, M.A., Rueckert, D., 2020. A deep learning approach to segmentation of the developing cortex in fetal brain MRI with minimal manual labeling, in: Arbel, T., Ben Ayed, I., de Bruijne, M., Descoteaux, M., Lombaert, H., Pal, C. (Eds.), Proceedings of the Third Conference on Medical Imaging with Deep Learning, Proceedings of Machine Learning Research. PMLR, pp. 241–261.

Fidon, L., Aertsen, M., Emam, D., Mufti, N., Guffens, F., Deprest, T., Demaerel, P., David, A.L., Melbourne, A., Ourselin, S., Deprest, J., Vercauteren, T., 2021. Label-Set Loss Functions for Partial Supervision: Application to Fetal Brain 3D MRI Parcellation, in: de Bruijne, M., Cattin, P.C., Cotin, S., Padoy, N., Speidel, S., Zheng, Y., Essert, C. (Eds.), Medical Image Computing and Computer Assisted Intervention – MICCAI 2021. Springer International Publishing, Cham, pp. 647–657. https://doi.org/10.1007/978-3-030-87196-3_60

Gholipour, A., Estroff, J.A., Barnewolt, C.E., Robertson, R.L., Grant, P.E., Gagoski, B., Warfield, S.K., Afacan, O., Connolly, S.A., Neil, J.J., Wolfberg, A., Mulkern, R.V., 2014. Fetal MRI: A Technical Update with Educational Aspirations. Concepts Magn Reson Part A Bridg Educ Res 43, 237–266. https://doi.org/10.1002/cmr.a.21321

Gholipour, A., Rollins, C., Velasco-Annis, C., Ouaalam, A., Akhondi-Asl, A., Afacan, O., Ortinau, C., Clancy, S., Limperopoulos, C., Yang, E., Estroff, J., Warfield, S., 2017. A normative spatiotemporal MRI atlas of the fetal brain for automatic segmentation and analysis of early brain growth. Scientific Reports 7. https://doi.org/10.1038/s41598-017-00525-w

Glenn, O.A., 2010. MR imaging of the fetal brain. Pediatr Radiol 40, 68–81. https://doi.org/10.1007/s00247-009-1459-3

Gowland, P., 2011. Safety of Fetal MRI Scanning, in: Prayer, D. (Ed.), Fetal MRI. Springer, Berlin, Heidelberg, pp. 49–54. https://doi.org/10.1007/174_2010_122

Griffiths, P.D., Bradburn, M., Campbell, M.J., Cooper, C.L., Embleton, N., Graham, R., Hart, A.R., Jarvis, D., Kilby, M.D., Lie, M., Mason, G., Mandefield, L., Mooney, C., Pennington, R., Robson, S.C., Wailoo, A., 2019. MRI in the diagnosis of fetal developmental brain abnormalities: the MERIDIAN diagnostic accuracy study. Health Technol Assess 23, 1–144. https://doi.org/10.3310/hta23490

Gwet, K.L., 2019. irrCAC: Computing Chance-Corrected Agreement Coefficients (CAC), R Package version 1.0.

Hart, A.R., Embleton, N.D., Bradburn, M., Connolly, D.J.A., Mandefield, L., Mooney, C., Griffiths, P.D., 2020. Accuracy of in-utero MRI to detect fetal brain abnormalities and prognosticate developmental outcome: postnatal follow-up of the MERIDIAN cohort. The Lancet Child & Adolescent Health 4, 131–140. https://doi.org/10.1016/S2352-4642(19)30349-9

Hausdorff, F., 1991. Set theory. American Mathematical Society (RI).

Hong, J., Yun, H.J., Park, G., Kim, S., Laurentys, C.T., Siqueira, L.C., Tarui, T., Rollins, C.K., Ortinau, C.M., Grant, P.E., Lee, J.-M., Im, K., 2020. Fetal Cortical Plate Segmentation Using Fully Convolutional Networks With Multiple Plane Aggregation. Frontiers in Neuroscience 14, 1226. https://doi.org/10.3389/fnins.2020.591683

Hosny, I.A., Elghawabi, H.S., 2010. Ultrafast MRI of the fetus: an increasingly important tool in prenatal diagnosis of congenital anomalies. Magn Reson Imaging 28, 1431–1439. https://doi.org/10.1016/j.mri.2010.06.024

Isensee, F., Jaeger, P.F., Kohl, S.A.A., Petersen, J., Maier-Hein, K.H., 2021. nnU-Net: a self-configuring method for deep learning-based biomedical image segmentation. Nat Methods 18, 203–211. https://doi.org/10.1038/s41592-020-01008-z

Jakab, A., Payette, K., Mazzone, L., Schauer, S., Muller, C.O., Kottke, R., Ochsenbein-Kölble, N., Tuura, R., Moehrlen, U., Meuli, M., 2021. Emerging magnetic resonance imaging techniques in open spina bifida in utero. European Radiology Experimental 5, 23. https://doi.org/10.1186/s41747-021-00219-z

Jakab, A., Pogledic, I., Schwartz, E., Gruber, G., Mitter, C., Brugger, P.C., Langs, G., Schöpf, V., Kasprian, G., Prayer, D., 2015. Fetal Cerebral Magnetic Resonance Imaging Beyond Morphology. Semin. Ultrasound CT MR 36, 465–475. https://doi.org/10.1053/j.sult.2015.06.003

Jarvis, D.A., Finney, C.R., Griffiths, P.D., 2019. Normative volume measurements of the fetal intra-cranial compartments using 3D volume in utero MR imaging. Eur Radiol 29, 3488–3495. https://doi.org/10.1007/s00330-018-5938-5

Kasprian, G., Langs, G., Brugger, P.C., Bittner, M., Weber, M., Arantes, M., Prayer, D., 2011. The prenatal origin of hemispheric asymmetry: an in utero neuroimaging study. Cereb. Cortex 21, 1076–1083. https://doi.org/10.1093/cercor/bhq179

Khalili, N., Lessmann, N., Turk, E., Claessens, N., Heus, R. de, Kolk, T., Viergever, M.A., Benders, M.J.N.L., Išgum, I., 2019. Automatic brain tissue segmentation in fetal MRI using convolutional neural networks. Magn Reson Imaging. https://doi.org/10.1016/j.mri.2019.05.020

Kikinis, R., Pieper, S.D., Vosburgh, K.G., 2014. 3D Slicer: A Platform for Subject-Specific Image Analysis, Visualization, and Clinical Support, in: Intraoperative Imaging and Image-Guided Therapy. Springer, New York, NY, pp. 277–289. https://doi.org/10.1007/978-1-4614-7657-3_19

Klein, A., Tourville, J., 2012. 101 Labeled Brain Images and a Consistent Human Cortical Labeling Protocol. Frontiers in Neuroscience 6, 171. https://doi.org/10.3389/fnins.2012.00171

Kuklisova-Murgasova, M., Quaghebeur, G., Rutherford, M.A., Hajnal, J.V., Schnabel, J.A., 2012. Reconstruction of fetal brain MRI with intensity matching and complete outlier removal. Med Image Anal 16, 1550–1564. https://doi.org/10.1016/j.media.2012.07.004

Kyriakopoulou, V., Vatansever, D., Davidson, A., Patkee, P., Elkommos, S., Chew, A., Martinez-Biarge, M., Hagberg, B., Damodaram, M., Allsop, J., Fox, M., Hajnal, J.V., Rutherford, M.A., 2017. Normative biometry of the fetal brain using magnetic resonance imaging. Brain Struct Funct 222, 2295–2307. https://doi.org/10.1007/s00429-016-1342-6

Licandro, R., Langs, G., Kasprian, G.J., Sablatnig, R., Prayer, D., Schwartz, E., 2016. A Longitudinal Diffeomorphic Atlas-Based Tissue Labeling Framework for Fetal Brains using Geodesic Regression. Presented at the 21st Computer Vision Winter Workshop, Rimske Toplice, Slovenia.

Maier-Hein, L., Reinke, A., Kozubek, M., Martel, A.L., Arbel, T., Eisenmann, M., Hanbury, A., Jannin, P., Müller, H., Onogur, S., Saez-Rodriguez, J., van Ginneken, B., Kopp-Schneider, A., Landman, B.A., 2020. BIAS: Transparent reporting of biomedical image analysis challenges. Medical Image Analysis 66, 101796. https://doi.org/10.1016/j.media.2020.101796

Menze, B.H., Jakab, A., Bauer, S., Kalpathy-Cramer, J., Farahani, K., Kirby, J., Burren, Y., Porz, N., Slotboom, J., Wiest, R., Lanczi, L., Gerstner, E., Weber, M.-A., Arbel, T., Avants, B.B., Ayache, N., Buendia, P., Collins, D.L., Cordier, N., Corso, J.J., Criminisi, A., Das, T., Delingette, H., Demiralp, Ç., Durst, C.R., Dojat, M., Doyle, S., Festa, J., Forbes, F., Geremia, E., Glocker, B., Golland, P., Guo, X., Hamamci, A., Iftekharuddin, K.M., Jena, R., John, N.M., Konukoglu, E., Lashkari, D., Mariz, J.A., Meier, R., Pereira, S., Precup, D., Price, S.J., Raviv, T.R., Reza, S.M.S., Ryan, M., Sarikaya, D., Schwartz, L., Shin, H.-C., Shotton, J., Silva, C.A., Sousa, N., Subbanna, N.K., Szekely, G., Taylor, T.J., Thomas, O.M., Tustison, N.J., Unal, G., Vasseur, F., Wintermark, M., Ye, D.H., Zhao, L., Zhao, B., Zikic, D., Prastawa, M., Reyes, M., Van Leemput, K., 2015. The Multimodal Brain Tumor Image Segmentation Benchmark (BRATS). IEEE Trans Med Imaging 34, 1993–2024. https://doi.org/10.1109/TMI.2014.2377694

Meuli, M., Meuli-Simmen, C., Hutchins, G.M., Seller, M.J., Harrison, M.R., Adzick, N.S., 1997. The spinal cord lesion in human fetuses with myelomeningocele: Implications for fetal surgery. Journal of Pediatric Surgery, Papers Presented at the 43rd Annual International Congress 32, 448–452. https://doi.org/10.1016/S0022-3468(97)90603-5

Meuli, M., Moehrlen, U., 2013. Fetal surgery for myelomeningocele: a critical appraisal. Eur J Pediatr Surg 23, 103–109. https://doi.org/10.1055/s-0033-1343082

MONAI Consortium, 2020. MONAI: Medical Open Network for AI. https://doi.org/10.5281/zenodo.4323058

Nagaraj, U.D., Venkatesan, C., Bierbrauer, K.S., Kline-Fath, B.M., 2022. Value of pre- and postnatal magnetic resonance imaging in the evaluation of congenital central nervous system anomalies. Pediatr Radiol 52, 802–816. https://doi.org/10.1007/s00247-021-05137-1

Payette, K., de Dumast, P., Kebiri, H., Ezhov, I., Paetzold, J.C., Shit, S., Iqbal, A., Khan, R., Kottke, R., Grehten, P., Ji, H., Lanczi, L., Nagy, M., Beresova, M., Nguyen, T.D., Natalucci, G., Karayannis, T., Menze, B., Bach Cuadra, M., Jakab, A., 2021a. An automatic multi-tissue human fetal brain segmentation benchmark using the Fetal Tissue Annotation Dataset. Sci Data 8, 167. https://doi.org/10.1038/s41597-021-00946-3

Payette, K., Dumast, P. de, Jakab, A., Cuadra, M.B., Vasung, L., Licandro, R., Menze, B., Zurich), H.L., 2021b. Fetal Brain Tissue Annotation and Segmentation Challenge. https://doi.org/10.5281/zenodo.4573144

Payette, K., Jakab, A., 2021. Fetal Tissue Annotation Challenge - FeTA MICCAI 2021 [WWW Document]. URL http://dx.doi.org/10.7303/syn25649159 (accessed 2.22.22).

Pierre Deman, Sebastien Tourbier, Reto Meuli, Meritxell Bach Cuadra, 2020. meribach/mevislabFetalMRI: MEVISLAB MIAL Super-Resolution Reconstruction of Fetal Brain MRI v1.0. Zenodo. https://doi.org/10.5281/zenodo.3878564

Polat, A., Barlow, S., Ber, R., Achiron, R., Katorza, E., 2017. Volumetric MRI study of the intrauterine growth restriction fetal brain. Eur Radiol 27, 2110–2118. https://doi.org/10.1007/s00330-016-4502-4

Prayer, D., Kasprian, G., Krampl, E., Ulm, B., Witzani, L., Prayer, L., Brugger, P.C., 2006. MRI of normal fetal brain development. European Journal of Radiology, Fetal Imaging 57, 199–216. https://doi.org/10.1016/j.ejrad.2005.11.020

R Core Team, 2020. R: A Language and Environment for Statistical Computing. R Foundation for Statistical Computing, Vienna, Austria.

Rollins, C.K., Ortinau, C.M., Stopp, C., Friedman, K.G., Tworetzky, W., Gagoski, B., Velasco-Annis, C., Afacan, O., Vasung, L., Beaute, J.I., Rofeberg, V., Estroff, J.A., Grant, P.E., Soul, J.S., Yang, E., Wypij, D., Gholipour, A., Warfield, S.K., Newburger, J.W., 2021. Regional Brain Growth Trajectories in Fetuses with Congenital Heart Disease. Ann Neurol 89, 143–157. https://doi.org/10.1002/ana.25940

Ronneberger, O., Fischer, P., Brox, T., 2015. U-Net: Convolutional Networks for Biomedical Image Segmentation, in: Navab, N., Hornegger, J., Wells, W.M., Frangi, A.F. (Eds.), Medical Image Computing and Computer-Assisted Intervention – MICCAI 2015, Lecture Notes in Computer Science. Springer International Publishing, Cham, pp. 234–241. https://doi.org/10.1007/978-3-319-24574-4_28

Rüegger, C.M., Bartsch, C., Martinez, R.M., Ross, S., Bolliger, S.A., Koller, B., Held, L., Bruder, E., Bode, P.K., Caduff, R., Frey, B., Schäffer, L., Bucher, H.U., 2014. Minimally invasive, imaging guided virtual autopsy compared to conventional autopsy in foetal, newborn and infant cases: study protocol for the paediatric virtual autopsy trial. BMC Pediatr 14, 15. https://doi.org/10.1186/1471-2431-14-15

Sadhwani, A., Wypij, D., Rofeberg, V., Gholipour, A., Mittleman, M., Rohde, J., Velasco-Annis, C., Calderon, J., Friedman, K.G., Tworetzky, W., Grant, P.E., Soul, J.S., Warfield, S.K., Newburger, J.W., Ortinau, C.M., Rollins, C.K., 2022. Fetal Brain Volume Predicts Neurodevelopment in Congenital Heart Disease. Circulation. https://doi.org/10.1161/CIRCULATIONAHA.121.056305

Sanroma, G., Benkarim, O.M., Piella, G., Lekadir, K., Hahner, N., Eixarch, E., González Ballester, M.A., 2018. Learning to combine complementary segmentation methods for fetal and 6-month infant brain MRI segmentation. Computerized Medical Imaging and Graphics 69, 52–59. https://doi.org/10.1016/j.compmedimag.2018.08.007

Serag, A., Aljabar, P., Ball, G., Counsell, S.J., Boardman, J.P., Rutherford, M.A., Edwards, A.D., Hajnal, J.V., Rueckert, D., 2012. Construction of a consistent high-definition spatio-temporal atlas of the developing brain using adaptive kernel regression. Neuroimage 59, 2255–2265. https://doi.org/10.1016/j.neuroimage.2011.09.062

Skotting, M.B., Eskildsen, S.F., Ovesen, A.S., Fonov, V.S., Ringgaard, S., Hjortdal, V.E., Lauridsen, M.H., 2021. Infants with congenital heart defects have reduced brain volumes. Sci Rep 11, 4191. https://doi.org/10.1038/s41598-021-83690-3

Taha, A.A., Hanbury, A., 2015. Metrics for evaluating 3D medical image segmentation: analysis, selection, and tool. BMC Med Imaging 15. https://doi.org/10.1186/s12880-015-0068-x

Tourbier, S., Bresson, X., Hagmann, P., Meuli, R., Bach Cuadra, M., 2019. sebastientourbier/mialsuperresolutiontoolkit: MIAL Super-Resolution Toolkit v1.0. Zenodo. https://doi.org/10.5281/zenodo.2598448

Tourbier, S., Bresson, X., Hagmann, P., Thiran, J.-P., Meuli, R., Cuadra, M.B., 2015. An efficient total variation algorithm for super-resolution in fetal brain MRI with adaptive regularization. Neuroimage 118, 584–597. https://doi.org/10.1016/j.neuroimage.2015.06.018

van den Heuvel, M.I., Hect, J.L., Smarr, B.L., Qawasmeh, T., Kriegsfeld, L.J., Barcelona, J., Hijazi, K.E., Thomason, M.E., 2021. Maternal stress during pregnancy alters fetal cortico-cerebellar connectivity in utero and increases child sleep problems after birth. Sci Rep 11, 2228. https://doi.org/10.1038/s41598-021-81681-y

Vasung, L., Abaci Turk, E., Ferradal, S.L., Sutin, J., Stout, J.N., Ahtam, B., Lin, P.-Y., Ellen Grant, P., 2019. Exploring early human brain development with structural and physiological neuroimaging. Neuroimage 187, 226–254. https://doi.org/10.1016/j.neuroimage.2018.07.041

Vasung, L., Rollins, C.K., Yun, H.J., Velasco-Annis, C., Zhang, J., Wagstyl, K., Evans, A., Warfield, S.K., Feldman, H.A., Grant, P.E., Gholipour, A., 2020. Quantitative In vivo MRI Assessment of Structural Asymmetries and Sexual Dimorphism of Transient Fetal Compartments in the Human Brain. Cerebral Cortex 30, 1752–1767. https://doi.org/10.1093/cercor/bhz200

Wiesenfarth, M., Reinke, A., Landman, B.A., Eisenmann, M., Saiz, L.A., Cardoso, M.J., Maier-Hein, L., Kopp-Schneider, A., 2021. Methods and open-source toolkit for analyzing and visualizing challenge results. Sci Rep 11, 2369. https://doi.org/10.1038/s41598-021-82017-6

Wu, Y., Lu, Y.-C., Jacobs, M., Pradhan, S., Kapse, K., Zhao, L., Niforatos-Andescavage, N., Vezina, G., du Plessis, A.J., Limperopoulos, C., 2020. Association of Prenatal Maternal Psychological Distress With Fetal Brain Growth, Metabolism, and Cortical Maturation. JAMA Network Open 3, e1919940. https://doi.org/10.1001/jamanetworkopen.2019.19940

Zvi, E., Shemer, A., Toussia-Cohen, S., Zvi, D., Bashan, Y., Hirschfeld-Dicker, L., Oselka, N., Amitai, M.-M., Ezra, O., Bar-Yosef, O., Katorza, E., 2020. Fetal Exposure to MR Imaging: Long-Term Neurodevelopmental Outcome. AJNR Am J Neuroradiol 41, 1989–1992. https://doi.org/10.3174/ajnr.A6771

# Overview of Challenge Participants Algorithm

## 1.1  2Ai

### Team Members and Affiliations

Team Members: Helena R. Torres[1,2,3,4], Bruno Oliveira[1,2,3,4], Pedro Morais, Jaime C. Fonseca, João L. Vilaca[1]

Affiliations: [1]2Ai – School of Technology, IPCA, Barcelos, Portugal; [2]Algoritmi Center, School of Engineering, University of Minho, Guimarães, Portugal; [3]Life and Health Sciences Research Institute (ICVS), School of Medicine, University of Minho, Braga, Portugal; [4]ICVS/3B's - PT Government Associate Laboratory, Braga/Guimarães, Portugal

### Model Description

In-utero MRI has been emerging as an important tool to evaluate fetal neurological development. Here, segmentation of fetal brain structures can have an important role in the interpretation of the fetal MRI, allowing to aid in the quantification of the changing brain morphology and in the diagnosis of congenital disorders. Since manual segmentation is highly prone to observer variability and time-consuming, automatic methods for fetal tissue segmentation can bring added value to clinical practice. The main goal of the FeTA challenge is to potentiate the development of new methods for fetal tissue segmentation.

In the scope of the FeTA challenge, an automatic method to segment the different brain structures of the fetus was developed. As well-known, encoder-decoder deep convolutional neural networks have been showed the best performance for medical imaging segmentation. In this work, the U-Net method was adapted to perform fetal tissue segmentation [2, 4]. Concerning the model architecture, similar to U-NET, an encoder-decoder network was used. The encoding path is composed of down-sampling blocks and each block of the encoding path is composed of two convolutional layers, with each layer consisting of a convolution, followed by batch normalization and a leaky rectified linear unit (ReLU). The down-sampling is implemented using a stride convolution on the first layer of each encoding block. The initial number of feature maps is defined to be 32, which is double in each down-sampling stride convolution operation. The decoding path corresponds to a symmetric expanding path and skip-connections are used to allow that the encoder and decoder share information. Overall, the number of trainable parameters was 29,971,032. Moreover, a Sigmoid activation layer was added to the end of the network to achieve the segmentation, which was performed through a one-hot labelling scheme using the Dice metric as loss function. Finally, a Xavier initialization of the model parameters was applied, in order to initialize the weights such that the variance of the activations were the same across every layer.

Concerning the input of the network, all training images of the FeTA challenge were used. A pre-processing of the images was applied, where image normalization was implemented in order to achieve zero mean intensity with unit variance. Moreover, the pixel size of the images was normalized to be the median value of the pixel size of all the images and a patch-based strategy was used, were a 3D patch with size 128x128x128 was applied. To overcome overfitting problems during training, data augmentation techniques were applied to the training images, namely spatial and intensity-based transformations [1]. Spatial transformations included random flip, rotation, scaling, grid distortion, optical distortion, and elastic transformations, whereas intensity transformations included random gaussian noise, brightness, contrast, and gamma transformations.

The network was trained during 800 epochs with a mini-batch size of 4 and using the Adam optimizer with an initial learning rate of 0.001 and a learning rate decay following the polinomial learning rate policy (lr=1- ((1-epoch)∕(max_epoch) )^power) with the power of 0.9 [3]. The training time of the method was ≈6 days on a computer with a CPU: i7-7700HQ @3.56GHz, RAM: 16GBytes @1.2GHz, and GPU: GTX1070 @1.443GHz, using the PyTorch python library. Finally, the final segmentation was obtained by post-processing the output of the network by removing isolated segmented voxels.

### References

[1] Buslaev, A., Iglovikov, V. I., Khvedchenya, E., Parinov, A., Druzhinin, M., & Kalinin, A. A. (2020). Albumentations: Fast and flexible image augmentations. Information (Switzerland), 11(2), 1–20. https://doi.org/10.3390/info11020125

[2] Isensee, F., Jaeger, P. F., Kohl, S. A. A., Petersen, J., & Maier-Hein, K. H. (2021). nnU-Net: a self-configuring method for deep learning-based biomedical image segmentation. Nature Methods, 18(2), 203–211. https://doi.org/10.1038/s41592-020-01008-z

[3] Kingma, D. P., & Ba, J. L. (2015). Adam: A method for stochastic optimization. 3rd International Conference on Learning Representations, ICLR 2015 - Conference Track Proceedings, 1–15.

[4] Ronneberger, O., Fischer, P., & Brox, T. (2015). U-Net: Convolutional Networks for Biomedical Image Segmentation. Computer Vision and Pattern Recognition, 1–8.

## 1.2 A3

### Team Members and Affiliations

Team Members: Yunzhi Xu and Li Zhao

Affiliations: College of Biomedical Engineering & Instrument Science, Zhejiang University, Hangzhou, China

### Model Description

Input: 3D volume with a shape of 128x128x128. The input data were preprocessed as follows.

Pre-processing: To improve the data balance, the background of the original data was cropped according to the edge of the brain in 3D. The border of the brain was selected by the non-zero pixels. The cropped images were zero-padded to a matrix size of 192x192x192. Then the images were down-sampled to 128x128x128 to fit in the memory of GPUs. Images were normalized, so the mean value was zero and the standard deviation equals one.

Model Architecture: The model was implemented based on V-net. Compared to the standard U-Net, a parametric rectified linear unit (PReLU) activation function. Starting number of features was 16 and was doubled in each descending layer. There are three layers of feature extraction. Each layer was a res-net composed of two consecutive 3x3x3 convolutional kernels. Layer normalization was used instead of original batch normalization. A 2x2x2 max-pooling was performed instead of 3D convolution in the original V-net. A dropout rate of 0.5 was used before the first transpose convolution layer. The decoding path was symmetric with the encoding path with the halved number of filters for each layer.

### Training Description

FeTA data were mixed in the training with 80% in the training and 20% in the validation. No additional data were used. Data augmentation was performed by randomly shifting the image up to 5 pixels in 3 directions, rotating the image up to 15 degrees along three axes, and flipping the brain in the left-right direction. Binary-cross-entropy was used as the loss function. Adam optimizer was used with a learning rate of 1e-4 with auto rate schedule using ReduceLROnPlateau. The model was trained in 200 epochs. Each epoch had 32 steps and a batch size of two. The algorithm was implemented using Tensorflow 2.4. The training was performed with two Nvidia P100 GPUs, 16 CPUs, and 118G Memory on Alibaba Cloud.

### 1.3 BIT_LILAB

#### 1.3.1 Team Members and Affiliations
Team Members: Weibin Liao, Yi Lv, Xuesong Li
Affiliations: School of Computer Science and Technology, Beijing Institute of Technology, Beijing, China

#### 1.3.2 Model Architecture
Our model structure is shown in the figure below, mainly referring to TransUNet [1]. This is a CNN-Transformer hybrid model. The network is similar to a U-shaped structure, adding Transformer layers with Multi-head Self-Attention mechanism to the encoder to encode the feature representation of image patches. The motivation of using this hybrid architecture is that we can not only use the high-resolution CNN feature map in the decoding path, but also use the global context information encoded by Transformers [2].

In encoder, ResNet-50 [3] is used as the feature extractor of CNN to generate a feature map for the input image. Transformer is used encoding feature representations from decomposed image patches.

Decoder consists of multiple up-sampling block, where each block consists of a 2× upsampling operator, a 3×3 convolution layer, a ReLU layer and a Batch Normalization layer successively. The overall architecture of our proposal network can be found in Figure 1. The detailed network architecture can be found in Table 1.

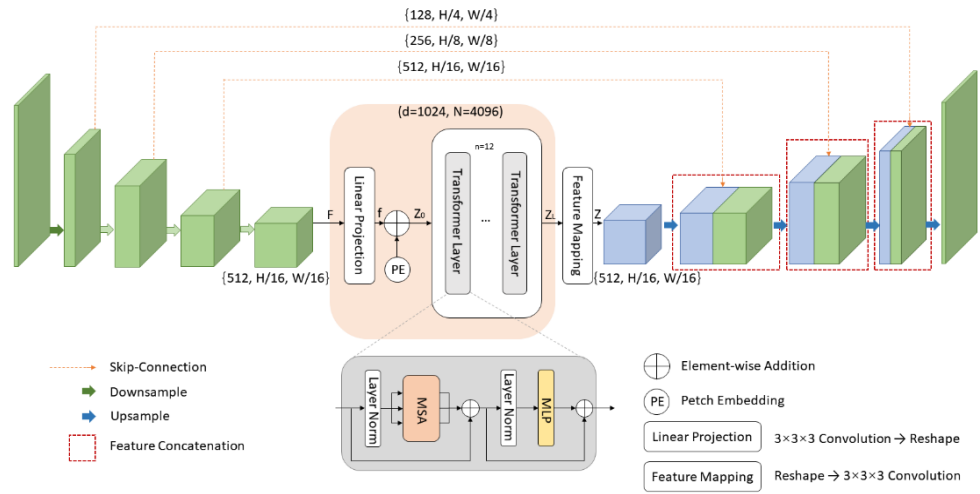

**Fig.1.** Overall architecture of our proposal network.

**Table 1.** The design details of our proposed network. $Conv\ 3 \times 3$ denotes a $3 \times 3$ convolutional layer. $GN$ denotes Group Normalization. $BN$ denotes Batch Normalization. Each ResBlock in encoder is a residual block. The size of input image is 3×256×256.

| Stage | Block name | Details | Output Size |
|---|---|---|---|
| CNN Encoder | InitConv | $Conv\ 7 \times 7, GN, ReLU$ | $64 \times 128 \times 128$ |
| | DownSample | $MaxPool$ | $64 \times 128 \times 128$ |
| | ResBlock1 | $\begin{cases} GN, Conv\ 1 \times 1, ReLU \\ GN, Conv\ 3 \times 3, ReLU \\ GN, Conv\ 1 \times 1, ReLU \end{cases} \times 3$ | $128 \times 64 \times 64$ |

|  | ResBlock2 | $\begin{cases} GN, Conv\ 1\times1, ReLU \\ GN, Conv\ 3\times3, ReLU \\ GN, Conv\ 1\times1, ReLU \end{cases} \times 4$ | $256\times32\times32$ |
|---|---|---|---|
|  | ResBlock3 | $\begin{cases} GN, Conv\ 1\times1, ReLU \\ GN, Conv\ 3\times3, ReLU \\ GN, Conv\ 1\times1, ReLU \end{cases} \times 9$ | $512\times16\times16$ |
| Transformer Encoder | Linear Projection | $Conv\ 3\times3, reshape$ | $1024\times4096$ |
|  | Transformer | $Transformer\ Layer \times 12$ | $1024\times4096$ |
| CNN Decoder | Feature Mapping | $reshape, Conv3\times3$ | $512\times16\times16$ |
|  | DeBlock1 | $Conv\ 3\times3, ReLU, BN \times 3$ | $512\times16\times16$ |
|  | UpSample1 | $Conv\ 3\times3, UpsamplingBilinear$ | $256\times32\times32$ |
|  | DeBlock2 | $Conv\ 3\times3, ReLU, BN \times 2$ | $256\times32\times32$ |
|  | UpSample2 | $Conv\ 3\times3, UpsamplingBilinear$ | $128\times64\times64$ |
|  | DeBlock3 | $Conv\ 3\times3, ReLU, BN \times 2$ | $128\times64\times64$ |
|  | UpSample3 | $Conv\ 3\times3, UpsamplingBilinear$ | $64\times128\times128$ |
|  | DeBlock4 | $Conv\ 3\times3, ReLU, BN \times 2$ | $64\times128\times128$ |
|  | EndConv | $Conv\ 1\times1, Softmax$ | $3\times256\times256$ |

### 1.3.3 Input and Output

Our model accepts 2D slices as input, and the 3D volume needs to be sliced in advance during model training. When testing, the 3D volume image can be directly input, and the program will automatically slice the volume image, predict each slice, and then stack it into a 3D volume image. The input 2D slice resolution is $256\times256$, and the image patch size is fixed as $16\times16$.

### 1.3.4 Dataset

We used all 80 data in the dataset. Due to the different reconstruction methods of data (the first 40 data are reconstructed by MialSRTK and the last 40 data are reconstructed by SimpleIRTK. When using data, we make a simple division based on two different reconstruction methods. 60 data (1-30, 41-70) are used as the training set, and 20 data (31-40, 71-80) are used as the test set. We don't differentiate the data for health conditions.

### 1.3.5 Preprocess and Data Augmentation

Due to we use 2D-input network, we need to preprocess 3D volume data into 2D slices. We adopt two simple data augmentation strategies: random rotation and flipping. Randomly rotate the 2D slice (-20, 20) degrees counterclockwise, or rotate it by 90 degrees, and then flip it up and down or left and right. The probability of data augmentation is 0.5.

### 1.3.6 Pre-training

ResNet-50 and ViT are combined in the hybrid encoder, we use the pre-training model of the backbone network provided by the author. We have also made some attempts on transfer learning: We first train the model on the Synapse multi-organ segmentation dataset, and then transfer the pre-training weights to the Fetal Tissue Annotation task. The results show that the model with pre-trained weights can converge faster on the Fetal Tissue Annotation dataset.

### 1.3.7 Implementation Details

The input resolution and patch size are set as $256\times256$ and 16. Models are trained with SGD optimizer with learning rate 0.01, momentum 0.9 and weight decay 1e-4. Batch size is 24, the max epochs are 150 and the number of training iterations are 30k. Loss function is a composite loss function of Cross Entropy and Dice loss. All experiments are conducted using four Nvidia GTX 1080Ti GPU. Training model takes about 10 hours at a time.

The python package version we use is as follows: torch==1.4.0, torchvision==0.5.0, numpy==1.20.3, tqdm==4.61.1, tensorboard==2.5.0, tensorboardX==2.2, ml-collections==0.1.0, medpy==0.4.0, SimpleITK==2.0.2, scipy==1.6.3, h5py==3.2.1.

### 1.3.8 Other Trained Models

We also tried other two baseline models on this task.

The first is nnUNet [4]. nnU-Net is the first standardized deep learning benchmark in biomedical segmentation. Without manual effort, researchers can compare their algorithms against nnU-Net on an arbitrary number of datasets to provide meaningful evidence for proposed improvements. We use 3D full resolution U-Net for training in nnUNet. We also used all 80 data, including 64 data as training data and 16 data as validation data. The test results show that nnUNet can reach the same level as TransUNet in DSC and HD95.

The other is CoTr [5]. This is also a framework combining CNN and transformer for medical image segmentation. Unlike TransUNet, CoTr can directly process 3D volume data. Unfortunately, the results we reproduced are far lower than TransUNet and nnUNet, and do not show better results compared with other CNN-based, Transformer-based and hybrid methods mentioned in this paper.

### 1.3.9 Reference

1. Chen, Jieneng, et al. "Transunet: Transformers make strong encoders for medical image segmentation." arXiv preprint arXiv:2102.04306 (2021).
2. Dosovitskiy, Alexey, et al. "An image is worth 16x16 words: Transformers for image recognition at scale." arXiv preprint arXiv:2010.11929 (2020).
3. He, Kaiming, et al. "Deep residual learning for image recognition." Proceedings of the IEEE conference on computer vision and pattern recognition. 2016.
4. Isensee, Fabian, et al. "nnU-Net: a self-configuring method for deep learning-based biomedical image segmentation." Nature methods 18.2 (2021): 203-211.
5. Xie, Yutong, et al. "CoTr: Efficiently Bridging CNN and Transformer for 3D Medical Image Segmentation." arXiv preprint arXiv:2103.03024 (2021).

## 1.4 Davoodkarimi

### 1.4.1 Team Members and Affiliations
Team Members: Davood Karimi, Ali Gholipour
Affiliations: Boston Children's Hospital, Harvard Medical School

### 1.4.2 Model Description
An encoder-decoded fully-convolutional neural network. The backbone is similar to UNet. Additional dense connections as well as short and long skip connections are included between the different stages of the encoder and the decoder. All layers are followed by ReLU.

The model works on $128^3$-voxel patches. During training, patches are selected from training images at random locations. During test, a sliding-window strategy is used to cover the entire volume. We use a novel loss function derived from mean-absolute-error loss [4]. Additionally, we have a loss term to constrain the relative volume of each of the segments/labels.

### 1.4.3 Training Method
All FeTA images (regardless of quality) were used in the same way. No additional datasets were used in any way for training. We used 15 of the FeTA images (selected to be representative of image quality and subject age distribution) for validation/testing. Once the final architecture/training hyper-parameters were decided upon, the final model was trained on all 80 FeTA images.

The only pre-processing applied consisted of intensity normalization. Specifically, each image was normalized via dividing by the image standard deviation.

Extensive image augmentation was used. Including simple geometric flips and rotations, as well as elastic deformations [2]. In addition, label perturbation and smoothing were used. Test-time image augmentation is also performed, but only includes flips; at test time no rotations or elastic deformations are used. Only a single model is trained. On a test image, the flips generate 8 labels that can be combined using simple averaging or STAPLE [5] to obtain a final label prediction.

We initialize all parameters using He's method [1] and optimize using Adam optimizer [3] with an initial learning rate of 10-4, with a batch size of 1. Learning rate was reduced by a factor of 0.9 after every 2000 training iterations if the validation loss did not decrease. If the validation loss did not decrease after 3 consecutive evaluations, training was stopped. No pre-training is used. The model training takes approximately $10^5$ iterations, equivalent to 24 hours on a single GPU. All implementation is done in TensorFlow 1.14. For data augmentation and other processing we used ITK.

### 1.4.4 References
1. He, K., Zhang, X., Ren, S., Sun, J.: Delving deep into rectifers: Surpassing human-level performance on imagenet classication. In: Proceedings of the IEEE international conference on computer vision. pp. 1026-1034 (2015)
2. Karimi, D., Samei, G., Kesch, C., Nir, G., Salcudean, S.E.: Prostate segmentation in mri using a convolutional neural network architecture and training strategy based on statistical shape models. International journal of computer assisted radiology and surgery 13(8), 1211-1219 (2018)
3. Kingma, D.P., Ba, J.: Adam: A method for stochastic optimization. arXiv preprint arXiv:1412.6980 (2014)
4. Wang, X., Hua, Y., Kodirov, E., Robertson, N.M.: Imae for noise-robust learning: Mean absolute error does not treat examples equally and gradient magnitude's variance matters. arXiv preprint arXiv:1903.12141 (2019)

5. Warfield, S.K., Zou, K.H., Wells, W.M.: Simultaneous truth and performance level estimation (staple): an algorithm for the validation of image segmentation. IEEE transactions on medical imaging 23(7), 903-921 (2004)

## 1.5 FeVer

### 1.5.1 Team Members and Affiliations

Team Members: KuanLun Liao[1,3], YiXuan Wu[2,3], and JinTai Chen[1,3]
Affiliations: [1]Zhejiang University, College of Computer Science and Technology, Road 38 West Lake District, Hangzhou 310058, P.R. China; [2]Zhejiang University, School of Medicine, 866 Yuhangtang Rd, Hangzhou 310058, P.R. China; [3]Real Doctor AI Research Centre, Zhejiang University, Hangzhou, China

### 1.5.2 Data Processing

Data source: We use all 80 T2-weighted fetal brain reconstructions data [5]. Besides, we don't use any other datasets. In order to preserve the internal structure and locality information of images, we use the original 3D images as input.

Data pre-processing: For pre-processing, we first limit the pixel values to the range of (-200, 1000) to removing the unused pixels. Then, since the spacing of each CT image is different, we resample the slice spaces for all CT images to the same scale by using spline interpolation. What's more, in order to utilize the data effectively, we remove the slices which corresponding labels only have background information.

Data augmentation: For data augmentation, we randomly crop each image with slices of 48, and randomly flip each image in the direction of up-and-down and left-and-right; besides, we also utilize the method of mixup [7], which convexly combines random pairs of images and their associated labels.

### 1.5.3 Model Description

We adopt a modified 3D Res-UNet model architecture (https://github.com/pykao/Modified-3D-UNet-Pytorch), which has a context pathway, a localization pathway and residual connections each with five layers.

Context path: In the context pathway, each layer contains three 3 × 3 × 3 convolutions each followed by a instance normalization, and then a leaky rectified linear unit (LeakyReLu); and every layer has a dropout layer with the ratio equals to 0.3. The shortcut path with residual connection brings the features of the nearby lower-level layers, alleviating the degradation problem and giving better performance.

Localization path: In the localization pathway, each layer consists of an up-sampling block with multiplier of spatial size 2, in the following of two sequence of a 3×3×3 convolution, a 3d-InstanceNorm, and then a leaky rectified linear unit (LeakyReLu). Besides, the upsample step is followed by a 3×3×3 convolution, a 3d-InstanceNorm, and then a leaky rectified linear unit (LeakyReLu).

Concat connection: The concat connections from layers of equal resolution in the context path provide the essential high-resolution features to the localization path. What's more, the concat connections are also utilized between different layers of the localization pathway to get the final predict.

InstanceNorm: In this work, we apply Instance Normalization [6] over a 5D input (a mini-batch of 3D inputs with additional channel dimension), which give normalization to H and W and D at the dimension of pixel value. The mean and standard-deviation are calculated per-dimension separately for each object in a mini-batch, giving a better outcome for the theme of semantic segmentation.

Leaky ReLU: Unlike the architecture in original U-Net, we utilize LeakyReLu instead of ReLu to be the activation function in our network. Leaky ReLU [3] is a type of activation function based on a ReLU, but it has a small slope for negative values instead of a flat slope.

Optimizer: For optimization, we choose the QHAdam [2] to be the optimizer with default configurations except for the weight decay rate 0.00001, which performs best in our task and settings.

Scheduler: For learning rate, we start with the value of 0.005, and choose the CosineAnnealingLR [1] as the learning rate scheduler, setting the period at 50 iterations and the minimum learning rate at 0.0005.

Loss function: For loss function, we apply Dice-Loss function [4], a popular loss function for image segmentation tasks based on the Dice coefficient, which is essentially a measure of overlap between two samples. This measure ranges from 0 to 1 where a Dice coefficient of 1 denotes perfect and complete overlap.

### 1.5.4 Training details

Parameter initialization: We initial the model parameters in a random way.

Parameter number: The architecture has 2,369,496 parameters in total.

Hyperparameter tuning: The hyperparameters are tuned by grid search.
- batch size: {1, 2, 4}
- learning rate: {0.01, 0.001, 0.005, 0.0001}
- weight decay rate: {0.0001, 0.00001}

Splits of the datasets: We randomly divide 95% of the dataset as the training set and the remaining 5% as the validation set.

Batch size: We set the batch size as 4.

Ensemble of models: We divide the dataset as training set and validation set randomly for five times, and use them to train different models respectively. Then, we choose the best two models among these five models and utilize average value of these two models to be the final prediction for labels.

Training time: 40 hours per model on RTX 3090

Software libraries and packages:
- Pytorch 1.8.0
- Qhoptim 1.1.0
- SimpleITK 2.0.0
- Torchvision 0.9.0
- Numpy 1.19.1
- Pandas 1.1.3
- Scipy 1.4.1
- Scikit-learn 0.23.2
- Tensorboard 2.2.2

### 1.5.5 References

1. Loshchilov, I., Hutter, F.: Sgdr: Stochastic gradient descent with warm restarts. arXiv preprint arXiv:1608.03983 (2016)
2. Ma, J., Yarats, D.: Quasi-hyperbolic momentum and adam for deep learning. arXiv preprint arXiv:1810.06801 (2018)
3. Maas, A.L., Hannun, A.Y., Ng, A.Y., et al.: Rectifier nonlinearities improve neural network acoustic models. In: Proc. icml. vol. 30, p. 3. Citeseer (2013)
4. Milletari, F., Navab, N., Ahmadi, S.A.: V-net: Fully convolutional neural networks for volumetric medical image segmentation. In: 2016 fourth international conference on 3D vision (3DV). pp. 565-571. IEEE (2016)
5. Payette, K., de Dumast, P., Kebiri, H., Ezhov, I., Paetzold, J.C., Shit, S., Iqbal, A., Khan, R., Kottke, R., Grehten, P., et al.: An automatic multi-tissue human fetal brain segmentation benchmark using the fetal tissue annotation dataset. Scientific Data 8(1), 1-14 (2021)
6. Ulyanov, D., Vedaldi, A., Lempitsky, V.: Instance normalization: The missing ingredient for fast stylization. arXiv preprint arXiv:1607.08022 (2016)
7. Zhang, H., Cisse, M., Dauphin, Y.N., Lopez-Paz, D.: mixup: Beyond empirical risk minimization. arXiv preprint arXiv:1710.09412 (2017)

### 1.6 Hilab

#### 1.6.1 Team Members and Affiliations
Team Members: Guiming Dong, Hao Fu, and Guotai Wang
Affiliations: School of Mechanical and Electrical Engineering, University of Electronic Science and Technology of China, Chengdu, China

#### 1.6.2 Model Description
Our method is based on nnU Net [1], and more details are described in the following. The network architecture is showed in Figure 1.

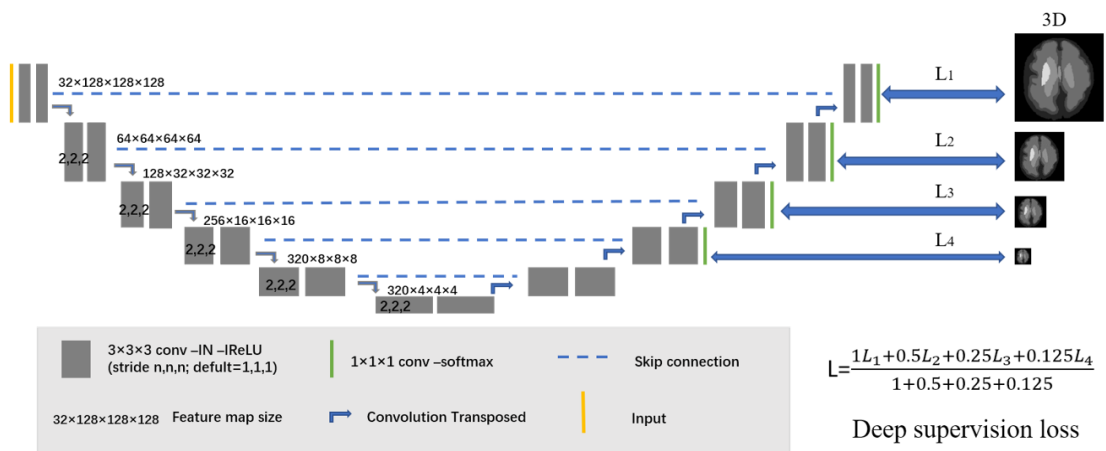

**Figure 1:** Network Architecture as generated by nnU Net.

We use instance normalization [2] and Leaky ReLU [3] following every convolution layer. And then, SGD optimizer with Nesterov momentum (μ = 0.99) is selected. Both cross entropy loss and Dice loss are used for segmentation training.

Training inputs are patches which are randomly sampled from training cases and the patch size of 128x128x128 is selected. Besides, the input data only has one modality.

Before fed into the network, all the training images need to be cropped according to the foreground. We also use Z score (mean subtraction and division by standard deviation) per image followed by cropping for data normalization.

Random initialization is selected for our model. Our framework is based on nn UNet [1], which is publically available at https://github.com/MIC DKFZ/nnUNet

#### 1.6.3 Training Method
During the training, we use all cases of FeTA dataset. We don't use any external dataset. We split the dataset into 5 folds so that we can run a 5 fold cross validation. During the validation, we found that performance of the model on pathological cases is worse than that one on neurotypical cases. So pathological cases are copied triple during the training. And then, an initial learning rate of 0.01 is used for learning network weights. The learning rate is decayed with following schedule [4]: $(1 - epoch/epoch\_max)^{0.9}$. Batch size is set to 2.

The data augmentation techniques are as follows: Rotation and Scaling, Gaussian noise, Gaussian blur, Brightness, Contrast, Simulation of low resolution, Gamma augmentation and Mirroring. More details can be seen in [1].

For FeTA challenge, we have trained 5 models via a 5 fold cross validation. And all the 5 models were selected to run ensembling.

Other details:
- System version: Ubuntu 20.04
- PyTorch version: pytorch1.6.0_py3.8_cuda10.2.89_cudnn7.6.5_0
- Devices: NVIDIA GeForce RTX 2080 Ti
- Training runs 400 epochs for each fold, and one epoch is defined as 250 iterations. Each epoch costs about 460s.

### 1.6.4 References

1. Isensee, F., Jaeger, P.F., Kohl, S.A.A. et al. "nnU Net: a self configuring method for deep learning based biomedical image segmentation." Nat Methods (2020).
2. D. Ulyanov, A. Vedaldi, and V. Lempitsky, "Instance normalization: The missing ingredient for fast stylization," arXiv preprint arXiv:1607.08022, 2016.
3. A. L. Maas, A. Y. Hannun, and A. Y. Ng, "Rectifier nonlinearities improve neural network acoustic models," in Proc. icml, vol. 30, no. 1, 2013, p. 3.
4. C. Chen, G. Papandreou, I. Kokkinos, K. Murphy, and A. L. Yuille, "Deeplab: Semantic image segmentation with deep convolutional nets, atrous convolution, and fully connected crfs," IEEE transactions on pattern analysis and machine.

## 1.7 Ichilove-ax and Ichilove-combi

### 1.7.1 Team Members and Affiliations

Team Members: Netanell Avisdris[1,2], Ori Ben-Zvi[2,3], Bella Fadida-Specktor, Prof. Leo Joskowicz, Prof. Dafna Ben Bashat[2,3,4]
Affiliations: [1]School of Computer Science and Engineering, The Hebrew University of Jerusalem, Israel; [2]Sagol Brain Institute, Tel Aviv Sourasky Medical Center, Israel, [3]Sagol School of Neuroscience, Tel Aviv University, Israel; [4]Sackler Faculty of Medicine, Tel Aviv University, Israel

Note: Only ichilove-ax and ichilove-combi were entries in the FeTA Challenge

### 1.7.2 Model Description

The approach taken in our algorithm included: reformat the brain into three orthogonal plans of 2D slice-volumes (axial, sagittal and coronal); The algorithm composed of two stages: (1) detecting the fetal brain ROI using 2D U-net (only on the axial slices); (2) performing 2D multi-class segmentation on the cropped ROI using 2D U-net; and finally, merging all resulting segmentations into one, based on majority voting.

We submitted four submissions, three where the segmentation obtained separately on each plane, and one which merge all segmentations.

- Ichilove-ax: First stage + segmentation in the axial plane of cropped ROI input volume
- Ichilove-sag: First stage + segmentation in the sagittal plane of cropped ROI input volume
- Ichilove-cor: First stage + segmentation in the coronal plane of cropped ROI input volume
- Ichilove-combi : First stage + segmentation in three planes of cropped ROI input volume, each segmentation performed twice, on the original volume, and on flipped volume, thus resulting in six segmentations. The merging strategy between segmentation is voxel-wise majority voting.

Figure 1 shows schematic overview of the four submissions.

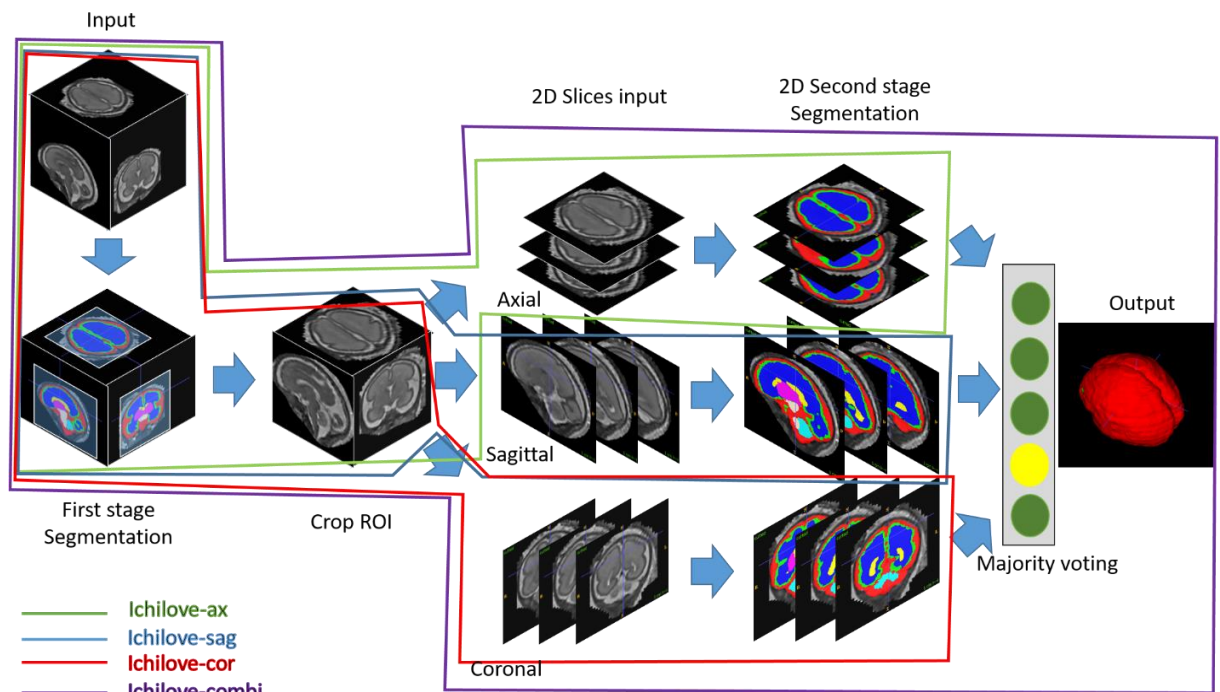

**Figure 17:** Schematic overview of 4 submissions: ichilove-{ax,sag,cor,combi}.

### 1.7.3 Networks

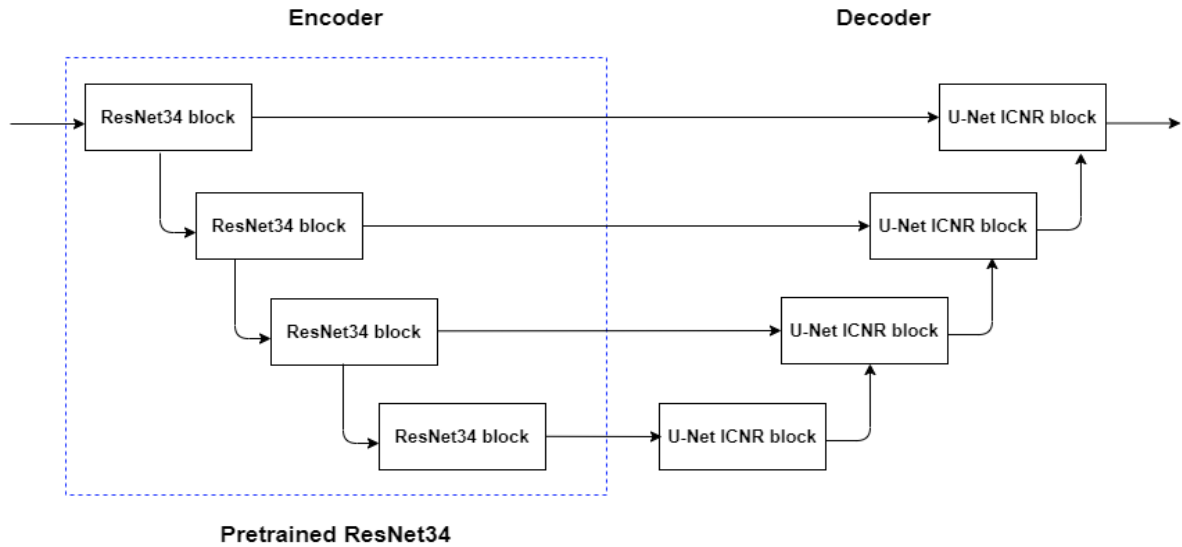

**Figure 18**: Architecture of 2D Segmentation network

For all segmentation steps, we use modified 2D U-net with similar architecture to the one used in [1] for fetal brain segmentation, where the encoder part composed of pre-trained ResNet34 network blocks and the decoder consists of U-Net PixelShuffle ICNR blocks for each scale described in [2]. The encoder and decoder are connected with the fast.ai Dynamic U-Net [3].

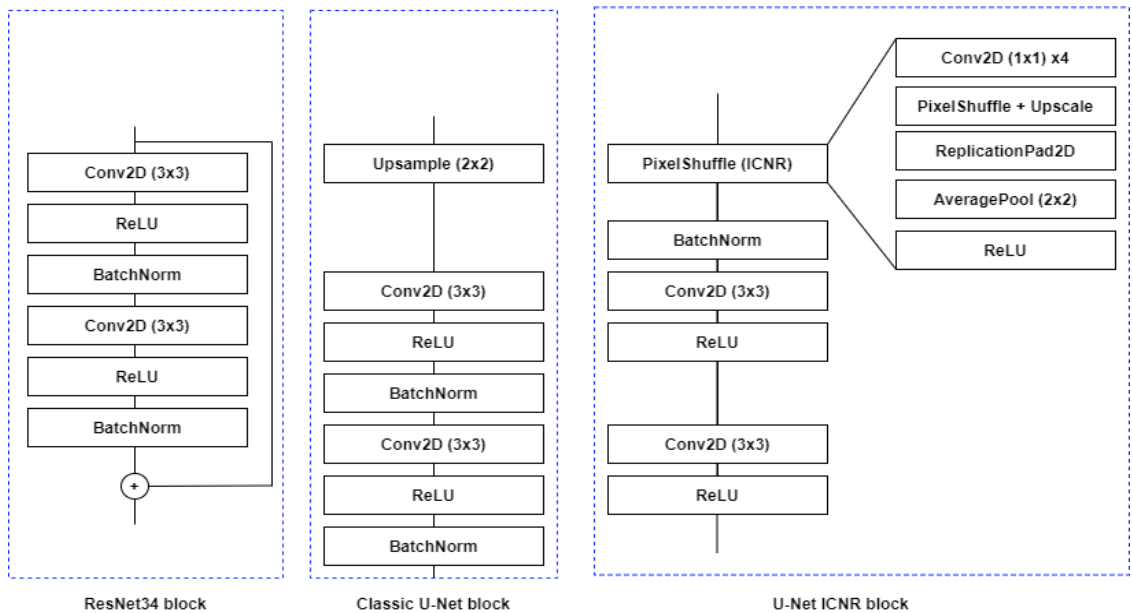

**Figure 3:** Blocks of the segmentation network: a ResNet34 block for the first scale of the encoder (left), a U-Net ICNR block for the decoder (right) and a comparison block between the classic U-Net and the U-Net ICNR blocks (middle).

To train the network, we used the OneCycle optimizer in [4] with a learning rate of 1e-3 for 30 epochs twice - the first time only the decoder layers (U-Net ICNR blocks), and the second time for the entire network with batch size of 16 slices. The network was trained with Lovasz-Softmax loss [5].

For training, we used four augmentations: 1) intensity with a contrast in the range 0.3-1.7 and brightness in the range 0.3-0.9 for training only; 2) scaling to size NxN and normalization in the range [0,1] for both training and inference, where for the first stage N=256, and for second stage N=160; 3) rotation in the range of 0-90 degrees, and; 4) Intensity Inhomogeneity Augmentation [6]. For the infer-ence, the slice output segmentation is computed by nearest-neighbor interpolation followed by zero-padding (background class) to obtain the original slice size. From 60 provided volumes, we use 50 volumes for train 5 for validation and 5 for test.

### 1.7.4 References

1. Avisdris, N., et al., Automatic linear measurements of the fetal brain on MRI with deep neural networks. International Journal of Computer Assisted Radiology and Surgery, 2021.
2. Aitken, A., et al., Checkerboard artifact free sub-pixel convolution: A note on sub-pixel convolution, resize convolution and convolution resize. arXiv preprint arXiv:1707.02937, 2017.
3. Howard, J. and S. Gugger, fastai: A Layered API for Deep Learning. Information, 2020. 11(2): p. 108.
4. Smith, L.N., A disciplined approach to neural network hyper-parameters: Part 1--learning rate, batch size, momentum, and weight decay. arXiv preprint arXiv:1803.09820, 2018.
5. Berman, M., A.R. Triki, and M.B. Blaschko. The lovász-softmax loss: A tractable surrogate for the optimization of the intersection-over-union measure in neural networks. in Proceedings of the IEEE Conference on Computer Vision and Pattern Recognition. 2018.
6. Khalili, N., et al., Automatic brain tissue segmentation in fetal MRI using convolutional neural networks. Magnetic resonance imaging, 2019. 64: p. 77-89.

### 1.8 Anonymous

One team participated in the FeTA Challenge, but declined to be a part of the paper. Therefore their scores are included in all rankings for completeness, but the algorithm description is not included.

### 1.9 MIAL

#### 1.9.1 Team Members and Affiliations

Team Members: Priscille de Dumast[2,1], Meritxell Bach Cuadra[1,2]

Affiliations: [1]CIBM Center for Biomedical Imaging; [2]Department of Radiology, Lausanne University Hospital and University of Lausanne

Note: This team contains challenge organizers. While included in all rankings, this team was ineligible for prizes.

#### 1.9.2 Model Description

Our segmentation method is based on the well-established 2D U-Net [2] image segmentation method that has been used in the past for 2D fetal brain MRI [1]. In our approach, two networks, with identical architecture, are trained separately: one network is trained on the MIALSRTK reconstructed images, and one network is trained on the IRTK reconstructed images.

We adopt a patch-based approach, feeding our networks with sub-image of 64x64 voxel size. The 2D U-Net architecture is composed of encoding and de-coding paths. The encoding path consists of 5 repetitions of the followings: two 3x3 convolutional layers, followed by a rectified linear unit (ReLu) activation function and a 2x2 max-pooling down-sampling layer. Feature maps are hence doubled from 32 to 512. In the expanding path, 2x2 up-sampled encoded features concatenated with the corresponding encoding path are 3x3 convolved and passed through ReLu. All convolutional layers output are batch normalized. The network prediction is computed with a final 1x1 convolution.

Neural networks are implemented in TensorFlow (version 2.3).

#### 1.9.3 Data

We proceed to a reconstruction-aware method. Each network was trained on 40 training cases, according to the reconstruction method used, either IRTK or MIALSRTK.
Input sub-image patches are extracted using a 2.5D approach: overlapping patches from the three orientations (axial, coronal and sagittal). Patches with tissue (positive patches) are duplicated once with data augmentation performed by randomly flipping and rotating patches (by $n$ x 90°, n ∈ [0;3]). Most negative patches (with positive intensities but no fetal brain tissue represented) were discarded as only 1/8 are kept. Intensities of all image patches are standardized to have mean 0 and variance 1. Final estimation of the segmentation is reconstructed using a majority voting approach from all probability maps.

#### 1.9.4 Training

Both networks are trained using a hybrid loss $\mathcal{L}_{hybrid}$ combining two terms:

$$\mathcal{L}_{hybrid} = (1 - \lambda)\,\mathcal{L}_{cce} + \lambda \mathcal{L}_{dice}$$

where $\mathcal{L}_{cce}$ is the categorical cross-entropy loss function, $\mathcal{L}_{dice}$ is based on the dice similarity coefficient and $\lambda$ balances the contribution of the two terms of the loss. $\lambda$ was set to 0.5.
Both networks were trained with an initial learning rate (LR) of 1 x 10$^{-3}$. Mialsrtk network was trained for 100 epochs with learning rate decay at epoch [23, 45] and irtk network was trained for 100 epochs with learning rate decay at [24,44].

A 5-folds cross-validation approach was used to determine the hyperparameters (initial LR, epochs for LR decay, and total number of epochs).

#### 1.9.5 References

[1] Khalili, N., et al.: Automatic brain tissue segmentation in fetal MRI using convolutional neural networks. Magnetic Resonance Imaging 64, 77-89 (Dec 2019). https://doi.org/10.1016/j.mri.2019.05.020

[2] Ronneberger, O., et al.: U-net: Convolutional networks for biomedical image segmentation. In: Medical Image Computing and Computer-Assisted Intervention - MICCAI 2015. pp. 234-241. Springer International Publishing, Cham (2015)

## 1.10 Moona Mazher

### 1.10.1 Team Members and Affiliations

Team Members: Moona Mazher[1], Abdul Qayyum[2], Abdesslam BENZINOU[2], Mohamed Abdel-Nasser[1,3] and Domenec Puig[1]

Affiliations: [1]Department of Computer Engineering and Mathematics, University Rovira i Virgili, Spain; [2]ENIB, UMR CNRS 6285 LabSTICC, Brest, 29238, France; [3]Department of Electrical Engineering, Faculty of Engineering, Aswan University, Egypt

### 1.10.2 Model Description

Main Contribution for Feta Segmentation Task:

1. We developed 2D Densely-based encoder network from scratch without using any pre-trained module and it is composed of a different number of layers with feature reuse capability. The efficient and lightweight 2D convolutional layer blocks with 2D up-sampling layers have been built on the decoder side of the proposed model (see detail in Figure 1). The proposed dense blocks used a successive number of feature maps in each encoder layer of the proposed model.
2. Initially, we have used axial 2D slices for training our proposed model and models did not perform well. Later, the same proposed model is trained on three different views (axial, coronal, and sagittal) of the 2D slices of the feta dataset and then combined these three output segmentation maps to develop 3D segmentation volume (see detail in Figure 3).

Proposed Method Description: The different deep learning models based on ResNet, MobileNet, efficientNet have been used to construct the encoder for the Feta challenge segmentation task. We have tried different deep leaning based modules such as ResNet [1], MobileNet [2], efficientNet[3] for Feta brain segmentation, and finally, the DenseNet [4] based module produced a better performance as compared to other deep learning modules. Our proposed model is based on an encoder and decoder module with a skip connection.

We have proposed a simple DenseNet module with a small number of blocks on the encoder side and some efficient simple 2D convolutional layers module with 2D up-sampling layers including regularization layers proposed at the decoder side. The DenseNet based deep learning with additional tricks produced better segmentation performance on the Feta challenge dataset.

In the DenseNet module, the feature maps of all preceding layers are used as inputs, and their feature maps are used as inputs into all subsequent layers. The dense module has many advantages such as alleviates the vanishing-gradient problem, strengthen feature propagation, encourage feature reuse, and substantially reduce the number of parameters. The proposed Dense feature maps module adding only a small set of feature maps to the collective knowledge of the network and keeps the remaining feature maps unchanged—and the final layer makes a decision based on all feature-maps in that are reused in the model. The proposed Dense module used a fewer number of parameters and improved the flow of information and gradients throughout the network, which makes them easy to train. Each layer has direct access to the gradients from the loss function and the original input signal, leading to implicit deep supervision. This helps training of deeper network architectures. Further, we also observe those dense connections have a regularizing effect, which reduces overfitting on tasks with smaller training set sizes.

A detailed description of the proposed model with proposed layers is shown in Figure 1. We have trained all layers' weights from scratch using the Fetal Brain Tissue Annotation and Segmentation Challenge (FeTA) dataset. Pre-trained models and extra datasets are not used in this our proposed solution.

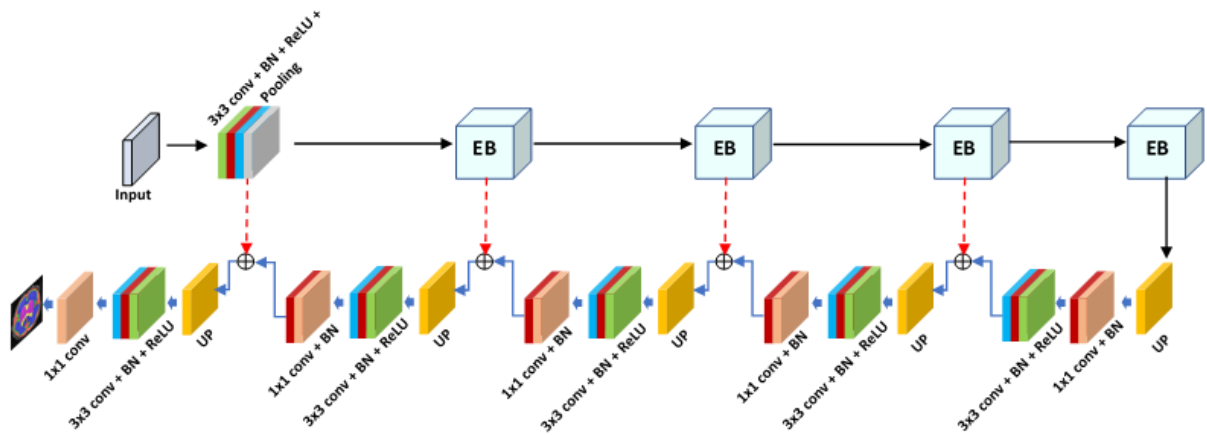

**Figure 1:** The proposed model configuration for Feta Segmentation (EB: Encoder Block).

The Dense-Layer (DL) consisted of two convolutional layers (Conv-BN-ReLU) with batch normalization (BN) layer and ReLU activation function. The feature maps are concatenated and reuse between each convolutional layer. The first convolutional layer used a 1x1 kernel and the second convolutional layer used a 3x3 kernel with a different number of feature maps. The feature maps of each layer are concatenated to form the Dense layer is shown in Figure 2. The six number of Dense-layers have been used to construct the encoder module of the proposed model with transition layer. We refer to layers between encoder blocks as transition layers, which do convolution and pooling. The transition layers used in our experiments consist of a batch normalization layer and a 1x1 convolutional layer followed by a 2x2 average pooling layer. The transition layer helps to reduce the spatial size in each encoder block in the proposed model. The transition layers have been used after each dense block in the encoder module inside the proposed model. The transition layer with six dense layers is shown in Figure 2.

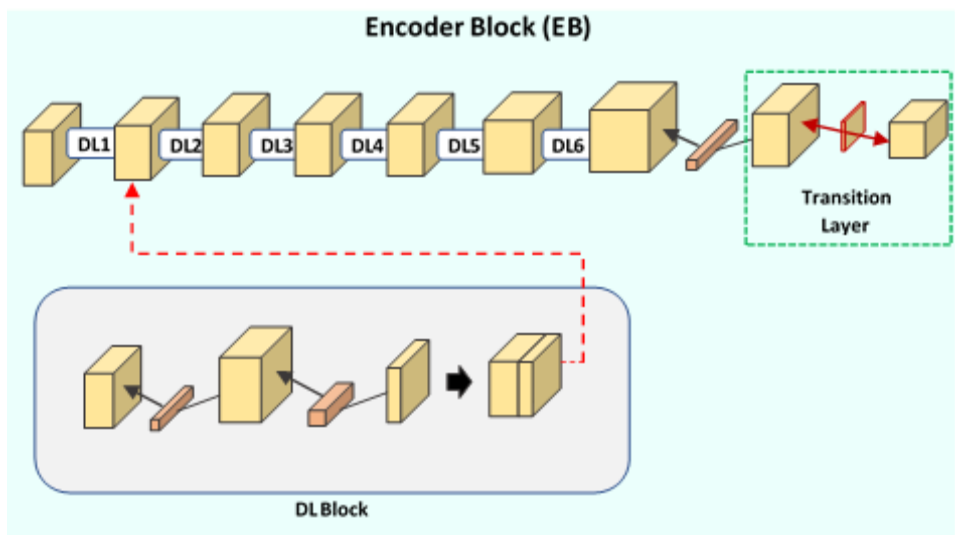

**Figure 2:** The encoder module based on proposed DensNet blocks (DL: Dense Layer).

On the decoder side, the first up-sampling 2D layer is used with a simple and efficient 2D convolutional layers block (1x1conv, 3x3conv, BN, and ReLU). The feature maps between encoder and decoder module are concatenated using skip connection based on second, third, and fourth up-blocks and proposed 2D-convolutional layer block (3x3conv, BN, ReLU) to reconstruct the segmentation from input images. The input features' maps that are obtained from every encoder block are concatenated with every decoder block feature's map to reconstruct the semantic information. The regularization layers (BN, ReLU) have been used after the concatenation of feature maps between 3x3 2D convolutional and up-sampled layer feature maps for smooth training and optimization of the training process of the proposed model. In the end, the 1x1 convolutional layer with a sigmoid layer is used to construct the segmentation map. Each dense block used several features maps (96, 768, 2112, and 2208).

First, we used the proposed model for axial 2D slices of the Feta dataset, and the model produced the worst dice score especially for Ventricles, Cerebellum, Deep Grey Matter, Brainstem classes. When we trained our model on coronal slices, the model provided optimal performance in terms of validation Dice score for all classes in the dataset that's the main motivation and reason to use all three views 2D slices to trained the 2D deep learning model. We have trained the proposed model for individual views of the 3D input volume and combined the output of the trained model on each view (axial, coronal, and sagittal) to construct a 3D segmentation map. The combined output developed by the proposed model with all three 2D views produced a better performance as compared to individual 2D views. The proposed solution is shown in Figure 3. Note. We also trained three different models on each view and could not get optimal performance on validation data as compared to the trained single model on each view.

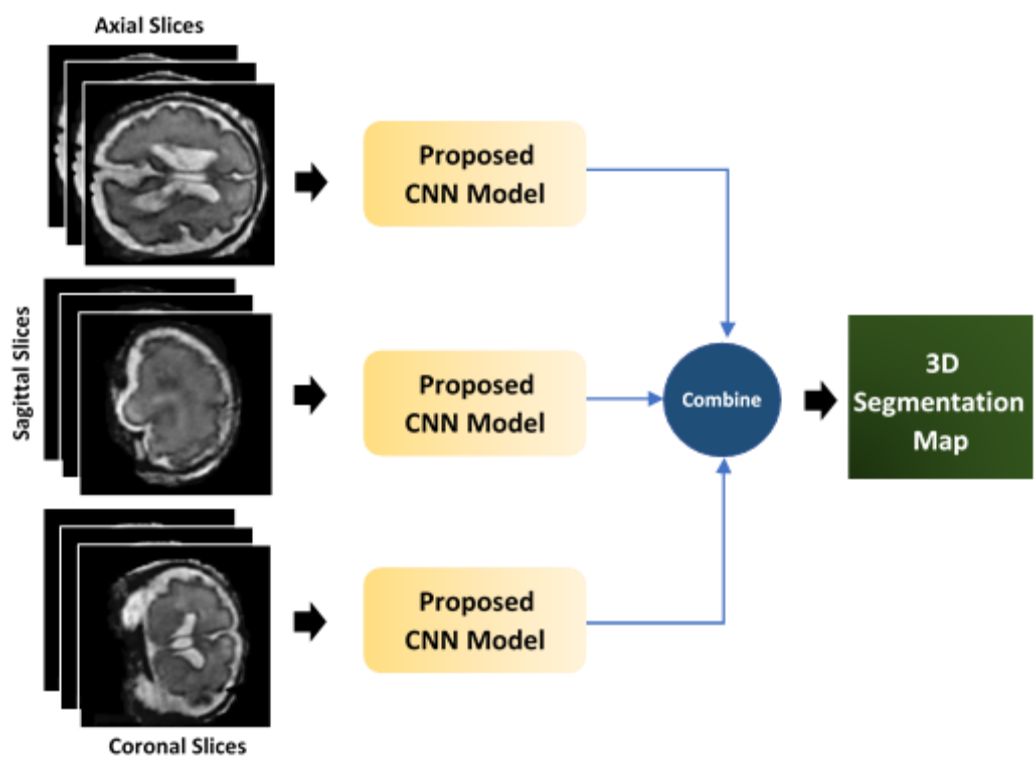

**Figure 3:** The proposed model using two separate views for Feta segmentation.

### 1.10.3 Training and Data Distribution used in Proposed Model

The Adam optimizers with a 0.0003 learning rate have been used for the training of our proposed model. In our experiments, 24 batch sizes with 1000 epochs using early stop criteria have been used for the training and optimization of the proposed model. First, the proposed model trained without data augmentation and got a good score, and later we did some data augmentation techniques such as CenterCrop, HorizontalFlip (p=0.5), VerticalFlip (p=0.5), RandomBrightnessContrast (p=0.8), and RandomGamma (p=0.8) during training the model that helps to improve a little performance. The binary cross-entropy loss function is used to compute the loss between ground-truth and predicted segmentation mask.

The one-hot encoded ground-truth segmentation map was used to compute the binary cross-entropy loss between ground truth and the predicted segmentation map.
Albumentations PyTorch segmentation library [5] used for data augmentation. The Pytorch library [6] has been used for all model implementation, training, validation, and testing. The V 100 Tesla machine with a single 12 GB GPU memory is used for training the proposed and state-of-the-art deep learning models. The proposed model contains the total number of trainable parameters (49,510,728) and total FLOPs (301846528).

The Feta dataset consisted of 80 subjects with a ground truth segmentation map. 80 % (64 subjects) data has been for training and 20% (16 subjects) for testing. The 5k fold cross-validation technique was used for random splitting the data into training and validation. The 16384 (256*64) number of images and masks in axial, coronal, and sagittal views are used for training the model 4096 (256*16) no. of images used for validation the proposed model. The standard normalization method to normalize the dataset. The pandas, skimage, OpenCV, nibabel, simpleITK python-based libraries are used for data preprocessing for training, validation, and testing the proposed model.
Note: There is no pre-trained model and an additional dataset has been used in our experiments.

### 1.10.4    References

[1] He, Kaiming, Xiangyu Zhang, Shaoqing Ren, and Jian Sun. "Identity mappings in deep residual networks." In European conference on computer vision, pp. 630-645. Springer, Cham, 2016.
[2] Sandler, Mark, Andrew Howard, Menglong Zhu, Andrey Zhmoginov, and Liang-Chieh Chen. "Mobilenetv2: Inverted residuals and linear bottlenecks." In Proceedings of the IEEE conference on computer vision and pattern recognition, pp. 4510-4520. 2018.
[3] Tan, Mingxing, and Quoc Le. "Efficientnet: Rethinking model scaling for convolutional neural networks." In International Conference on Machine Learning, pp. 6105-6114. PMLR, 2019.
[4] Huang, Gao, Zhuang Liu, Laurens Van Der Maaten, and Kilian Q. Weinberger. "Densely connected convolutional networks." In Proceedings of the IEEE conference on computer vision and pattern recognition, pp. 4700-4708. 2017.
[5] https://albumentations.ai/docs/
[6] https://pytorch.org/

## 1.11 muw_dsobotka

### 1.11.1 Team Members and Affiliations Model Description
Team Members: Daniel Sobotka, Georg Langs
Affiliations: Computational Imaging Research Lab, Department of Biomedical Imaging and Image-guided Therapy, Medical University of Vienna, Vienna, Austria

### 1.11.2 Model Description

The proposed multi-task learning model is building on a 3D U-Net architecture [1] with a shared encoder and two task specific decoders. The segmentation task TS generates the fetal tissue annotations, where the image reconstruction task TR reconstructs the input image. In contrast to other multi-task learning approaches, layer-wise feature fusion [2] is used to share features between the image segmentation and image reconstruction decoders. The encoder of the network consists of three down-sampling blocks and each task specific decoder of three up-sampling blocks. Each down-sampling block contains two 3 x 3 x 3 convolutions followed by Rectified Linear Units (ReLU) and Group Normalizations (GN) [6], as well as a 2 x 2 x 2 max pooling operation. The up-sampling path for both, image segmentation and image reconstruction tasks uses nearest neighbor interpolation and is symmetric to the down-sampling path. Neural Discriminative Dimensionality Reduction (NDDR) layers [2] are used to utilize features learned from $T_S$ and $T_R$ and to learn layerwise feature fusion between the two task specific decoders. There, features with the same spatial resolution from both, the image segmentation and image reconstruction decoders, are concatenated followed by a 1 x 1 x 1 convolution, ReLu and GN. The loss function $L$ uses homoscedastic uncertainty [3] to balance the single-task losses for the weights $W$ of the network:

$$L(W, \sigma_S, \sigma_R) = \frac{1}{2\sigma_S^2} L_S(W) + \frac{1}{2\sigma_R^2} L_R(W) + \log(\sigma_S) + \log(\sigma_R)$$

where $L_S(W)$ denotes the cross entropy loss for $T_S$. $L_R(W)$ is the mean squared error loss for $T_R$. Based on preliminary results we set $\sigma_S$ and $\sigma_R$ to fixed values with a ratio 1:8. The network consists of 6491385 trainable parameters and uses 3D patches of size 128 x 96 x 96 as input with stride between patches of size 64 x 48 x 48 and predicts 3D patches of size 128 x 96 x 96 combined at inference to an fetal tissue annotation image of size 256 x 256 x 256. All the input patches were Z-score normalized and the initialization of the model parameters was random. As baseline implementation the 3D-UNet from [5] was used.

### 1.11.3 Training Method

The proposed network model was trained on all available 80 FeTa subjects with-out validation and testing splitting and without additional external data. Further, the network was optimized using Adam [4] with an initial learning rate of 0:001, batch size of 1 and a total of 100 epochs. Data augmentation included elastic deformation with spline order 3, random flipping, random rotation by 90 degrees, random rotation by ±15 degrees, random contrast, Gaussian noise and Poisson noise. As software library PyTorch 1.3.1 with Python 3.7.3 was used and training of the network lasted around one week. For the final selection of the here described model approximately 5 different models with different values of $\sigma_S$ and $\sigma_R$ have been trained.

### 1.11.4 References

1. Cicek, O., Abdulkadir, A., Lienkamp, S.S., Brox, T., Ronneberger, O.: 3d u-net: learning dense volumetric segmentation from sparse annotation. In: International conference on medical image computing and computer-assisted intervention. pp. 424-432. Springer (2016)

2. Gao, Y., Ma, J., Zhao, M., Liu, W., Yuille, A.L.: Nddr-cnn: Layerwise feature fusing in multi-task cnns by neural discriminative dimensionality reduction. In: Proceedings of the IEEE/CVF Conference on Computer Vision and Pattern Recognition. pp. 3205-3214 (2019)

3. Kendall, A., Gal, Y., Cipolla, R.: Multi-task learning using uncertainty to weigh losses for scene geometry and semantics. In: Proceedings of the IEEE conference on computer vision and pattern recognition. pp. 7482-7491 (2018)

4. Kingma, D.P., Ba, J.: Adam: A method for stochastic optimization. arXiv preprint arXiv:1412.6980 (2014)

5. Wolny, A., Cerrone, L., Vijayan, A., Tofanelli, R., Barro, A.V., Louveaux, M.,Wenzl, C., Strauss, S., Wilson-Sanchez, D., Lymbouridou, R., et al.: Accurate and versatile 3d segmentation of plant tissues at cellular resolution. Elife 9, e57613 (2020)

6. Wu, Y., He, K.: Group normalization. In: Proceedings of the European conference on computer vision (ECCV). pp. 3-19 (2018)

## 1.12 Neurophet

### 1.12.1 Team Members and Affiliations
Team Members: ZunHyan Rieu[1], Donghyeon Kim[1], Hyun Gi Kim[2]
Affiliations: [1]NEUROPHET, Republic of Korea; [2]The Catholic University of Korea, Eunpyeong St. Mary's Hospital, Republic of Korea

### 1.12.2 Model Description

The base segmentation architecture we used is U-Net, and we designed it to receive its input patch 3D with the size of (64, 64, 64). We used patch-based training since the actual brain size MRI is pretty small compared to its MRI dimensions (256, 256, 256). Our U-Net design consists of encoding blocks of 5 followed by residual blocks. We applied a probability-based sampling method to avoid our model being trained for non-label patches.

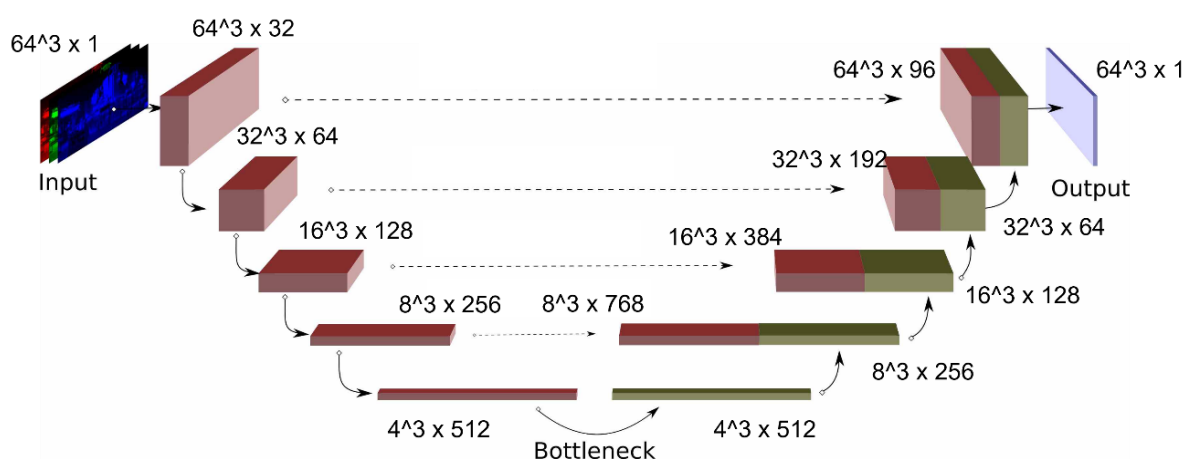

As for the loss function, we used the sum of cross-entropy loss function and dice loss function. In cross-entropy loss function, specifically, we applied different weights in each class to define the vague segmentation boundaries between 'External Cerebrospinal Fluid' and 'Grey Matter, and 'Brain-stem'. For the optimizer, we applied AdamW with the learning rate of 1E-5. To normalize the intensity range of the given MRI dataset, we performed intensity normalization from 0 to 1 with a percentile cutoff of (0, 99.8).

### 1.12.3 Training Method
Neither an additional dataset nor a pre-trained model was used for training. All case of FeTA dataset was used for training. However, we performed a visual inspection to define the poor and good quality images and distributed them equally to the train and validation. For the train and validation dataset, we divided 80 subjects with a split ratio of 0.8 (training:64, validation:16).

To reduce the dependency on the training dataset, we performed the data augmentation (random affine, random blur). For the hyper-parameter tuning, we applied the inferences for multiple patch overlap combinations (8 x 8 x 8, 16 x 16 x 16, 32 x 32 x 32, 48 x 48 x 48) and chose the best out of it.
Once the output is generated, we removed the mislabeled outer blobs using the connected component-based label noise removal method.

All preprocessing and architecture modeling, we used torchIO and scikit-image library.

Inference Example

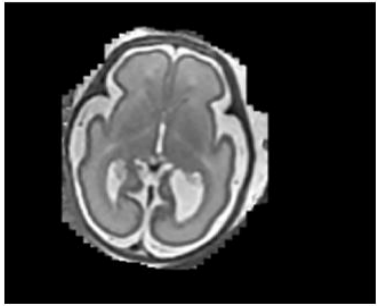 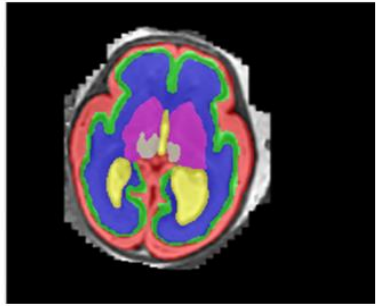 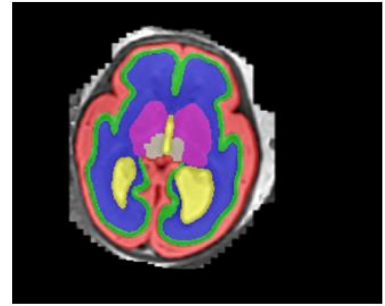

### 1.13 NVAUTO

#### 1.13.1 Team Members and Affiliations
Team Members: Md Mahfuzur Rahman Siddiquee[1], Andriy Myronenko[2], Daguang Xu[2]
Affiliations: [1]Arizona State University; [2]NVIDIA, Santa Clara, CA

#### 1.13.2 Model Description
We implemented our approach with MONAI (https://github.com/Project-MONAI/MONAI) [1]. We use the encoder-decoder backbone based on [4] with an asymmetrically larger encoder to extract image features and a smaller decoder to reconstruct the segmentation mask [6, 8, 9]. We have also utilized the OCR module from [7].

Encoder part: The encoder part uses ResNet [2] blocks. We have used 5 stage of down-sampling, each stage have 1, 2, 2, 4, and 4 convolutional blocks, respectively. We have used batch normalization and ReLU. Each block's output is followed by additive identity skip connection. We follow a common CNN approach to progressively downsize image dimensions by 2 and simultaneously increase feature size by 2. For downsizing we use strided convolutions. All convolutions are 3x3x3 with initial number of filters equal to 32. The encoder is trained with
224x224x144 input region.

Decoder part: The decoder structure is similar to the encoder one, but with a single block per each spatial level. Each decoder level begins with upsizing with transposed convolution: reducing the number of features by a factor of 2 and doubling the spatial dimension, followed by an addition of encoder output of the equivalent spatial level. The end of the decoder has the same spatial size as the original image, and the number of features equal to the initial input feature size, followed by 1x1x1 convolution into 8 channels and a softmax.

#### 1.13.3 Training Method
Dataset: We have used the FeTA dataset [5] only for training the model. We have randomly split the entire dataset into 5-folds and trained a model for each.

Loss: We have used Dice loss for training [3].

Optimization: We use AdamW optimizer with initial learning rate of 2E-4 and decrease it to zero at the end of final epoch using Cosine annealing scheduler. We have used a batch-size of 4. The model is trained of 4 GPUs, each GPU optimizing for batch-size of 1. However, we have calculated batch normalization across all the GPUs. We have ensembled 10 models for the submission: 5 of these models contain OCR module [7] and the rest do not. All the models were trained for 300 epochs.

Regularization: We use L2 norm regularization on the convolutional kernel parameters with a weight of 1E-5.

Data preprocessing and augmentation: We normalize all input images to have zero mean and unit std (based on nonzero voxels only). We have applied random flip on each axis, random rotation, and random zoom with probability of 0.5. We have also applied random contrast adjustment, random Gaussian noise, and random Gaussian smoothing with probability of 0.2.

### 1.13.4 Results on Cross-Validation

Our cross-validation results on the 5-folds can be found in Table 1.

**Table 1**. Average DICE among classes using 5-fold cross-validation.

| Fold 1 | Fold 2 | Fold 3 | Fold 4 | Fold 5 | Average |
|---|---|---|---|---|---|
| 0.8531 | 0.8515 | 0.8304 | 0.8284 | 0.8481 | 0.8423 |

### 1.13.5 References

1. Project-monai/monai, https://doi.org/10.5281/zenodo.5083813
2. He, K., Zhang, X., Ren, S., Sun, J.: Identity mappings in deep residual networks. In: European conference on computer vision. pp. 630-645. Springer (2016)
3. Milletari, F., Navab, N., Ahmadi, S.A.: V-net: Fully convolutional neural networks for volumetric medical image segmentation. In: 2016 fourth international conference on 3D vision (3DV). pp. 565-571. IEEE (2016)
4. Myronenko, A.: 3D MRI brain tumor segmentation using autoencoder regularization. In: International MICCAI Brainlesion Workshop. pp. 311-320. Springer (2018)
5. Payette, K., de Dumast, P., Kebiri, H., Ezhov, I., Paetzold, J.C., Shit, S., Iqbal, A., Khan, R., Kottke, R., Grehten, P., et al.: An automatic multi-tissue human fetal brain segmentation benchmark using the fetal tissue annotation dataset. Scientific Data 8(1), 1-14 (2021)
6. Ronneberger, O., Fischer, P., Brox, T.: U-net: Convolutional networks for biomedical image segmentation. In: International Conference on Medical image computing and computer-assisted intervention. pp. 234-241. Springer (2015)
7. Yuan, Y., Chen, X., Wang, J.: Object-contextual representations for semantic segmentation. In: Computer Vision - ECCV 2020: 16th European Conference, Glasgow, UK, August 23-28, 2020, Proceedings, Part VI 16. pp. 173-190. Springer (2020)
8. Zhou, Z., Siddiquee, M.M.R., Tajbakhsh, N., Liang, J.: Unet++: A nested u-net architecture for medical image segmentation. In: Deep learning in medical image analysis and multimodal learning for clinical decision support, pp. 3-11. Springer (2018)
9. Zhou, Z., Siddiquee, M.M.R., Tajbakhsh, N., Liang, J.: Unet++: Redesigning skip connections to exploit multiscale features in image segmentation. IEEE transactions on medical imaging 39(6), 1856-1867 (2019)

## 1.14 pengyy

### 1.14.1 Team Members and Affiliations

Team Members: Ying Peng[1], Juanying Xie[1], Huiquan Zhang[1]

Affiliations: [1]School of Computer Science, Shaanxi Normal University, Xi'an 710119, PR China

### 1.14.2 Model Description

Our network is based on the residual 3D U-Net. The architecture of our network is shown in Figure 1. Residual blocks in the encoder consist of two convolution blocks and a shortcut connection. If the number of input/output channels is different in a residual block, a non-linear projection is performed by adding the 1×1×1 convolutional block to the shortcut connection, so as to match the dimensions. Down-sampling in the encoder is done by a specific stride convolution. The decoder of the network consists of a stack of Conv-IN-Leaky ReLU blocks. The 2×2×2 transposed convolutions are used to realize up-sampling the feature maps at each layer of decoder. There are 32 channels of feature map in our network initially. The feature map channels will be doubled by each down-sampling operation in the encoder, until 320 feature maps at last. The feature maps will be halved by each transposed convolution in the decoder. At last the size at the end of decoder is same as the spatial size in the input end. Then is a convolution layer with 8 channels and the 1×1×1 convolution kernel size and a Softmax function.

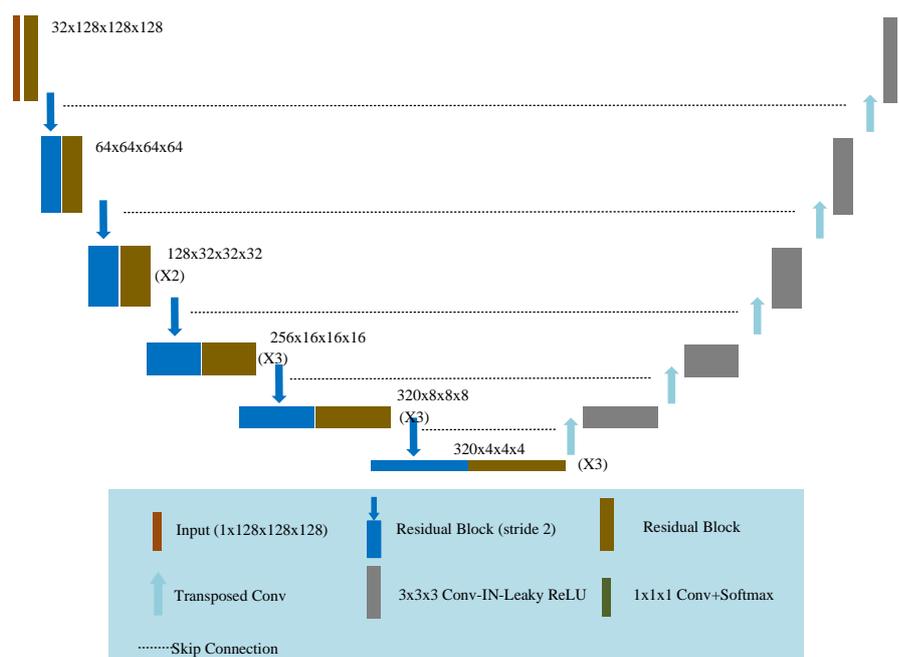

**Figure 1:** The architecture of 3D ResU-Net model.

### 1.14.3 Data Preprocessing

Since the images in the FeTA dataset have different scanning orientations, we reorient all images to the RAI orientation. At the same time we also resample the images to the isotropic resolution of 0.5mm×0.5mm×0.5mm. The detail is that the third order spline interpolation and linear interpolation are used to the images and labels, respectively. Finally, the Z-score (mean subtraction and division by standard deviation) is used to normalize each image. We only use the official FeTA dataset to evaluate the proposed method without using any other datasets.

### 1.14.4 Implementation Details

The loss function used in our network is the one combining the cross-entropy and dice loss. The optimizer is SGD (Stochastic Gradient Descent). The initial learning rate is 0.01, and the nesterov momentum is 0.99. The mini-batch size is 2, and the number of epochs is 1000. The patch size is 128 × 128 × 128. 10-fold cross validation experiments are carried out by randomly partitioning the training set into 10 folds. All our models are trained by starting from scratch, without using any pre-trained model. To avoid overfitting, standard data augmentation techniques are used during training procedure, such as random rotation, random scaling, random elastic deformation, mirroring, adding Gaussian noise, and Gamma correction. The program is implemented on the basis of nnU-Net [1] and the experimental environment is Python 3.7 with Pytorch 1.8. The nnU-Net repository is available at: https://github.com/mic-dkfz/nnunet. All the experiments were conducted on GeForce RTX 3090 GPU with 24 GB memory. Training each fold costs about 24 hours.

There are 10 models obtained by using 10-fold cross validation experiments. These 10 models are ensemble together to predict the images in the test set. In particular, test time augmentation technique is employed by mirroring along all axes.

Our method is different from the original nnU-Net, due to the data are re-rotated and training data are partitioned using different way.

### 1.14.5 References

[1] Isensee, F., Jaeger, P. F., Kohl, S. A., Petersen, J., & Maier-Hein, K. H. (2020). nnU-Net: a self-configuring method for deep learning-based biomedical image segmentation. Nature Methods, 1-9.

Author's Original Version / manuscript under review### 1.15 Physense-UPF Team

#### 1.15.1 Team Members and Affiliations

Team Members: Mireia Alenyà[1], Maria Inmaculada Villanueva[2], Mateus Riva, Oscar Camara[1]
Affiliations: [1]BCN-MedTech, Department of Information and Communications Technologies, Universitat Pompeu Fabra, Barcelona, Spain; [2]Department of Information and Communications Technologies, Universitat Pompeu Fabra, Barcelona, Spain; [3]Institut d'Investigacions Biomèdiques August Pi i Sunyer, Barcelona, Spain

#### 1.15.2 Model Description

The deep learning model used for this challenge was the nnU-Net, publicly available https://github.com/MIC-DKFZ/nnUNet [1]. The network features that have improved the performance the most have been the generalised Dice [2], which accounts for the volume of segmented structures and the implementation of data augmentation. These two features are described in more detail in the following sections.

Additionally, several attempts have been made in order to improve the segmentation accuracy, whether in the pre-processing stage or in the computation of the loss function. Those are described in the section *Report of all tested models*.

The submitted model presents a modification in the computation of the loss function. It still uses Dice and cross entropy terms as the original network does, but the former has been modified to be a generalised Dice as presented in [2]. This new Dice metric assesses multilabel segmentations with a unique score, assigning a different weight for each structure according to its volume.
It is calculated as follows:

$$DICE_{ml} = \frac{2TC_{ml}}{TC_{ml}+1}, TC_{ml} = \frac{\sum_{labels,l} \propto_l \sum_{voxels,i} \min(GT_{li}, X_{li})}{\sum_{labels,l} \propto_l \sum_{voxels,i} \propto_l \max(GT_{li}, X_{li})}, \quad \propto_l = \frac{1}{V_l}$$

where $TC_{ml}$ is the multilabel Tanimoto coefficient [2]; $GT_{li}$ is the value of voxel $i$ for label $l$ in the ground-truth segmentation; $X_{li}$ is the analogous for the predicted one; $\propto_l$ is the label-specific weighting factor that affects how much each structure $l$ contributes to the overlap accumulated over all labels; and $V_l$ is the volume of each label $l$.

To run five folds (switching subset of samples for training and validation in each of them), implement cross validation and data augmentation are possible options already defined in the nnU-Net codes.

Model architecture: The model architecture is based on an encoder–decoder with skip-connection ('U-Net-like') and instance normalization, leaky ReLU and deep supervision [1]. Its scheme is shown in Figure 1.

Author's Original Version / 07. Apr 2022 / Manuscript under review.    55

**Figure 1:** nnU-Net architecture. Each box corresponds to a multi-channel feature map. The arrows denote different operations. The number of channels is denoted on top of the box. The x-y-z-size is provided at the lower left edge of the box.

The loss function is based on Dice and cross-entropy, while the optimizer is the stochastic gradient descent (SGD) with Nesterov momentum ($\mu = 0.99$). Conv3d layers are of kernel size [3, 3, 3], stride [1, 1, 1] or [2, 2, 2] and padding [1, 1, 1]. LeakyReLU is applied with negative slope = 0.01 and Instance Normalization is applied with parameters: eps=1e-05, momentum=0.1, affine=True.

ConvTranspose3d layers present the following characteristics, depending on the level in which they are applied:
(0): ConvTranspose3d(320, 320, kernel_size=[2, 2, 2], stride=[2, 2, 2], bias=False)
(1): ConvTranspose3d(320, 256, kernel_size=[2, 2, 2], stride=[2, 2, 2], bias=False)
(2): ConvTranspose3d(256, 128, kernel_size=[2, 2, 2], stride=[2, 2, 2], bias=False)
(3): ConvTranspose3d(128, 64, kernel_size=[2, 2, 2], stride=[2, 2, 2], bias=False)
(4): ConvTranspose3d(64, 32, kernel_size=[2, 2, 2], stride=[2, 2, 2], bias=False)

Inputs/outputs: Inputs and outputs have been FeTA images and labels in 3D. The used network is 3D-nnU-Net with full resolution. The patch size has been: [128 128 128].

Pre-processing of inputs: Pre-processing done on the inputs is cropping, resampling (if necessary) and data normalisation.

The parameters' configuration is:
- batch_size: 2
- num pool per axis: [5, 5, 5]
- patch size: array([128, 128, 128])
- spacing: array([0.5, 0.5, 0.5])

- pool kernel sizes: [[2, 2, 2], [2, 2, 2], [2, 2, 2], [2, 2, 2], [2, 2, 2]]
- convolution kernel sizes: [[3, 3, 3], [3, 3, 3], [3, 3, 3], [3, 3, 3], [3, 3, 3], [3, 3, 3]]

Initialization of model parameters: The initialization of model parameters is fully optimized by the network after extracting the dataset fingerprint (a set of dataset-specific properties such as image sizes, voxel spacings, intensity information etc). The initial loss rate is 0.01.

### 1.15.3 Training method:

To train the network, cases from 1 to 60 have been used while the 20 remaining have served to test the network performance.

In the final version of the model no additional datasets have been used. The main reason for this is that no public fetal datasets were known and neither were found.

Optimization: splits of the datasets (training, validation, testing), learning rate/batch size:
The submitted model includes the independent run of five different folds, in each of which different samples have been used for training and validation (48 - 12) during training. Then, cross validation between those five folds is applied.

Data augmentation strategies: The following data augmentation features have been applied:
- Rotation along each axis, range (-15º,15º)
- Elastic deformation
- Scaling, range (0.85, 1.25)
- Add Gaussian noise, range (0, 0.1)
- Add Gaussian blur, range (0.5, 1)
- Gamma Transform, range (0.7, 1.5)
- Mirror along all axes
- Additive brightness transform, range (0.75, 1.25)
- Contrast transform, range (0.75, 1.25)
- Simulate low resolution transform, zoom range (0.5, 1)

### 1.15.4 Report of all tested models:

Several models with different approaches have been trained, and then used to predict segmentations, whose results were evaluated with the following metrics: Hausdorff Distance, Dice coefficient and volume similarity index.

Changes in the network configuration:
- Single label models (segmenting all 7 tissues separately) have been compared with multilabel models, obtaining better results with the latter. Not only in terms of accuracy but also in terms of computing time and memory optimization.
- Options of network architecture: 3D-UNet with full resolution, 2D-UNet, 3D-UNet Cascade have been tested. Best results were obtained with the 3D-UNet full resolution configuration.

Changes in the pre-processing stage:
- Registration of FeTA images and labels to the publicly available Gholipour ATLAS has been performed using the Advanced Normalization Tools (ANTs) software. FeTA data of subjects of GA higher than 27 has been registered to the ATLAS of GA 31 weeks, while FeTA data of subjects of less or equal than 27 weeks has been registered to the ATLAS of 25 weeks. This allows us to have all the input data in the same position and orientation by applying rigid transformations. Results obtained with this implementation are worse than the ones with the original images.
- Addition of gestational age as an extra input channel in the medical images has been implemented, but it has not supposed any improvements in the model performance.

Modifying the loss function:
- The Dice loss term computation has been changed for the generalised Dice. This implementation led to an improvement in the performance of the model, so it is included in the version submitted.
- Spatial constraints by relational graphs [3] have been applied. To do so, the centroids of label structures have been computed on ground-truth data and the L2 distances between the labels of adjacent structures have been defined as prior distances. Differences between priors and distances between predicted labels have been computed during training and have been included as an additional term in the loss function.
    - Centroids and respective L2 distances have been computed on the whole brain (centroids of 7 labels computed).
    - Centroids and distances have been computed on separated hemispheres (centroids of 14 labels computed). To separate the labels according to hemispheres: (1) FeTA data has been registered to the Gholipour ATLAS (Fetal Brain Atlas (harvard.edu)) by applying a rigid transform using ANTs (GitHub - ANTsX/ANTs: Advanced Normalization Tools (ANTs)); (2) After having all images aligned, labels of the right hemisphere have been defined as *label_number*+7; (3) centroids have been computed on those 14 labels.
    - None of those two approaches lead to an improvement of the model performance, so they have not been included in the presented version.

Others:
- Curriculum learning [4] methods impose a gradual and systematic way of learning, first training with best quality images before considering the low-quality ones. A qualitative classification of the quality of the medical images was made between high=1, medium=2 and low=3 by different members of the team. The data loader of the nnU-Net was modified accordingly, in order to start training with the annotated high-quality images (easier to segment correctly), continue with the medium and ending with the low-quality images (presenting more difficulty or uncertainty). This implementation does not improve the performance of the model either.
- Other details: software libraries and packages used, including version, training time:
    - Python 3.8 with an Anaconda virtual environment has been used. The nnU-Net runs in the HPC cluster of Universitat Pompeu Fabra (UPF) using a GPU and the computing platform is CUDA 10.2.
    - The average training time was of ~48 hours to run 100 epochs of the model.
    - Required Libraries:
        - nibabel 3.2.1
        - glob 0.7
        - os 0.1.4
        - numpy 1.20.3

- argparse 1.4.0
- Pytorch 1.6

### 1.15.5   References

[1] Isensee, F., Jaeger, P. F., Kohl, S. A., Petersen, J., & Maier-Hein, K. H. (2020). nnU-Net: a self-configuring method for deep learning-based biomedical image segmentation. Nature Methods, 1-9.
[2] Crum, W. R., Camara, O. & Hill, D. L. G. (2006). Generalized overlap measures for evaluation and validation in medical image analysis. IEEE Trans. Med. Imaging 25, 1451–1461.
[3] Riva, M. (2018) A new calibration approach to graph-based semantic segmentation. Master's thesis, Institute of Mathematics and Statistics.
[4] Y. Bengio, J. Louradour, R. Collobert, and J. Weston, "Curriculum learning," in Proceedings of the 26th International Conference on Machine Learning, ser. ICML 2009. New York, NY, USA: ACM, 2009, pp. 41–48.

### 1.16 SingleNets

#### 1.16.1 Team Members and Affiliations
Team Members: Bella Specktor Fadida[1], Leo Joskowicz[1], Dafna Ben Bashat[2,3], Netanell Avisdris[1,2]
Affiliations: [1]School of Computer Science and Engineering, The Hebrew University of Jerusalem, Israel; [2]Sagol Brain Institute, Tel Aviv Sourasky Medical Center, Israel; [3]Sackler Faculty of Medicine & Sagol School of Neuroscience, Tel Aviv University, Israel

#### 1.16.2 Background
Fetal brain structures segmentation is important for quantitative brain evaluation. Usually, works on brain segmentation focus on a single brain structure or a general brain segmentation. A question remains whether a method that is used for a single structure can be effectively used for multi-structure segmentation.

Multiple works have been proposed for single structure fetal brain segmentation including a work by Dudovitch et al [1] that reached a Dice score of 0.96 for whole brain segmentation using only 9 training cases.

Isensee et al. [2] observed that different frameworks should be used for segmenting different structures. The authors propose a single network framework, a single network on downscaled data framework and a cascaded framework of detection followed by segmentation. In [3] a novel contour dice loss function was proposed for segmentation of fuzzy boundaries that was shown to be effective for placenta segmentation.

#### 1.16.3 Model Description
For brain structure multi-class prediction, we used 3D networks trained separately for each brain sub-structure and for the background ("Skull" network). The network architecture was the same for all networks and is similar to Dudovitch et al [1] (Figure 1). To combine the predictions, we used the maximum network response for each voxel. Figure 2 illustrates our method.

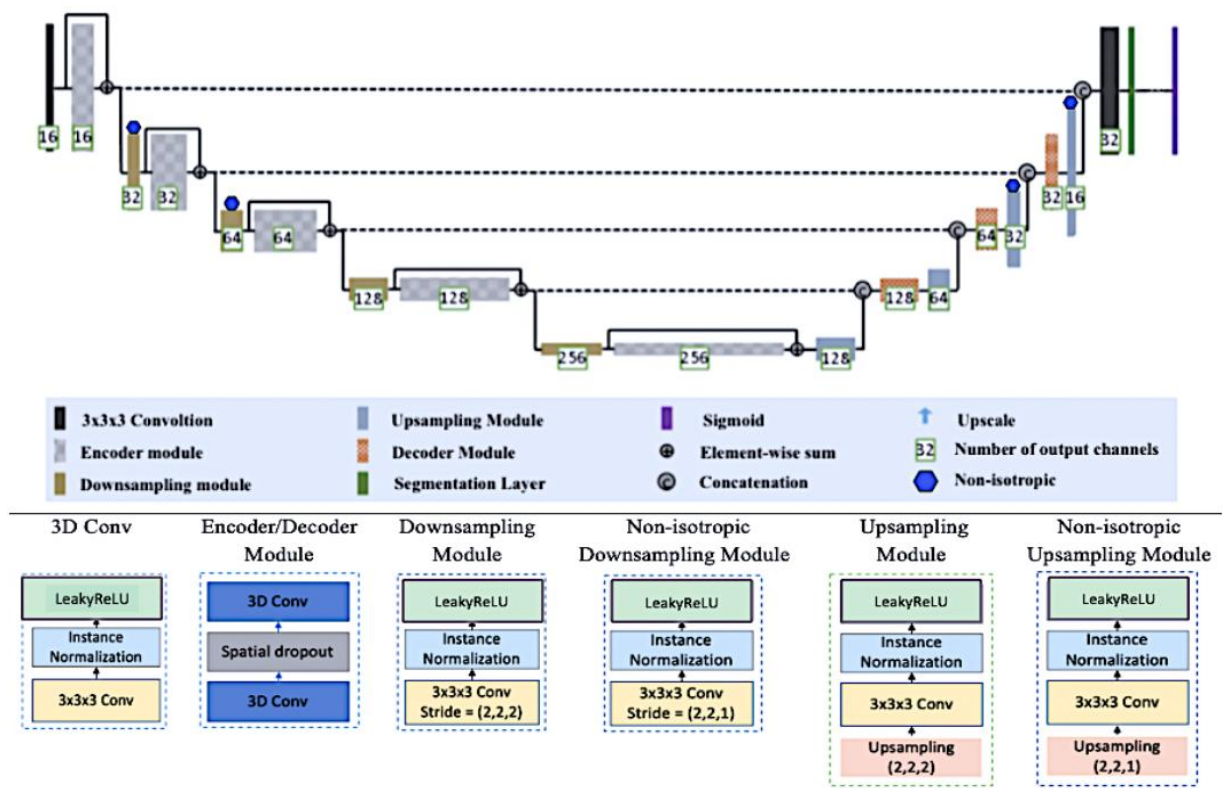

**Fig. 1.** Top: Architecture of 3D segmentation network. The number of output channels of each unit is indicated next to it. Bottom: (a-f) network modules details.

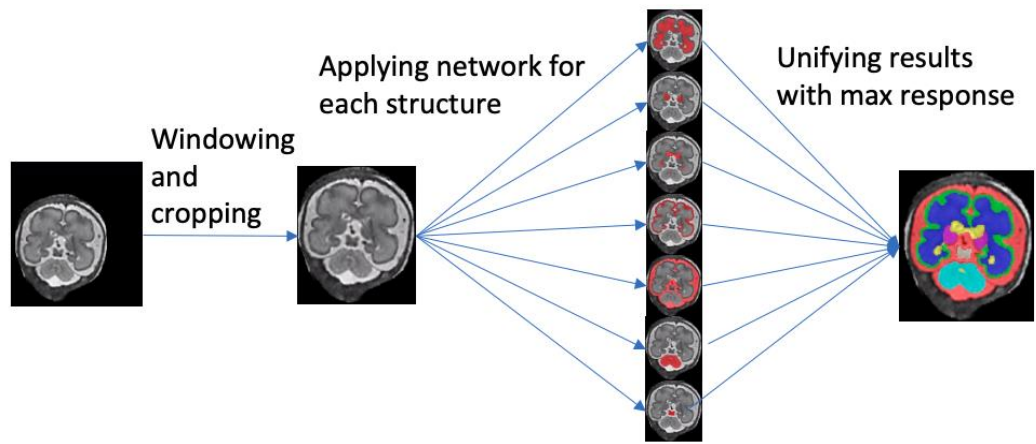

**Fig. 2** Illustration of multi-class fetal brain structure segmentation method using networks for each structure

Segmentation Frameworks: Single downscaled network for most structures. Data was downscaled by a factor of 0.5 in all planes. Cascaded framework for CBM (Cerebellum) and BS (Brainstem). ROI extraction was performed by a segmentation network on downscaled data by a factor of 0.5 and segmentation network was trained on the original scale. Most networks used soft dice loss and some used contour dice loss in combination with soft dice based on validation results. Table 1 summarizes network configurations for each structure.

**Table 1.** Frameworks and loss functions for each one of the structures

|  | Network framework | Loss Function |
|---|---|---|
| Skull | Single | Contour Dice + Soft Dice |
| CSF | Single | Soft Dice |
| GM | Single | Soft Dice |
| WM | Single | Soft Dice |
| LV | Single | Contour Dice + Soft Dice |
| CBM | Cascade | Soft Dice for both detection and segmentation networks |
| SGM | Single | Soft Dice |
| BS | Cascade | Soft dice for detection, Contour Dice + Soft Dice for segmentation |

Preprocessing:
1. Bounding box around the data with voxel intensities above th=2.
2. Windowing on intensity values so that the top and bottom 1% intensities were discarded.
3. Data normalization to zero mean and std of 1.
4. Downscaling by 0.5 in all planes for increased input ROI (block size of [96,96,96] was used due to GPU memory limitations).

Postprocessing: Voxels that were predicted as background by the Skull network were set to 0.

### 1.16.4  Contour Dice (CD) Loss Method – new formulation

We implemented a new version of Contour Dice (CD) loss which is closer to the definition of Contour Dice metric used in previous works [4,5]. Formally, let $\partial T$ and $\partial S$ be the extracted surfaces of the ground truth delineation and the network results, respectively and let $B_{\partial T}$ and $B_{\partial S}$ be their respective offset bands. The contour Dice loss of the offset bands is:

$$\text{Contour Dice Loss (T,S)} = - \frac{|\partial T \cap B_{\partial S}| + |\partial S \cap B_{\partial T}|}{|\partial T| + |\partial S|}$$

Although the Contour Dice is a non-differentiable function, in practice the contours $\partial T$ and $\partial S$ have a width, and $|\partial T|$ and $|\partial T|$ are summations over contour voxels. The formula can therefore be directly used as a loss function.

### 1.16.5  Training Details

For development, we randomly split the data to 55/25 training and validation examples. For deployment, best network setups were retrained on 75 training cases.

We did not use metadata information at all during training. Initial learning rate of 0.005 was used with learning rate drop. Models were trained with early stopping of 20 epochs. For all networks we used block size of 96×96×96, batch size of 2 and the augmentations: flipping, rotation, translation, scaling, Poisson noise, contrast, and intensity multiplication.

For Skull network we used CD loss with band size of 0 and for LV and BS segmentation networks we used band size of 3. Networks with Soft Dice loss were fine-tuned from previous best networks.

Software Packages: For inference, we used cuda=9.0, tensorflow_gpu=1.8.0, Keras=2.1.5, nibabel=2.5.0, scipy=1.3.3, numpy=1.17.4, h5py=2.0, keras_contrib=2.0.8, SimpleITK=2.1.0 and tqdm=4.48.2 packages. The training was performed on a cluster environment with cuda=10.0 and tensorflow_gpu=1.14.0.

### 1.16.6  Results
To combine the networks, we tested two different approaches against one another:

1. Stacking results one on top of the other where structures order is CSF<GM<WM<LV<CBM<SGM<BS. Holes were filled with maximum network response.
2. Maximum networks response for each voxel.

Preliminary results on 55/25 training/validation split are depicted in Table 2. Figure 3 shows illustrative results examples using each one of the approaches. Combining segmentation results of the structures with maximum response approach resulted in better segmentation results compared to structures stacking for almost all brain structures.

**Table 2.** Results comparison between stacking and max response

|       |              | Dice  | VOD   | Hausdorff 95 2D Avg | ASSD 2D Avg |
|-------|--------------|-------|-------|---------------------|-------------|
| *Skull* | Stacking     | 0.971 | 0.057 | 2.88                | 0.79        |
|       | Max Response | 0.971 | 0.057 | 2.88                | 0.79        |
| *CSF*   | Stacking     | 0.806 | 0.307 | 4.39                | 1.08        |
|       | Max Response | 0.811 | 0.298 | 4.36                | 1.11        |
| *GM*    | Stacking     | 0.717 | 0.436 | 3.54                | 0.77        |
|       | Max Response | 0.723 | 0.429 | 3.45                | 0.76        |
| *WM*    | Stacking     | 0.895 | 0.189 | 4.53                | 0.99        |
|       | Max Response | 0.896 | 0.186 | 4.65                | 1.03        |
| *LV*    | Stacking     | 0.796 | 0.326 | 7.22                | 1.59        |
|       | Max Response | 0.808 | 0.311 | 7.17                | 1.52        |
| *CBM*   | Stacking     | 0.818 | 0.283 | 4.06                | 1.21        |
|       | Max Response | 0.834 | 0.265 | 4.01                | 1.24        |
| *SGM*   | Stacking     | 0.777 | 0.349 | 5.79                | 1.88        |
|       | Max Response | 0.793 | 0.328 | 5.61                | 1.86        |
| *BS*    | Stacking     | 0.723 | 0.418 | 5.53                | 1.62        |
|       | Max Response | 0.736 | 0.400 | 5.19                | 1.54        |

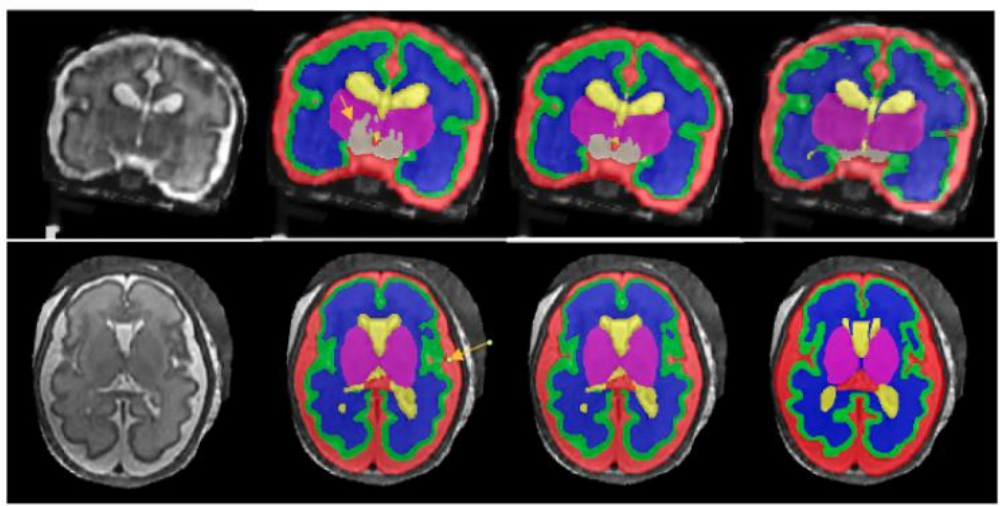

**Figure 3.** Results comparison between stacking approach (2nd column) and maximum response (3rd column). 4th column is the ground truth

### 1.16.7 References

1. Dudovitch G, Link-Sourani D, Sira LB, Miller E, Bashat DB, Joskowicz L. Deep Learning Automatic Fetal Structures Segmentation in MRI Scans with Few Annotated Datasets. Proc. Int. Conf. Medical Image Computing and Computer-Assisted Interventions pp365-74, 2020.

2. Isensee F, Jaeger PF, Kohl SA, Petersen J, Maier-Hein KH. nnU-Net: a self-configuring method for deep learning-based biomedical image segmentation. Nature Methods 18(2):203-11, 2021.
3. Specktor Fadida B, Link Sourani D, Ferster-Kveller S, Ben Sira L, Miller E, Ben Bashat D, Joskowicz L. A bootstrap self-training method for sequence transfer: State-of-the-art placenta segmentation in fetal MRI. Perinatal, Preterm and Pediatric Image Analysis MICCAI workshop 2021.
4. Nikolov S, Blackwell S, Mendes R, De Fauw J, Meyer C, Hughes C, Askham H, Romera- Paredes B, Karthikesalingam A, Chu C, Carnell D. Deep learning to achieve clinically applicable segmentation of head and neck anatomy for radiotherapy. arXiv preprint arXiv:1809.04430, Sep 12, 2018.
5. Moltz JH, Hänsch A, Lassen-Schmidt B, Haas B, Genghi A, Schreier J, Morgas T, Klein J. Learning a loss function for segmentation: a feasibility study. In Proc. IEEE 17th Int. Symp. on Biomedical Imaging, pp. 357-360), 2020.

### 1.17 SJTU_EIEE_2-426Lab

#### 1.17.1 Team Members and Affiliations
Team Members Hao Liu, Yuchen Pei, Huai Chen, and Lisheng Wang
Affiliations: Team SJTU_EIEE_2-426LAB, Shanghai Jiao Tong University

#### 1.17.2 Model Description
Data: For the task of fetal brain tissue segmentation, we propose a cumbersome coarse-to-fine segmentation framework inspired by [1], which divides the segmentation process into two stages. In the first stage, the coarse model segments all the 7 tissues at a time. Then in the second stage, with the regions of interest (ROIs) determined by the coarse result from the first step, the segmentation of each tissue is refined by a separate model, respectively. Specifically, we use nnU-Net [1] as well as a vanilla 3D U-Net [2] with residual architecture [3] (referred to below as Res-U-Net) for the coarse stage, and then we train 5 3D Res-U-Net models separately to refine the segmentation of white matter, ventricles, cerebellum, deep grey matter, and brainstem.

All the 3D Res-U-Net models we use share the similarly architecture. The detailed structure of Res-U-Net in the coarse stage is shown in Figure 1.

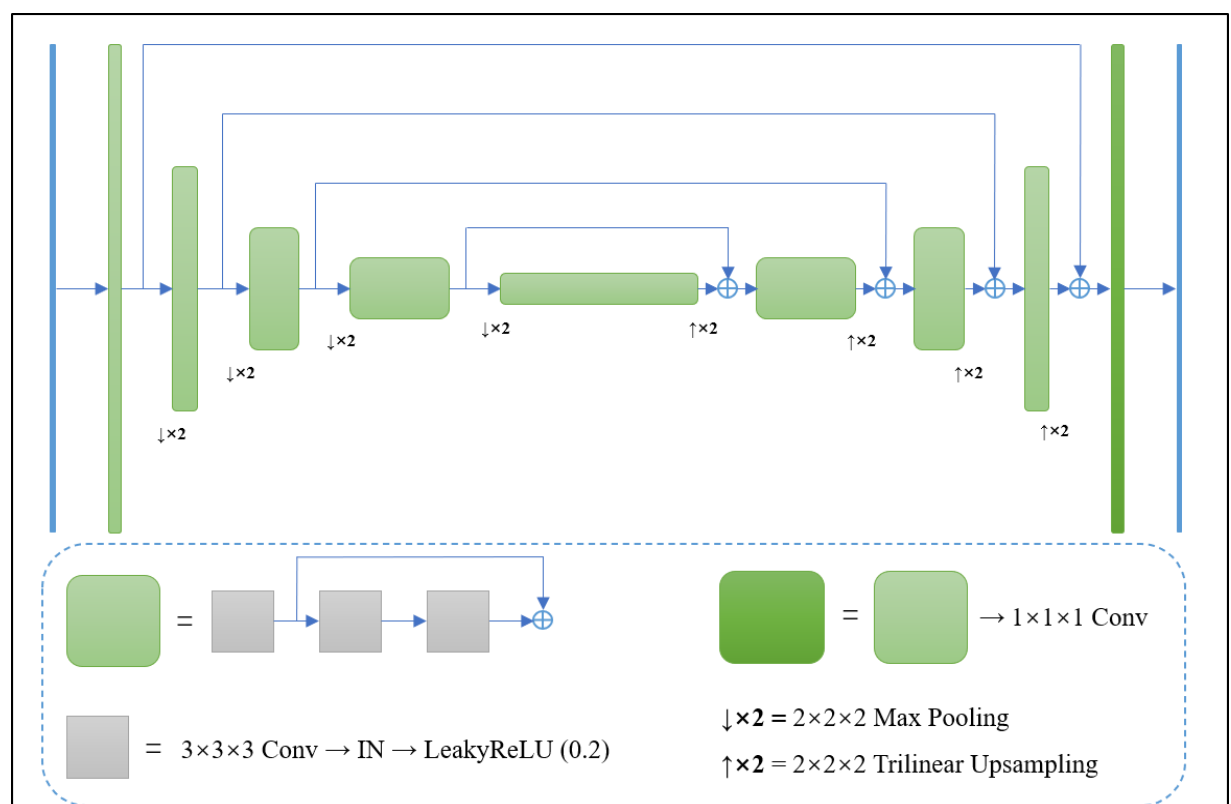

**Figure 1:** The network architecture of the Res-U-Net for coarse segmentation. There are 8 filters in the first convolution layer with a ratio of 2 between subsequent spatial levels and the last layer of the encoder has 128 filters.

Since we use small ROIs of corresponding tissues as the input of refinement networks, the width (i.e., the number of filters in intermediate convolutional layers) of the refinement networks is allowed to be larger. For the model to refine the segmentation of white matter, the width of the network is two times than the coarse network and there is one more spatial level where convolution layers have 256 filters. The other refinement networks are four times wider than the coarse network and also have an additional spatial level where convolution layers have 512 filters.
For the nnU-Net used in the first stage, we follow the default setting.

Loss Function: In the first stage, we use the combination of cross-entropy and dice loss to train the multi-class segmentation networks. The weights of two losses are both 1. In the second stage, since the 95th percentile Hausdorff Coefficient (HD95) is one of the evaluation metrics, we replace the cross-entropy loss with a distance transform-based HD loss function [4] to reduce HD directly. At the end of each training epoch, the weight for HD loss is modified to the ratio of the mean of the dice loss term on the training data to that of the HD loss term.

Pre-processing: In the first stage, we first crop the nonzero regions of each fetal brain reconstructions and then pad them into the size of 256×256×256. Then, they are normalized to zero mean and unit variance and used as the input data of the 3D Res-U-Net. The preprocessing for nnU-Net is still left to its default. In the second stage, we crop the ROIs of tissues according to the segmentation results in the first stage. Specifically, the ROIs of white matter and ventricles are determined by the segmentation results of nnU-Net, while the ROIs of cerebellum and deep grey matter are selected on the basis of the segmentation by Res-U-Net. With regard to brainstem, the union of the two coarse segmentation masks is used to crop the ROIs. Note that we do not refine the segmentation of external cerebrospinal fluid and grey matter. All the cropped ROIs are slightly bigger than the bounding box and the size of each dimension is evenly divisible by 32 since the input data will go through 5 downsampling layers. And in order to inherit the results in the first stage, we concatenate the cropped ROIs and their corresponding segmentation predictions from the coarse models, which are used as the input of the refinement networks in the second stage.

### 1.17.3 Training Method

Data: We split all the provided data randomly into 64 training cases and 16 test cases. And since our coarse models in the first stage works poorly on the Case 5 and Case 7, we remove them from test dataset and use the remaining 14 cases to select refinement models. We do not use any additional datasets.

The data augmentations used for training the Res-U-Net in the first stage includes random rotation, random scale and random flip in each dimension. And we use only random flip in the second stage.

Training Details: All the training details mentioned below are for our Res-U-Net models and the nnU-Net is trained with default setting. The optimizer of network parameters of all the Res-U-Net is performed via Adam optimizer. All the model parameters are initialized randomly. Learning rate is initialized as 1e-3 in the first stage and 1e-4 in the second stage. The weight decay is set as 1e-5 and the batch size is 1. We train the network in the first stage for 500 epochs and the refinement networks in the second stage for 1000 epochs at most. We only save the model with highest dice coefficient on the test dataset in the first stage, while in the second stage, the models with the highest dice coefficient or the lowest HD are both saved. Our algorithm is implemented in Python 3.8.5 using Pytorch 1.7.1 framework. Experiments are performed on one NVIDIA RTX 3090 with 24GB of RAM memory.

### 1.17.4 Inference Strategy and Post-processing

We use two inference strategies in the refinement stage to improve the quality of segmentation. First, for each input volume, the result is produced by the mean of predictions by the two saved models. Second, for each case, we predict the results of 8 variants obtained through flips in each dimension, and then take the average of them as the final result.

The mask of refinement segmentation is generated by the threshold of 0.5. And if there is overlap of segmentation masks for different tissues, the tissue with the maximum average volume on the whole dataset is treated as the predicted label for the overlapping regions.

### 1.17.5 References

1. Huai, Chen & Wang, Xiuying & Huang, Yijie & Wu, Xiyi & Yu, Yizhou & Wang, Lisheng. (2019). Harnessing 2D Networks and 3D Features for Automated Pancreas Segmentation from Volumetric CT Images. 339-347. 10.1007/978-3-030-32226-7_38.
2. Isensee, Fabian & Jaeger, Paul & Kohl, Simon & Petersen, Jens & Maier-Hein, Klaus. (2021). nnU-Net: a self-configuring method for deep learning-based biomedical image segmentation. Nature Methods. 18. 1-9. 10.1038/s41592-020-01008-z.
3. Ronneberger, Olaf & Fischer, Philipp & Brox, Thomas. (2015). U-Net: Convolutional Networks for Biomedical Image Segmentation. LNCS. 9351. 234-241. 10.1007/978-3-319-24574-4_28.
4. He, Kaiming & Zhang, Xiangyu & Ren, Shaoqing & Sun, Jian. (2016). Deep Residual Learning for Image Recognition. 770-778. 10.1109/CVPR.2016.90.
5. Karimi, Davood & Salcudean, Septimiu. (2019). Reducing the Hausdorff Distance in Medical Image Segmentation with Convolutional Neural Networks. IEEE Transactions on Medical Imaging. PP. 1-1. 10.1109/TMI.2019.2930068.

### 1.18   TRABIT

#### 1.18.1   Team Members and Affiliations

Team Members: Lucas Fidon[1], Michael Aertsen[2], Suprosanna Shit[3], Philippe Demaerel[2], Sébastien Ourselin[1], Jan Deprest[2,4,5], and Tom Vercauteren[1]

Affiliations: [1]School of Biomedical Engineering & Imaging Sciences, King's College London, UK; [2]Department of Radiology, University Hospitals Leuven, Belgium; [3]Technical University of Munich, Germany; [4]Institute for Women's Health, University College London, UK; [5]Department of Obstetrics and Gynaecology, University Hospitals Leuven, Belgium

#### 1.18.2   Objective

The Fetal Brain Tissue Annotation and Segmentation Challenge (FeTA) aims at comparing algorithms for multi-class automatic segmentation of fetal brain 3D T2 MRI. Seven tissue types are considered [13]:
1. extra-axial cerebrospinal fluid
2. cortical gray matter
3. white matter
4. ventricular system
5. cerebellum
6. deep gray matter
7. brainstem

This paper describes our method for our participation in the FeTA challenge 2021 (team name: TRABIT).

The performance of convolutional neural networks for medical image segmentation is thought to correlate positively with the number of training data [1]. The FeTA challenge does not restrict participants to using only the provided training data but also allows for using other publicly available sources. Yet, open access fetal brain data remains limited. An advantageous strategy could thus be to expand the training data to cover broader perinatal brain imaging sources. Perinatal brain MRIs, other than the FeTA challenge data, that are currently publicly available, span normal and pathological fetal atlases as well as neonatal scans [4,5,7,17]. However, perinatal brain MRIs segmented in different datasets typically come with different annotation protocols. This makes it challenging to combine those datasets to train a deep neural network.

We recently proposed a family of loss functions, the label-set loss functions [3], for partially supervised learning. Label-set loss functions allow to train deep neural networks with partially segmented images, i.e. segmentations in which some classes may be grouped into super-classes. We propose to use label-set loss functions [3] to improve the segmentation performance of a state-of-the-art deep learning pipeline for multi-class fetal brain segmentation by merging several publicly available datasets. To promote generalisability, our approach does not introduce any additional hyper-parameters tuning.

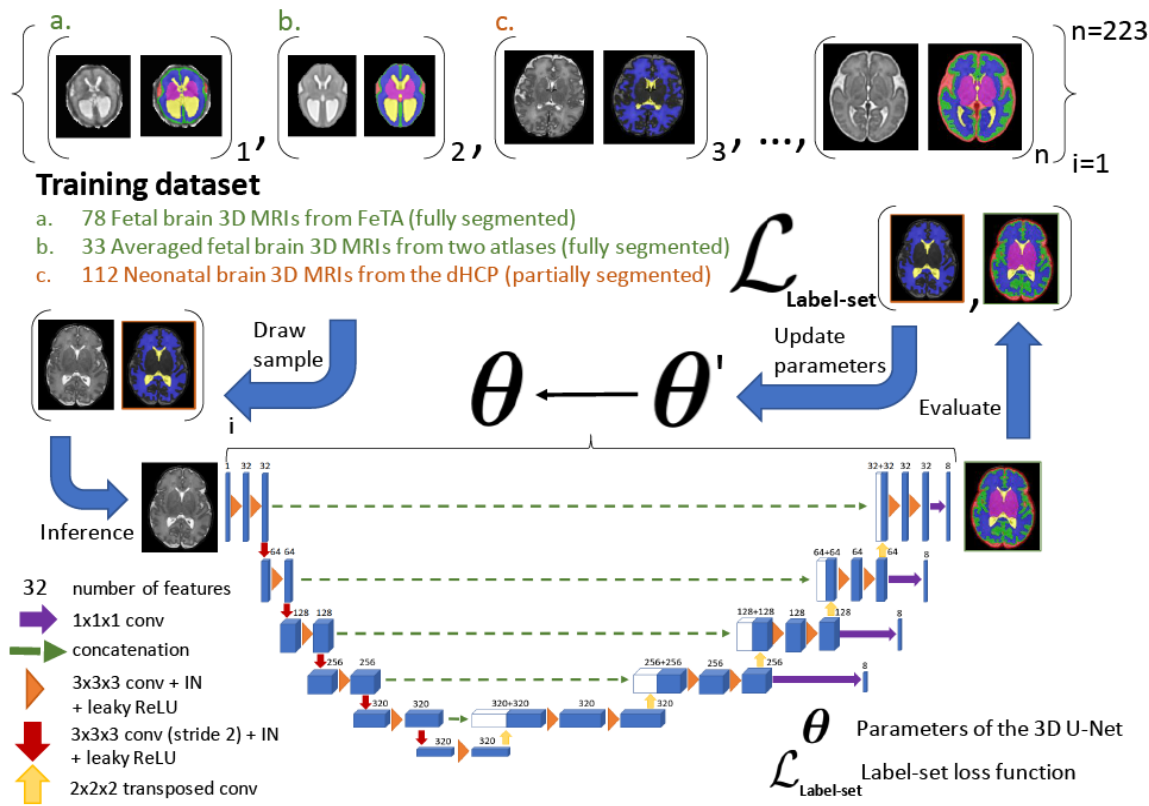

**Figure 1:** Overview of the training process with partial supervision. The 3D MRIs from datasets a. [13] and b. [4,5] are fully-segmented while the 3D MRIs from dataset c. [7] were only segmented manually for white matter, ventricular system, cerebellum, and the full brain. We propose to use a label-set loss function [3] to allow to train a 3D U-Net [2] with fully and partially segmented 3D MRIs.

### 1.18.3 Materials and Methods

In this section, we give the detail of our segmentation pipeline and the data used for training the deep neural networks. Our segmentation software will be publicly available.

<u>FeTA challenge amended training data:</u> The original FeTA challenge training data provides 80 fetal brain 3D T2 MRIs with manual segmentations of all 7 target tissue types [13]. 40 fetal brain 3D MRIs were reconstructed using MIAL [14] and 40 fetal brain 3D MRIs were reconstructed using Simple IRTK [8].

For the 40 MIAL 3D MRIs, corrections of the segmentations were performed by authors MA, LF, and PD using ITK-SNAP [18] to reduce the variability against the published segmentation guidelines that was released with the FeTA dataset [13]. Those corrections were performed as part of our previous work [3] and are publicly available (DOI: 10.5281/zenodo.5148611). Two spina bifida cases were excluded (sub-feta007 and sub-feta009) because we considered that the image quality did not allow to segment them reliably for all the tissue types. Only the remaining 78 3D MRIs were used for training.

<u>Other public training data:</u> We also included 18 average fetal brain 3D T2 MRIs from a neurotypical fetal brain atlas (http://crl.med.harvard.edu/research/fetal_brain_atlas/) [5], 15 average fetal brain 3D T2 MRIs from a spina bifida fetal brain atlas (https://www.synapse.org/#!Synapse:syn25887675/wiki/611424) [4]. Segmentations for all 7 tissue types are available for all the atlas data. In addition, we used 112 neonatal brain MRIs from the developing human connectome project [7] (dHCP data release 2). We excluded the brain MRIs of babies with a gestational age higher than 38 weeks. We started from the brain masks and the automatic segmentations publicly available for the neonatal brain MRIs for white matter, ventricular system, and cerebellum [9] and we modified them manually to match the annotation protocol of the FeTA dataset [13] using ITK-SNAP [18]. Ground-truth segmentations for the other tissue types in the dHCP data were not available for our training.

Pre-processing: A brain mask for the fetal brain 3D MRI is computed by affine registration of template volumes from two fetal brain atlases [4,5]. We use all the template volumes with a gestational age that does not differ to the gestation age of fetal brain by more than 1:5 weeks. The affine registrations are computed using a symmetric block-matching approach [10] as implemented in NiftyReg [11]. The affine transformations are initialized by a translation that aligns the centre of gravity of the non-zero intensity regions of the two volumes. The brain mask is obtained by averaging the warped brain mask and thresholding at 0.5.

After a brain mask has been computed, the fetal brain 3D MRI is registered rigidly to a neurotypical fetal brain atlas [5] and the 3D MRI resampled to a resolution of 0.8 mm isotropic. The rigid registration is computed using NiftyReg [10,11] and the transformation is initialized by a translation that aligns the centre of gravity of the brain masks.

Deep learning pipeline: We used an ensemble of 10 3D U-Nets [2]. We used the DynU-Net of MONAI [12] to implement a 3D U-Net with one input block, 4 down-sampling blocks, one bottleneck block, 5 up-sampling blocks, 32 features in the first level, instance normalization [15], and leaky-ReLU with slope 0.01. An illustration of the architecture is provided in Fig. 1. The CNN used has 31195784 trainable parameters. The patch size was set to 128 x 160 x 128. All the pre-processed 3D MRIs on which the pipeline was tested fitted inside a patch of size 128 x 160 x 128. Every input volume is skull stripped after dilating the brain mask by 5 voxels and cropped or padded with zeros to fit the patch size. The non-zeros image intensity values are clipped for the values above percentile 99.9, and normalized to zeros mean and unit variance. Test-time augmentation [16] with all the combinations of flipping along the three spatial dimensions is performed (8 predictions). The 8 score map predictions are averaged to obtain the output of each CNN. Ensembling is obtained by averaging the softmax predictions of the 10 CNNs. The deep learning pipeline was implemented using MONAI v5.2.0 by authors LF and SS.

Loss function: We used the sum of two label-set loss functions as loss function: the Leaf-Dice loss [3] and the marginalized cross entropy loss [3].

Optimization: For each network in the ensemble, the training dataset was split into 90% training and 10% validation at random. The random initialization of the 3D U-Net weights was performed using He initialization [6]. We used SGD with Nesterov momentum, batch size 2, weight decay $3 \times 10^{-5}$, initial learning rate 0.01, and polynomial learning rate decay with power 0.9 for a total of 2200 epochs. The CNN parameters used at inference corresponds to the last epoch. We used deep supervision with 4 levels during training. Training each 3D U-Net required 12GB of GPU memory and took on average 3 days. We have trained exactly 10 CNNs and used all of them for the ensemble submitted to the challenge.

Data augmentation: We used random zoom (zoom ratio range [0.7, 1.5] drawn uniformly at random; probability of augmentation 0:3), random rotation (rotation angle range [-15°; 15°] for all dimensions drawn uniformly at random; probability of augmentation 0.3), random additive Gaussian noise (mean 0, standard deviation 0.1; probability of augmentation 0.3), random Gaussian spatial smoothing (standard deviation range [0.5; 1.5] in voxels for all dimensions drawn uniformly at random; probability of augmentation 0.2), random gamma augmentation (gamma range [0.7; 1.5] drawn uniformly at random; probability of augmentation 0.3), and random flip along all dimension (probability of augmentation 0.5 for each dimension).

Post-processing: The mean softmax prediction is resampled to the original 3D MRI using the inverse of the rigid transformation computed in the pre-processing step to register the 3D MRI to the template space. This image registration is computed using NiftyReg [11] with an interpolation order equal to 1. After resampling, the final multi-class segmentation prediction is obtained by taking the argmax of the mean softmax.

### 1.18.4 Conclusion

Partially supervised learning can be used to train deep neural networks using multiple publicly available perinatal brain 3D MRI datasets that have different level of segmentations available. We used label-set loss functions [3] to train an ensemble of 3D U-Nets using four publicly available datasets [4,5,7,13]. We have submitted our segmentation algorithm to the FeTA challenge 2021.

### 1.18.5 Funding Sources

This project has received funding from the European Union's Horizon 2020 research and innovation program under the Marie Sklodowska-Curie grant agreement TRABIT No 765148. This work was supported by core and project funding from the Wellcome [203148/Z/16/Z; 203145Z/16/Z; WT101957], and EP-SRC [NS/A000049/1; NS/A000050/1; NS/A000027/1]. TV is supported by a Medtronic / RAEng Research Chair [RCSRF1819\7\34].

### 1.18.6 References

1. Bakas, S., Reyes, M., Jakab, A., Bauer, S., Remper, M., Crimi, A., Shinohara, R.T., Berger, C., Ha, S.M., Rozycki, M., et al.: Identifying the best machine learning algorithms for brain tumor segmentation, progression assessment, and overall survival prediction in the brats challenge. arXiv preprint arXiv:1811.02629 (2018)
2. Cicek, O., Abdulkadir, A., Lienkamp, S.S., Brox, T., Ronneberger, O.: 3d u-net: learning dense volumetric segmentation from sparse annotation. In: International conference on medical image computing and computer-assisted intervention. pp. 424-432. Springer (2016)
3. Fidon, L., Aertsen, M., Emam, D., Mufti, N., Guffens, F., Deprest, T., Demaerel, P., David, A.L., Melbourne, A., Ourselin, S., et al.: Label-set loss functions for partial supervision: Application to fetal brain 3d mri parcellation. arXiv preprint arXiv:2107.03846 (2021)
4. Fidon, L., Viola, E., Mufti, N., David, A., Melbourne, A., Demaerel, P., Ourselin, S., Vercauteren, T., Deprest, J., Aertsen, M.: A spatio-temporal atlas of the developing fetal brain with spina bifida aperta. Open Research Europe (2021)
5. Gholipour, A., Rollins, C.K., Velasco-Annis, C., Ouaalam, A., Akhondi-Asl, A., Afacan, O., Ortinau, C.M., Clancy, S., Limperopoulos, C., Yang, E., et al.: A normative spatiotemporal MRI atlas of the fetal brain for automatic segmentation and analysis of early brain growth. Scientific reports 7(1), 1-13 (2017)
6. He, K., Zhang, X., Ren, S., Sun, J.: Delving deep into rectifiers: Surpassing human-level performance on imagenet classification. In: Proceedings of the IEEE international conference on computer vision. pp. 1026-1034 (2015)

7. Hughes, E.J., Winchman, T., Padormo, F., Teixeira, R., Wurie, J., Sharma, M., Fox, M., Hutter, J., Cordero-Grande, L., Price, A.N., et al.: A dedicated neonatal brain imaging system. Magnetic resonance in medicine 78(2), 794-804 (2017)

8. Kuklisova-Murgasova, M., Quaghebeur, G., Rutherford, M.A., Hajnal, J.V., Schnabel, J.A.: Reconstruction of fetal brain mri with intensity matching and complete outlier removal. Medical image analysis 16(8), 1550-1564 (2012)

9. Makropoulos, A., Robinson, E.C., Schuh, A., Wright, R., Fitzgibbon, S., Bozek, J., Counsell, S.J., Steinweg, J., Vecchiato, K., Passerat-Palmbach, J., et al.: The developing human connectome project: A minimal processing pipeline for neonatal cortical surface reconstruction. Neuroimage 173, 88-112 (2018)

10. Modat, M., Cash, D.M., Daga, P., Winston, G.P., Duncan, J.S., Ourselin, S.: Global image registration using a symmetric block-matching approach. Journal of Medical Imaging 1(2), 024003 (2014)

11. Modat, M., Ridgway, G.R., Taylor, Z.A., Lehmann, M., Barnes, J., Hawkes, D.J., Fox, N.C., Ourselin, S.: Fast free-form deformation using graphics processing units. Computer methods and programs in biomedicine 98(3), 278-284 (2010)

12. MONAI Consortium: MONAI: Medical open network for AI (3 2020). https://doi.org/10.5281/zenodo.4323058, https://github.com/Project-MONAI/MONAI

13. Payette, K., de Dumast, P., Kebiri, H., Ezhov, I., Paetzold, J.C., Shit, S., Iqbal, A., Khan, R., Kottke, R., Grehten, P., et al.: An automatic multi-tissue human fetal brain segmentation benchmark using the fetal tissue annotation dataset. Scientific Data 8(1), 1-14 (2021)

14. Tourbier, S., Bresson, X., Hagmann, P., Thiran, J.P., Meuli, R., Cuadra, M.B.: An efficient total variation algorithm for super-resolution in fetal brain mri with adaptive regularization. NeuroImage 118, 584-597 (2015)

15. Ulyanov, D., Vedaldi, A., Lempitsky, V.: Instance normalization: The missing ingredient for fast stylization. arXiv preprint arXiv:1607.08022 (2016)

16. Wang, G., Li, W., Aertsen, M., Deprest, J., Ourselin, S., Vercauteren, T.: Aleatoric uncertainty estimation with test-time augmentation for medical image segmentation with convolutional neural networks. Neurocomputing 338, 34-45 (2019)

17. Wu, J., Sun, T., Yu, B., Li, Z., Wu, Q., Wang, Y., Qian, Z., Zhang, Y., Jiang, L., Wei, H.: Age-specific structural fetal brain atlases construction and cortical development quantification for chinese population. NeuroImage p. 118412 (2021)

18. Yushkevich, P.A., Gao, Y., Gerig, G.: ITK-SNAP: an interactive tool for semiautomatic segmentation of multi-modality biomedical images. In: 2016 38th Annual International Conference of the IEEE Engineering in Medicine and Biology Society (EMBC). pp. 3342-3345. IEEE (2016)

## 1.19 Xlab

### 1.19.1 Team Members and Affiliations
Team Members: Yang Lin

Affiliations: Department of Computer Science, Hong Kong University of Science and Technology

### 1.19.2 Model description

The model uses the architecture similar to U-net [1]. The contracting path consists of five repetition of two 3*3 convolutions, a rectified linear unit and a 2*2 max pooling with stride 2. In each down sampling, the number of feature channels is doubled. The expansive path consists of an up sampling of the feature map, a 2x2 convolution, a concatenation with the correspondingly cropped feature map from the contracting path, and two 3x3 convolutions, each followed by a ReLU. The final layer is a 1x1 convolution mapping each 64- component feature vector to the desired number of classes. The difference is that padding is added in the convolutions, so that the size of the feature maps won't change inside each repetition, as a result, more information will be transferred to the next layer. And batch norm is used instead of instance norm.

The preprocessing step is similar to nnUNet [2]. But more pre-process on the input images are designed after observing the features of FeTA images.

### 1.19.3 Training method

All cases of FeTA are used in the training progress. The loss function is the combination of dice loss and cross entropy loss. The datasets are split to training set and validation and 5-fold cross validation is used. 5 models are generated in the 5-fold cross validation. Strategies includes mirroring, random rotation, random scaling, gamma correction and random elastic transformation are used.
The training in done with CUDA 11.0, Ubuntu 20, python 3.8 and pytorch 1.8 and takes about 5 days and has 1000 epochs.

### 1.19.4 Reference
[1] Olaf Ronneberger, Philipp Fischer, Thomas Brox. U-Net: Convolutional Networks for Biomedical Image Segmentation.
[2] Isensee, F., Jaeger, P. F., Kohl, S. A., Petersen, J., & Maier-Hein, K. H. (2020). nnU-Net: a self-configuring method for deep learning-based biomedical image segmentation. Nature Methods, 1-9.

## 1.20 ZJUWULAB

### 1.20.1 Team Members and Affiliations
Team Members: Zelin Zhang[1], Xinyi Xu[1], Dan Wu[1]
Affiliations: [1]Key Laboratory for Biomedical Engineering of Ministry of Education, Department of Biomedical Engineering, College of Biomedical Engineering & Instrument Science, Zhejiang University, Yuquan Campus, Hangzhou, China

### 1.20.2 Model Description
Our segmentation network adopts a standard 2D Unet framework, but we replace all 2x max pool down sampling layers with 2x convolution down sampling layers. Our loss function consists of two parts: L1 regularization loss function and feature matching loss function. In the feature matching loss function, we cascade the predicted output and corresponding label of the model with the original input image respectively and input them into the pretrained vgg19 network to calculate the L1 loss from the deep features output by each convolution layers, and finally carry out back propagation training model after weighted average of all feature matching losses.

The input to our model is a series of 2D slices with the size of 256x256.

The input preprocessing includes 2 steps: 1) Normalize all voxel values of one case to the interval [0, 255]; 2) Convert class mapping of 8 categories into color mapping (Each category corresponds to the combination of three channels in the RGB color map, in which each channel has only 0 and 1.); 3) Extract all 2D slices and convert to RGB images.

The initialization of the model adopts the method of random initialization.

The whole model is implemented under the framework of pytoch 1.0.

### 1.20.3 Training method
We used all cases to optimize our model. We did not use any additional public datasets. The whole dataset is divided into training set, validation set and test set with the ratio of 7:2:1. We use the Adam optimizer, the initialization learning rate is 0.002, and the batch size of training is 32. We did not use any data augmentation methods. According to the divided validation set and test set, we select the checkpoint with the best accuracy in 100 epochs as the final submitted model.

Other details: Pytorch 1.0; CUDA 10.1; SimpleITK 1.2.4; Nvidia GPU RTX 3080 ti x 4; Training time: 6.8h.

## 2. Benchmarking report for multiTaskChallengeDice_combined

created by challengeR v1.0.2

17 December, 2021

This document presents a systematic report on the benchmark study "multiTaskChallengeDice_combined". Input data comprises raw metric values for all algorithms and cases. Generated plots are:

- Visualization of assessment data: Dot- and boxplot, podium plot and ranking heatmap
- Visualization of ranking stability: Blob plot, violin plot and significance map, line plot

Details can be found in Wiesenfarth et al. (2021).

### 2.1 Ranking

Algorithms within a task are ranked according to the following ranking scheme:

*aggregate using function ("mean") then rank*

The analysis is based on 21 algorithms and 280 cases. 0 missing cases have been found in the data set.

Ranking:

|  | Dice_mean | rank |
|---|---|---|
| NVAUTO | 0.7858525 | 1 |
| SJTU_EIEE_2-426Lab | 0.7753286 | 2 |
| Neurophet | 0.7745291 | 3 |
| Pengyy | 0.7744333 | 4 |
| Hilab | 0.7735193 | 5 |
| davoodkarimi | 0.7709845 | 6 |
| Xlab | 0.7709073 | 7 |
| TRABIT | 0.7685622 | 8 |
| Physense-UPF Team | 0.7674980 | 9 |
| 2Ai | 0.7669346 | 10 |
| ichilove-ax | 0.7660280 | 11 |
| muw_dsobotka | 0.7648522 | 12 |
| ichilove-combi | 0.7619306 | 13 |
| Moona_Mazher | 0.7550563 | 14 |
| BIT_LILAB | 0.7518695 | 15 |
| SingleNets | 0.7478066 | 16 |
| MIAL | 0.7402952 | 17 |
| ZJUWULAB | 0.7026920 | 18 |
| FeVer | 0.6826219 | 19 |
| Anonymous | 0.6207273 | 20 |
| A3 | 0.5335483 | 21 |

## 2.2 Visualization of raw assessment data

### 2.2.1 Dot- and boxplot

*Dot- and boxplots* for visualizing raw assessment data separately for each algorithm. Boxplots representing descriptive statistics over all cases (median, quartiles and outliers) are combined with horizontally jittered dots representing individual cases.

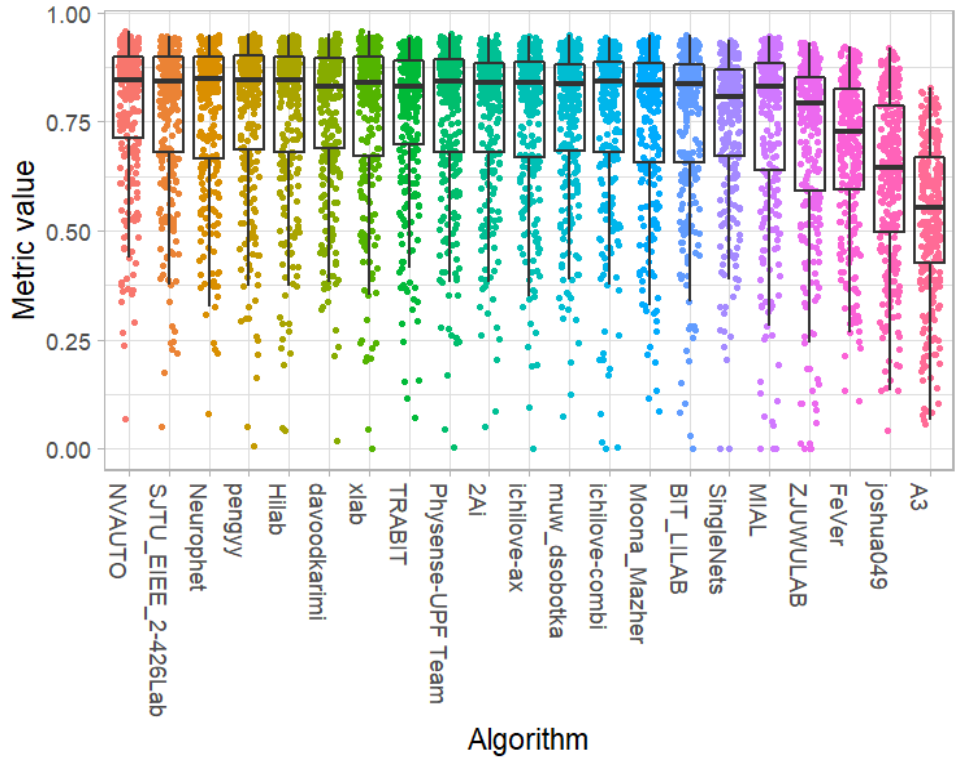

**2.2.2 Podium plot**

*Podium plots* (see also Eugster et al., 2008) for visualizing raw assessment data. Upper part (spaghetti plot): Participating algorithms are color-coded, and each colored dot in the plot represents a metric value achieved with the respective algorithm. The actual metric value is encoded by the y-axis. Each podium (here: $p$=21) represents one possible rank, ordered from best (1) to last (here: 21). The assignment of metric values (i.e. colored dots) to one of the podiums is based on the rank that the respective algorithm achieved on the corresponding case. Note that the plot part above each podium place is further subdivided into $p$ "columns", where each column represents one participating algorithm (here: $p = 21$). Dots corresponding to identical cases are connected by a line, leading to the shown spaghetti structure. Lower part: Bar charts represent the relative frequency for each algorithm to achieve the rank encoded by the podium place.

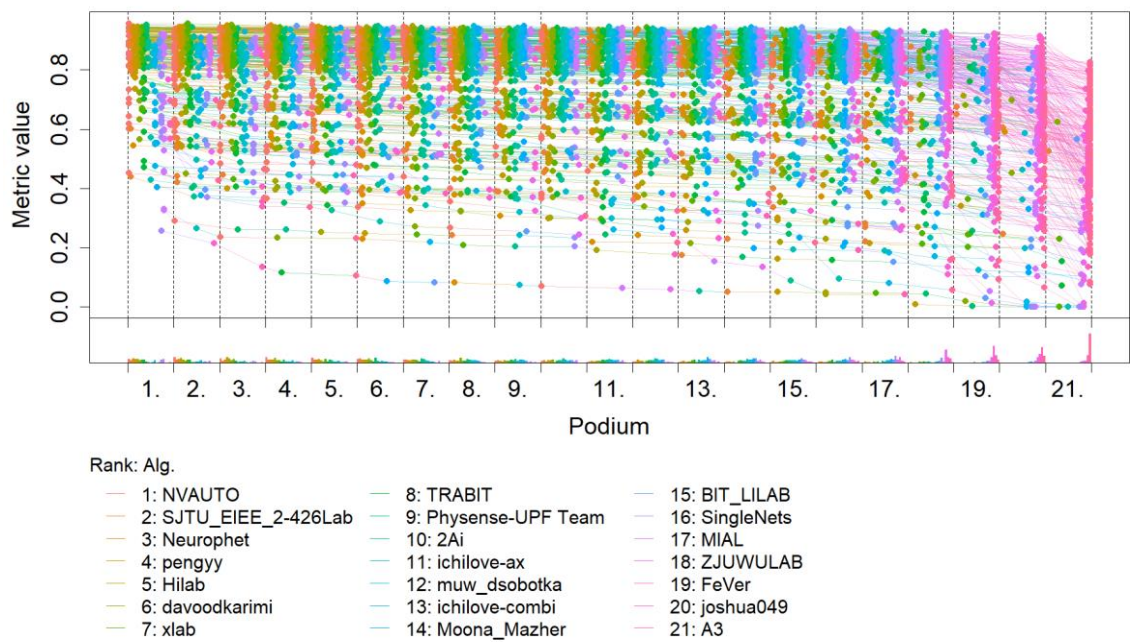

### 2.2.3 Ranking heatmap

*Ranking heatmaps* for visualizing raw assessment data. Each cell $(i, A_j)$ shows the absolute frequency of cases in which algorithm $A_j$ achieved rank $i$.

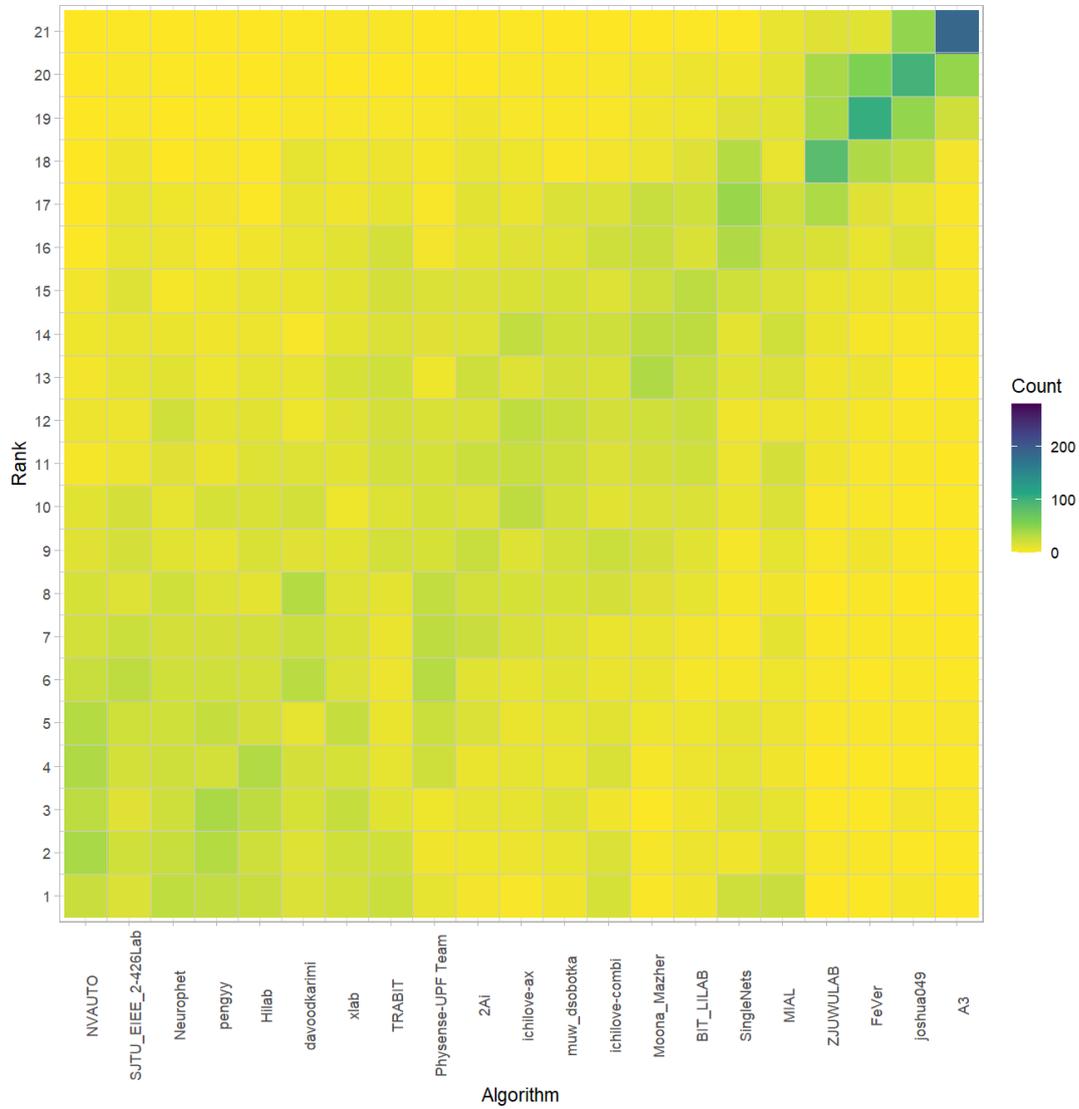

## 2.3 Visualization of ranking stability

### 2.3.1 *Blob plot* for visualizing ranking stability based on bootstrap sampling

Algorithms are color-coded, and the area of each blob at position ($A_i$, rank $j$) is proportional to the relative frequency $A_i$ achieved rank $j$ across $b = 1000$ bootstrap samples. The median rank for each algorithm is indicated by a black cross. 95% bootstrap intervals across bootstrap samples are indicated by black lines.

```
## Warning: `guides(<scale> = FALSE)` is deprecated. Please use `guides(<scale> =
## "none")` instead.
```

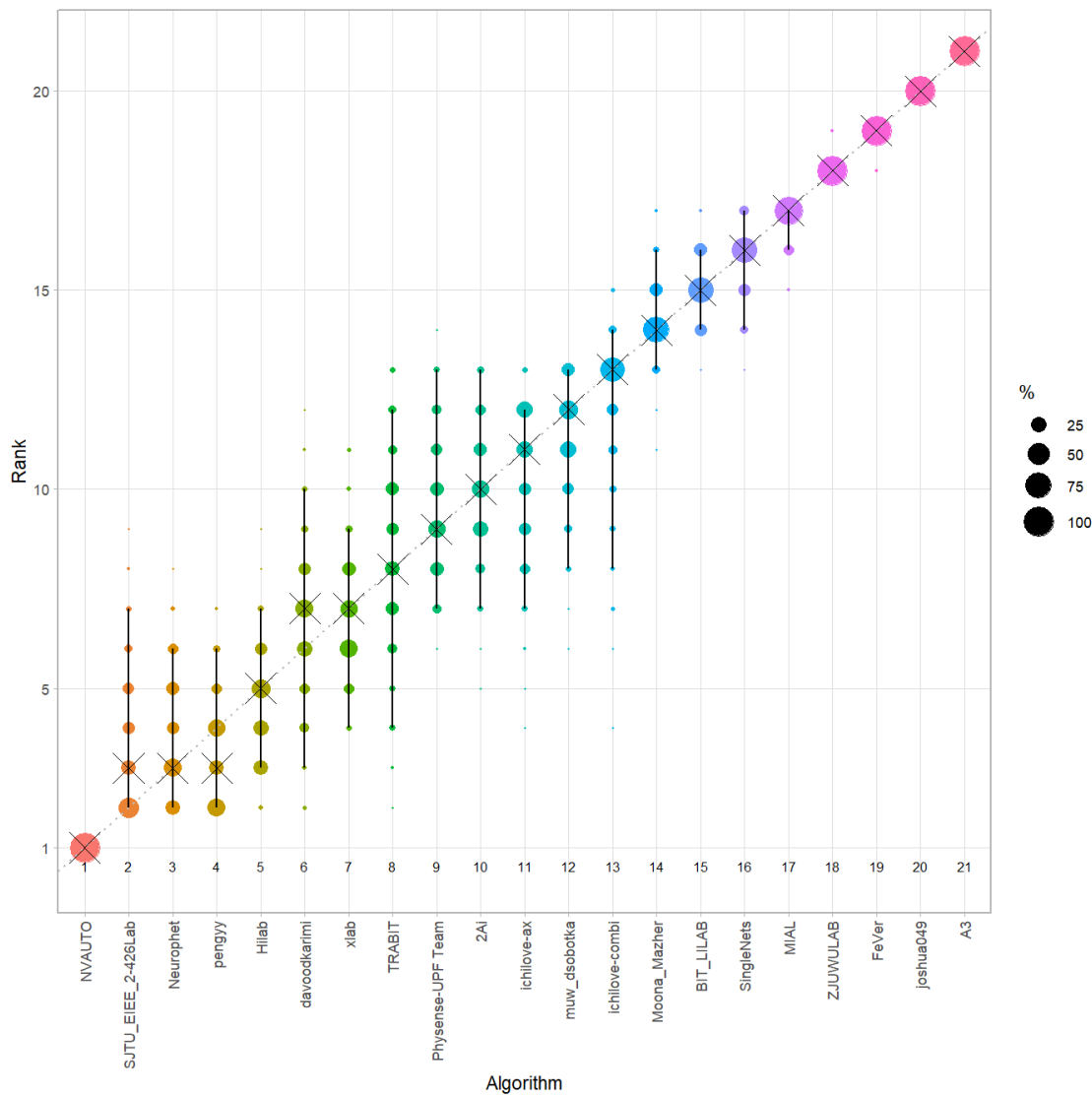

### 2.3.2 *Violin plot* for visualizing ranking stability based on bootstrapping

The ranking list based on the full assessment data is pairwise compared with the ranking lists based on the individual bootstrap samples (here $b = 1000$ samples). For each pair of rankings, Kendall's $\tau$ correlation is computed. Kendall's $\tau$ is a scaled index determining the correlation between the lists. It is computed by evaluating the number of pairwise concordances and discordances between ranking lists and produces values between $-1$ (for inverted order) and 1 (for identical order). A violin plot, which simultaneously depicts a boxplot and a density plot, is generated from the results.

Summary Kendall's tau:

| Task | mean | median | q25 | q75 |
|---|---|---|---|---|
| dummyTask | 0.9217905 | 0.9238095 | 0.8952381 | 0.952381 |

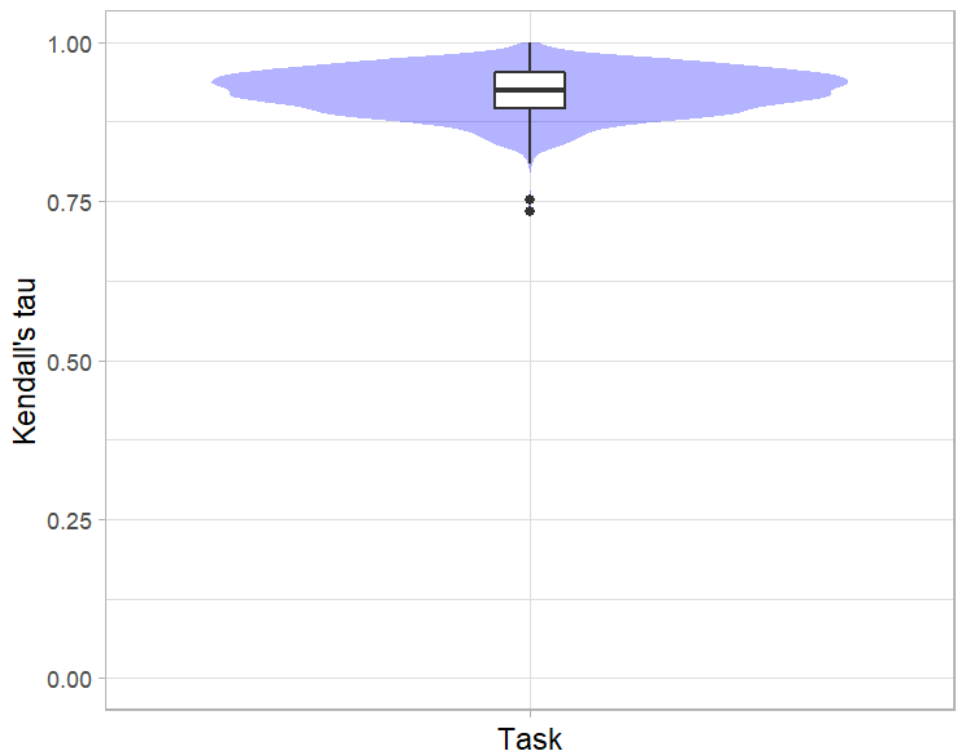

### 2.3.3 *Significance maps* for visualizing ranking stability based on statistical significance

*Significance maps* depict incidence matrices of pairwise significant test results for the one-sided Wilcoxon signed rank test at a 5% significance level with adjustment for multiple testing according to Holm. Yellow shading indicates that metric values from the algorithm on the x-axis were significantly superior to those from the algorithm on the y-axis, blue color indicates no significant difference.

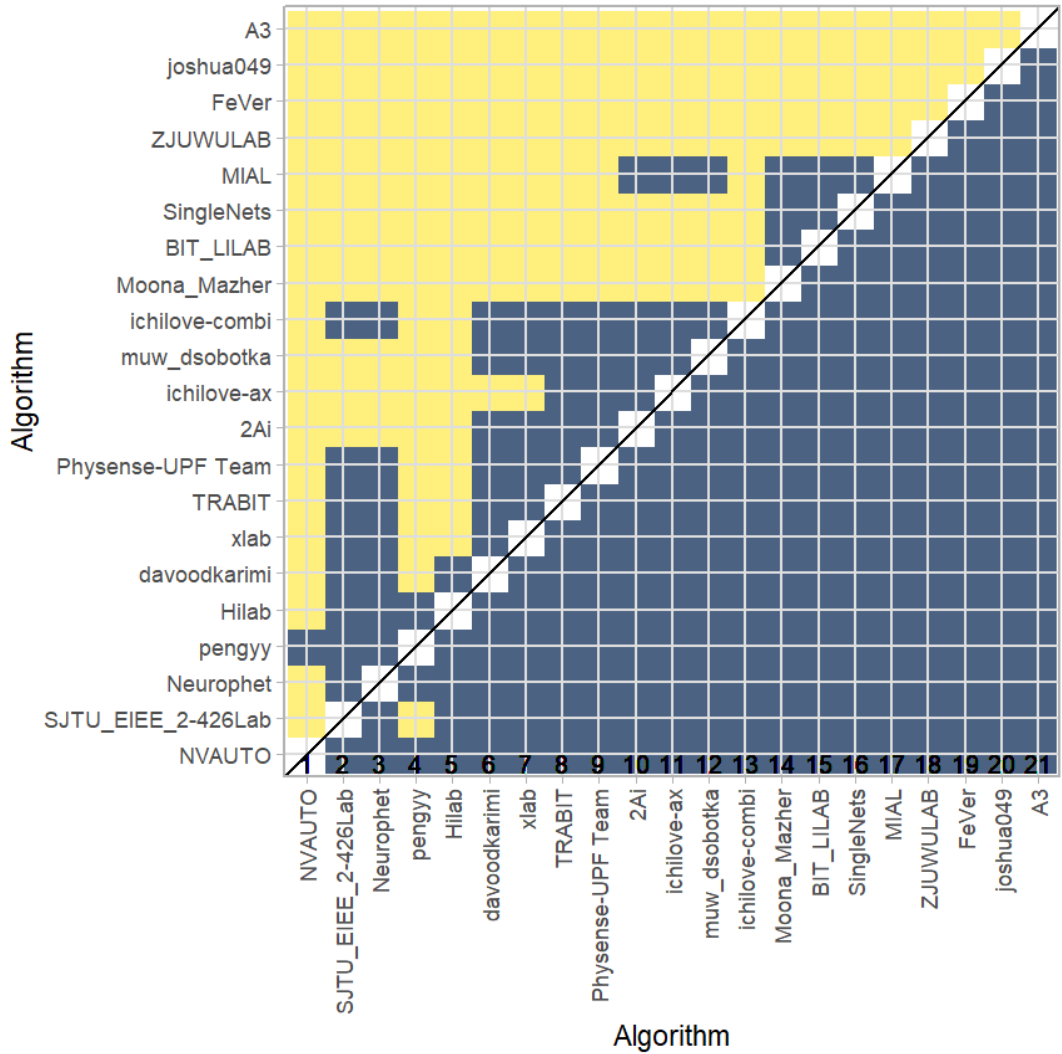

**2.3.4 Ranking robustness to ranking methods**

*Line plots* for visualizing ranking robustness across different ranking methods. Each algorithm is represented by one colored line. For each ranking method encoded on the x-axis, the height of the line represents the corresponding rank. Horizontal lines indicate identical ranks for all methods.

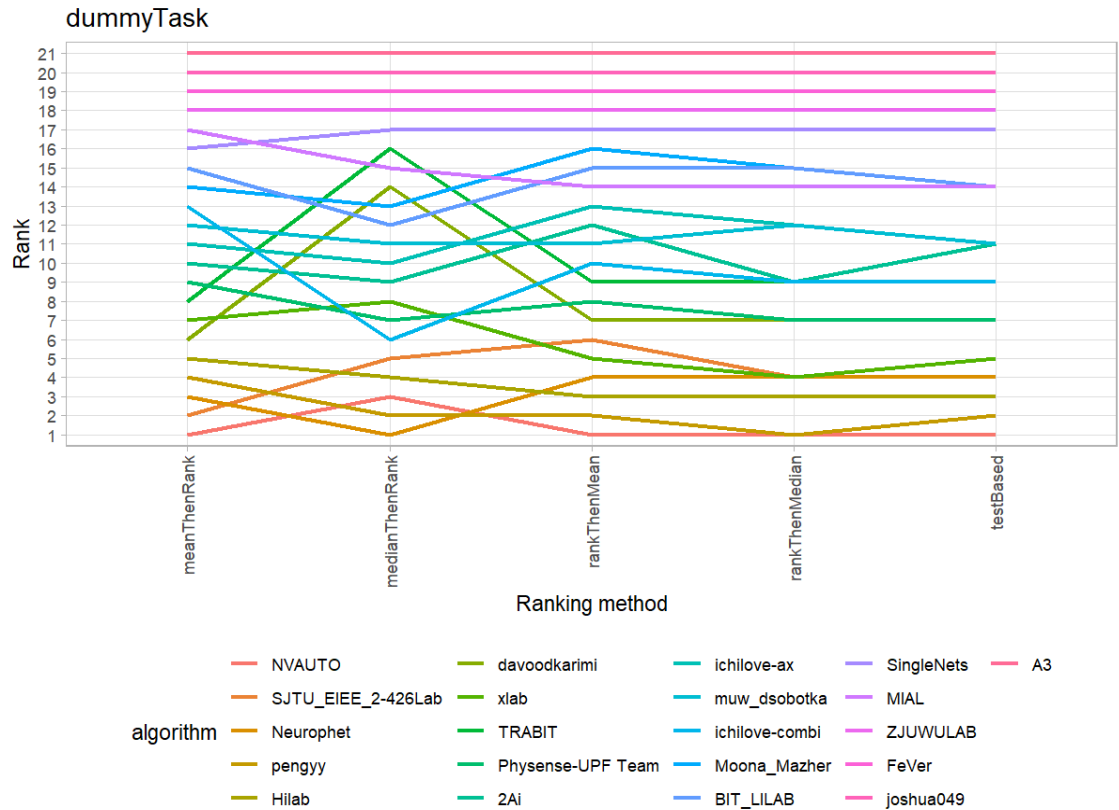

## 2.4 References

Wiesenfarth, M., Reinke, A., Landman, B.A., Eisenmann, M., Aguilera Saiz, L., Cardoso, M.J., Maier-Hein, L. and Kopp-Schneider, A. Methods and open-source toolkit for analyzing and visualizing challenge results. *Sci Rep* **11**, 2369 (2021). https://doi.org/10.1038/s41598-021-82017-6

M. J. A. Eugster, T. Hothorn, and F. Leisch, "Exploratory and inferential analysis of benchmark experiments," Institut fuer Statistik, Ludwig-Maximilians-Universitaet Muenchen, Germany, Technical Report 30, 2008. [Online]. Available: http://epub.ub.uni-muenchen.de/4134/.

## 3. Benchmarking report for multiTaskChallengeHD_combined

created by challengeR v1.0.2

17 December, 2021

This document presents a systematic report on the benchmark study "multiTaskChallengeHD_combined". Input data comprises raw metric values for all algorithms and cases. Generated plots are:

- Visualization of assessment data: Dot- and boxplot, podium plot and ranking heatmap
- Visualization of ranking stability: Blob plot, violin plot and significance map, line plot

Details can be found in Wiesenfarth et al. (2021).

### 3.1 Ranking

Algorithms within a task are ranked according to the following ranking scheme:

*aggregate using function ("mean") then rank*

The analysis is based on 21 algorithms and 280 cases. 0 missing cases have been found in the data set.

Ranking:

|  | Hausdorff_mean | rank |
|---|---|---|
| NVAUTO | 14.01218 | 1 |
| Hilab | 14.56878 | 2 |
| 2Ai | 14.62533 | 3 |
| SJTU_EIEE_2-426Lab | 14.67063 | 4 |
| pengyy | 14.69852 | 5 |
| TRABIT | 14.90064 | 6 |
| Physense-UPF Team | 15.01775 | 7 |
| xlab | 15.26159 | 8 |
| Neurophet | 15.37497 | 9 |
| ichilove-combi | 16.03872 | 10 |
| davoodkarimi | 16.75480 | 11 |
| muw_dsobotka | 17.15931 | 12 |
| BIT_LILAB | 18.16190 | 13 |
| Moona_Mazher | 18.54792 | 14 |
| ichilove-ax | 21.32919 | 15 |
| MIAL | 25.10674 | 16 |
| SingleNets | 26.12090 | 17 |
| ZJUWULAB | 27.94799 | 18 |
| FeVer | 34.41890 | 19 |
| Anonymous | 37.38550 | 20 |
| A3 | 39.60763 | 21 |

### 3.2 Visualization of raw assessment data

#### 3.2.1 Dot- and boxplot

*Dot- and boxplots* for visualizing raw assessment data separately for each algorithm. Boxplots representing descriptive statistics over all cases (median, quartiles and outliers) are combined with horizontally jittered dots representing individual cases.

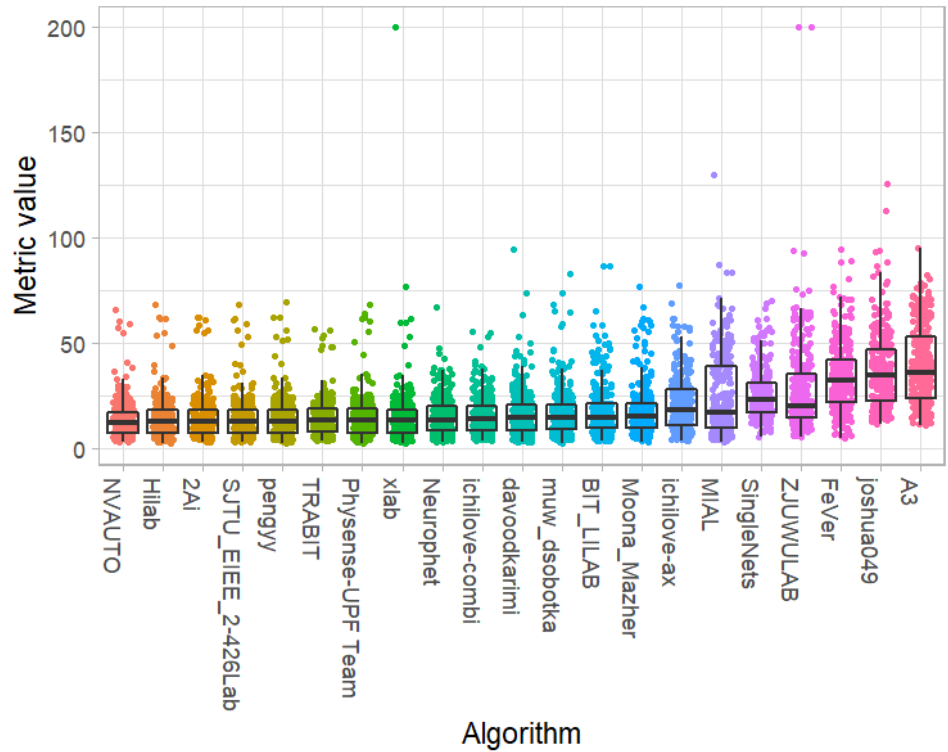

### 3.2.2 Podium plot

*Podium plots* (see also Eugster et al., 2008) for visualizing raw assessment data. Upper part (spaghetti plot): Participating algorithms are color-coded, and each colored dot in the plot represents a metric value achieved with the respective algorithm. The actual metric value is encoded by the y-axis. Each podium (here: $p$=21) represents one possible rank, ordered from best (1) to last (here: 21). The assignment of metric values (i.e. colored dots) to one of the podiums is based on the rank that the respective algorithm achieved on the corresponding case. Note that the plot part above each podium place is further subdivided into $p$ "columns", where each column represents one participating algorithm (here: $p = 21$). Dots corresponding to identical cases are connected by a line, leading to the shown spaghetti structure. Lower part: Bar charts represent the relative frequency for each algorithm to achieve the rank encoded by the podium place.

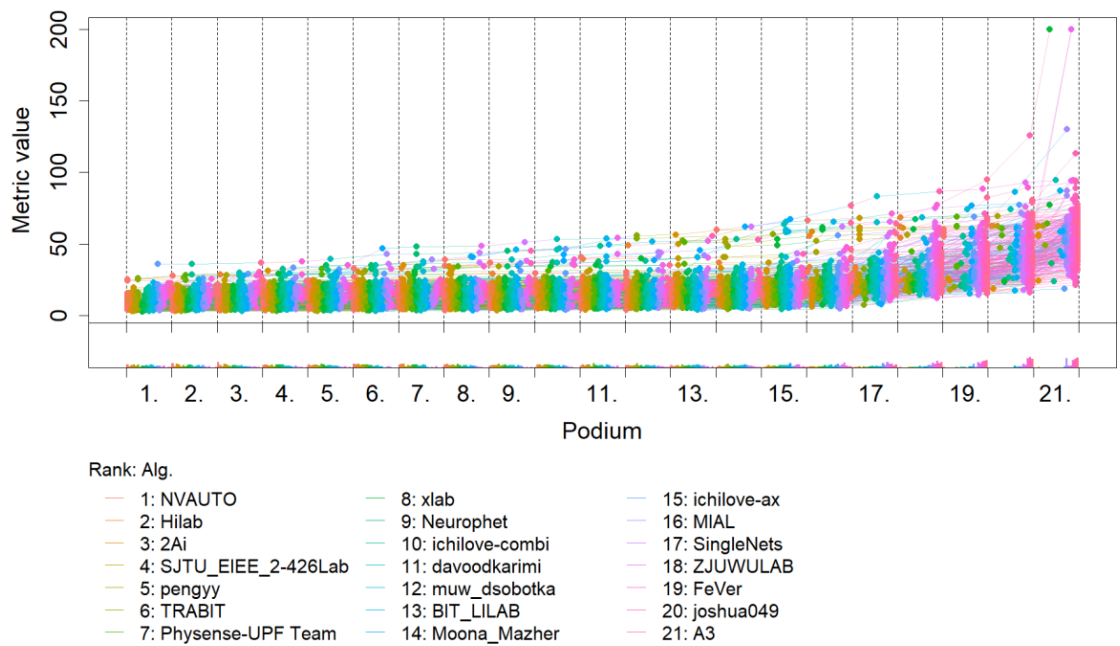

### 3.2.3 Ranking heatmap

*Ranking heatmaps* for visualizing raw assessment data. Each cell $(i, A_j)$ shows the absolute frequency of cases in which algorithm $A_j$ achieved rank $i$.

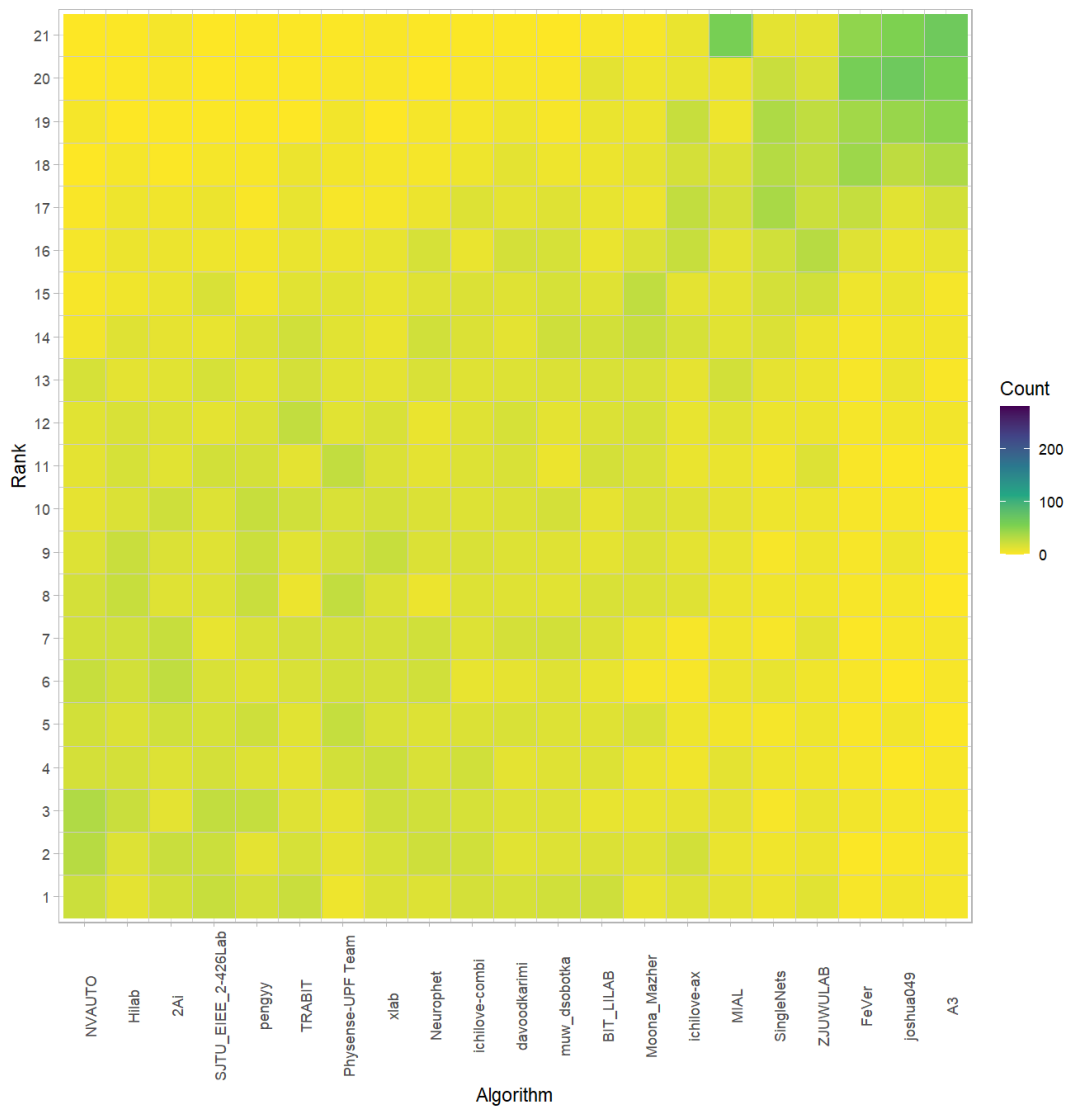

### 3.3 Visualization of ranking stability

#### 3.3.1 *Blob plot* for visualizing ranking stability based on bootstrap sampling

Algorithms are color-coded, and the area of each blob at position $(A_i, \text{rank } j)$ is proportional to the relative frequency $A_i$ achieved rank $j$ across $b = 1000$ bootstrap samples. The median rank for each algorithm is indicated by a black cross. 95% bootstrap intervals across bootstrap samples are indicated by black lines.

```
## Warning: `guides(<scale> = FALSE)` is deprecated. Please use
`guides(<scale> =
## "none")` instead.
```

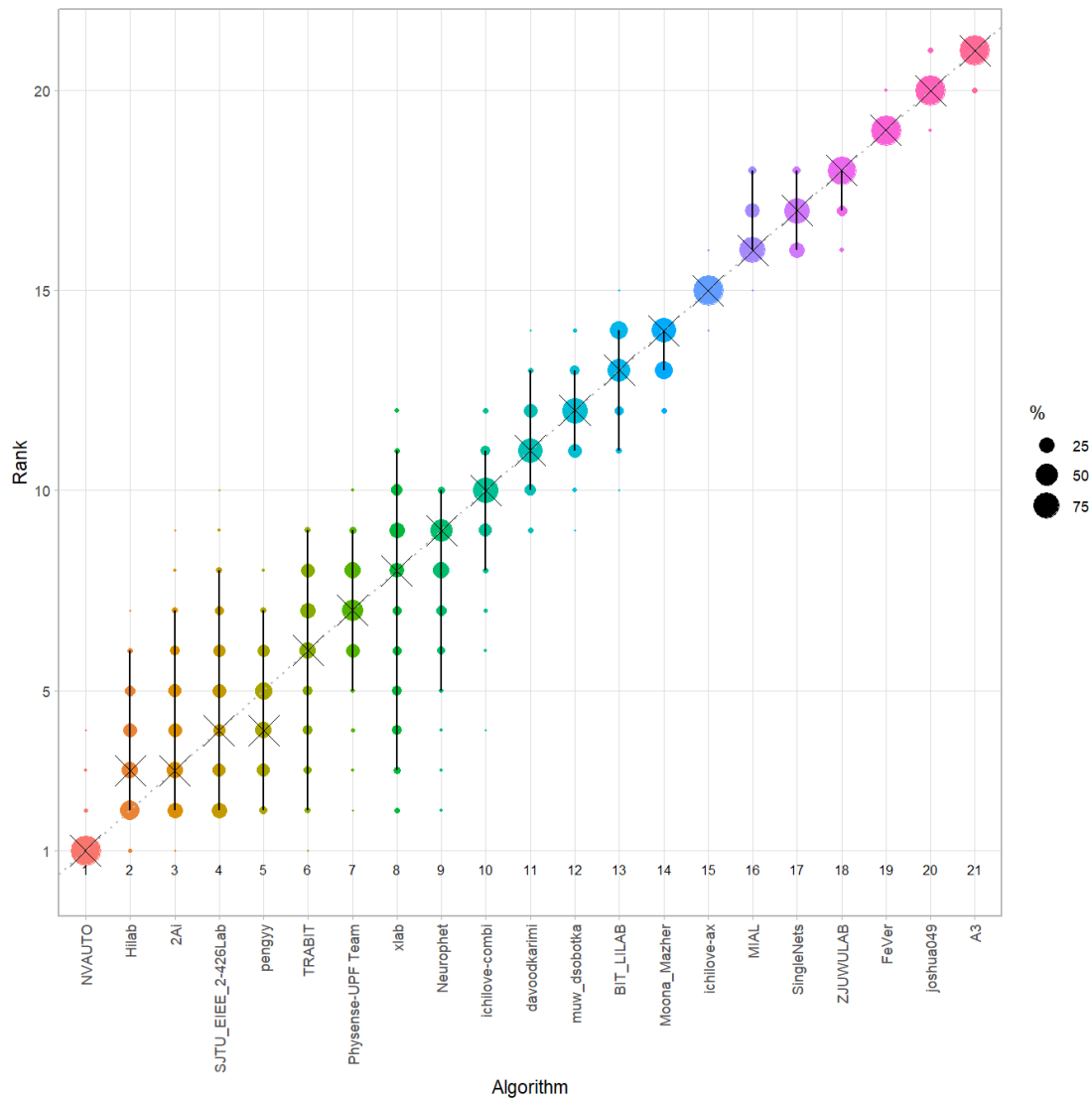

### 3.3.2 *Violin plot* for visualizing ranking stability based on bootstrapping

The ranking list based on the full assessment data is pairwise compared with the ranking lists based on the individual bootstrap samples (here $b = 1000$ samples). For each pair of rankings, Kendall's $\tau$ correlation is computed. Kendall's $\tau$ is a scaled index determining the correlation between the lists. It is computed by evaluating the number of pairwise concordances and discordances between ranking lists and produces values between $-1$ (for inverted order) and $1$ (for identical order). A violin plot, which simultaneously depicts a boxplot and a density plot, is generated from the results.

Summary Kendall's tau:

| Task      | mean      | median    | q25       | q75      |
|-----------|-----------|-----------|-----------|----------|
| dummyTask | 0.9334476 | 0.9333333 | 0.9142857 | 0.952381 |

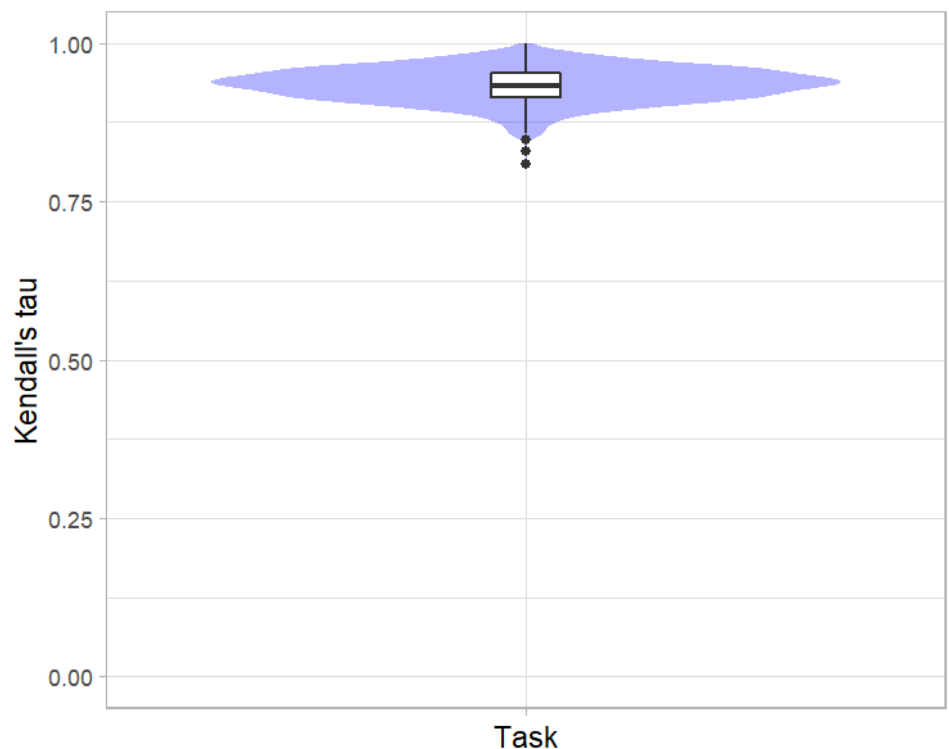

### 3.3.3 *Significance maps* for visualizing ranking stability based on statistical significance

*Significance maps* depict incidence matrices of pairwise significant test results for the one-sided Wilcoxon signed rank test at a 5% significance level with adjustment for multiple testing according to Holm. Yellow shading indicates that metric values from the algorithm on the x-axis were significantly superior to those from the algorithm on the y-axis, blue color indicates no significant difference.

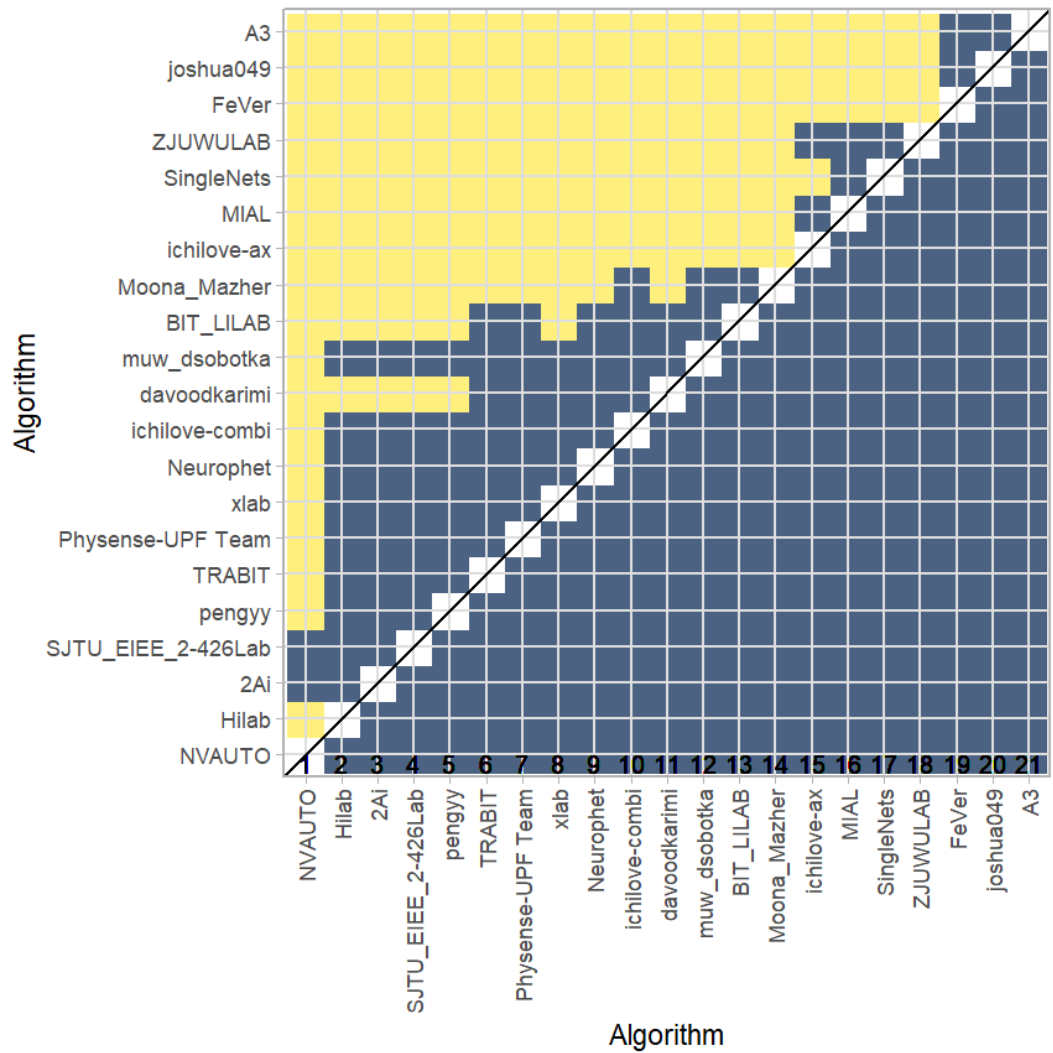

### 3.3.4 Ranking robustness to ranking methods

*Line plots* for visualizing ranking robustness across different ranking methods. Each algorithm is represented by one colored line. For each ranking method encoded on the x-axis, the height of the line represents the corresponding rank. Horizontal lines indicate identical ranks for all methods.

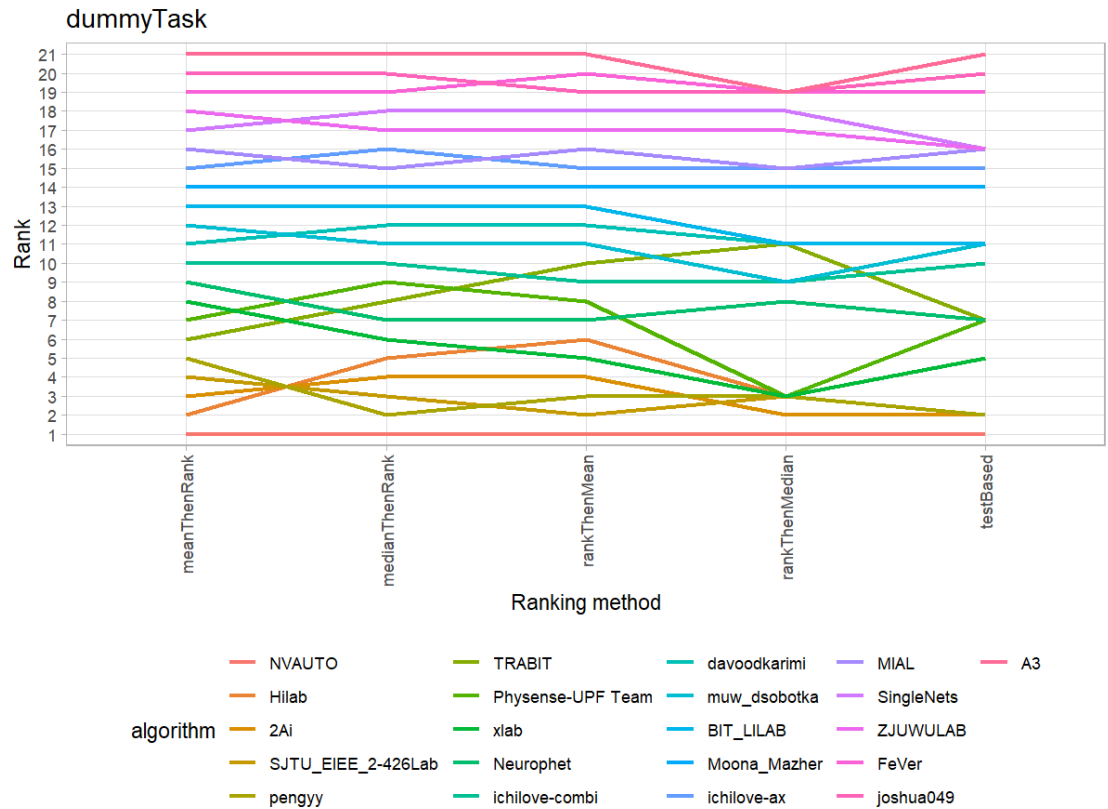

### 3.4 References

Wiesenfarth, M., Reinke, A., Landman, B.A., Eisenmann, M., Aguilera Saiz, L., Cardoso, M.J., Maier-Hein, L. and Kopp-Schneider, A. Methods and open-source toolkit for analyzing and visualizing challenge results. *Sci Rep* **11**, 2369 (2021). https://doi.org/10.1038/s41598-021-82017-6

M. J. A. Eugster, T. Hothorn, and F. Leisch, "Exploratory and inferential analysis of benchmark experiments," Institut fuer Statistik, Ludwig-Maximilians-Universitaet Muenchen, Germany, Technical Report 30, 2008. [Online]. Available: http://epub.ub.uni-muenchen.de/4134/.

# 4. Benchmarking report for multiTaskChallengeVolSim_combined

created by challengeR v1.0.2

17 December, 2021

This document presents a systematic report on the benchmark study "multiTaskChallengeVolSim_combined". Input data comprises raw metric values for all algorithms and cases. Generated plots are:

- Visualization of assessment data: Dot- and boxplot, podium plot and ranking heatmap
- Visualization of ranking stability: Blob plot, violin plot and significance map, line plot

Details can be found in Wiesenfarth et al. (2021).

## 4.1 Ranking

Algorithms within a task are ranked according to the following ranking scheme:

*aggregate using function ("mean") then rank*

The analysis is based on 21 algorithms and 280 cases. 0 missing cases have been found in the data set.

Ranking:

|                    | Volume_Similarity_mean | rank |
|--------------------|------------------------|------|
| ichilove-ax        | 0.8879231              | 1    |
| NVAUTO             | 0.8849992              | 2    |
| SJTU_EIEE_2-426Lab | 0.8829732              | 3    |
| davoodkarimi       | 0.8817364              | 4    |
| Neurophet          | 0.8767972              | 5    |
| SingleNets         | 0.8758778              | 6    |
| pengyy             | 0.8746371              | 7    |
| muw_dsobotka       | 0.8736648              | 8    |
| ichilove-combi     | 0.8732283              | 9    |
| Hilab              | 0.8732049              | 10   |
| xlab               | 0.8731572              | 11   |
| BIT_LILAB          | 0.8675333              | 12   |
| 2Ai                | 0.8671930              | 13   |
| TRABIT             | 0.8657770              | 14   |
| Moona_Mazher       | 0.8657734              | 15   |
| Physense-UPF Team  | 0.8633648              | 16   |
| MIAL               | 0.8447757              | 17   |
| ZJUWULAB           | 0.8353674              | 18   |
| FeVer              | 0.8278280              | 19   |
| Anonymous | 0.8012755 | 20 |
| A3 | 0.7909992 | 21 |

## 4.2 Visualization of raw assessment data

### 4.2.1 Dot- and boxplot

*Dot- and boxplots* for visualizing raw assessment data separately for each algorithm. Boxplots representing descriptive statistics over all cases (median, quartiles and outliers) are combined with horizontally jittered dots representing individual cases.

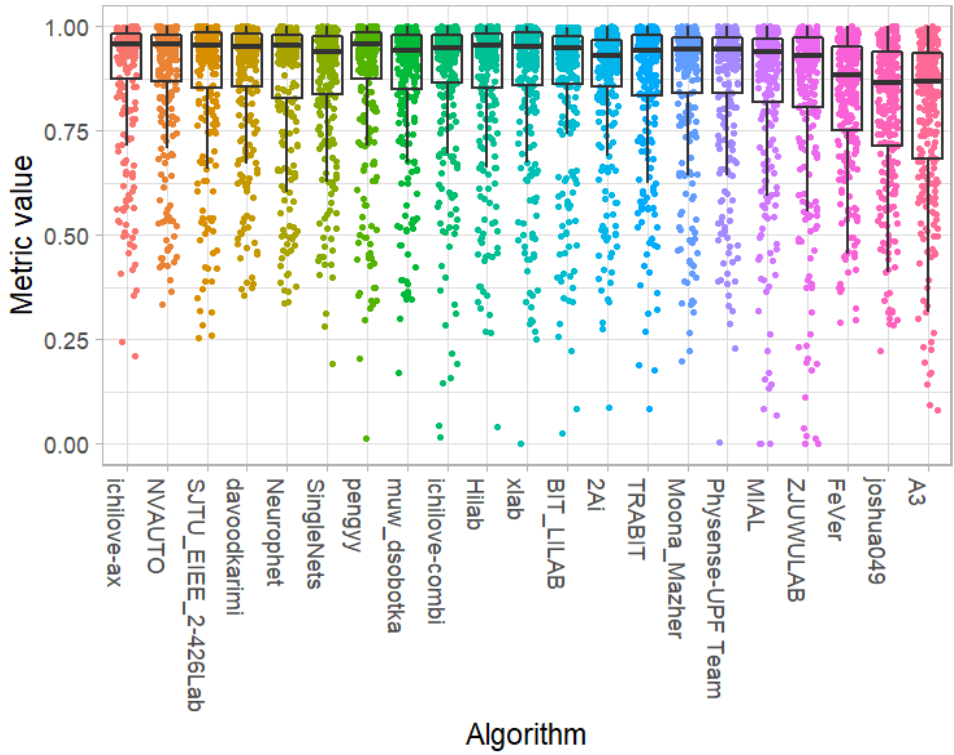

**4.2.2 Podium plot**

*Podium plots* (see also Eugster et al., 2008) for visualizing raw assessment data. Upper part (spaghetti plot): Participating algorithms are color-coded, and each colored dot in the plot represents a metric value achieved with the respective algorithm. The actual metric value is encoded by the y-axis. Each podium (here: $p$=21) represents one possible rank, ordered from best (1) to last (here: 21). The assignment of metric values (i.e. colored dots) to one of the podiums is based on the rank that the respective algorithm achieved on the corresponding case. Note that the plot part above each podium place is further subdivided into $p$ "columns", where each column represents one participating algorithm (here: $p = 21$). Dots corresponding to identical cases are connected by a line, leading to the shown spaghetti structure. Lower part: Bar charts represent the relative frequency for each algorithm to achieve the rank encoded by the podium place.

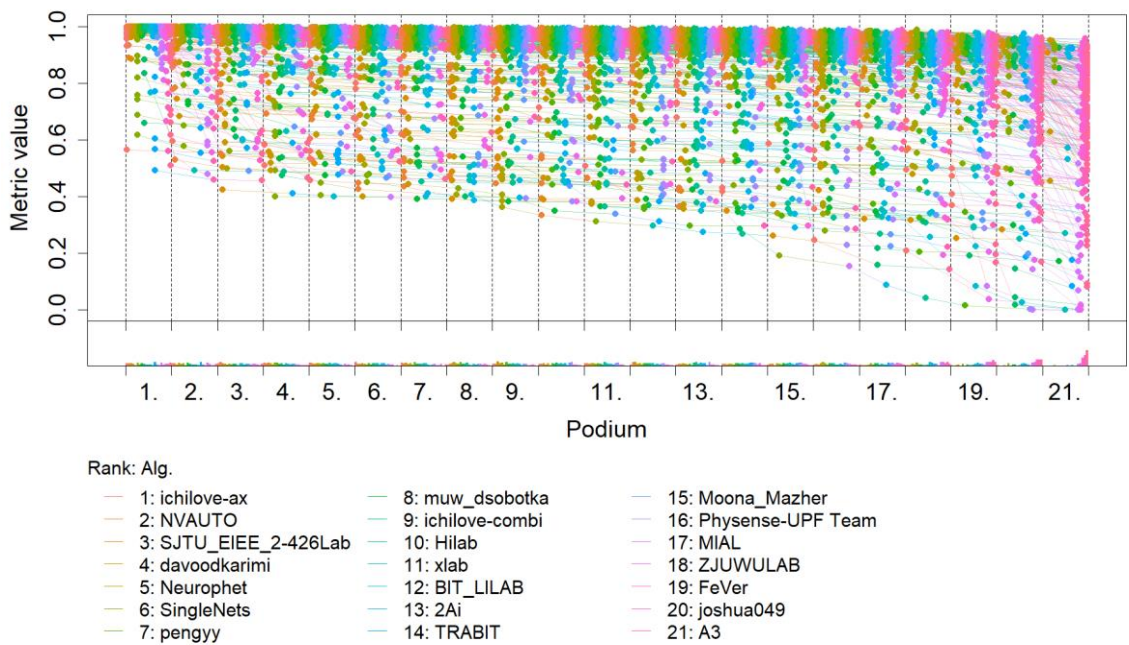

**4.2.3 Ranking heatmap**

*Ranking heatmaps* for visualizing raw assessment data. Each cell $(i, A_j)$ shows the absolute frequency of cases in which algorithm $A_j$ achieved rank $i$.

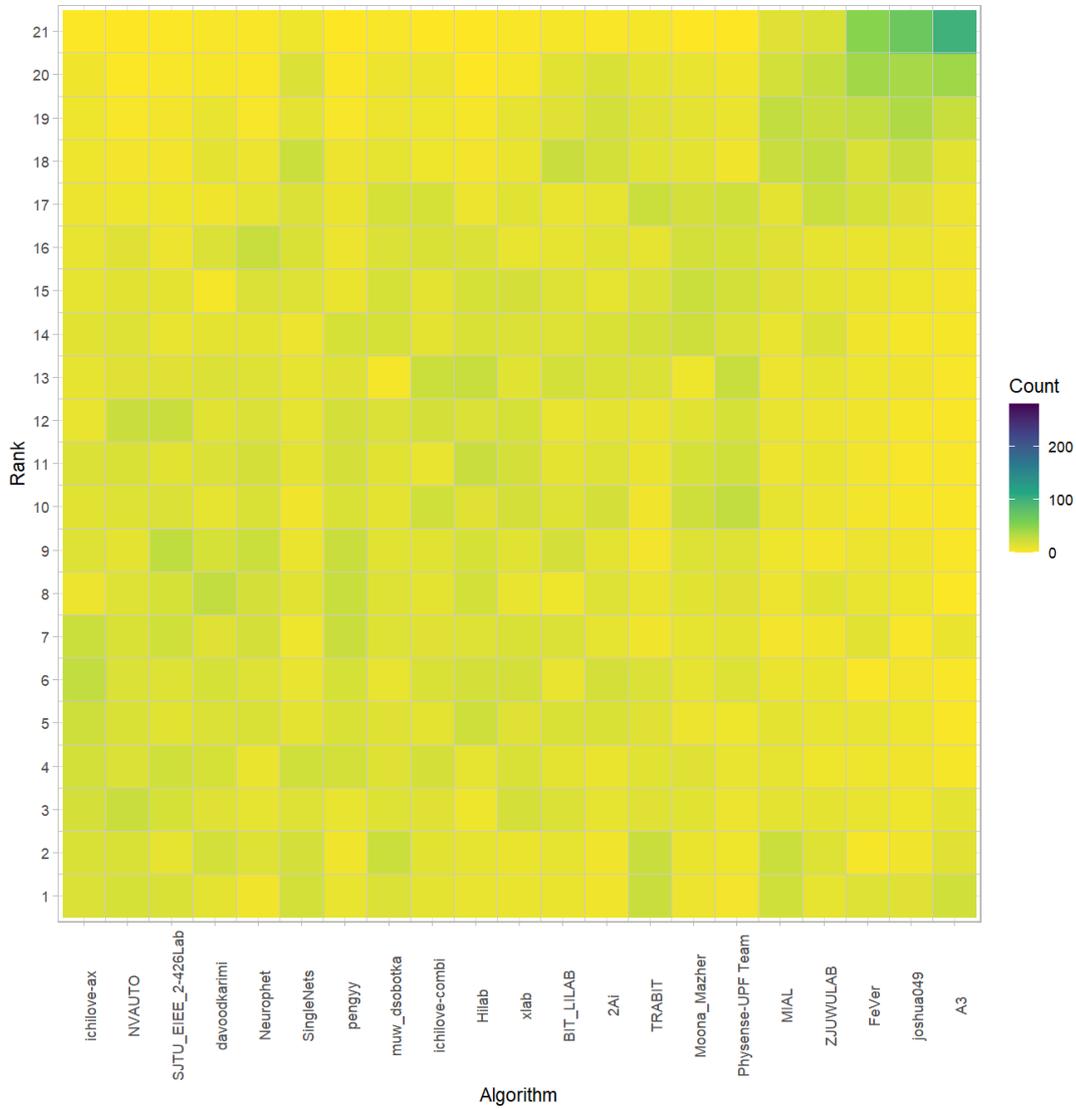

## 4.3 Visualization of ranking stability

### 4.3.1 *Blob plot* for visualizing ranking stability based on bootstrap sampling

Algorithms are color-coded, and the area of each blob at position ($A_i$, rank $j$) is proportional to the relative frequency $A_i$ achieved rank $j$ across $b = 1000$ bootstrap samples. The median rank for each algorithm is indicated by a black cross. 95% bootstrap intervals across bootstrap samples are indicated by black lines.

```
## Warning: `guides(<scale> = FALSE)` is deprecated. Please use `guides(<scale> =
## "none")` instead.
```

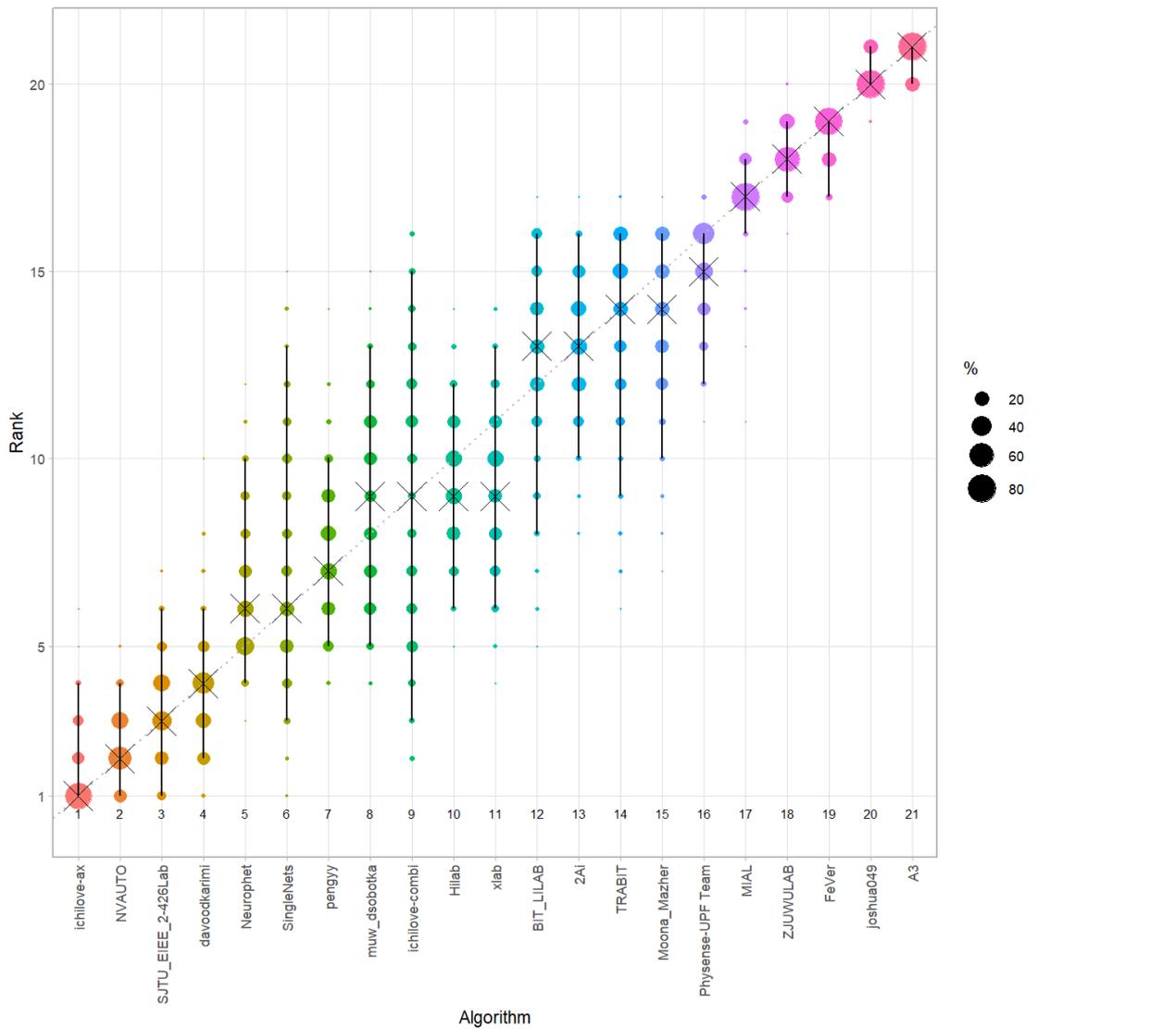

### 4.3.2 *Violin plot* for visualizing ranking stability based on bootstrapping

The ranking list based on the full assessment data is pairwise compared with the ranking lists based on the individual bootstrap samples (here $b = 1000$ samples). For each pair of rankings, Kendall's $\tau$ correlation is computed. Kendall's $\tau$ is a scaled index determining the correlation between the lists. It is computed by evaluating the number of pairwise concordances and discordances between ranking lists and produces values between $-1$ (for inverted order) and $1$ (for identical order). A violin plot, which simultaneously depicts a boxplot and a density plot, is generated from the results.

Summary Kendall's tau:

| Task | mean | median | q25 | q75 |
|---|---|---|---|---|
| dummyTask | 0.8657238 | 0.8666667 | 0.8380952 | 0.8952381 |

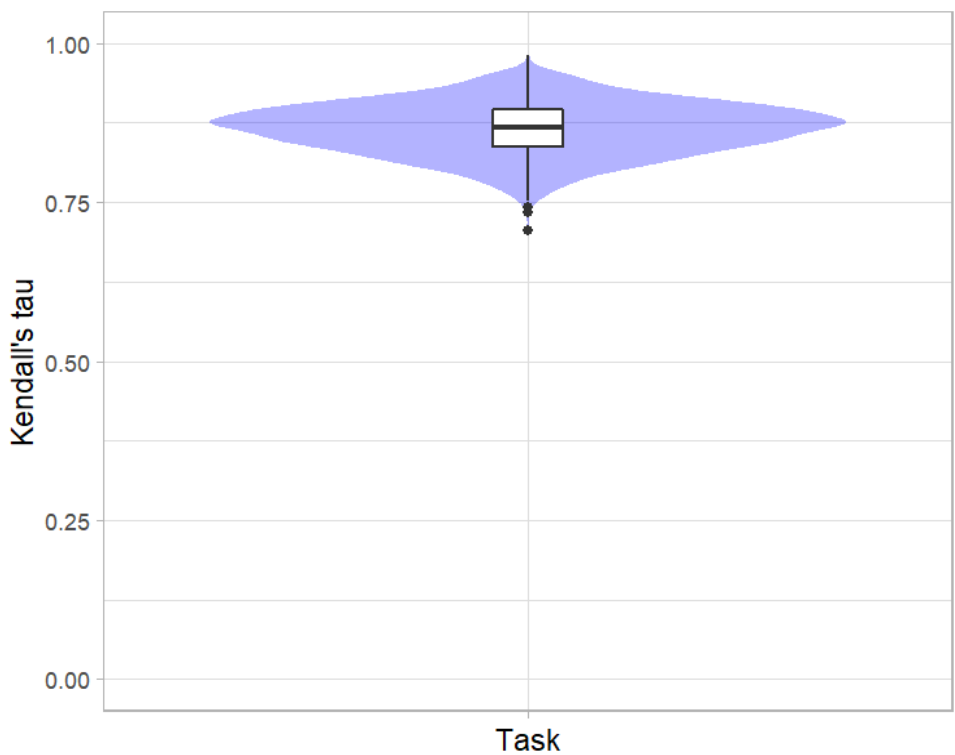

### 4.3.3 *Significance maps* for visualizing ranking stability based on statistical significance

*Significance maps* depict incidence matrices of pairwise significant test results for the one-sided Wilcoxon signed rank test at a 5% significance level with adjustment for multiple testing according to Holm. Yellow shading indicates that metric values from the algorithm on the x-axis were significantly superior to those from the algorithm on the y-axis, blue color indicates no significant difference.

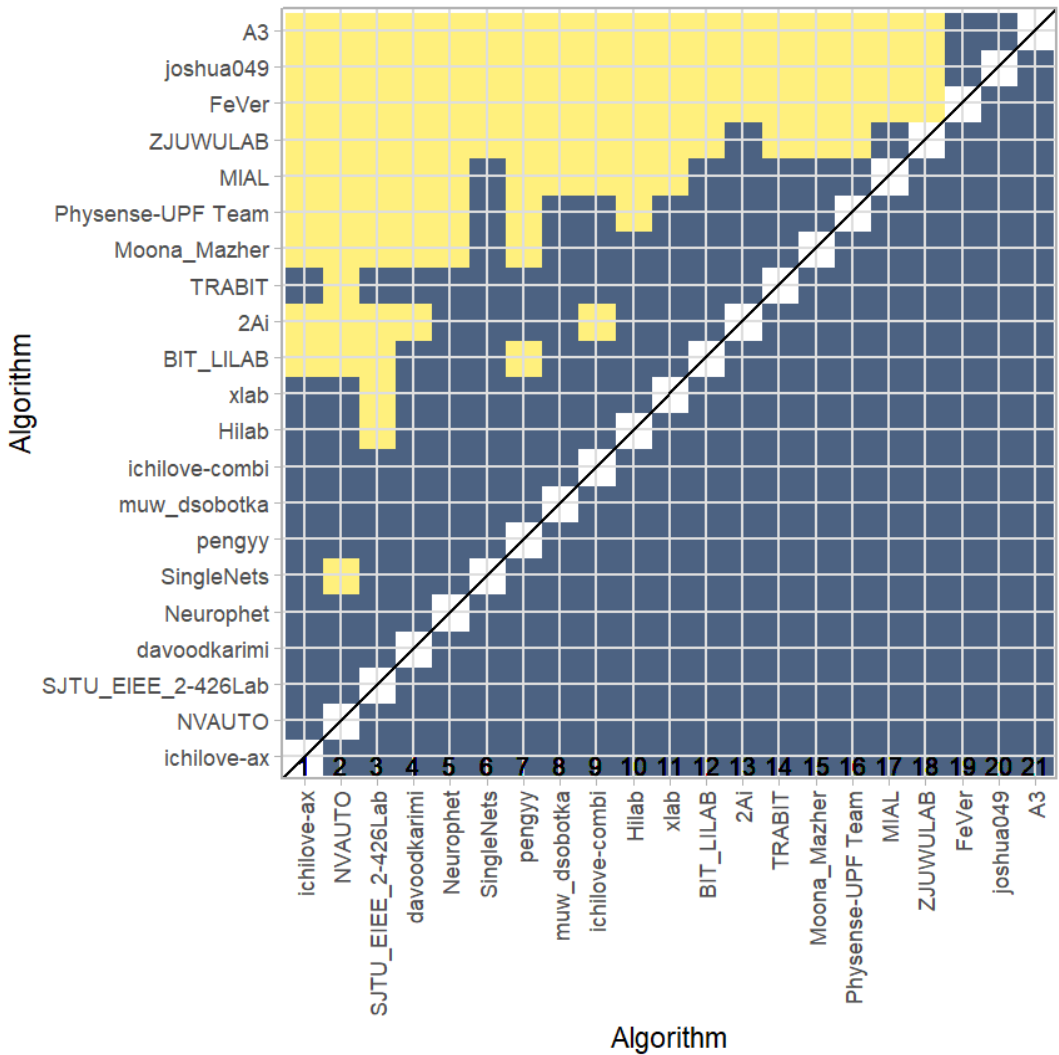

**4.3.4 Ranking robustness to ranking methods**

*Line plots* for visualizing ranking robustness across different ranking methods. Each algorithm is represented by one colored line. For each ranking method encoded on the x-axis, the height of the line represents the corresponding rank. Horizontal lines indicate identical ranks for all methods.

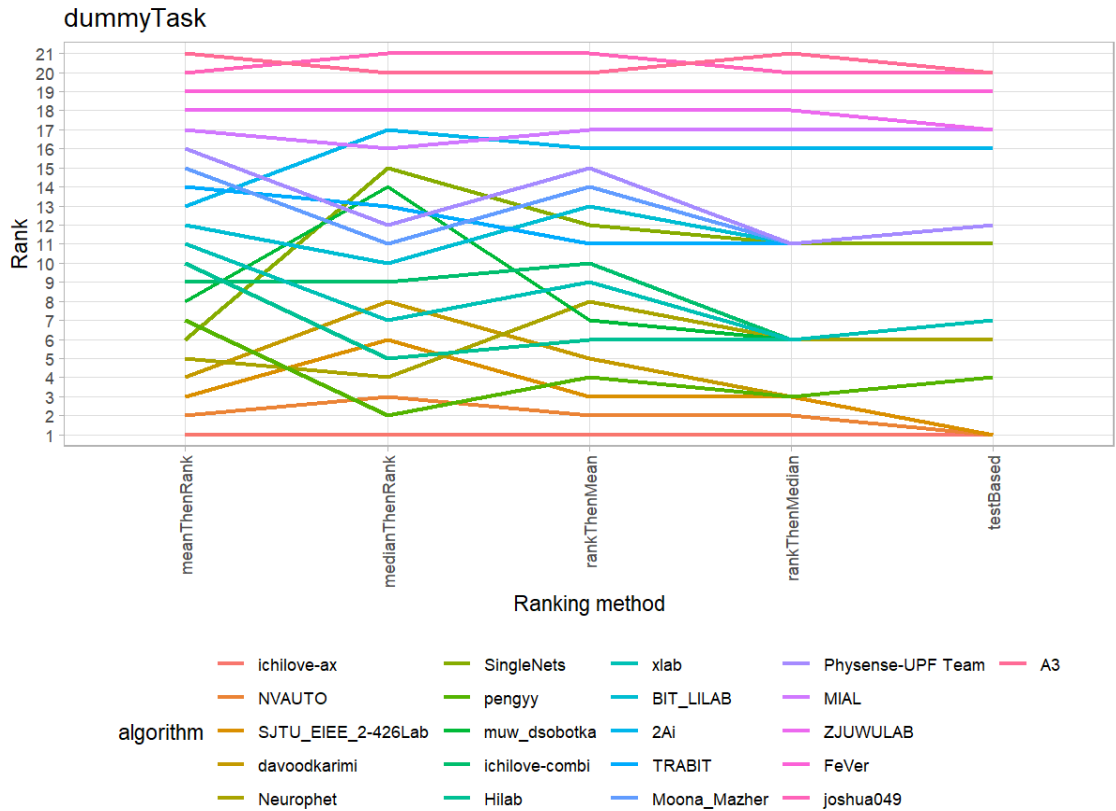

## 4.4 References

Wiesenfarth, M., Reinke, A., Landman, B.A., Eisenmann, M., Aguilera Saiz, L., Cardoso, M.J., Maier-Hein, L. and Kopp-Schneider, A. Methods and open-source toolkit for analyzing and visualizing challenge results. *Sci Rep* **11**, 2369 (2021). https://doi.org/10.1038/s41598-021-82017-6

M. J. A. Eugster, T. Hothorn, and F. Leisch, "Exploratory and inferential analysis of benchmark experiments," Institut fuer Statistik, Ludwig-Maximilians-Universitaet Muenchen, Germany, Technical Report 30, 2008. [Online]. Available: http://epub.ub.uni-muenchen.de/4134/.

Appendix

Table of Contents

# 1. Overview of Challenge Participants Algorithm

## 1.1 2Ai

### 1.1.1 Team Members and Affiliations

Team Members: Helena R. Torres[1,2,3,4], Bruno Oliveira[1,2,3,4], Pedro Morais, Jaime C. Fonseca, João L. Vilaca[1]

Affiliations: [1]2Ai – School of Technology, IPCA, Barcelos, Portugal; [2]Algoritmi Center, School of Engineering, University of Minho, Guimarães, Portugal; [3]Life and Health Sciences Research Institute (ICVS), School of Medicine, University of Minho, Braga, Portugal; [4]ICVS/3B's - PT Government Associate Laboratory, Braga/Guimarães, Portugal

### 1.1.2 Model Description

In-utero MRI has been emerging as an important tool to evaluate fetal neurological development. Here, segmentation of fetal brain structures can have an important role in the interpretation of the fetal MRI, allowing to aid in the quantification of the changing brain morphology and in the diagnosis of congenital disorders. Since manual segmentation is highly prone to observer variability and time-consuming, automatic methods for fetal tissue segmentation can bring added value to clinical practice. The main goal of the FeTA challenge is to potentiate the development of new methods for fetal tissue segmentation.

In the scope of the FeTA challenge, an automatic method to segment the different brain structures of the fetus was developed. As well-known, encoder-decoder deep convolutional neural networks have been showed the best performance for medical imaging segmentation. In this work, the U-Net method was adapted to perform fetal tissue segmentation [2, 4]. Concerning the model architecture, similar to U-NET, an encoder-decoder network was used. The encoding path is composed of down-sampling blocks and each block of the encoding path is composed of two convolutional layers, with each layer consisting of a convolution, followed by batch normalization and a leaky rectified linear unit (ReLU). The down-sampling is implemented using a stride convolution on the first layer of each encoding block. The initial number of feature maps is defined to be 32, which is double in each down-sampling stride convolution operation. The decoding path corresponds to a symmetric expanding path and skip-connections are used to allow that the encoder and decoder share information. Overall, the number of trainable parameters was 29,971,032. Moreover, a Sigmoid activation layer was added to the end of the network to achieve the segmentation, which was performed through a one-hot labelling scheme using the Dice metric as loss function. Finally, a Xavier initialization of the model parameters was applied, in order to initialize the weights such that the variance of the activations were the same across every layer.

Concerning the input of the network, all training images of the FeTA challenge were used. A pre-processing of the images was applied, where image normalization was implemented in order to achieve zero mean intensity with unit variance. Moreover, the pixel size of the images was normalized to be the median value of the pixel size of all the images and a patch-based strategy was used, were a 3D patch with size 128x128x128 was applied. To overcome overfitting problems during training, data augmentation techniques were applied to the training images, namely spatial and intensity-based transformations [1]. Spatial transformations included random flip, rotation, scaling, grid distortion, optical distortion, and elastic transformations, whereas intensity transformations included random gaussian noise, brightness, contrast, and gamma transformations.

The network was trained during 800 epochs with a mini-batch size of 4 and using the Adam optimizer with an initial learning rate of 0.001 and a learning rate decay following the polinomial learning rate policy (lr=1- ((1-epoch)/(max_epoch) )^power) with the power of 0.9 [3]. The training time of the method was ≈6 days on a computer with a CPU: i7-7700HQ @3.56GHz, RAM: 16GBytes @1.2GHz, and GPU: GTX1070 @1.443GHz, using the PyTorch python library. Finally, the final segmentation was obtained by post-processing the output of the network by removing isolated segmented voxels.

### 1.1.3 References

[1] Buslaev, A., Iglovikov, V. I., Khvedchenya, E., Parinov, A., Druzhinin, M., & Kalinin, A. A. (2020). Albumentations: Fast and flexible image augmentations. *Information (Switzerland)*, *11*(2), 1–20. https://doi.org/10.3390/info11020125
[2] Isensee, F., Jaeger, P. F., Kohl, S. A. A., Petersen, J., & Maier-Hein, K. H. (2021). nnU-Net: a self-configuring method for deep learning-based biomedical image segmentation. *Nature Methods*, *18*(2), 203–211. https://doi.org/10.1038/s41592-020-01008-z
[3] Kingma, D. P., & Ba, J. L. (2015). Adam: A method for stochastic optimization. *3rd International Conference on Learning Representations, ICLR 2015 - Conference Track Proceedings*, 1–15.
[4] Ronneberger, O., Fischer, P., & Brox, T. (2015). U-Net: Convolutional Networks for Biomedical Image Segmentation. *Computer Vision and Pattern Recognition*, 1–8.

## 1.2 A3

### 1.2.1 Team Members and Affiliations
Team Members: Yunzhi Xu and Li Zhao
Affiliations: College of Biomedical Engineering & Instrument Science, Zhejiang University, Hangzhou, China

### 1.2.2 Model Description
Input: 3D volume with a shape of 128x128x128. The input data were preprocessed as follows.
Pre-processing: To improve the data balance, the background of the original data was cropped according to the edge of the brain in 3D. The border of the brain was selected by the non-zero pixels. The cropped images were zero-padded to a matrix size of 192x192x192. Then the images were down-sampled to 128x128x128 to fit in the memory of GPUs. Images were normalized, so the mean value was zero and the standard deviation equals one.

Model Architecture: The model was implemented based on V-net. Compared to the standard U-Net, a parametric rectified linear unit (PReLU) activation function. Starting number of features was 16 and was doubled in each descending layer. There are three layers of feature extraction. Each layer was a res-net composed of two consecutive 3x3x3 convolutional kernels. Layer normalization was used instead of original batch normalization. A 2x2x2 max-pooling was performed instead of 3D convolution in the original V-net. A dropout rate of 0.5 was used before the first transpose convolution layer. The decoding path was symmetric with the encoding path with the halved number of filters for each layer.

### 1.2.3 Training Description
FeTA data were mixed in the training with 80% in the training and 20% in the validation. No additional data were used. Data augmentation was performed by randomly shifting the image up to 5 pixels in 3 directions, rotating the image up to 15 degrees along three axes, and flipping the brain in the left-right direction. Binary-cross-entropy was used as the loss function. Adam optimizer was used with a learning rate of 1e-4 with auto rate schedule using ReduceLROnPlateau. The model was trained in 200 epochs. Each epoch had 32 steps and a batch size of two. The algorithm was implemented using Tensorflow 2.4. The training was performed with two Nvidia P100 GPUs, 16 CPUs, and 118G Memory on Alibaba Cloud.

## 1.3 BIT_LILAB

### 1.3.1 Team Members and Affiliations
Team Members: Weibin Liao, Yi Lv, Xuesong Li
Affiliations: School of Computer Science and Technology, Beijing Institute of Technology, Beijing, China

### 1.3.2 Model Architecture

Our model structure is shown in the figure below, mainly referring to TransUNet [1]. This is a CNN-Transformer hybrid model. The network is similar to a U-shaped structure, adding Transformer layers with Multi-head Self-Attention mechanism to the encoder to encode the feature representation of image patches. The motivation of using this hybrid architecture is that we can not only use the high-resolution CNN feature map in the decoding path, but also use the global context information encoded by Transformers [2].

In encoder, ResNet-50 [3] is used as the feature extractor of CNN to generate a feature map for the input image. Transformer is used encoding feature representations from decomposed image patches.
Decoder consists of multiple up-sampling block, where each block consists of a 2× upsampling operator, a 3×3 convolution layer, a ReLU layer and a Batch Normalization layer successively. The overall architecture of our proposal network can be found in Figure 1. The detailed network architecture can be found in Table 1.

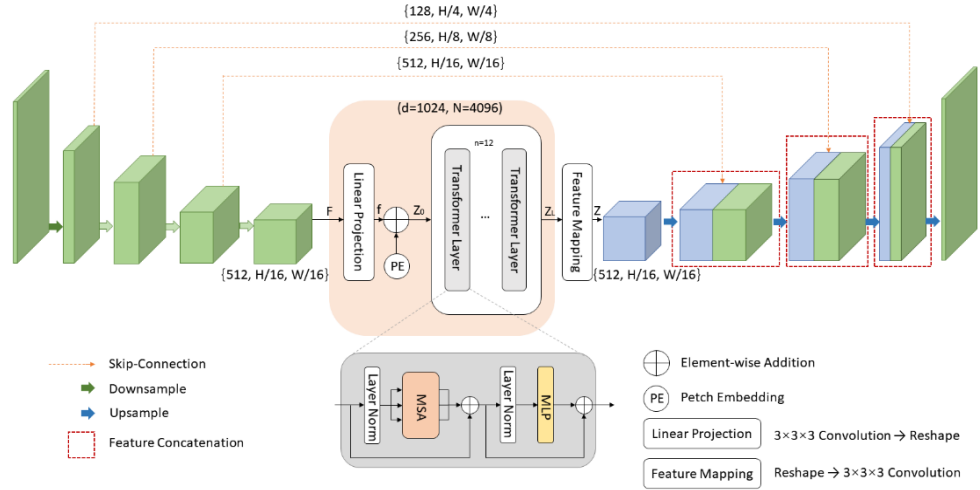

**Fig.1.** Overall architecture of our proposal network.

**Table 1.** The design details of our proposed network. $Conv\ 3\times 3$ denotes a $3\times 3$ convolutional layer. $GN$ denotes Group Normalization. $BN$ denotes Batch Normalization. Each ResBlock in encoder is a residual block. The size of input image is 3×256×256.

| Stage | Block name | Details | Output Size |
|---|---|---|---|
| CNN Encoder | InitConv | $Conv\ 7\times 7, GN, ReLU$ | $64\times 128\times 128$ |
| | DownSample | $MaxPool$ | $64\times 128\times 128$ |
| | ResBlock1 | $\begin{cases} GN, Conv\ 1\times 1, ReLU \\ GN, Conv\ 3\times 3, ReLU \\ GN, Conv\ 1\times 1, ReLU \end{cases} \times 3$ | $128\times 64\times 64$ |
| | ResBlock2 | $\begin{cases} GN, Conv\ 1\times 1, ReLU \\ GN, Conv\ 3\times 3, ReLU \\ GN, Conv\ 1\times 1, ReLU \end{cases} \times 4$ | $256\times 32\times 32$ |
| | ResBlock3 | $\begin{cases} GN, Conv\ 1\times 1, ReLU \\ GN, Conv\ 3\times 3, ReLU \\ GN, Conv\ 1\times 1, ReLU \end{cases} \times 9$ | $512\times 16\times 16$ |
| Transformer Encoder | Linear Projection | $Conv\ 3\times 3, reshape$ | $1024\times 4096$ |
| | Transformer | $Transformer\ Layer \times 12$ | $1024\times 4096$ |
| CNN Decoder | Feature Mapping | $reshape, Conv 3\times 3$ | $512\times 16\times 16$ |
| | DeBlock1 | $Conv\ 3\times 3, ReLU, BN \times 3$ | $512\times 16\times 16$ |
| | UpSample1 | $Conv\ 3\times 3, UpsamplingBilinear$ | $256\times 32\times 32$ |
| | DeBlock2 | $Conv\ 3\times 3, ReLU, BN \times 2$ | $256\times 32\times 32$ |
| | UpSample2 | $Conv\ 3\times 3, UpsamplingBilinear$ | $128\times 64\times 64$ |
| | DeBlock3 | $Conv\ 3\times 3, ReLU, BN \times 2$ | $128\times 64\times 64$ |
| | UpSample3 | $Conv\ 3\times 3, UpsamplingBilinear$ | $64\times 128\times 128$ |

| | DeBlock4 | $Conv\ 3\times 3, ReLU, BN \times 2$ | $64 \times 128 \times 128$ |
| | EndConv | $Conv\ 1\times 1, Softmax$ | $3 \times 256 \times 256$ |

#### 1.3.3 Input and Output

Our model accepts 2D slices as input, and the 3D volume needs to be sliced in advance during model training. When testing, the 3D volume image can be directly input, and the program will automatically slice the volume image, predict each slice, and then stack it into a 3D volume image. The input 2D slice resolution is 256 × 256, and the image patch size is fixed as 16 × 16.

#### 1.3.4 Dataset

We used all 80 data in the dataset. Due to the different reconstruction methods of data (the first 40 data are reconstructed by MialSRTK and the last 40 data are reconstructed by SimpleIRTK. When using data, we make a simple division based on two different reconstruction methods. 60 data (1-30, 41-70) are used as the training set, and 20 data (31-40, 71-80) are used as the test set. We don't differentiate the data for health conditions.

#### 1.3.5 Preprocess and Data Augmentation

Due to we use 2D-input network, we need to preprocess 3D volume data into 2D slices. We adopt two simple data augmentation strategies: random rotation and flipping. Randomly rotate the 2D slice (-20, 20) degrees counterclockwise, or rotate it by 90 degrees, and then flip it up and down or left and right. The probability of data augmentation is 0.5.

#### 1.3.6 Pre-training

ResNet-50 and ViT are combined in the hybrid encoder, we use the pre-training model of the backbone network provided by the author. We have also made some attempts on transfer learning: We first train the model on the Synapse multi-organ segmentation dataset, and then transfer the pre-training weights to the Fetal Tissue Annotation task. The results show that the model with pre-trained weights can converge faster on the Fetal Tissue Annotation dataset.

#### 1.3.7 Implementation Details

The input resolution and patch size are set as 256 × 256 and 16. Models are trained with SGD optimizer with learning rate 0.01, momentum 0.9 and weight decay 1e-4. Batch size is 24, the max epochs are 150 and the number of training iterations are 30k. Loss function is a composite loss function of Cross Entropy and Dice loss. All experiments are conducted using four Nvidia GTX 1080Ti GPU. Training model takes about 10 hours at a time.

The python package version we use is as follows: torch==1.4.0, torchvision==0.5.0, numpy==1.20.3, tqdm==4.61.1, tensorboard==2.5.0, tensorboardX==2.2, ml-collections==0.1.0, medpy==0.4.0, SimpleITK==2.0.2, scipy==1.6.3, h5py==3.2.1.

#### 1.3.8 Other Trained Models

We also tried other two baseline models on this task.

The first is nnUNet [4]. nnU-Net is the first standardized deep learning benchmark in biomedical segmentation. Without manual effort, researchers can compare their algorithms against nnU-Net on an arbitrary number of datasets to provide meaningful evidence for proposed improvements. We use 3D full resolution U-Net for training in nnUNet. We also used all 80 data, including 64 data as training data and 16 data as validation data. The test results show that nnUNet can reach the same level as TransUNet in DSC and HD95.

The other is CoTr [5]. This is also a framework combining CNN and transformer for medical image segmentation. Unlike TransUNet, CoTr can directly process 3D volume data. Unfortunately, the results we reproduced are far lower than TransUNet and nnUNet, and do not show better results compared with other CNN-based, Transformer-based and hybrid methods mentioned in this paper.

#### 1.3.9 Reference

1. Chen, Jieneng, et al. "Transunet: Transformers make strong encoders for medical image segmentation." arXiv preprint arXiv:2102.04306 (2021).
2. Dosovitskiy, Alexey, et al. "An image is worth 16x16 words: Transformers for image recognition at scale." arXiv preprint arXiv:2010.11929 (2020).
3. He, Kaiming, et al. "Deep residual learning for image recognition." Proceedings of the IEEE conference on computer vision and pattern recognition. 2016.
4. Isensee, Fabian, et al. "nnU-Net: a self-configuring method for deep learning-based biomedical image segmentation." Nature methods 18.2 (2021): 203-211.
5. Xie, Yutong, et al. "CoTr: Efficiently Bridging CNN and Transformer for 3D Medical Image Segmentation." arXiv preprint arXiv:2103.03024 (2021).

## 1.4 Davoodkarimi

### 1.4.1 Team Members and Affiliations
Team Members: Davood Karimi, Ali Gholipour
Affiliations: Boston Children's Hospital, Harvard Medical School

### 1.4.2 Model Description

An encoder-decoded fully-convolutional neural network. The backbone is similar to UNet. Additional dense connections as well as short and long skip connections are included between the different stages of the encoder and the decoder. All layers are followed by ReLU.

The model works on $128^3$-voxel patches. During training, patches are selected from training images at random locations. During test, a sliding-window strategy is used to cover the entire volume.

We use a novel loss function derived from mean-absolute-error loss [4]. Additionally, we have a loss term to constrain the relative volume of each of the segments/labels.

### 1.4.3 Training Method

All FeTA images (regardless of quality) were used in the same way. No additional datasets were used in any way for training. We used 15 of the FeTA images (selected to be representative of image quality and subject age distribution) for validation/testing. Once the final architecture/training hyper-parameters were decided upon, the final model was trained on all 80 FeTA images.

The only pre-processing applied consisted of intensity normalization. Specifically, each image was normalized via dividing by the image standard deviation.

Extensive image augmentation was used. Including simple geometric flips and rotations, as well as elastic deformations [2]. In addition, label perturbation and smoothing were used. Test-time image augmentation is also performed, but only includes flips; at test time no rotations or elastic deformations are used. Only a single model is trained. On a test image, the flips generate 8 labels that can be combined using simple averaging or STAPLE [5] to obtain a final label prediction.

We initialize all parameters using He's method [1] and optimize using Adam optimizer [3] with an initial learning rate of 10-4, with a batch size of 1. Learning rate was reduced by a factor of 0.9 after every 2000 training iterations if the validation loss did not decrease. If the validation loss did not decrease after 3 consecutive evaluations, training was stopped. No pre-training is used. The model training takes approximately $10^5$ iterations, equivalent to 24 hours on a single GPU. All implementation is done in TensorFlow 1.14. For data augmentation and other processing we used ITK.

### 1.4.4 References

1. He, K., Zhang, X., Ren, S., Sun, J.: Delving deep into rectifers: Surpassing human-level performance on imagenet classication. In: Proceedings of the IEEE international conference on computer vision. pp. 1026-1034 (2015)
2. Karimi, D., Samei, G., Kesch, C., Nir, G., Salcudean, S.E.: Prostate segmentation in mri using a convolutional neural network architecture and training strategy based on statistical shape models. International journal of computer assisted radiology and surgery 13(8), 1211-1219 (2018)
3. Kingma, D.P., Ba, J.: Adam: A method for stochastic optimization. arXiv preprint arXiv:1412.6980 (2014)
4. Wang, X., Hua, Y., Kodirov, E., Robertson, N.M.: Imae for noise-robust learning: Mean absolute error does not treat examples equally and gradient magnitude's variance matters. arXiv preprint arXiv:1903.12141 (2019)
5. Warfield, S.K., Zou, K.H., Wells, W.M.: Simultaneous truth and performance level estimation (staple): an algorithm for the validation of image segmentation. IEEE transactions on medical imaging 23(7), 903-921 (2004)

## 1.5 FeVer

### 1.5.1 Team Members and Affiliations
Team Members: KuanLun Liao[1,3], YiXuan Wu[2,3], and JinTai Chen[1,3]
Affiliations: [1]Zhejiang University, College of Computer Science and Technology, Road 38 West
Lake District, Hangzhou 310058, P.R. China; [2]Zhejiang University, School of Medicine, 866 Yuhangtang Rd, Hangzhou 310058, P.R. China;
[3]Real Doctor AI Research Centre, Zhejiang University, Hangzhou, China

### 1.5.2 Data Processing
Data source: We use all 80 T2-weighted fetal brain reconstructions data [5]. Besides, we don't use any other datasets. In order to preserve the internal structure and locality information of images, we use the original 3D images as input.

Data pre-processing: For pre-processing, we first limit the pixel values to the range of (-200, 1000) to removing the unused pixels. Then, since the spacing of each CT image is different, we resample the slice spaces for all CT images to the same scale by using spline interpolation. What's more, in order to utilize the data effectively, we remove the slices which corresponding labels only have background information.

Data augmentation: For data augmentation, we randomly crop each image with slices of 48, and randomly flip each image in the direction of up-and-down and left-and-right; besides, we also utilize the method of mixup [7], which convexly combines random pairs of images and their associated labels.

### 1.5.3 Model Description
We adopt a modified 3D Res-UNet model architecture (https://github.com/pykao/Modified-3D-UNet-Pytorch), which has a context pathway, a localization pathway and residual connections each with five layers.

Context path: In the context pathway, each layer contains three 3 × 3 × 3 convolutions each followed by a instance normalization, and then a leaky rectified linear unit (LeakyReLu); and every layer has a dropout layer with the ratio equals to 0.3. The shortcut path with residual connection brings the features of the nearby lower-level layers, alleviating the degradation problem and giving better performance.

Localization path: In the localization pathway, each layer consists of an up-sampling block with multiplier of spatial size 2, in the following of two sequence of a 3×3×3 convolution, a 3d-InstanceNorm, and then a leaky rectified linear unit (LeakyReLu). Besides, the upsample step is followed by a 3×3×3 convolution, a 3d-InstanceNorm, and then a leaky rectified linear unit (LeakyReLu).

Concat connection: The concat connections from layers of equal resolution in the context path provide the essential high-resolution features to the localization path. What's more, the concat connections are also utilized between different layers of the localization pathway to get the final predict.

InstanceNorm: In this work, we apply Instance Normalization [6] over a 5D input (a mini-batch of 3D inputs with additional channel dimension), which give normalization to H and W and D at the dimension of pixel value. The mean and standard-deviation are calculated per-dimension separately for each object in a mini-batch, giving a better outcome for the theme of semantic segmentation.

Leaky ReLU: Unlike the architecture in original U-Net, we utilize LeakyReLu instead of ReLu to be the activation function in our network. Leaky ReLU [3] is a type of activation function based on a ReLU, but it has a small slope for negative values instead of a flat slope.

Optimizer: For optimization, we choose the QHAdam [2] to be the optimizer with default configurations except for the weight decay rate 0.00001, which performs best in our task and settings.

Scheduler: For learning rate, we start with the value of 0.005, and choose the CosineAnnealingLR [1] as the learning rate scheduler, setting the period at 50 iterations and the minimum learning rate at 0.0005.

Loss function: For loss function, we apply Dice-Loss function [4], a popular loss function for image segmentation tasks based on the Dice coefficient, which is essentially a measure of overlap between two samples. This measure ranges from 0 to 1 where a Dice coefficient of 1 denotes perfect and complete overlap.

### 1.5.4 Training details
Parameter initialization: We initial the model parameters in a random way.

Parameter number: The architecture has 2,369,496 parameters in total.

Hyperparameter tuning: The hyperparameters are tuned by grid search.
- batch size: {1, 2, 4}

- learning rate: {0.01, 0.001, 0.005, 0.0001}
- weight decay rate: {0.0001, 0.00001}

<u>Splits of the datasets:</u> We randomly divide 95% of the dataset as the training set and the remaining 5% as the validation set.

<u>Batch size:</u> We set the batch size as 4.

<u>Ensemble of models:</u> We divide the dataset as training set and validation set randomly for five times, and use them to train different models respectively. Then, we choose the best two models among these five models and utilize average value of these two models to be the final prediction for labels.

<u>Training time:</u> 40 hours per model on RTX 3090

<u>Software libraries and packages:</u>
- Pytorch 1.8.0
- Qhoptim 1.1.0
- SimpleITK 2.0.0
- Torchvision 0.9.0
- Numpy 1.19.1
- Pandas 1.1.3
- Scipy 1.4.1
- Scikit-learn 0.23.2
- Tensorboard 2.2.2

### 1.5.5 References

1. Loshchilov, I., Hutter, F.: Sgdr: Stochastic gradient descent with warm restarts. arXiv preprint arXiv:1608.03983 (2016)
2. Ma, J., Yarats, D.: Quasi-hyperbolic momentum and adam for deep learning. arXiv preprint arXiv:1810.06801 (2018)
3. Maas, A.L., Hannun, A.Y., Ng, A.Y., et al.: Rectifier nonlinearities improve neural network acoustic models. In: Proc. icml. vol. 30, p. 3. Citeseer (2013)
4. Milletari, F., Navab, N., Ahmadi, S.A.: V-net: Fully convolutional neural networks for volumetric medical image segmentation. In: 2016 fourth international conference on 3D vision (3DV). pp. 565-571. IEEE (2016)
5. Payette, K., de Dumast, P., Kebiri, H., Ezhov, I., Paetzold, J.C., Shit, S., Iqbal, A., Khan, R., Kottke, R., Grehten, P., et al.: An automatic multi-tissue human fetal brain segmentation benchmark using the fetal tissue annotation dataset. Scientific Data 8(1), 1-14 (2021)
6. Ulyanov, D., Vedaldi, A., Lempitsky, V.: Instance normalization: The missing ingredient for fast stylization. arXiv preprint arXiv:1607.08022 (2016)
7. Zhang, H., Cisse, M., Dauphin, Y.N., Lopez-Paz, D.: mixup: Beyond empirical risk minimization. arXiv preprint arXiv:1710.09412 (2017)

## 1.6 Hilab

### 1.6.1 Team Members and Affiliations
Team Members: Guiming Dong, Hao Fu, and Guotai Wang
Affiliations: School of Mechanical and Electrical Engineering, University of Electronic Science and
Technology of China, Chengdu, China

### 1.6.2 Model Description
Our method is based on nnU Net [1], and more details are described in the following. The network architecture is showed in Figure 1.

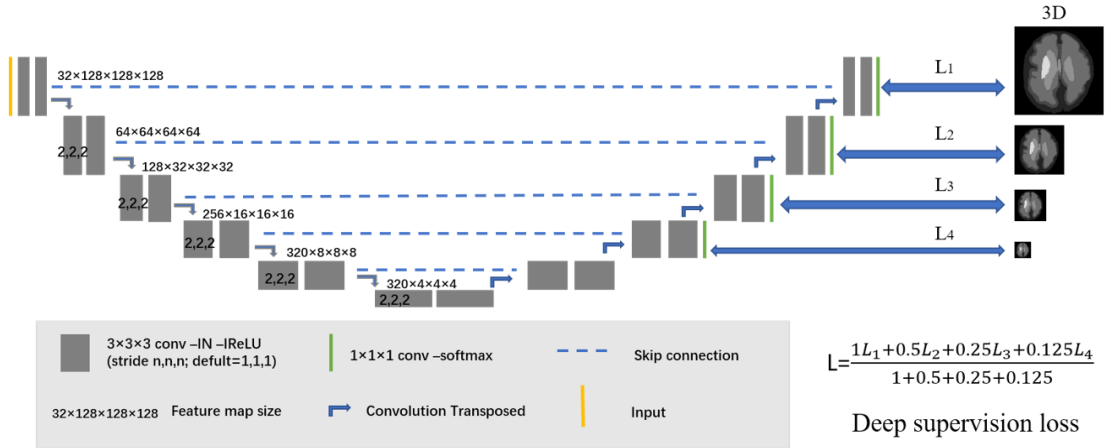

Figure 1: Network Architecture as generated by nnU Net.

We use instance normalization [2] and Leaky ReLU [3] following every convolution layer. And then, SGD optimizer with Nesterov momentum (μ = 0.99) is selected. Both cross entropy loss and Dice loss are used for segmentation training.

Training inputs are patches which are randomly sampled from training cases and the patch size of 128x128x128 is selected. Besides, the input data only has one modality.

Before fed into the network, all the training images need to be cropped according to the foreground. We also use Z score (mean subtraction and division by standard deviation) per image followed by cropping for data normalization.

Random initialization is selected for our model. Our framework is based on nn UNet [1], which is publically available at https://github.com/MIC DKFZ/nnUNet

### 1.6.3 Training Method
During the training, we use all cases of FeTA dataset. We don't use any external dataset. We split the dataset into 5 folds so that we can run a 5 fold cross validation. During the validation, we found that performance of the model on pathological cases is worse than that one on neurotypical cases. So pathological cases are copied triple during the training. And then, an initial learning rate of 0.01 is used for learning network weights. The learning rate is decayed with following schedule [4]:
$(1 - epoch/epoch\_max)^{0.9}$. Batch size is set to 2.

The data augmentation techniques are as follows: Rotation and Scaling, Gaussian noise, Gaussian blur, Brightness, Contrast, Simulation of low resolution, Gamma augmentation and Mirroring. More details can be seen in [1].

For FeTA challenge, we have trained 5 models via a 5 fold cross validation. And all the 5 models were selected to run ensembling.

Other details:
- System version: Ubuntu 20.04
- PyTorch version: pytorch1.6.0_py3.8_cuda10.2.89_cudnn7.6.5_0
- Devices: NVIDIA GeForce RTX 2080 Ti
- Training runs 400 epochs for each fold, and one epoch is defined as 250 iterations. Each epoch costs about 460s.

### 1.6.4 References
1. Isensee, F., Jaeger, P.F., Kohl, S.A.A. et al. "nnU Net: a self configuring method for deep learning based biomedical image segmentation." Nat Methods (2020).
2. D. Ulyanov, A. Vedaldi, and V. Lempitsky, "Instance normalization: The missing ingredient for fast stylization," arXiv preprint arXiv:1607.08022, 2016.

3. A. L. Maas, A. Y. Hannun, and A. Y. Ng, "Rectifier nonlinearities improve neural network acoustic models," in Proc. icml, vol. 30, no. 1, 2013, p. 3.
4. C. Chen, G. Papandreou, I. Kokkinos, K. Murphy, and A. L. Yuille, "Deeplab: Semantic image segmentation with deep convolutional nets, atrous convolution, and fully connected crfs," IEEE transactions on pattern analysis and machine.

## 1.7 Ichilove-ax and Ichilove-combi

### 1.7.1 Team Members and Affiliations
Team Members: Netanell Avisdris[1,2], Ori Ben-Zvi[2,3], Bella Fadida-Specktor, Prof. Leo Joskowicz, Prof. Dafna Ben Bashat[2,3,4]
Affiliations: [1]School of Computer Science and Engineering, The Hebrew University of Jerusalem, Israel; [2]Sagol Brain Institute, Tel Aviv Sourasky Medical Center, Israel, [3]Sagol School of Neuroscience, Tel Aviv University, Israel; [4]Sackler Faculty of Medicine, Tel Aviv University, Israel
Note: Only ichilove-ax and ichilove-combi were entries in the FeTA Challenge

### 1.7.2 Model Description
The approach taken in our algorithm included: reformat the brain into three orthogonal plans of 2D slice-volumes (axial, sagittal and coronal); The algorithm composed of two stages: (1) detecting the fetal brain ROI using 2D U-net (only on the axial slices); (2) performing 2D multi-class segmentation on the cropped ROI using 2D U-net; and finally, merging all resulting segmentations into one, based on majority voting.

We submitted four submissions, three where the segmentation obtained separately on each plane, and one which merge all segmentations.
- Ichilove-ax: First stage + segmentation in the axial plane of cropped ROI input volume
- Ichilove-sag: First stage + segmentation in the sagittal plane of cropped ROI input volume
- Ichilove-cor: First stage + segmentation in the coronal plane of cropped ROI input volume
- Ichilove-combi : First stage + segmentation in three planes of cropped ROI input volume, each segmentation performed twice, on the original volume, and on flipped volume, thus resulting in six segmentations. The merging strategy between segmentation is voxel-wise majority voting.

Figure 1 shows schematic overview of the four submissions.

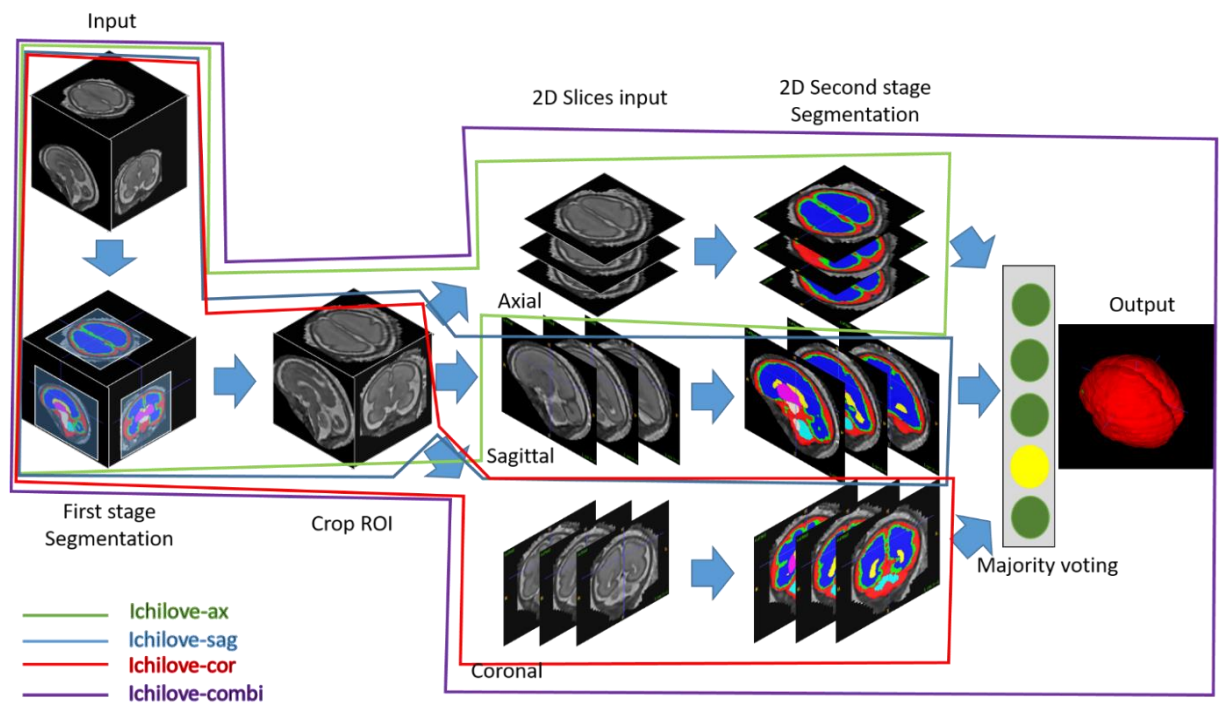

**Figure 1:** Schematic overview of 4 submissions: ichilove-{ax,sag,cor,combi}.

### 1.7.3 Networks

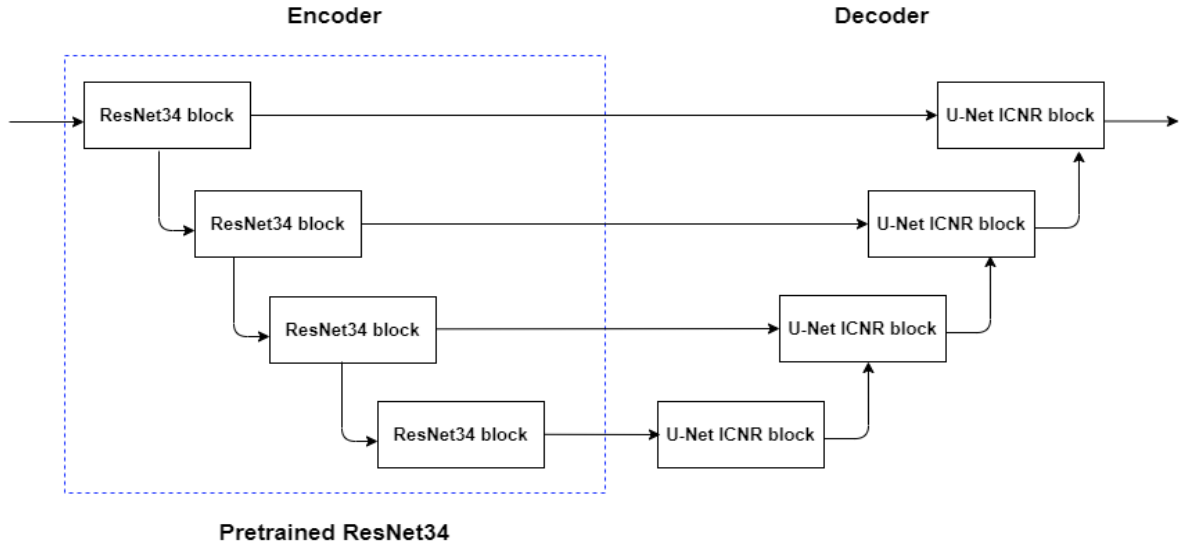

**Figure 2**: Architecture of 2D Segmentation network

For all segmentation steps, we use modified 2D U-net with similar architecture to the one used in [1] for fetal brain segmentation, where the encoder part composed of pre-trained ResNet34 network blocks and the decoder consists of U-Net PixelShuffle ICNR blocks for each scale described in [2]. The encoder and decoder are connected with the fast.ai Dynamic U-Net [3].

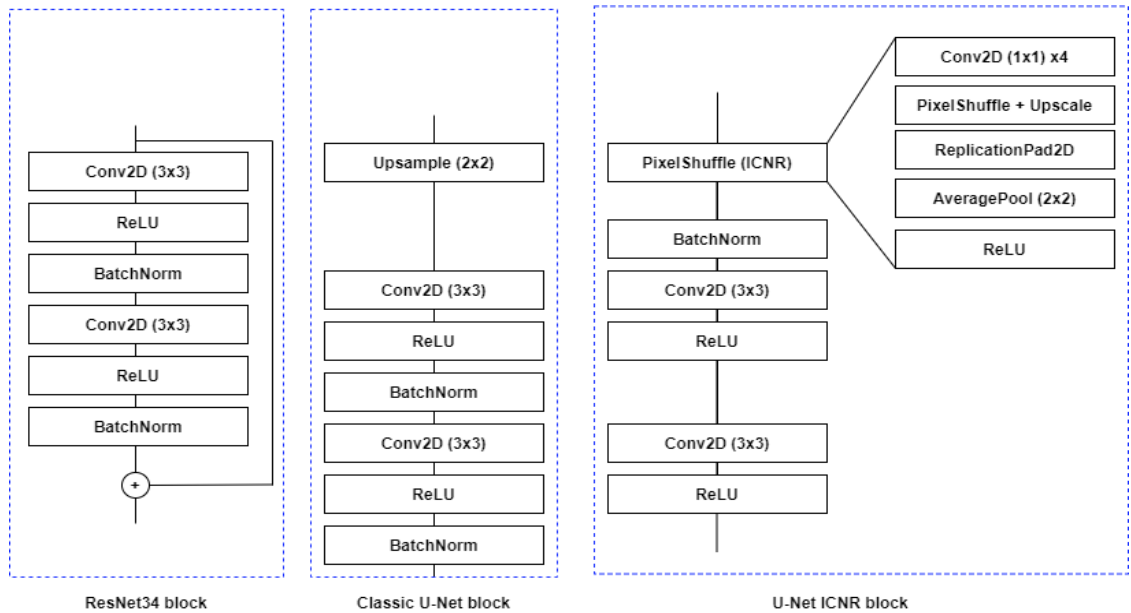

**Figure 3:** Blocks of the segmentation network: a ResNet34 block for the first scale of the encoder (left), a U-Net ICNR block for the decoder (right) and a comparison block between the classic U-Net and the U-Net ICNR blocks (middle).

To train the network, we used the OneCycle optimizer in [4] with a learning rate of 1e-3 for 30 epochs twice - the first time only the decoder layers (U-Net ICNR blocks), and the second time for the entire network with batch size of 16 slices. The network was trained with Lovasz-Softmax loss [5].

For training, we used four augmentations: 1) intensity with a contrast in the range 0.3-1.7 and brightness in the range 0.3-0.9 for training only; 2) scaling to size NxN and normalization in the range [0,1] for both training and inference, where for the first stage N=256, and for second stage N=160; 3) rotation in the range of 0-90 degrees, and; 4) Intensity Inhomogeneity Augmentation [6]. For the inference, the slice output segmentation is computed by nearest-neighbor interpolation followed by zero-padding (background class) to obtain the original slice size. From 60 provided volumes, we use 50 volumes for train 5 for validation and 5 for test.

### 1.7.4 References

1. Avisdris, N., et al., Automatic linear measurements of the fetal brain on MRI with deep neural networks. International Journal of Computer Assisted Radiology and Surgery, 2021.

2. Aitken, A., et al., Checkerboard artifact free sub-pixel convolution: A note on sub-pixel convolution, resize convolution and convolution resize. arXiv preprint arXiv:1707.02937, 2017.

3. Howard, J. and S. Gugger, fastai: A Layered API for Deep Learning. Information, 2020. 11(2): p. 108.

4. Smith, L.N., A disciplined approach to neural network hyper-parameters: Part 1--learning rate, batch size, momentum, and weight decay. arXiv preprint arXiv:1803.09820, 2018.

5. Berman, M., A.R. Triki, and M.B. Blaschko. The lovász-softmax loss: A tractable surrogate for the optimization of the intersection-over-union measure in neural networks. in Proceedings of the IEEE Conference on Computer Vision and Pattern Recognition. 2018.

6. Khalili, N., et al., Automatic brain tissue segmentation in fetal MRI using convolutional neural networks. Magnetic resonance imaging, 2019. 64: p. 77-89.

### 1.8 Anonymous

One team participated in the FeTA Challenge, but declined to be a part of the paper. Therefore their scores are included in all rankings for completeness, but the algorithm description is not included.

## 1.9 MIAL

### 1.9.1 Team Members and Affiliations

Team Members: Priscille de Dumast[2,1], Meritxell Bach Cuadra[1,2]

Affiliations: [1]CIBM Center for Biomedical Imaging; [2]Department of Radiology, Lausanne University Hospital and University of Lausanne
Note: This team contains challenge organizers. While included in all rankings, this team was ineligible for prizes.

### 1.9.2 Model Description

Our segmentation method is based on the well-established 2D U-Net [2] image segmentation method that has been used in the past for 2D fetal brain MRI [1]. In our approach, two networks, with identical architecture, are trained separately: one network is trained on the MIALSRTK reconstructed images, and one network is trained on the IRTK reconstructed images.

We adopt a patch-based approach, feeding our networks with sub-image of 64x64 voxel size. The 2D U-Net architecture is composed of encoding and de-coding paths. The encoding path consists of 5 repetitions of the followings: two 3x3 convolutional layers, followed by a rectified linear unit (ReLu) activation function and a 2x2 max-pooling down-sampling layer. Feature maps are hence doubled from 32 to 512. In the expanding path, 2x2 up-sampled encoded features concatenated with the corresponding encoding path are 3x3 convolved and passed through ReLu. All convolutional layers output are batch normalized. The network prediction is computed with a final 1x1 convolution.

Neural networks are implemented in TensorFlow (version 2.3).

### 1.9.3 Data

We proceed to a reconstruction-aware method. Each network was trained on 40 training cases, according to the reconstruction method used, either IRTK or MIALSRTK.
Input sub-image patches are extracted using a 2.5D approach: overlapping patches from the three orientations (axial, coronal and sagittal). Patches with tissue (positive patches) are duplicated once with data augmentation performed by randomly flipping and rotating patches (by $n$ x 90°, n ∈ [0;3]). Most negative patches (with positive intensities but no fetal brain tissue represented) were discarded as only 1/8 are kept. Intensities of all image patches are standardized to have mean 0 and variance 1.
Final estimation of the segmentation is reconstructed using a majority voting approach from all probability maps.

### 1.9.4 Training

Both networks are trained using a hybrid loss $\mathcal{L}_{hybrid}$ combining two terms:
$$\mathcal{L}_{hybrid} = (1 - \lambda)\,\mathcal{L}_{cce} + \lambda \mathcal{L}_{dice}$$
where $\mathcal{L}_{cce}$ is the categorical cross-entropy loss function, $\mathcal{L}_{dice}$ is based on the dice similarity coefficient and $\lambda$ balances the contribution of the two terms of the loss. $\lambda$ was set to 0.5.
Both networks were trained with an initial learning rate (LR) of $1 \times 10^{-3}$. Mialsrtk network was trained for 100 epochs with learning rate decay at epoch [23, 45] and irtk network was trained for 100 epochs with learning rate decay at [24,44].

A 5-folds cross-validation approach was used to determine the hyperparameters (initial LR, epochs for LR decay, and total number of epochs).

### 1.9.5 References

[1] Khalili, N., et al.: Automatic brain tissue segmentation in fetal MRI using convolutional neural networks. Magnetic Resonance Imaging 64, 77-89 (Dec 2019). https://doi.org/10.1016/j.mri.2019.05.020
[2] Ronneberger, O., et al.: U-net: Convolutional networks for biomedical image segmentation. In: Medical Image Computing and Computer-Assisted Intervention - MICCAI 2015. pp. 234-241. Springer International Publishing, Cham (2015)

## 1.10 Moona Mazher

### 1.10.1 Team Members and Affiliations

Team Members: Moona Mazher[1], Abdul Qayyum[2], Abdesslam BENZINOU[2], Mohamed Abdel-Nasser[1,3] and Domenec Puig[1]
Affiliations: [1]Department of Computer Engineering and Mathematics, University Rovira i Virgili, Spain; [2]ENIB, UMR CNRS 6285 LabSTICC, Brest, 29238, France; [3]Department of Electrical Engineering, Faculty of Engineering, Aswan University, Egypt

### 1.10.2 Model Description

Main Contribution for Feta Segmentation Task:

1. We developed 2D Densely-based encoder network from scratch without using any pre-trained module and it is composed of a different number of layers with feature reuse capability. The efficient and lightweight 2D convolutional layer blocks with 2D up-sampling layers have been built on the decoder side of the proposed model (see detail in Figure 1). The proposed dense blocks used a successive number of feature maps in each encoder layer of the proposed model.
2. Initially, we have used axial 2D slices for training our proposed model and models did not perform well. Later, the same proposed model is trained on three different views (axial, coronal, and sagittal) of the 2D slices of the feta dataset and then combined these three output segmentation maps to develop 3D segmentation volume (see detail in Figure 3).

Proposed Method Description: The different deep learning models based on ResNet, MobileNet, efficientNet have been used to construct the encoder for the Feta challenge segmentation task. We have tried different deep leaning based modules such as ResNet [1], MobileNet [2], efficientNet[3] for Feta brain segmentation, and finally, the DenseNet [4] based module produced a better performance as compared to other deep learning modules. Our proposed model is based on an encoder and decoder module with a skip connection.

We have proposed a simple DenseNet module with a small number of blocks on the encoder side and some efficient simple 2D convolutional layers module with 2D up-sampling layers including regularization layers proposed at the decoder side. The DenseNet based deep learning with additional tricks produced better segmentation performance on the Feta challenge dataset.

In the DenseNet module, the feature maps of all preceding layers are used as inputs, and their feature maps are used as inputs into all subsequent layers. The dense module has many advantages such as alleviates the vanishing-gradient problem, strengthen feature propagation, encourage feature reuse, and substantially reduce the number of parameters. The proposed Dense feature maps module adding only a small set of feature maps to the collective knowledge of the network and keeps the remaining feature maps unchanged—and the final layer makes a decision based on all feature-maps in that are reused in the model. The proposed Dense module used a fewer number of parameters and improved the flow of information and gradients throughout the network, which makes them easy to train. Each layer has direct access to the gradients from the loss function and the original input signal, leading to implicit deep supervision. This helps training of deeper network architectures. Further, we also observe those dense connections have a regularizing effect, which reduces overfitting on tasks with smaller training set sizes.

A detailed description of the proposed model with proposed layers is shown in Figure 1. We have trained all layers' weights from scratch using the Fetal Brain Tissue Annotation and Segmentation Challenge (FeTA) dataset. Pre-trained models and extra datasets are not used in this our proposed solution.

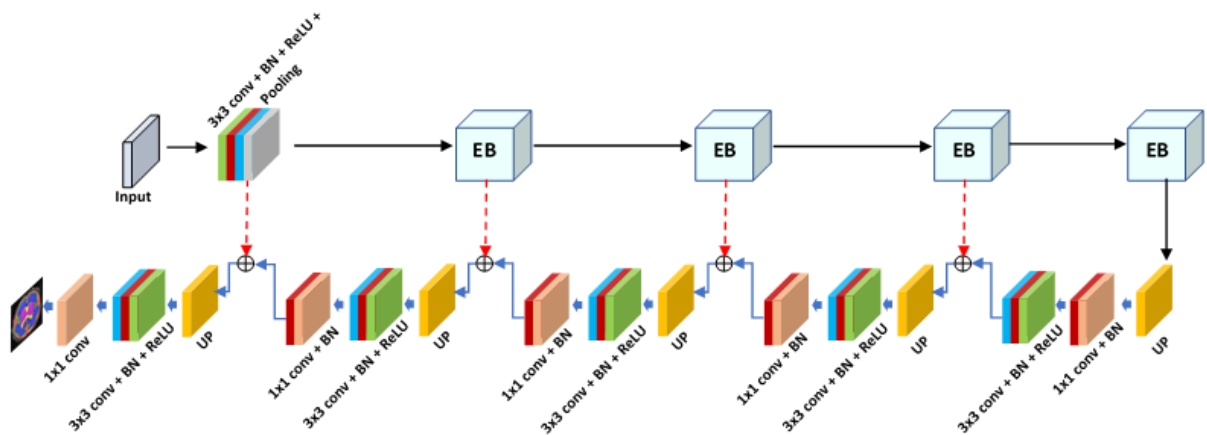

**Figure 1:** The proposed model configuration for Feta Segmentation (EB: Encoder Block).

The Dense-Layer (DL) consisted of two convolutional layers (Conv-BN-ReLU) with batch normalization (BN) layer and ReLU activation function. The feature maps are concatenated and reuse between each convolutional layer. The first convolutional layer used a 1x1 kernel and the second convolutional layer used a 3x3 kernel with a different number of feature maps. The feature maps of each layer are concatenated to form the Dense layer is shown in Figure 2. The six number of Dense-layers have been used to construct the encoder module of the proposed model with transition layer. We refer to layers between encoder blocks as transition layers, which do convolution and pooling. The transition layers used in our experiments consist of a batch normalization layer and a 1x1 convolutional layer followed by a 2x2 average pooling layer. The transition layer helps to reduce the spatial size in each encoder block in the proposed model. The transition layers have been used after each dense block in the encoder module inside the proposed model. The transition layer with six dense layers is shown in Figure 2.

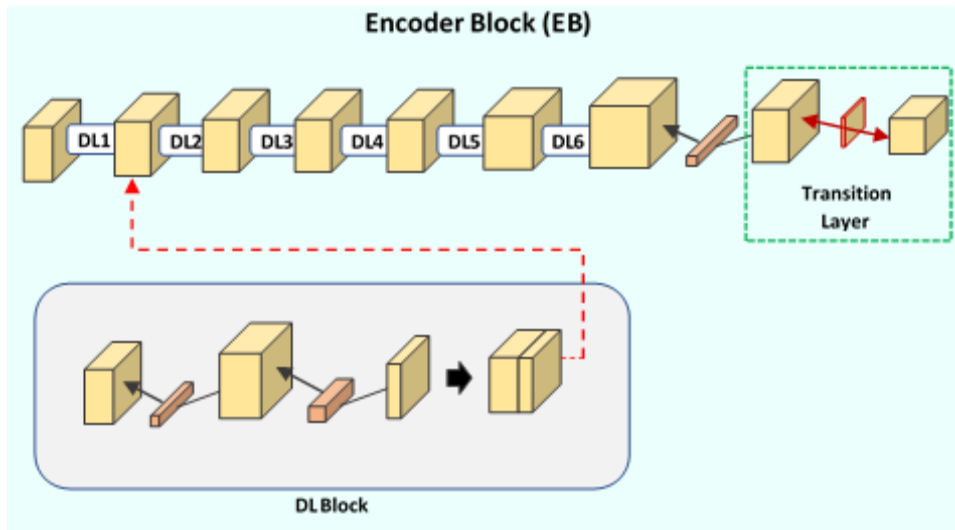

**Figure 2:** The encoder module based on proposed DensNet blocks (DL: Dense Layer).

On the decoder side, the first up-sampling 2D layer is used with a simple and efficient 2D convolutional layers block (1x1conv,3x3conv, BN, and ReLU). The feature maps between encoder and decoder module are concatenated using skip connection based on second, third, and fourth up-blocks and proposed 2D-convolutional layer block (3x3conv, BN, ReLU) to reconstruct the segmentation from input images. The input features' maps that are obtained from every encoder block are concatenated with every decoder block feature's map to reconstruct the semantic information. The regularization layers (BN, ReLU) have been used after the concatenation of feature maps between 3x3 2D convolutional and up-sampled layer feature maps for smooth training and optimization of the training process of the proposed model. In the end, the 1x1 convolutional layer with a sigmoid layer is used to construct the segmentation map. Each dense block used several features maps (96, 768, 2112, and 2208).

First, we used the proposed model for axial 2D slices of the Feta dataset, and the model produced the worst dice score especially for Ventricles, Cerebellum, Deep Grey Matter, Brainstem classes. When we trained our model on coronal slices, the model provided optimal performance in terms of validation Dice score for all classes in the dataset that's the main motivation and reason to use all three views 2D slices to trained the 2D deep learning model. We have trained the proposed model for individual views of the 3D input volume and combined the output of the trained model on each view (axial, coronal, and sagittal) to construct a 3D segmentation map. The combined output developed by the proposed model with all three 2D views produced a better performance as compared to individual 2D views. The proposed solution is shown in Figure 3. Note. We also trained three different models on each view and could not get optimal performance on validation data as compared to the trained single model on each view.

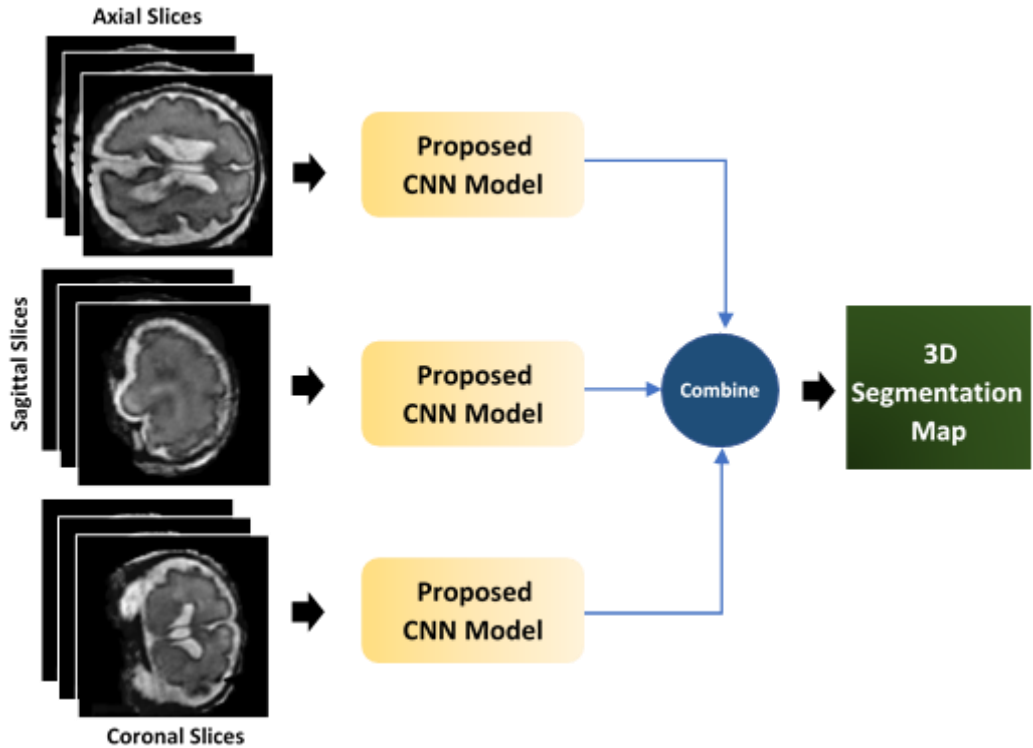

**Figure 3:** The proposed model using two separate views for Feta segmentation.

### 1.10.3 Training and Data Distribution used in Proposed Model

The Adam optimizers with a 0.0003 learning rate have been used for the training of our proposed model. In our experiments, 24 batch sizes with 1000 epochs using early stop criteria have been used for the training and optimization of the proposed model. First, the proposed model trained without data augmentation and got a good score, and later we did some data augmentation techniques such as CenterCrop, HorizontalFlip (p=0.5), VerticalFlip (p=0.5), RandomBrightnessContrast (p=0.8), and RandomGamma (p=0.8) during training the model that helps to improve a little performance. The binary cross-entropy loss function is used to compute the loss between ground-truth and predicted segmentation mask.

The one-hot encoded ground-truth segmentation map was used to compute the binary cross-entropy loss between ground truth and the predicted segmentation map.
Albumentations PyTorch segmentation library [5] used for data augmentation. The Pytorch library [6] has been used for all model implementation, training, validation, and testing. The V 100 Tesla machine with a single 12 GB GPU memory is used for training the proposed and state-of-the-art deep learning models. The proposed model contains the total number of trainable parameters (49,510,728) and total FLOPs (301846528).

The Feta dataset consisted of 80 subjects with a ground truth segmentation map. 80 % (64 subjects) data has been for training and 20% (16 subjects) for testing. The 5k fold cross-validation technique was used for random splitting the data into training and validation. The 16384 (256*64) number of images and masks in axial, coronal, and sagittal views are used for training the model 4096 (256*16) no. of images used for validation the proposed model. The standard normalization method to normalize the dataset. The pandas, skimage, OpenCV, nibabel, simpleITK python-based libraries are used for data preprocessing for training, validation, and testing the proposed model. Note: There is no pre-trained model and an additional dataset has been used in our experiments.

### 1.10.4 References

[1] He, Kaiming, Xiangyu Zhang, Shaoqing Ren, and Jian Sun. "Identity mappings in deep residual networks." In European conference on computer vision, pp. 630-645. Springer, Cham, 2016.
[2] Sandler, Mark, Andrew Howard, Menglong Zhu, Andrey Zhmoginov, and Liang-Chieh Chen. "Mobilenetv2: Inverted residuals and linear bottlenecks." In Proceedings of the IEEE conference on computer vision and pattern recognition, pp. 4510-4520. 2018.
[3] Tan, Mingxing, and Quoc Le. "Efficientnet: Rethinking model scaling for convolutional neural networks." In International Conference on Machine Learning, pp. 6105-6114. PMLR, 2019.
[4] Huang, Gao, Zhuang Liu, Laurens Van Der Maaten, and Kilian Q. Weinberger. "Densely connected convolutional networks." In Proceedings of the IEEE conference on computer vision and pattern recognition, pp. 4700-4708. 2017.
[5] https://albumentations.ai/docs/
[6] https://pytorch.org/

## 1.11 muw_dsobotka

### 1.11.1 Team Members and Affiliations Model Description
Team Members: Daniel Sobotka, Georg Langs
Affiliations: Computational Imaging Research Lab, Department of Biomedical Imaging and
Image-guided Therapy, Medical University of Vienna, Vienna, Austria

### 1.11.2 Model Description
The proposed multi-task learning model is building on a 3D U-Net architecture [1] with a shared encoder and two task specific decoders. The segmentation task TS generates the fetal tissue annotations, where the image reconstruction task TR reconstructs the input image. In contrast to other multi-task learning approaches, layer-wise feature fusion [2] is used to share features between the image segmentation and image reconstruction decoders. The encoder of the network consists of three down-sampling blocks and each task specific decoder of three up-sampling blocks. Each down-sampling block contains two 3 x 3 x 3 convolutions followed by Rectified Linear Units (ReLU) and Group Normalizations (GN) [6], as well as a 2 x 2 x 2 max pooling operation. The up-sampling path for both, image segmentation and image reconstruction tasks uses nearest neighbor interpolation and is symmetric to the down-sampling path. Neural Discriminative Dimensionality Reduction (NDDR) layers [2] are used to utilize features learned from $T_S$ and $T_R$ and to learn layerwise feature fusion between the two task specific decoders. There, features with the same spatial resolution from both, the image segmentation and image reconstruction decoders, are concatenated followed by a 1 x 1 x 1 convolution, ReLu and GN. The loss function $L$ uses homoscedastic uncertainty [3] to balance the single-task losses for the weights $W$ of the network:

$$L(W, \sigma_S, \sigma_R) = \frac{1}{2\sigma_S^2} L_S(W) + \frac{1}{2\sigma_R^2} L_R(W) + \log(\sigma_S) + \log(\sigma_R)$$

where $L_S(W)$ denotes the cross entropy loss for $T_S$. $L_R(W)$ is the mean squared error loss for $T_R$. Based on preliminary results we set $\sigma_S$ and $\sigma_R$ to fixed values with a ratio 1:8. The network consists of 6491385 trainable parameters and uses 3D patches of size 128 x 96 x 96 as input with stride between patches of size 64 x 48 x 48 and predicts 3D patches of size 128 x 96 x 96 combined at inference to an fetal tissue annotation image of size 256 x 256 x 256. All the input patches were Z-score normalized and the initialization of the model parameters was random. As baseline implementation the 3D-UNet from [5] was used.

### 1.11.3 Training Method
The proposed network model was trained on all available 80 FeTa subjects with-out validation and testing splitting and without additional external data. Further, the network was optimized using Adam [4] with an initial learning rate of 0:001, batch size of 1 and a total of 100 epochs. Data augmentation included elastic deformation with spline order 3, random flipping, random rotation by 90 degrees, random rotation by ±15 degrees, random contrast, Gaussian noise and Poisson noise. As software library PyTorch 1.3.1 with Python 3.7.3 was used and training of the network lasted around one week. For the final selection of the here described model approximately 5 different models with different values of $\sigma_S$ and $\sigma_R$ have been trained.

### 1.11.4 References

1. Cicek, O., Abdulkadir, A., Lienkamp, S.S., Brox, T., Ronneberger, O.: 3d u-net: learning dense volumetric segmentation from sparse annotation. In: International conference on medical image computing and computer-assisted intervention. pp. 424-432. Springer (2016)
2. Gao, Y., Ma, J., Zhao, M., Liu, W., Yuille, A.L.: Nddr-cnn: Layerwise feature fusing in multi-task cnns by neural discriminative dimensionality reduction. In: Proceedings of the IEEE/CVF Conference on Computer Vision and Pattern Recognition. pp. 3205-3214 (2019)
3. Kendall, A., Gal, Y., Cipolla, R.: Multi-task learning using uncertainty to weigh losses for scene geometry and semantics. In: Proceedings of the IEEE conference on computer vision and pattern recognition. pp. 7482-7491 (2018)
4. Kingma, D.P., Ba, J.: Adam: A method for stochastic optimization. arXiv preprint arXiv:1412.6980 (2014)
5. Wolny, A., Cerrone, L., Vijayan, A., Tofanelli, R., Barro, A.V., Louveaux, M.,Wenzl, C., Strauss, S., Wilson-Sanchez, D., Lymbouridou, R., et al.: Accurate and versatile 3d segmentation of plant tissues at cellular resolution. Elife 9, e57613 (2020)
6. Wu, Y., He, K.: Group normalization. In: Proceedings of the European conference on computer vision (ECCV). pp. 3-19 (2018)

## 1.12 Neurophet

### 1.12.1 Team Members and Affiliations
Team Members: ZunHyan Rieu[1], Donghyeon Kim[1], Hyun Gi Kim[2]
Affiliations: [1]NEUROPHET, Republic of Korea; [2]The Catholic University of Korea, Eunpyeong St. Mary's Hospital, Republic of Korea

### 1.12.2 Model Description

The base segmentation architecture we used is U-Net, and we designed it to receive its input patch 3D with the size of (64, 64, 64). We used patch-based training since the actual brain size MRI is pretty small compared to its MRI dimensions (256, 256, 256). Our U-Net design consists of encoding blocks of 5 followed by residual blocks. We applied a probability-based sampling method to avoid our model being trained for non-label patches.

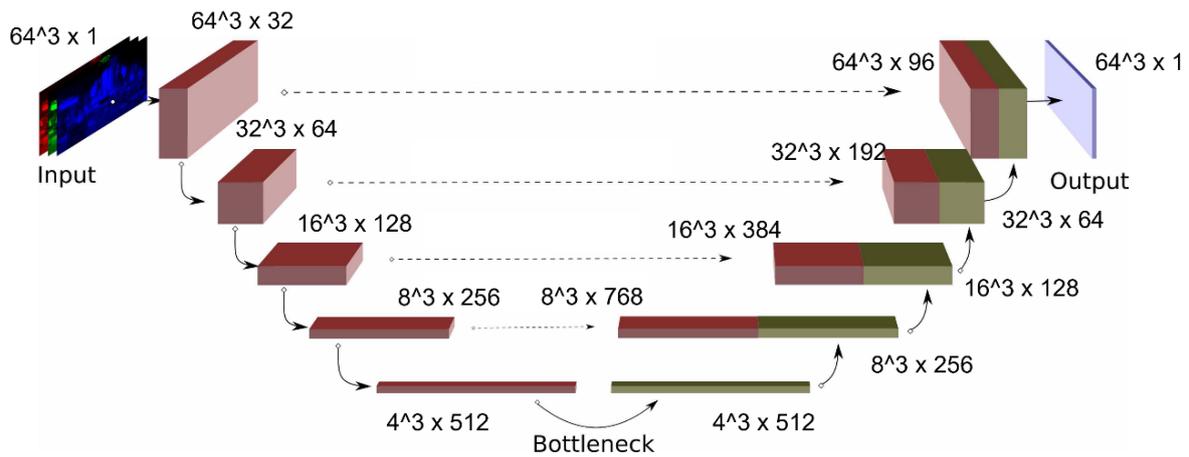

As for the loss function, we used the sum of cross-entropy loss function and dice loss function. In cross-entropy loss function, specifically, we applied different weights in each class to define the vague segmentation boundaries between 'External Cerebrospinal Fluid' and 'Grey Matter, and 'Brainstem'. For the optimizer, we applied AdamW with the learning rate of 1E-5. To normalize the intensity range of the given MRI dataset, we performed intensity normalization from 0 to 1 with a percentile cutoff of (0, 99.8).

### 1.12.3 Training Method
Neither an additional dataset nor a pre-trained model was used for training. All case of FeTA dataset was used for training. However, we performed a visual inspection to define the poor and good quality images and distributed them equally to the train and validation. For the train and validation dataset, we divided 80 subjects with a split ratio of 0.8 (training:64, validation:16).

To reduce the dependency on the training dataset, we performed the data augmentation (random affine, random blur). For the hyper-parameter tuning, we applied the inferences for multiple patch overlap combinations (8 x 8 x 8, 16 x 16 x 16, 32 x 32 x 32, 48 x 48 x 48) and chose the best out of it.

Once the output is generated, we removed the mislabeled outer blobs using the connected component-based label noise removal method.

All preprocessing and architecture modeling, we used torchIO and scikit-image library.

Inference Example

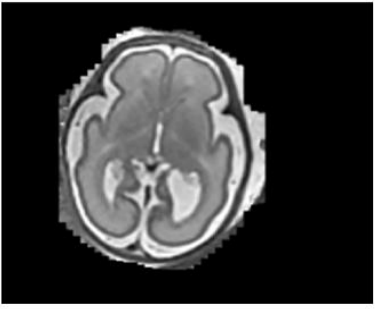 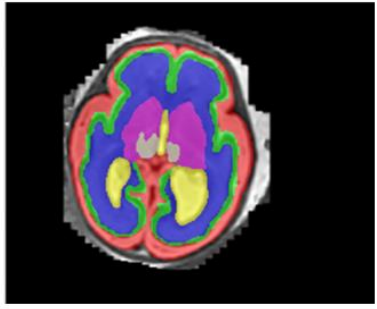 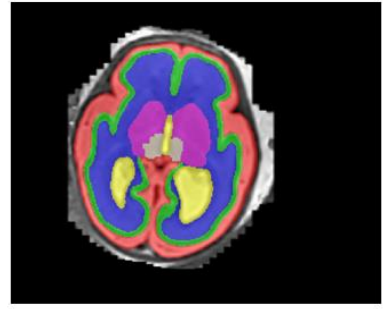

| MRI | Ground-truth | Prediction |

## 1.13 NVAUTO

### 1.13.1 Team Members and Affiliations
Team Members: Md Mahfuzur Rahman Siddiquee[1], Andriy Myronenko[2], Daguang Xu[2]
Affiliations: [1]Arizona State University; [2]NVIDIA, Santa Clara, CA

### 1.13.2 Model Description
We implemented our approach with MONAI (https://github.com/Project-MONAI/MONAI) [1]. We use the encoder-decoder backbone based on [4] with an asymmetrically larger encoder to extract image features and a smaller decoder to reconstruct the segmentation mask [6, 8, 9]. We have also utilized the OCR module from [7].

Encoder part: The encoder part uses ResNet [2] blocks. We have used 5 stage of down-sampling, each stage have 1, 2, 2, 4, and 4 convolutional blocks, respectively. We have used batch normalization and ReLU. Each block's output is followed by additive identity skip connection. We follow a common CNN approach to progressively downsize image dimensions by 2 and simultaneously increase feature size by 2. For downsizing we use strided convolutions. All convolutions are 3x3x3 with initial number of filters equal to 32. The encoder is trained with
224x224x144 input region.

Decoder part: The decoder structure is similar to the encoder one, but with a single block per each spatial level. Each decoder level begins with upsizing with transposed convolution: reducing the number of features by a factor of 2 and doubling the spatial dimension, followed by an addition of encoder output of the equivalent spatial level. The end of the decoder has the same spatial size as the original image, and the number of features equal to the initial input feature size, followed by 1x1x1 convolution into 8 channels and a softmax.

### 1.13.3 Training Method
Dataset: We have used the FeTA dataset [5] only for training the model. We have randomly split the entire dataset into 5-folds and trained a model for each.

Loss: We have used Dice loss for training [3].

Optimization: We use AdamW optimizer with initial learning rate of 2E-4 and decrease it to zero at the end of final epoch using Cosine annealing scheduler. We have used a batch-size of 4. The model is trained of 4 GPUs, each GPU optimizing for batch-size of 1. However, we have calculated batch normalization across all the GPUs. We have ensembled 10 models for the submission: 5 of these models contain OCR module [7] and the rest do not. All the models were trained for 300 epochs.

Regularization: We use L2 norm regularization on the convolutional kernel parameters with a weight of 1E-5.

Data preprocessing and augmentation: We normalize all input images to have zero mean and unit std (based on nonzero voxels only). We have applied random flip on each axis, random rotation, and random zoom with probability of 0.5. We have also applied random contrast adjustment, random Gaussian noise, and random Gaussian smoothing with probability of 0.2.

### 1.13.4 Results on Cross-Validation
Our cross-validation results on the 5-folds can be found in Table 1.

Table 1. Average DICE among classes using 5-fold cross-validation.

| Fold 1 | Fold 2 | Fold 3 | Fold 4 | Fold 5 | Average |
|---|---|---|---|---|---|
| 0.8531 | 0.8515 | 0.8304 | 0.8284 | 0.8481 | 0.8423 |

### 1.13.5 References

1. Project-monai/monai, https://doi.org/10.5281/zenodo.5083813
2. He, K., Zhang, X., Ren, S., Sun, J.: Identity mappings in deep residual networks. In: European conference on computer vision. pp. 630-645. Springer (2016)
3. Milletari, F., Navab, N., Ahmadi, S.A.: V-net: Fully convolutional neural networks for volumetric medical image segmentation. In: 2016 fourth international conference on 3D vision (3DV). pp. 565-571. IEEE (2016)
4. Myronenko, A.: 3D MRI brain tumor segmentation using autoencoder regularization. In: International MICCAI Brainlesion Workshop. pp. 311-320. Springer (2018)
5. Payette, K., de Dumast, P., Kebiri, H., Ezhov, I., Paetzold, J.C., Shit, S., Iqbal, A., Khan, R., Kottke, R., Grehten, P., et al.: An automatic multi-tissue human fetal brain segmentation benchmark using the fetal tissue annotation dataset. Scientific Data 8(1), 1-14 (2021)

6. Ronneberger, O., Fischer, P., Brox, T.: U-net: Convolutional networks for biomedical image segmentation. In: International Conference on Medical image computing and computer-assisted intervention. pp. 234-241. Springer (2015)
7. Yuan, Y., Chen, X., Wang, J.: Object-contextual representations for semantic segmentation. In: Computer Vision - ECCV 2020: 16th European Conference, Glasgow, UK, August 23-28, 2020, Proceedings, Part VI 16. pp. 173-190. Springer (2020)
8. Zhou, Z., Siddiquee, M.M.R., Tajbakhsh, N., Liang, J.: Unet++: A nested u-net architecture for medical image segmentation. In: Deep learning in medical image analysis and multimodal learning for clinical decision support, pp. 3-11. Springer (2018)
9. Zhou, Z., Siddiquee, M.M.R., Tajbakhsh, N., Liang, J.: Unet++: Redesigning skip connections to exploit multiscale features in image segmentation. IEEE transactions on medical imaging 39(6), 1856-1867 (2019)

## 1.14 pengyy

### 1.14.1 Team Members and Affiliations

Team Members: Ying Peng[1], Juanying Xie[1], Huiquan Zhang[1]
Affiliations: [1]School of Computer Science, Shaanxi Normal University, Xi'an 710119, PR China

### 1.14.2 Model Description

Our network is based on the residual 3D U-Net. The architecture of our network is shown in Figure 1. Residual blocks in the encoder consist of two convolution blocks and a shortcut connection. If the number of input/output channels is different in a residual block, a non-linear projection is performed by adding the 1×1×1 convolutional block to the shortcut connection, so as to match the dimensions. Down-sampling in the encoder is done by a specific stride convolution. The decoder of the network consists of a stack of Conv-IN-Leaky ReLU blocks. The 2×2×2 transposed convolutions are used to realize up-sampling the feature maps at each layer of decoder. There are 32 channels of feature map in our network initially. The feature map channels will be doubled by each down-sampling operation in the encoder, until 320 feature maps at last. The feature maps will be halved by each transposed convolution in the decoder. At last the size at the end of decoder is same as the spatial size in the input end. Then is a convolution layer with 8 channels and the 1×1×1 convolution kernel size and a Softmax function.

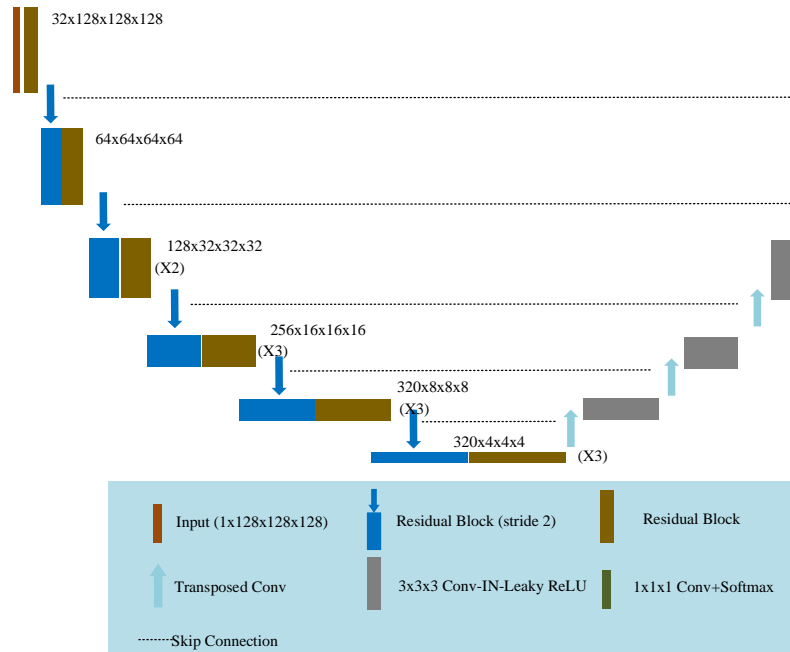

**Figure 1:** The architecture of 3D ResU-Net model.

### 1.14.3 Data Preprocessing

Since the images in the FeTA dataset have different scanning orientations, we reorient all images to the RAI orientation. At the same time we also resample the images to the isotropic resolution of 0.5mm×0.5mm×0.5mm. The detail is that the third order spline interpolation and linear interpolation are used to the images and labels, respectively. Finally, the Z-score (mean subtraction and division by standard deviation) is used to normalize each image. We only use the official FeTA dataset to evaluate the proposed method without using any other datasets.

### 1.14.4 Implementation Details

The loss function used in our network is the one combining the cross-entropy and dice loss. The optimizer is SGD (Stochastic Gradient Descent). The initial learning rate is 0.01, and the nesterov momentum is 0.99. The mini-batch size is 2, and the number of epochs is 1000. The patch size is 128 × 128 × 128. 10-fold cross validation experiments are carried out by randomly partitioning the training set into 10 folds. All our models are trained by starting from scratch, without using any pre-trained model. To avoid overfitting, standard data augmentation techniques are used during training procedure, such as random rotation, random scaling, random elastic deformation, mirroring, adding Gaussian noise, and Gamma correction. The program is implemented on the basis of nnU-Net [1] and the experimental environment is Python 3.7 with Pytorch 1.8. The nnU-Net repository is available at: https://github.com/mic-dkfz/nnunet. All the experiments were conducted on GeForce RTX 3090 GPU with 24 GB memory. Training each fold costs about 24 hours.

There are 10 models obtained by using 10-fold cross validation experiments. These 10 models are ensemble together to predict the images in the test set. In particular, test time augmentation technique is employed by mirroring along all axes.

Our method is different from the original nnU-Net, due to the data are re-rotated and training data are partitioned using different way.

### 1.14.5 References

[1] Isensee, F., Jaeger, P. F., Kohl, S. A., Petersen, J., & Maier-Hein, K. H. (2020). nnU-Net: a self-configuring method for deep learning-based biomedical image segmentation. Nature Methods, 1-9.

## 1.15 Physense-UPF Team

### 1.15.1 Team Members and Affiliations

Team Members: Mireia Alenyà[1], Maria Inmaculada Villanueva[2], Mateus Riva, Oscar Camara[1]
Affiliations: [1]BCN-MedTech, Department of Information and Communications Technologies, Universitat Pompeu Fabra, Barcelona, Spain; [2]Department of Information and Communications Technologies, Universitat Pompeu Fabra, Barcelona, Spain; [3]Institut d'Investigacions Biomèdiques August Pi i Sunyer, Barcelona, Spain

### 1.15.2 Model Description

The deep learning model used for this challenge was the nnU-Net, publicly available https://github.com/MIC-DKFZ/nnUNet [1]. The network features that have improved the performance the most have been the generalised Dice [2], which accounts for the volume of segmented structures and the implementation of data augmentation. These two features are described in more detail in the following sections.

Additionally, several attempts have been made in order to improve the segmentation accuracy, whether in the pre-processing stage or in the computation of the loss function. Those are described in the section *Report of all tested models*.

The submitted model presents a modification in the computation of the loss function. It still uses Dice and cross entropy terms as the original network does, but the former has been modified to be a generalised Dice as presented in [2]. This new Dice metric assesses multilabel segmentations with a unique score, assigning a different weight for each structure according to its volume.
It is calculated as follows:

$$DICE_{ml} = \frac{2TC_{ml}}{TC_{ml} + 1}, TC_{ml} = \frac{\sum_{labels,l} \propto_l \sum_{voxels,i} \min(GT_{li}, X_{li})}{\sum_{labels,l} \propto_l \sum_{voxels,i} \propto_l \max(GT_{li}, X_{li})}, \propto_l = \frac{1}{V_l}$$

where $TC_{ml}$ is the multilabel Tanimoto coefficient [2]; $GT_{li}$ is the value of voxel $i$ for label $l$ in the ground-truth segmentation; $X_{li}$ is the analogous for the predicted one; $\propto_l$ is the label-specific weighting factor that affects how much each structure $l$ contributes to the overlap accumulated over all labels; and $V_l$ is the volume of each label $l$.

To run five folds (switching subset of samples for training and validation in each of them), implement cross validation and data augmentation are possible options already defined in the nnU-Net codes.

Model architecture: The model architecture is based on an encoder–decoder with skip-connection ('U-Net-like') and instance normalization, leaky ReLU and deep supervision [1]. Its scheme is shown in Figure 1.

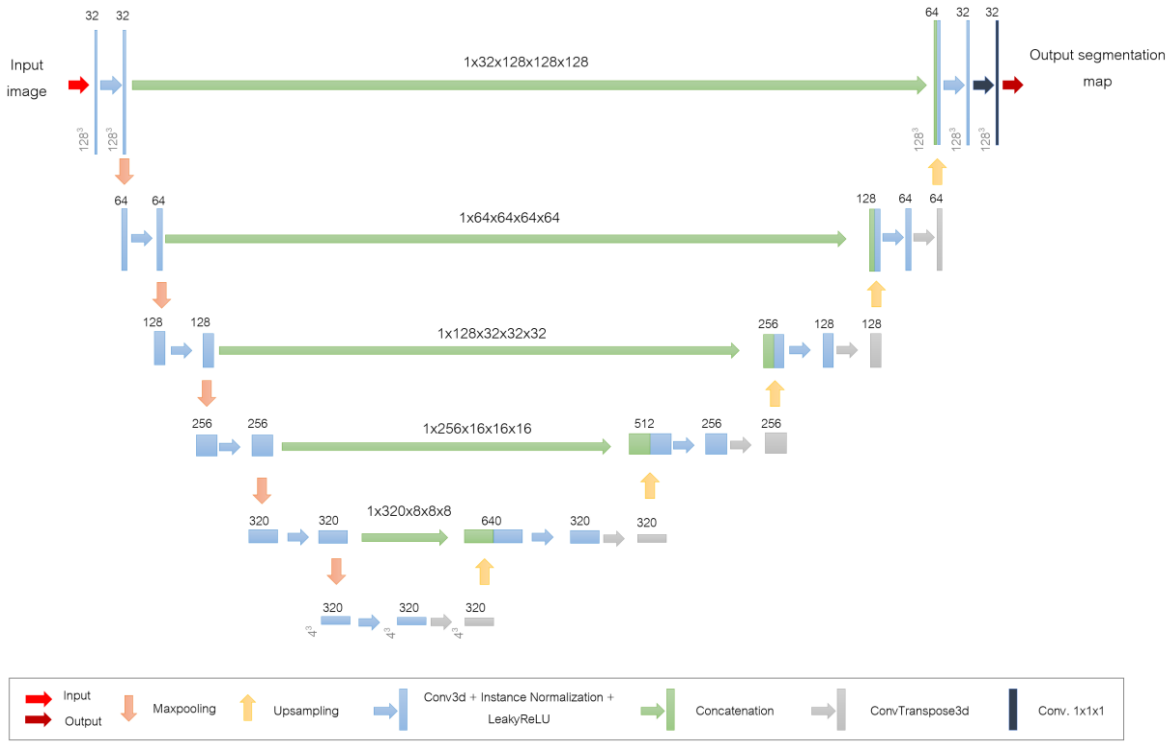

**Figure 1:** nnU-Net architecture. Each box corresponds to a multi-channel feature map. The arrows denote different operations. The number of channels is denoted on top of the box. The x-y-z-size is provided at the lower left edge of the box.

The loss function is based on Dice and cross-entropy, while the optimizer is the stochastic gradient descent (SGD) with Nesterov momentum (μ = 0.99). Conv3d layers are of kernel size [3, 3, 3], stride [1, 1, 1] or [2, 2, 2] and padding [1, 1, 1]. LeakyReLU is applied with negative slope = 0.01 and Instance Normalization is applied with parameters: eps=1e-05, momentum=0.1, affine=True.

ConvTranspose3d layers present the following characteristics, depending on the level in which they are applied:
(0): ConvTranspose3d(320, 320, kernel_size=[2, 2, 2], stride=[2, 2, 2], bias=False)
(1): ConvTranspose3d(320, 256, kernel_size=[2, 2, 2], stride=[2, 2, 2], bias=False)
(2): ConvTranspose3d(256, 128, kernel_size=[2, 2, 2], stride=[2, 2, 2], bias=False)
(3): ConvTranspose3d(128, 64, kernel_size=[2, 2, 2], stride=[2, 2, 2], bias=False)
(4): ConvTranspose3d(64, 32, kernel_size=[2, 2, 2], stride=[2, 2, 2], bias=False)

Inputs/outputs: Inputs and outputs have been FeTA images and labels in 3D. The used network is 3D-nnU-Net with full resolution. The patch size has been: [128 128 128].

Pre-processing of inputs: Pre-processing done on the inputs is cropping, resampling (if necessary) and data normalisation.

The parameters' configuration is:
- batch_size: 2
- num pool per axis: [5, 5, 5]
- patch size: array([128, 128, 128])
- spacing: array([0.5, 0.5, 0.5])
- pool kernel sizes: [[2, 2, 2], [2, 2, 2], [2, 2, 2], [2, 2, 2], [2, 2, 2]]
- convolution kernel sizes: [[3, 3, 3], [3, 3, 3], [3, 3, 3], [3, 3, 3], [3, 3, 3], [3, 3, 3]]

Initialization of model parameters: The initialization of model parameters is fully optimized by the network after extracting the dataset fingerprint (a set of dataset-specific properties such as image sizes, voxel spacings, intensity information etc). The initial loss rate is 0.01.

### 1.15.3 Training method:

To train the network, cases from 1 to 60 have been used while the 20 remaining have served to test the network performance.
In the final version of the model no additional datasets have been used. The main reason for this is that no public fetal datasets were known and neither were found.

Optimization: splits of the datasets (training, validation, testing), learning rate/batch size:
The submitted model includes the independent run of five different folds, in each of which different samples have been used for training and validation (48 - 12) during training. Then, cross validation between those five folds is applied.

Data augmentation strategies: The following data augmentation features have been applied:
- Rotation along each axis, range (-15º,15º)
- Elastic deformation
- Scaling, range (0.85, 1.25)
- Add Gaussian noise, range (0, 0.1)
- Add Gaussian blur, range (0.5, 1)
- Gamma Transform, range (0.7, 1.5)
- Mirror along all axes
- Additive brightness transform, range (0.75, 1.25)
- Contrast transform, range (0.75, 1.25)
- Simulate low resolution transform, zoom range (0.5, 1)

### 1.15.4 Report of all tested models:

Several models with different approaches have been trained, and then used to predict segmentations, whose results were evaluated with the following metrics: Hausdorff Distance, Dice coefficient and volume similarity index.

Changes in the network configuration:
- Single label models (segmenting all 7 tissues separately) have been compared with multilabel models, obtaining better results with the latter. Not only in terms of accuracy but also in terms of computing time and memory optimization.

- Options of network architecture: 3D-UNet with full resolution, 2D-UNet, 3D-UNet Cascade have been tested. Best results were obtained with the 3D-UNet full resolution configuration.

Changes in the pre-processing stage:
- Registration of FeTA images and labels to the publicly available Gholipour ATLAS has been performed using the Advanced Normalization Tools (ANTs) software. FeTA data of subjects of GA higher than 27 has been registered to the ATLAS of GA 31 weeks, while FeTA data of subjects of less or equal than 27 weeks has been registered to the ATLAS of 25 weeks. This allows us to have all the input data in the same position and orientation by applying rigid transformations. Results obtained with this implementation are worse than the ones with the original images.
- Addition of gestational age as an extra input channel in the medical images has been implemented, but it has not supposed any improvements in the model performance.

Modifying the loss function:
- The Dice loss term computation has been changed for the generalised Dice. This implementation led to an improvement in the performance of the model, so it is included in the version submitted.
- Spatial constraints by relational graphs [3] have been applied. To do so, the centroids of label structures have been computed on ground-truth data and the L2 distances between the labels of adjacent structures have been defined as prior distances. Differences between priors and distances between predicted labels have been computed during training and have been included as an additional term in the loss function.
    - Centroids and respective L2 distances have been computed on the whole brain (centroids of 7 labels computed).
    - Centroids and distances have been computed on separated hemispheres (centroids of 14 labels computed). To separate the labels according to hemispheres: (1) FeTA data has been registered to the Gholipour ATLAS (Fetal Brain Atlas (harvard.edu)) by applying a rigid transform using ANTs (GitHub - ANTsX/ANTs: Advanced Normalization Tools (ANTs)); (2) After having all images aligned, labels of the right hemisphere have been defined as *label_number*+7; (3) centroids have been computed on those 14 labels.
    - None of those two approaches lead to an improvement of the model performance, so they have not been included in the presented version.

Others:
- Curriculum learning [4] methods impose a gradual and systematic way of learning, first training with best quality images before considering the low-quality ones. A qualitative classification of the quality of the medical images was made between high=1, medium=2 and low=3 by different members of the team. The data loader of the nnU-Net was modified accordingly, in order to start training with the annotated high-quality images (easier to segment correctly), continue with the medium and ending with the low-quality images (presenting more difficulty or uncertainty). This implementation does not improve the performance of the model either.
- Other details: software libraries and packages used, including version, training time:
    - Python 3.8 with an Anaconda virtual environment has been used. The nnU-Net runs in the HPC cluster of Universitat Pompeu Fabra (UPF) using a GPU and the computing platform is CUDA 10.2.
    - The average training time was of ~48 hours to run 100 epochs of the model.
    - Required Libraries:
        - nibabel 3.2.1
        - glob 0.7
        - os 0.1.4
        - numpy 1.20.3
        - argparse 1.4.0
        - Pytorch 1.6

### 1.15.5 References

[1] Isensee, F., Jaeger, P. F., Kohl, S. A., Petersen, J., & Maier-Hein, K. H. (2020). nnU-Net: a self-configuring method for deep learning-based biomedical image segmentation. Nature Methods, 1-9.
[2] Crum, W. R., Camara, O. & Hill, D. L. G. (2006). Generalized overlap measures for evaluation and validation in medical image analysis. IEEE Trans. Med. Imaging 25, 1451–1461.
[3] Riva, M. (2018) A new calibration approach to graph-based semantic segmentation. Master's thesis, Institute of Mathematics and Statistics.
[4] Y. Bengio, J. Louradour, R. Collobert, and J. Weston, "Curriculum learning," in Proceedings of the 26th International Conference on Machine Learning, ser. ICML 2009. New York, NY, USA: ACM, 2009, pp. 41–48.

## 1.16 SingleNets

### 1.16.1 Team Members and Affiliations

Team Members: Bella Specktor Fadida[1], Leo Joskowicz[1], Dafna Ben Bashat[2,3], Netanell Avisdris[1,2]
Affiliations: [1]School of Computer Science and Engineering, The Hebrew University of Jerusalem, Israel; [2]Sagol Brain Institute, Tel Aviv Sourasky Medical Center, Israel; [3]Sackler Faculty of Medicine & Sagol School of Neuroscience, Tel Aviv University, Israel

### 1.16.2 Background

Fetal brain structures segmentation is important for quantitative brain evaluation. Usually, works on brain segmentation focus on a single brain structure or a general brain segmentation. A question remains whether a method that is used for a single structure can be effectively used for multi-structure segmentation.

Multiple works have been proposed for single structure fetal brain segmentation including a work by Dudovitch et al [1] that reached a Dice score of 0.96 for whole brain segmentation using only 9 training cases.

Isensee et al. [2] observed that different frameworks should be used for segmenting different structures. The authors propose a single network framework, a single network on downscaled data framework and a cascaded framework of detection followed by segmentation. In [3] a novel contour dice loss function was proposed for segmentation of fuzzy boundaries that was shown to be effective for placenta segmentation.

### 1.16.3 Model Description

For brain structure multi-class prediction, we used 3D networks trained separately for each brain sub-structure and for the background ("Skull" network). The network architecture was the same for all networks and is similar to Dudovitch et al [1] (Figure 1). To combine the predictions, we used the maximum network response for each voxel. Figure 2 illustrates our method.

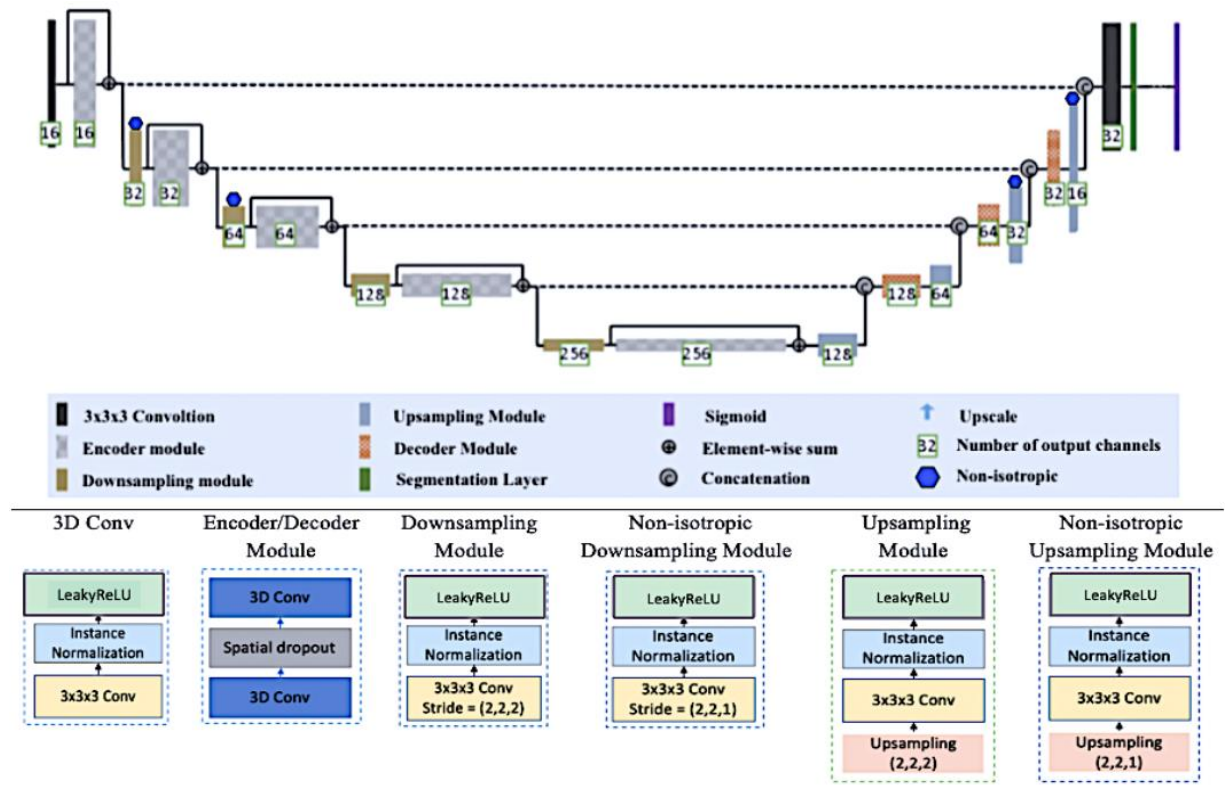

**Fig. 1.** Top: Architecture of 3D segmentation network. The number of output channels of each unit is indicated next to it. Bottom: (a-f) network modules details.

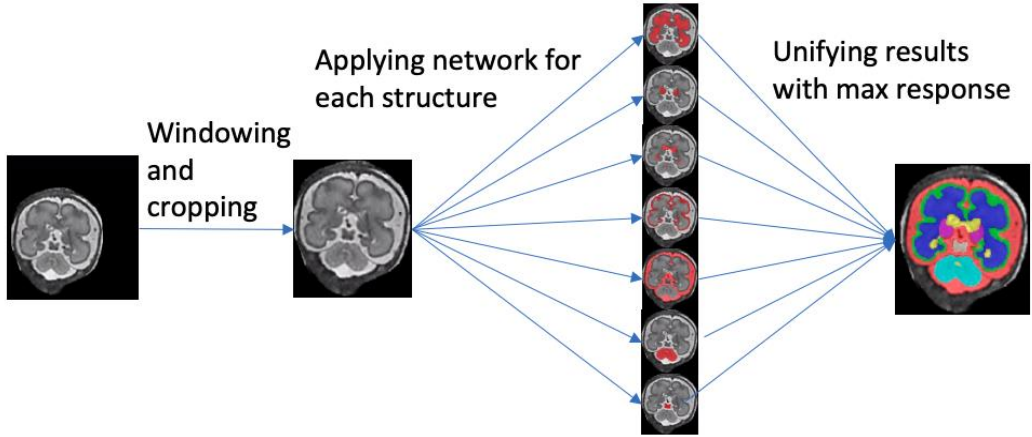

**Fig. 2** Illustration of multi-class fetal brain structure segmentation method using networks for each structure

Segmentation Frameworks: Single downscaled network for most structures. Data was downscaled by a factor of 0.5 in all planes. Cascaded framework for CBM (Cerebellum) and BS (Brainstem). ROI extraction was performed by a segmentation network on downscaled data by a factor of 0.5 and segmentation network was trained on the original scale. Most networks used soft dice loss and some used contour dice loss in combination with soft dice based on validation results. Table 1 summarizes network configurations for each structure.

**Table 1.** Frameworks and loss functions for each one of the structures

|  | Network framework | Loss Function |
| --- | --- | --- |
| Skull | Single | Contour Dice + Soft Dice |
| CSF | Single | Soft Dice |
| GM | Single | Soft Dice |
| WM | Single | Soft Dice |
| LV | Single | Contour Dice + Soft Dice |
| CBM | Cascade | Soft Dice for both detection and segmentation networks |
| SGM | Single | Soft Dice |
| BS | Cascade | Soft dice for detection, Contour Dice + Soft Dice for segmentation |

Preprocessing:
1. Bounding box around the data with voxel intensities above th=2.
2. Windowing on intensity values so that the top and bottom 1% intensities were discarded.
3. Data normalization to zero mean and std of 1.
4. Downscaling by 0.5 in all planes for increased input ROI (block size of [96,96,96] was used due to GPU memory limitations).

Postprocessing: Voxels that were predicted as background by the Skull network were set to 0.

### 1.16.4 Contour Dice (CD) Loss Method – new formulation

We implemented a new version of Contour Dice (CD) loss which is closer to the definition of Contour Dice metric used in previous works [4,5]. Formally, let $\partial T$ and $\partial S$ be the extracted surfaces of the ground truth delineation and the network results, respectively and let $B_{\partial T}$ and $B_{\partial S}$ be their respective offset bands. The contour Dice loss of the offset bands is:

$$\text{Contour Dice Loss (T,S)} = -\frac{|\partial T \cap B_{\partial S}| + |\partial S \cap B_{\partial T}|}{|\partial T| + |\partial S|}$$

Although the Contour Dice is a non-differentiable function, in practice the contours $\partial T$ and $\partial S$ have a width, and $|\partial T|$ and $|\partial T|$ are summations over contour voxels. The formula can therefore be directly used as a loss function.

### 1.16.5 Training Details

For development, we randomly split the data to 55/25 training and validation examples. For deployment, best network setups were retrained on 75 training cases.

We did not use metadata information at all during training. Initial learning rate of 0.005 was used with learning rate drop. Models were trained with early stopping of 20 epochs. For all networks we used block size of 96×96×96, batch size of 2 and the augmentations: flipping, rotation, translation, scaling, Poisson noise, contrast, and intensity multiplication.

For Skull network we used CD loss with band size of 0 and for LV and BS segmentation networks we used band size of 3. Networks with Soft Dice loss were fine-tuned from previous best networks.

Software Packages: For inference, we used cuda=9.0, tensorflow_gpu=1.8.0, Keras=2.1.5, nibabel=2.5.0, scipy=1.3.3, numpy=1.17.4, h5py=2.0, keras_contrib=2.0.8, SimpleITK=2.1.0 and tqdm=4.48.2 packages. The training was performed on a cluster environment with cuda=10.0 and tensorflow_gpu=1.14.0.

### 1.16.6 Results

To combine the networks, we tested two different approaches against one another:
1. Stacking results one on top of the other where structures order is CSF<GM<WM<LV<CBM<SGM<BS. Holes were filled with maximum network response.
2. Maximum networks response for each voxel.

Preliminary results on 55/25 training/validation split are depicted in Table 2. Figure 3 shows illustrative results examples using each one of the approaches. Combining segmentation results of the structures with maximum response approach resulted in better segmentation results compared to structures stacking for almost all brain structures.

**Table 2.** Results comparison between stacking and max response

|       |              | Dice  | VOD   | Hausdorff 95 2D Avg | ASSD 2D Avg |
|-------|--------------|-------|-------|---------------------|-------------|
| Skull | Stacking     | 0.971 | 0.057 | 2.88                | 0.79        |
|       | Max Response | 0.971 | 0.057 | 2.88                | 0.79        |
| CSF   | Stacking     | 0.806 | 0.307 | 4.39                | 1.08        |
|       | Max Response | 0.811 | 0.298 | 4.36                | 1.11        |
| GM    | Stacking     | 0.717 | 0.436 | 3.54                | 0.77        |
|       | Max Response | 0.723 | 0.429 | 3.45                | 0.76        |
| WM    | Stacking     | 0.895 | 0.189 | 4.53                | 0.99        |
|       | Max Response | 0.896 | 0.186 | 4.65                | 1.03        |
| LV    | Stacking     | 0.796 | 0.326 | 7.22                | 1.59        |
|       | Max Response | 0.808 | 0.311 | 7.17                | 1.52        |
| CBM   | Stacking     | 0.818 | 0.283 | 4.06                | 1.21        |
|       | Max Response | 0.834 | 0.265 | 4.01                | 1.24        |
| SGM   | Stacking     | 0.777 | 0.349 | 5.79                | 1.88        |
|       | Max Response | 0.793 | 0.328 | 5.61                | 1.86        |
| BS    | Stacking     | 0.723 | 0.418 | 5.53                | 1.62        |
|       | Max Response | 0.736 | 0.400 | 5.19                | 1.54        |

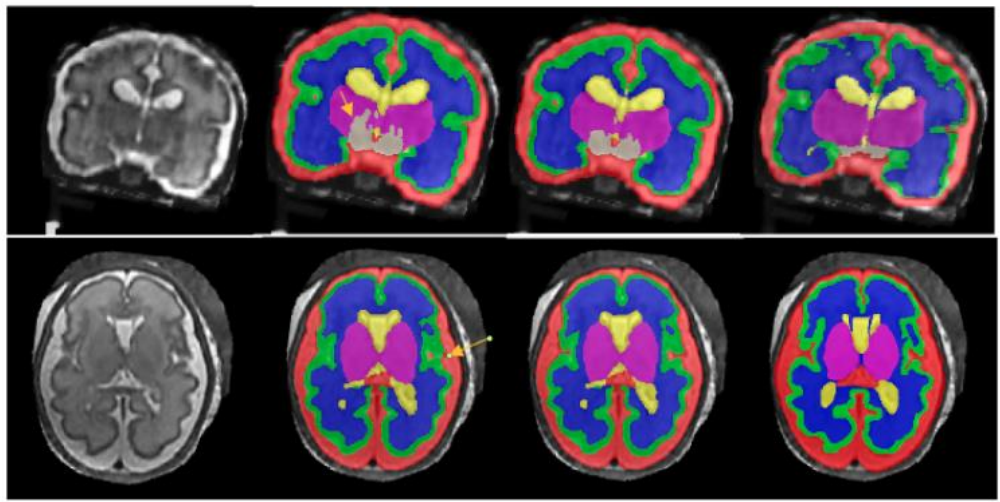

**Figure 3.** Results comparison between stacking approach (2nd column) and maximum response (3rd column). 4th column is the ground truth

## 1.16.7 References

1. Dudovitch G, Link-Sourani D, Sira LB, Miller E, Bashat DB, Joskowicz L. Deep Learning Automatic Fetal Structures Segmentation in MRI Scans with Few Annotated Datasets. Proc. Int. Conf. Medical Image Computing and Computer-Assisted Interventions pp365-74, 2020.
2. Isensee F, Jaeger PF, Kohl SA, Petersen J, Maier-Hein KH. nnU-Net: a self-configuring method for deep learning-based biomedical image segmentation. Nature Methods 18(2):203-11, 2021.
3. Specktor Fadida B, Link Sourani D, Ferster-Kveller S, Ben Sira L, Miller E, Ben Bashat D, Joskowicz L. A bootstrap self-training method for sequence transfer: State-of-the-art placenta segmentation in fetal MRI. Perinatal, Preterm and Pediatric Image Analysis MICCAI workshop 2021.
4. Nikolov S, Blackwell S, Mendes R, De Fauw J, Meyer C, Hughes C, Askham H, Romera- Paredes B, Karthikesalingam A, Chu C, Carnell D. Deep learning to achieve clinically applicable segmentation of head and neck anatomy for radiotherapy. arXiv preprint arXiv:1809.04430, Sep 12, 2018.
5. Moltz JH, Hänsch A, Lassen-Schmidt B, Haas B, Genghi A, Schreier J, Morgas T, Klein J. Learning a loss function for segmentation: a feasibility study. In Proc. IEEE 17th Int. Symp. on Biomedical Imaging, pp. 357-360), 2020.

## 1.17 SJTU_EIEE_2-426Lab

### 1.17.1 Team Members and Affiliations
Team Members Hao Liu, Yuchen Pei, Huai Chen, and Lisheng Wang
Affiliations: Team SJTU_EIEE_2-426LAB, Shanghai Jiao Tong University

### 1.17.2 Model Description
Data: For the task of fetal brain tissue segmentation, we propose a cumbersome coarse-to-fine segmentation framework inspired by [1], which divides the segmentation process into two stages. In the first stage, the coarse model segments all the 7 tissues at a time. Then in the second stage, with the regions of interest (ROIs) determined by the coarse result from the first step, the segmentation of each tissue is refined by a separate model, respectively. Specifically, we use nnU-Net [1] as well as a vanilla 3D U-Net [2] with residual architecture [3] (referred to below as Res-U-Net) for the coarse stage, and then we train 5 3D Res-U-Net models separately to refine the segmentation of white matter, ventricles, cerebellum, deep grey matter, and brainstem.

All the 3D Res-U-Net models we use share the similarly architecture. The detailed structure of Res-U-Net in the coarse stage is shown in Figure 1.

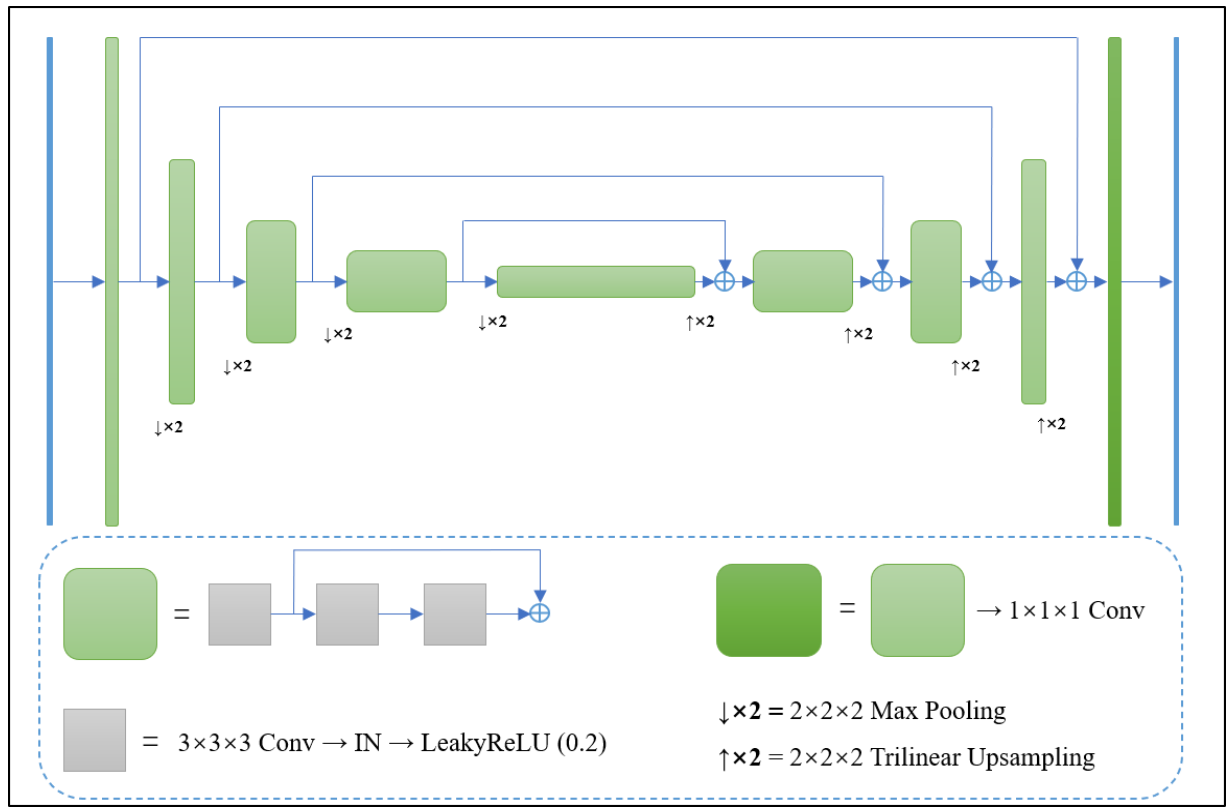

**Figure 1:** The network architecture of the Res-U-Net for coarse segmentation. There are 8 filters in the first convolution layer with a ratio of 2 between subsequent spatial levels and the last layer of the encoder has 128 filters.

Since we use small ROIs of corresponding tissues as the input of refinement networks, the width (i.e., the number of filters in intermediate convolutional layers) of the refinement networks is allowed to be larger. For the model to refine the segmentation of white matter, the width of the network is two times than the coarse network and there is one more spatial level where convolution layers have 256 filters. The other refinement networks are four times wider than the coarse network and also have an additional spatial level where convolution layers have 512 filters.
For the nnU-Net used in the first stage, we follow the default setting.

Loss Function: In the first stage, we use the combination of cross-entropy and dice loss to train the multi-class segmentation networks. The weights of two losses are both 1. In the second stage, since the 95th percentile Hausdorff Coefficient (HD95) is one of the evaluation metrics, we replace the cross-entropy loss with a distance transform-based HD loss function [4] to reduce HD directly. At the end of each training epoch, the weight for HD loss is modified to the ratio of the mean of the dice loss term on the training data to that of the HD loss term.

Pre-processing: In the first stage, we first crop the nonzero regions of each fetal brain reconstructions and then pad them into the size of 256×256×256. Then, they are normalized to zero mean and unit variance and used as the input data of the 3D Res-U-Net. The preprocessing for nnU-Net is still left to its default. In the second stage, we crop the ROIs of tissues according to the segmentation results in the first stage.

Specifically, the ROIs of white matter and ventricles are determined by the segmentation results of nnU-Net, while the ROIs of cerebellum and deep grey matter are selected on the basis of the segmentation by Res-U-Net. With regard to brainstem, the union of the two coarse segmentation masks is used to crop the ROIs. Note that we do not refine the segmentation of external cerebrospinal fluid and grey matter. All the cropped ROIs are slightly bigger than the bounding box and the size of each dimension is evenly divisible by 32 since the input data will go through 5 downsampling layers. And in order to inherit the results in the first stage, we concatenate the cropped ROIs and their corresponding segmentation predictions from the coarse models, which are used as the input of the refinement networks in the second stage.

### 1.17.3 Training Method

Data: We split all the provided data randomly into 64 training cases and 16 test cases. And since our coarse models in the first stage works poorly on the Case 5 and Case 7, we remove them from test dataset and use the remaining 14 cases to select refinement models. We do not use any additional datasets.

The data augmentations used for training the Res-U-Net in the first stage includes random rotation, random scale and random flip in each dimension. And we use only random flip in the second stage.

Training Details: All the training details mentioned below are for our Res-U-Net models and the nnU-Net is trained with default setting. The optimizer of network parameters of all the Res-U-Net is performed via Adam optimizer. All the model parameters are initialized randomly. Learning rate is initialized as 1e-3 in the first stage and 1e-4 in the second stage. The weight decay is set as 1e-5 and the batch size is 1. We train the network in the first stage for 500 epochs and the refinement networks in the second stage for 1000 epochs at most. We only save the model with highest dice coefficient on the test dataset in the first stage, while in the second stage, the models with the highest dice coefficient or the lowest HD are both saved. Our algorithm is implemented in Python 3.8.5 using Pytorch 1.7.1 framework. Experiments are performed on one NVIDIA RTX 3090 with 24GB of RAM memory.

### 1.17.4 Inference Strategy and Post-processing

We use two inference strategies in the refinement stage to improve the quality of segmentation. First, for each input volume, the result is produced by the mean of predictions by the two saved models. Second, for each case, we predict the results of 8 variants obtained through flips in each dimension, and then take the average of them as the final result.

The mask of refinement segmentation is generated by the threshold of 0.5. And if there is overlap of segmentation masks for different tissues, the tissue with the maximum average volume on the whole dataset is treated as the predicted label for the overlapping regions.

### 1.17.5 References

1. Huai, Chen & Wang, Xiuying & Huang, Yijie & Wu, Xiyi & Yu, Yizhou & Wang, Lisheng. (2019). Harnessing 2D Networks and 3D Features for Automated Pancreas Segmentation from Volumetric CT Images. 339-347. 10.1007/978-3-030-32226-7_38.
2. Isensee, Fabian & Jaeger, Paul & Kohl, Simon & Petersen, Jens & Maier-Hein, Klaus. (2021). nnU-Net: a self-configuring method for deep learning-based biomedical image segmentation. Nature Methods. 18. 1-9. 10.1038/s41592-020-01008-z.
3. Ronneberger, Olaf & Fischer, Philipp & Brox, Thomas. (2015). U-Net: Convolutional Networks for Biomedical Image Segmentation. LNCS. 9351. 234-241. 10.1007/978-3-319-24574-4_28.
4. He, Kaiming & Zhang, Xiangyu & Ren, Shaoqing & Sun, Jian. (2016). Deep Residual Learning for Image Recognition. 770-778. 10.1109/CVPR.2016.90.
5. Karimi, Davood & Salcudean, Septimiu. (2019). Reducing the Hausdorff Distance in Medical Image Segmentation with Convolutional Neural Networks. IEEE Transactions on Medical Imaging. PP. 1-1. 10.1109/TMI.2019.2930068.

## 1.18 TRABIT

### 1.18.1 Team Members and Affiliations

Team Members: Lucas Fidon[1], Michael Aertsen[2], Suprosanna Shit[3], Philippe Demaerel[2], Sébastien Ourselin[1], Jan Deprest[2,4,5], and Tom Vercauteren[1]

Affiliations: [1]School of Biomedical Engineering & Imaging Sciences, King's College London, UK; [2]Department of Radiology, University Hospitals Leuven, Belgium; [3]Technical University of Munich, Germany; [4]Institute for Women's Health, University College London, UK; [5]Department of Obstetrics and Gynaecology, University Hospitals Leuven, Belgium

### 1.18.2 Objective

The Fetal Brain Tissue Annotation and Segmentation Challenge (FeTA) aims at comparing algorithms for multi-class automatic segmentation of fetal brain 3D T2 MRI. Seven tissue types are considered [13]:
1. extra-axial cerebrospinal fluid
2. cortical gray matter
3. white matter
4. ventricular system
5. cerebellum
6. deep gray matter
7. brainstem

This paper describes our method for our participation in the FeTA challenge 2021 (team name: TRABIT).

The performance of convolutional neural networks for medical image segmentation is thought to correlate positively with the number of training data [1]. The FeTA challenge does not restrict participants to using only the provided training data but also allows for using other publicly available sources. Yet, open access fetal brain data remains limited. An advantageous strategy could thus be to expand the training data to cover broader perinatal brain imaging sources. Perinatal brain MRIs, other than the FeTA challenge data, that are currently publicly available, span normal and pathological fetal atlases as well as neonatal scans [4,5,7,17]. However, perinatal brain MRIs segmented in different datasets typically come with different annotation protocols. This makes it challenging to combine those datasets to train a deep neural network.

We recently proposed a family of loss functions, the label-set loss functions [3], for partially supervised learning. Label-set loss functions allow to train deep neural networks with partially segmented images, i.e. segmentations in which some classes may be grouped into super-classes. We propose to use label-set loss functions [3] to improve the segmentation performance of a state-of-the-art deep learning pipeline for multi-class fetal brain segmentation by merging several publicly available datasets. To promote generalisability, our approach does not introduce any additional hyper-parameters tuning.

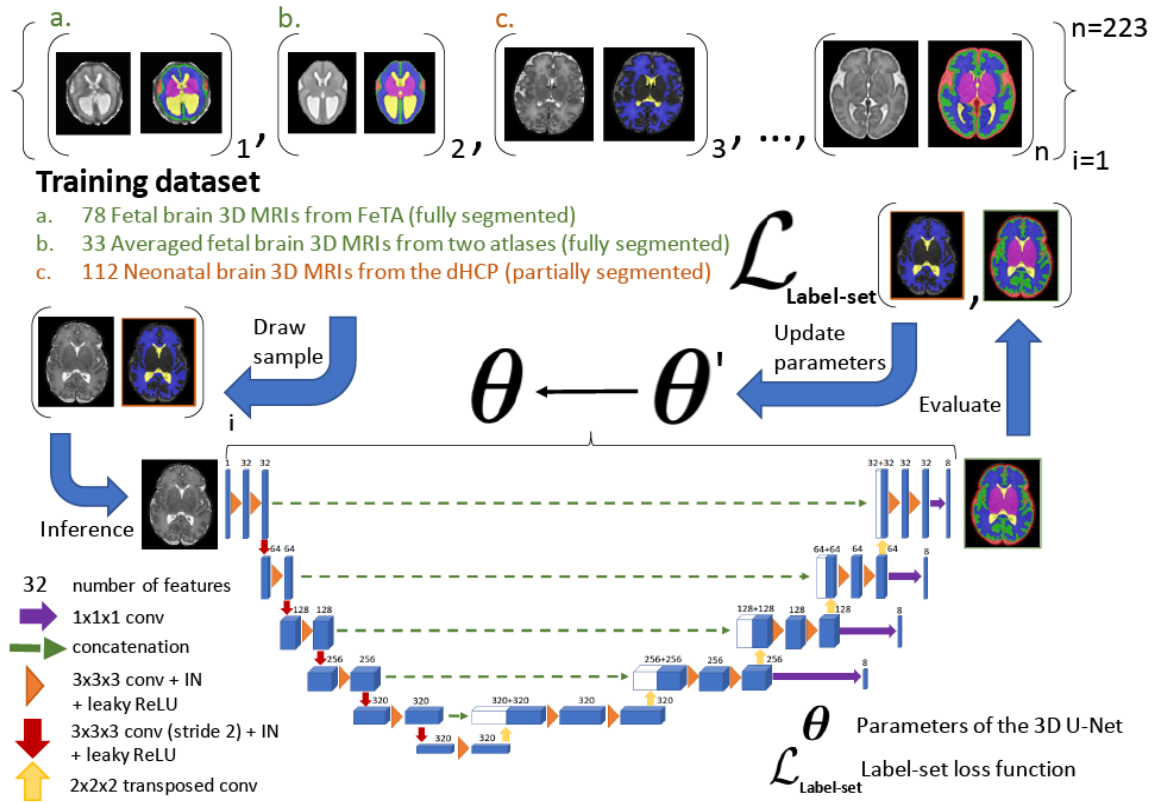

**Figure 1:** Overview of the training process with partial supervision. The 3D MRIs from datasets a. [13] and b. [4,5] are fully-segmented while the 3D MRIs from dataset c. [7] were only segmented manually for white matter, ventricular system, cerebellum, and the full brain. We propose to use a label-set loss function [3] to allow to train a 3D U-Net [2] with fully and partially segmented 3D MRIs.

1.18.3   **Materials and Methods**

In this section, we give the detail of our segmentation pipeline and the data used for training the deep neural networks. Our segmentation software will be publicly available.

<u>FeTA challenge amended training data:</u> The original FeTA challenge training data provides 80 fetal brain 3D T2 MRIs with manual segmentations of all 7 target tissue types [13]. 40 fetal brain 3D MRIs were reconstructed using MIAL [14] and 40 fetal brain 3D MRIs were reconstructed using Simple IRTK [8].

For the 40 MIAL 3D MRIs, corrections of the segmentations were performed by authors MA, LF, and PD using ITK-SNAP [18] to reduce the variability against the published segmentation guidelines that was released with the FeTA dataset [13]. Those corrections were performed as part of our previous work [3] and are publicly available (DOI: 10.5281/zenodo.5148611). Two spina bifida cases were excluded (sub-feta007 and sub-feta009) because we considered that the image quality did not allow to segment them reliably for all the tissue types. Only the remaining 78 3D MRIs were used for training.

<u>Other public training data:</u> We also included 18 average fetal brain 3D T2 MRIs from a neurotypical fetal brain atlas (http://crl.med.harvard.edu/research/fetal_brain_atlas/) [5], 15 average fetal brain 3D T2 MRIs from a spina bifida fetal brain atlas (https://www.synapse.org/#!Synapse:syn25887675/wiki/611424) [4]. Segmentations for all 7 tissue types are available for all the atlas data. In addition, we used 112 neonatal brain MRIs from the developing human connectome project [7] (dHCP data release 2). We excluded the brain MRIs of babies with a gestational age higher than 38 weeks. We started from the brain masks and the automatic segmentations publicly available for the neonatal brain MRIs for white matter, ventricular system, and cerebellum [9] and we modified them manually to match the annotation protocol of the FeTA dataset [13] using ITK-SNAP [18]. Ground-truth segmentations for the other tissue types in the dHCP data were not available for our training.

<u>Pre-processing:</u> A brain mask for the fetal brain 3D MRI is computed by affine registration of template volumes from two fetal brain atlases [4,5]. We use all the template volumes with a gestational age that does not differ to the gestation age of fetal brain by more than 1:5 weeks. The affine registrations are computed using a symmetric block-matching approach [10] as implemented in NiftyReg [11]. The affine transformations are initialized by a translation that aligns the centre of gravity of the non-zero intensity regions of the two volumes. The brain mask is obtained by averaging the warped brain mask and thresholding at 0.5.

After a brain mask has been computed, the fetal brain 3D MRI is registered rigidly to a neurotypical fetal brain atlas [5] and the 3D MRI resampled to a resolution of 0.8 mm isotropic. The rigid registration is computed using NiftyReg [10,11] and the transformation is initialized by a translation that aligns the centre of gravity of the brain masks.

Deep learning pipeline: We used an ensemble of 10 3D U-Nets [2]. We used the DynU-Net of MONAI [12] to implement a 3D U-Net with one input block, 4 down-sampling blocks, one bottleneck block, 5 up-sampling blocks, 32 features in the first level, instance normalization [15], and leaky-ReLU with slope 0.01. An illustration of the architecture is provided in Fig. 1. The CNN used has 31195784 trainable parameters. The patch size was set to 128 x 160 x 128. All the pre-processed 3D MRIs on which the pipeline was tested fitted inside a patch of size 128 x 160 x 128. Every input volume is skull stripped after dilating the brain mask by 5 voxels and cropped or padded with zeros to fit the patch size. The non-zeros image intensity values are clipped for the values above percentile 99.9, and normalized to zeros mean and unit variance. Test-time augmentation [16] with all the combinations of flipping along the three spatial dimensions is performed (8 predictions). The 8 score map predictions are averaged to obtain the output of each CNN. Ensembling is obtained by averaging the softmax predictions of the 10 CNNs. The deep learning pipeline was implemented using MONAI v5.2.0 by authors LF and SS.

Loss function: We used the sum of two label-set loss functions as loss function: the Leaf-Dice loss [3] and the marginalized cross entropy loss [3].

Optimization: For each network in the ensemble, the training dataset was split into 90% training and 10% validation at random. The random initialization of the 3D U-Net weights was performed using He initialization [6]. We used SGD with Nesterov momentum, batch size 2, weight decay $3 \times 10^{-5}$, initial learning rate 0.01, and polynomial learning rate decay with power 0.9 for a total of 2200 epochs. The CNN parameters used at inference corresponds to the last epoch. We used deep supervision with 4 levels during training. Training each 3D U-Net required 12GB of GPU memory and took on average 3 days. We have trained exactly 10 CNNs and used all of them for the ensemble submitted to the challenge.

Data augmentation: We used random zoom (zoom ratio range [0.7, 1.5] drawn uniformly at random; probability of augmentation 0:3), random rotation (rotation angle range [-15°; 15°] for all dimensions drawn uniformly at random; probability of augmentation 0.3), random additive Gaussian noise (mean 0, standard deviation 0.1; probability of augmentation 0.3), random Gaussian spatial smoothing (standard deviation range [0.5; 1.5] in voxels for all dimensions drawn uniformly at random; probability of augmentation 0.2), random gamma augmentation (gamma range [0.7; 1.5] drawn uniformly at random; probability of augmentation 0.3), and random flip along all dimension (probability of augmentation 0.5 for each dimension).

Post-processing: The mean softmax prediction is resampled to the original 3D MRI using the inverse of the rigid transformation computed in the pre-processing step to register the 3D MRI to the template space. This image registration is computed using NiftyReg [11] with an interpolation order equal to 1. After resampling, the final multi-class segmentation prediction is obtained by taking the argmax of the mean softmax.

### 1.18.4    Conclusion

Partially supervised learning can be used to train deep neural networks using multiple publicly available perinatal brain 3D MRI datasets that have different level of segmentations available. We used label-set loss functions [3] to train an ensemble of 3D U-Nets using four publicly available datasets [4,5,7,13]. We have submitted our segmentation algorithm to the FeTA challenge 2021.

### 1.18.5    Funding Sources

This project has received funding from the European Union's Horizon 2020 research and innovation program under the Marie Sklodowska-Curie grant agreement TRABIT No 765148. This work was supported by core and project funding from the Wellcome [203148/Z/16/Z; 203145Z/16/Z; WT101957], and EP-SRC [NS/A000049/1; NS/A000050/1; NS/A000027/1]. TV is supported by a Medtronic / RAEng Research Chair [RCSRF1819\7\34].

### 1.18.6    References

1. Bakas, S., Reyes, M., Jakab, A., Bauer, S., Remper, M., Crimi, A., Shinohara, R.T., Berger, C., Ha, S.M., Rozycki, M., et al.: Identifying the best machine learning algorithms for brain tumor segmentation, progression assessment, and overall survival prediction in the brats challenge. arXiv preprint arXiv:1811.02629 (2018)
2. Cicek, O., Abdulkadir, A., Lienkamp, S.S., Brox, T., Ronneberger, O.: 3d u-net: learning dense volumetric segmentation from sparse annotation. In: International conference on medical image computing and computer-assisted intervention. pp. 424-432. Springer (2016)
3. Fidon, L., Aertsen, M., Emam, D., Mufti, N., Guffens, F., Deprest, T., Demaerel, P., David, A.L., Melbourne, A., Ourselin, S., et al.: Label-set loss functions for partial supervision: Application to fetal brain 3d mri parcellation. arXiv preprint arXiv:2107.03846 (2021)
4. Fidon, L., Viola, E., Mufti, N., David, A., Melbourne, A., Demaerel, P., Ourselin, S., Vercauteren, T., Deprest, J., Aertsen, M.: A spatio-temporal atlas of the developing fetal brain with spina bifida aperta. Open Research Europe (2021)

5. Gholipour, A., Rollins, C.K., Velasco-Annis, C., Ouaalam, A., Akhondi-Asl, A., Afacan, O., Ortinau, C.M., Clancy, S., Limperopoulos, C., Yang, E., et al.: A normative spatiotemporal MRI atlas of the fetal brain for automatic segmentation and analysis of early brain growth. Scientific reports 7(1), 1-13 (2017)

6. He, K., Zhang, X., Ren, S., Sun, J.: Delving deep into rectifiers: Surpassing human-level performance on imagenet classification. In: Proceedings of the IEEE international conference on computer vision. pp. 1026-1034 (2015)

7. Hughes, E.J., Winchman, T., Padormo, F., Teixeira, R., Wurie, J., Sharma, M., Fox, M., Hutter, J., Cordero-Grande, L., Price, A.N., et al.: A dedicated neonatal brain imaging system. Magnetic resonance in medicine 78(2), 794-804 (2017)

8. Kuklisova-Murgasova, M., Quaghebeur, G., Rutherford, M.A., Hajnal, J.V., Schnabel, J.A.: Reconstruction of fetal brain mri with intensity matching and complete outlier removal. Medical image analysis 16(8), 1550-1564 (2012)

9. Makropoulos, A., Robinson, E.C., Schuh, A., Wright, R., Fitzgibbon, S., Bozek, J., Counsell, S.J., Steinweg, J., Vecchiato, K., Passerat-Palmbach, J., et al.: The developing human connectome project: A minimal processing pipeline for neonatal cortical surface reconstruction. Neuroimage 173, 88-112 (2018)

10. Modat, M., Cash, D.M., Daga, P., Winston, G.P., Duncan, J.S., Ourselin, S.: Global image registration using a symmetric block-matching approach. Journal of Medical Imaging 1(2), 024003 (2014)

11. Modat, M., Ridgway, G.R., Taylor, Z.A., Lehmann, M., Barnes, J., Hawkes, D.J., Fox, N.C., Ourselin, S.: Fast free-form deformation using graphics processing units. Computer methods and programs in biomedicine 98(3), 278-284 (2010)

12. MONAI Consortium: MONAI: Medical open network for AI (3 2020). https://doi.org/10.5281/zenodo.4323058, https://github.com/Project-MONAI/MONAI

13. Payette, K., de Dumast, P., Kebiri, H., Ezhov, I., Paetzold, J.C., Shit, S., Iqbal, A., Khan, R., Kottke, R., Grehten, P., et al.: An automatic multi-tissue human fetal brain segmentation benchmark using the fetal tissue annotation dataset. Scientific Data 8(1), 1-14 (2021)

14. Tourbier, S., Bresson, X., Hagmann, P., Thiran, J.P., Meuli, R., Cuadra, M.B.: An efficient total variation algorithm for super-resolution in fetal brain mri with adaptive regularization. NeuroImage 118, 584-597 (2015)

15. Ulyanov, D., Vedaldi, A., Lempitsky, V.: Instance normalization: The missing ingredient for fast stylization. arXiv preprint arXiv:1607.08022 (2016)

16. Wang, G., Li, W., Aertsen, M., Deprest, J., Ourselin, S., Vercauteren, T.: Aleatoric uncertainty estimation with test-time augmentation for medical image segmentation with convolutional neural networks. Neurocomputing 338, 34-45 (2019)

17. Wu, J., Sun, T., Yu, B., Li, Z., Wu, Q., Wang, Y., Qian, Z., Zhang, Y., Jiang, L., Wei, H.: Age-specific structural fetal brain atlases construction and cortical development quantification for chinese population. NeuroImage p. 118412 (2021)

18. Yushkevich, P.A., Gao, Y., Gerig, G.: ITK-SNAP: an interactive tool for semiautomatic segmentation of multi-modality biomedical images. In: 2016 38th Annual International Conference of the IEEE Engineering in Medicine and Biology Society (EMBC). pp. 3342-3345. IEEE (2016)

## 1.19 Xlab

### 1.19.1 Team Members and Affiliations
Team Members: Yang Lin

Affiliations: Department of Computer Science, Hong Kong University of Science and Technology

### 1.19.2 Model description
The model uses the architecture similar to U-net [1]. The contracting path consists of five repetition of two 3*3 convolutions, a rectified linear unit and a 2*2 max pooling with stride 2. In each down sampling, the number of feature channels is doubled. The expansive path consists of an up sampling of the feature map, a 2x2 convolution, a concatenation with the correspondingly cropped feature map from the contracting path, and two 3x3 convolutions, each followed by a ReLU. The final layer is a 1x1 convolution mapping each 64- component feature vector to the desired number of classes. The difference is that padding is added in the convolutions, so that the size of the feature maps won't change inside each repetition, as a result, more information will be transferred to the next layer. And batch norm is used instead of instance norm.

The preprocessing step is similar to nnUNet [2]. But more pre-process on the input images are designed after observing the features of FeTA images.

### 1.19.3 Training method
All cases of FeTA are used in the training progress. The loss function is the combination of dice loss and cross entropy loss. The datasets are split to training set and validation and 5-fold cross validation is used. 5 models are generated in the 5-fold cross validation. Strategies includes mirroring, random rotation, random scaling, gamma correction and random elastic transformation are used.
The training in done with CUDA 11.0, Ubuntu 20, python 3.8 and pytorch 1.8 and takes about 5 days and has 1000 epochs.

### 1.19.4 Reference

[1] Olaf Ronneberger, Philipp Fischer, Thomas Brox. U-Net: Convolutional Networks for Biomedical Image Segmentation.
[2] Isensee, F., Jaeger, P. F., Kohl, S. A., Petersen, J., & Maier-Hein, K. H. (2020). nnU-Net: a self-configuring method for deep learning-based biomedical image segmentation. Nature Methods, 1-9.

## 1.20 ZJUWULAB

### 1.20.1 Team Members and Affiliations

Team Members: Zelin Zhang[1], Xinyi Xu[1], Dan Wu[1]

Affiliations: [1]Key Laboratory for Biomedical Engineering of Ministry of Education, Department of Biomedical Engineering, College of Biomedical Engineering & Instrument Science, Zhejiang University, Yuquan Campus, Hangzhou, China

### 1.20.2 Model Description

Our segmentation network adopts a standard 2D Unet framework, but we replace all 2x max pool down sampling layers with 2x convolution down sampling layers. Our loss function consists of two parts: L1 regularization loss function and feature matching loss function. In the feature matching loss function, we cascade the predicted output and corresponding label of the model with the original input image respectively and input them into the pretrained vgg19 network to calculate the L1 loss from the deep features output by each convolution layers, and finally carry out back propagation training model after weighted average of all feature matching losses.

The input to our model is a series of 2D slices with the size of 256x256.

The input preprocessing includes 2 steps: 1) Normalize all voxel values of one case to the interval [0, 255]; 2) Convert class mapping of 8 categories into color mapping (Each category corresponds to the combination of three channels in the RGB color map, in which each channel has only 0 and 1.); 3) Extract all 2D slices and convert to RGB images.

The initialization of the model adopts the method of random initialization.

The whole model is implemented under the framework of pytoch 1.0.

### 1.20.3 Training method

We used all cases to optimize our model. We did not use any additional public datasets. The whole dataset is divided into training set, validation set and test set with the ratio of 7:2:1. We use the Adam optimizer, the initialization learning rate is 0.002, and the batch size of training is 32. We did not use any data augmentation methods. According to the divided validation set and test set, we select the checkpoint with the best accuracy in 100 epochs as the final submitted model.

Other details: Pytorch 1.0; CUDA 10.1; SimpleITK 1.2.4; Nvidia GPU RTX 3080 ti x 4; Training time: 6.8h.

## 2. Benchmarking report for multiTaskChallengeDice_combined

created by challengeR v1.0.2

12 April, 2022

This document presents a systematic report on the benchmark study "multiTaskChallengeDice_combined". Input data comprises raw metric values for all algorithms and cases. Generated plots are:

- Visualization of assessment data: Dot- and boxplot, podium plot and ranking heatmap
- Visualization of ranking stability: Blob plot, violin plot and significance map, line plot

Details can be found in Wiesenfarth et al. (2021).

### 2.1 Ranking

Algorithms within a task are ranked according to the following ranking scheme:

*aggregate using function ("mean") then rank*

The analysis is based on 21 algorithms and 280 cases. 0 missing cases have been found in the data set.

Ranking:

|                     | Dice_mean | rank |
|---------------------|-----------|------|
| NVAUTO              | 0.7858525 | 1    |
| SJTU_EIEE_2-426Lab  | 0.7753286 | 2    |
| Neurophet           | 0.7745291 | 3    |
| Pengyy              | 0.7744333 | 4    |
| Hilab               | 0.7735193 | 5    |
| davoodkarimi        | 0.7709845 | 6    |
| Xlab                | 0.7709073 | 7    |
| TRABIT              | 0.7685622 | 8    |
| Physense-UPF Team   | 0.7674980 | 9    |
| 2Ai                 | 0.7669346 | 10   |
| ichilove-ax         | 0.7660280 | 11   |
| muw_dsobotka        | 0.7648522 | 12   |
| ichilove-combi      | 0.7619306 | 13   |
| Moona_Mazher        | 0.7550563 | 14   |
| BIT_LILAB           | 0.7518695 | 15   |
| SingleNets          | 0.7478066 | 16   |
| MIAL                | 0.7402952 | 17   |
| ZJUWULAB            | 0.7026920 | 18   |
| FeVer               | 0.6826219 | 19   |
| Anonymous           | 0.6207273 | 20   |
| A3                  | 0.5335483 | 21   |

## 2.2 Visualization of raw assessment data

### 2.2.1 Dot- and boxplot

*Dot- and boxplots* for visualizing raw assessment data separately for each algorithm. Boxplots representing descriptive statistics over all cases (median, quartiles and outliers) are combined with horizontally jittered dots representing individual cases.

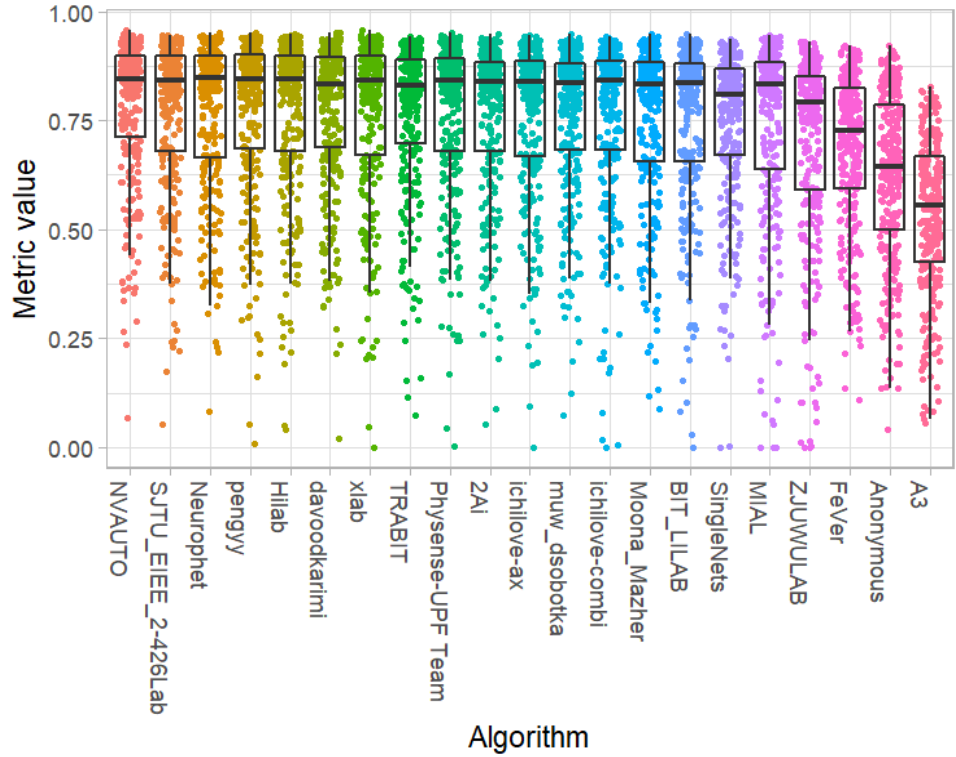

### 2.2.2 Podium plot

*Podium plots* (see also Eugster et al., 2008) for visualizing raw assessment data. Upper part (spaghetti plot): Participating algorithms are color-coded, and each colored dot in the plot represents a metric value achieved with the respective algorithm. The actual metric value is encoded by the y-axis. Each podium (here: $p$=21) represents one possible rank, ordered from best (1) to last (here: 21). The assignment of metric values (i.e. colored dots) to one of the podiums is based on the rank that the respective algorithm achieved on the corresponding case. Note that the plot part above each podium place is further subdivided into $p$ "columns", where each column represents one participating algorithm (here: $p = 21$). Dots corresponding to identical cases are connected by a line, leading to the shown spaghetti structure. Lower part: Bar charts represent the relative frequency for each algorithm to achieve the rank encoded by the podium place.

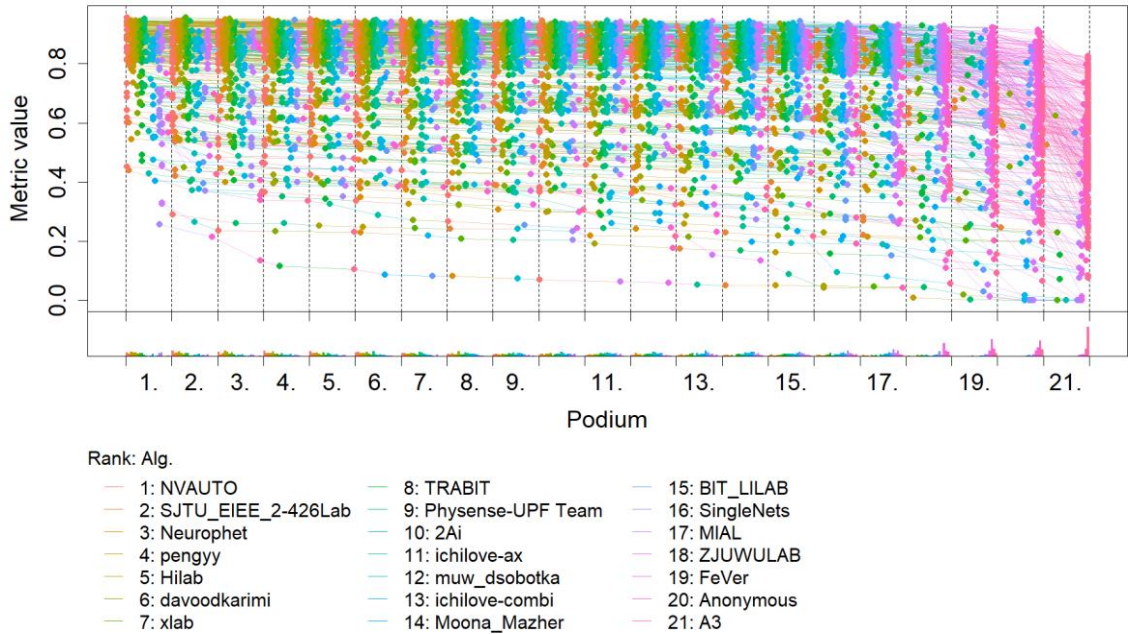

### 2.2.3 Ranking heatmap

*Ranking heatmaps* for visualizing raw assessment data. Each cell $(i, A_j)$ shows the absolute frequency of cases in which algorithm $A_j$ achieved rank $i$.

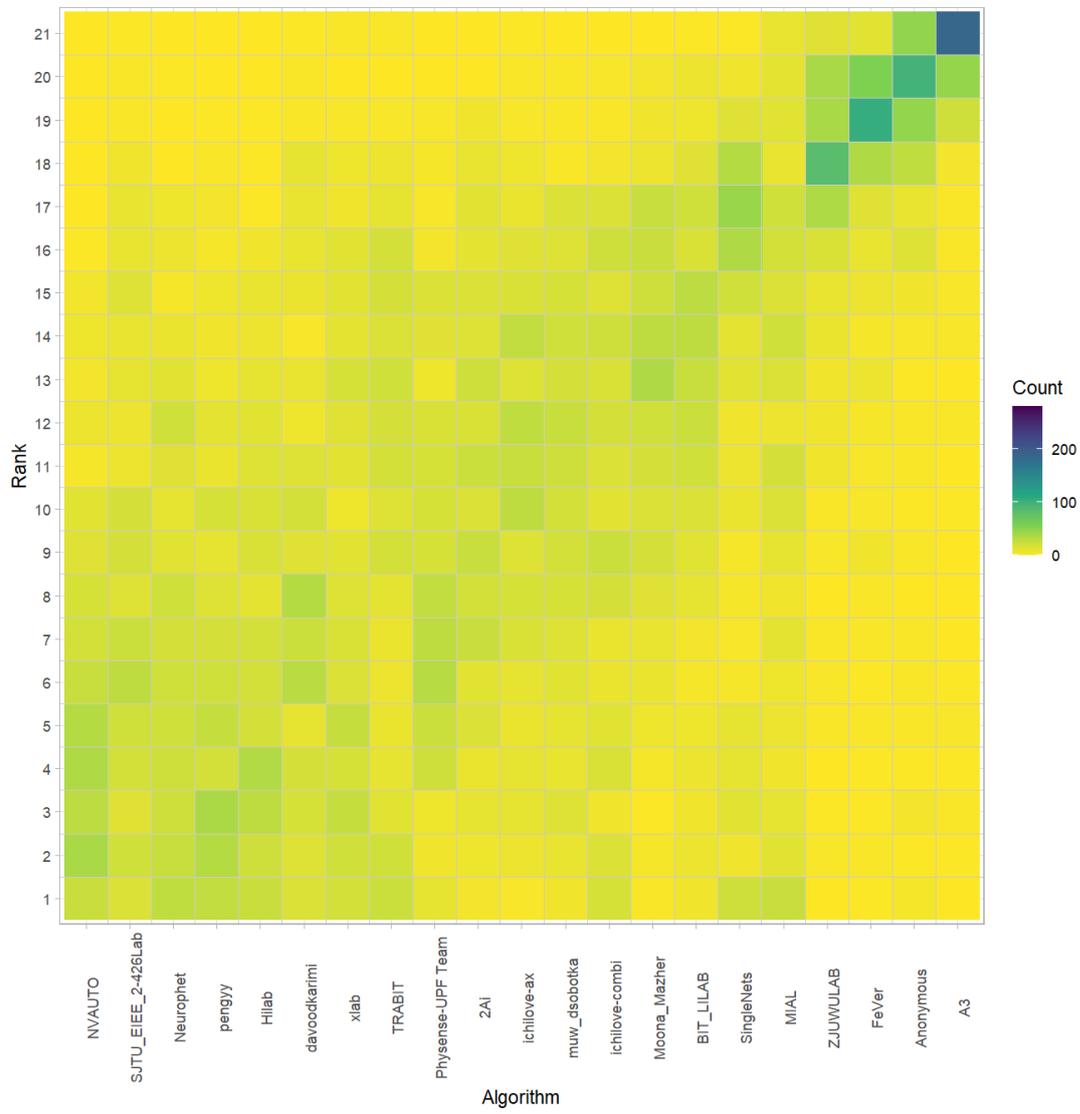

## 2.3 Visualization of ranking stability

### 2.3.1 *Blob plot* for visualizing ranking stability based on bootstrap sampling

Algorithms are color-coded, and the area of each blob at position $(A_i, \text{rank } j)$ is proportional to the relative frequency $A_i$ achieved rank $j$ across $b = 1000$ bootstrap samples. The median rank for each algorithm is indicated by a black cross. 95% bootstrap intervals across bootstrap samples are indicated by black lines.

```
## Warning: `guides(<scale> = FALSE)` is deprecated. Please use `guides(<scale> =
## "none")` instead.
```

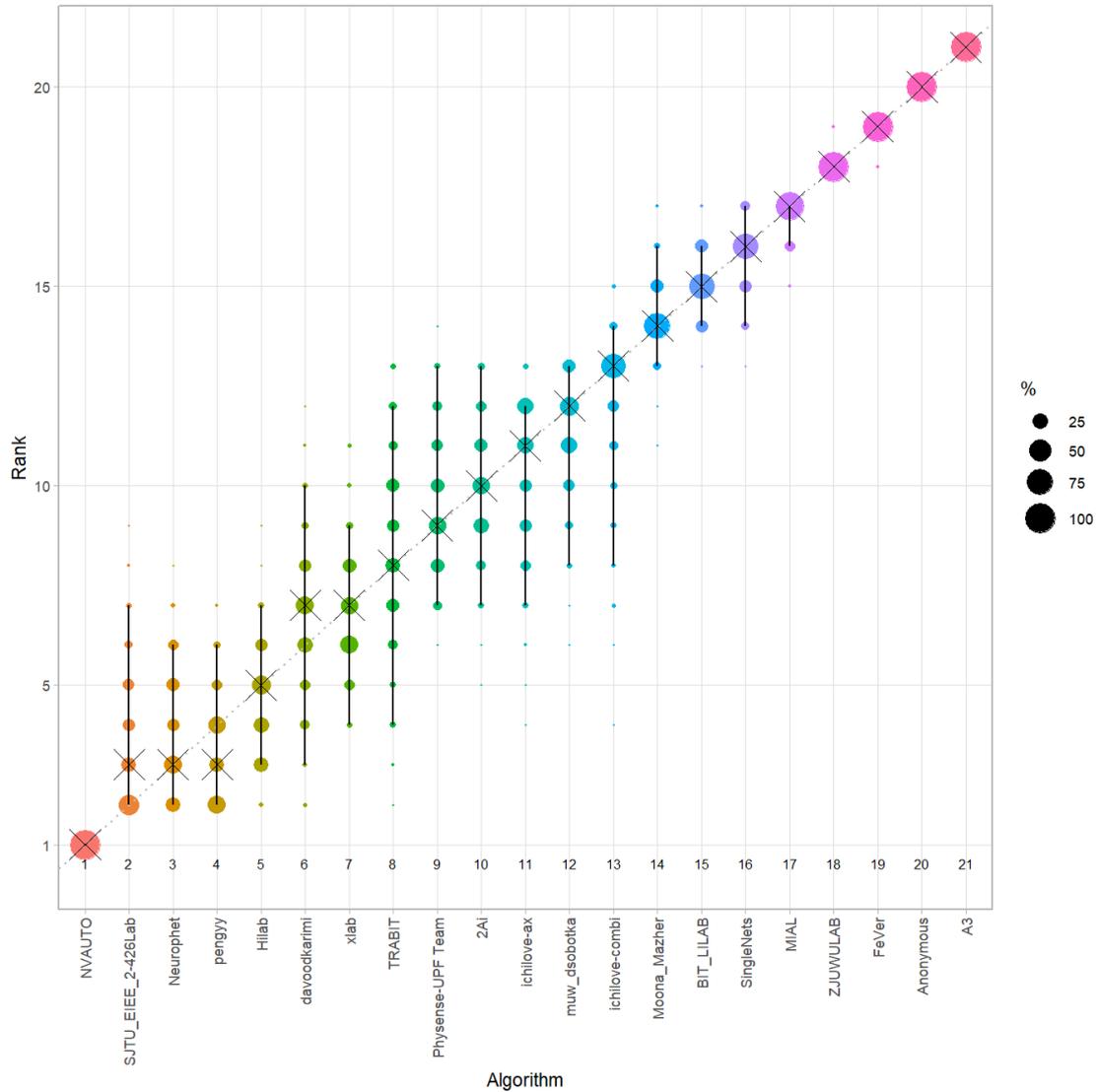

2.3.2 *Violin plot* for visualizing ranking stability based on bootstrapping

The ranking list based on the full assessment data is pairwise compared with the ranking lists based on the individual bootstrap samples (here $b = 1000$ samples). For each pair of rankings, Kendall's $\tau$ correlation is computed. Kendall's $\tau$ is a scaled index determining the correlation between the lists. It is computed by evaluating the number of pairwise concordances and discordances between ranking lists and produces values between $-1$ (for inverted order) and $1$ (for identical order). A violin plot, which simultaneously depicts a boxplot and a density plot, is generated from the results.

Summary Kendall's tau:

| Task | mean | median | q25 | q75 |
|---|---|---|---|---|
| dummyTask | 0.9217905 | 0.9238095 | 0.8952381 | 0.952381 |

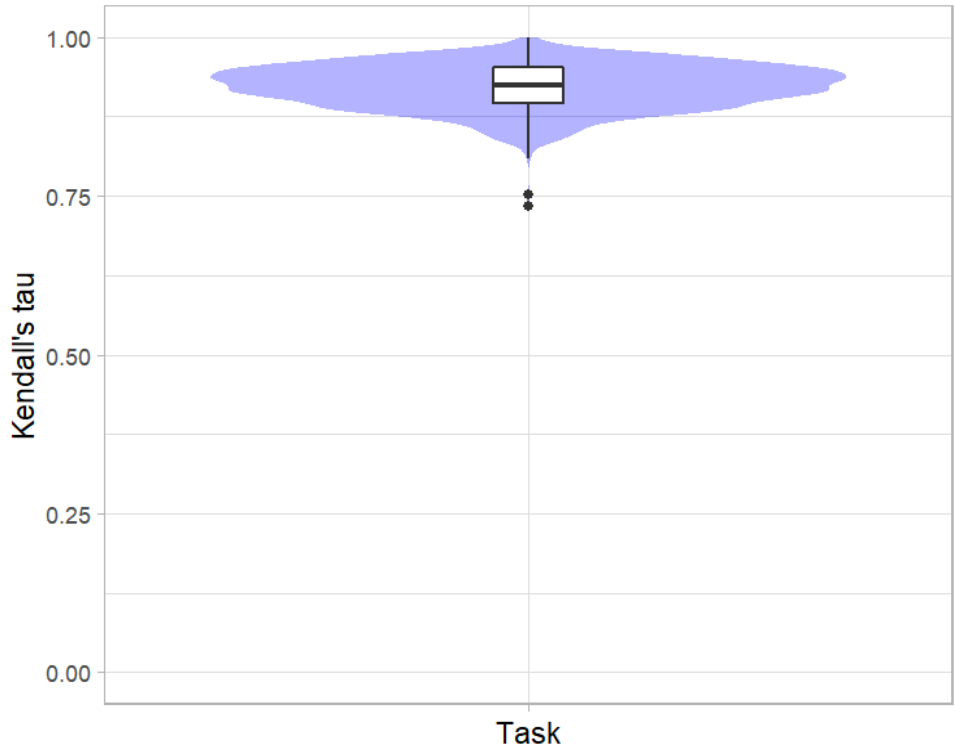

### 2.3.3 *Significance maps* for visualizing ranking stability based on statistical significance

*Significance maps* depict incidence matrices of pairwise significant test results for the one-sided Wilcoxon signed rank test at a 5% significance level with adjustment for multiple testing according to Holm. Yellow shading indicates that metric values from the algorithm on the x-axis were significantly superior to those from the algorithm on the y-axis, blue color indicates no significant difference.

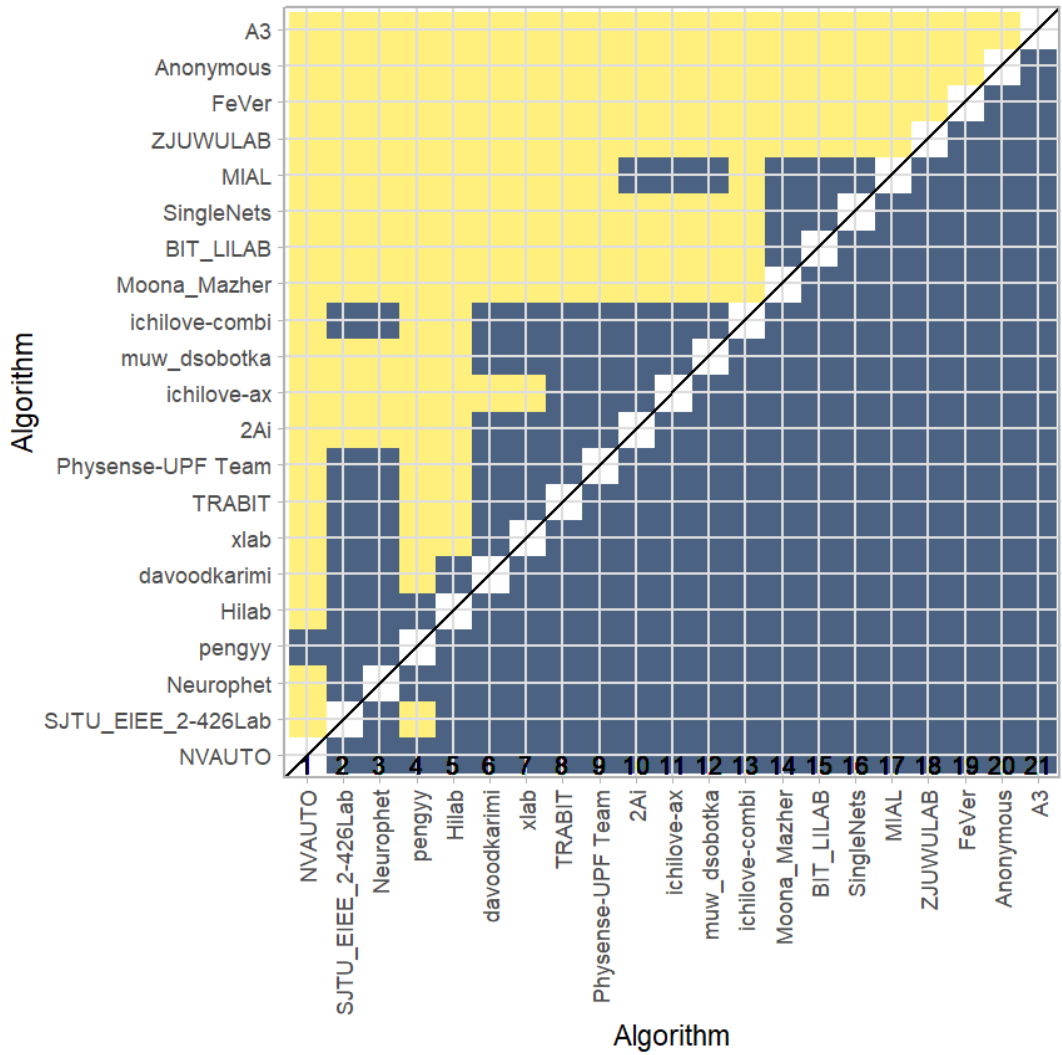

### 2.3.4 Ranking robustness to ranking methods

*Line plots* for visualizing ranking robustness across different ranking methods. Each algorithm is represented by one colored line. For each ranking method encoded on the x-axis, the height of the line represents the corresponding rank. Horizontal lines indicate identical ranks for all methods.

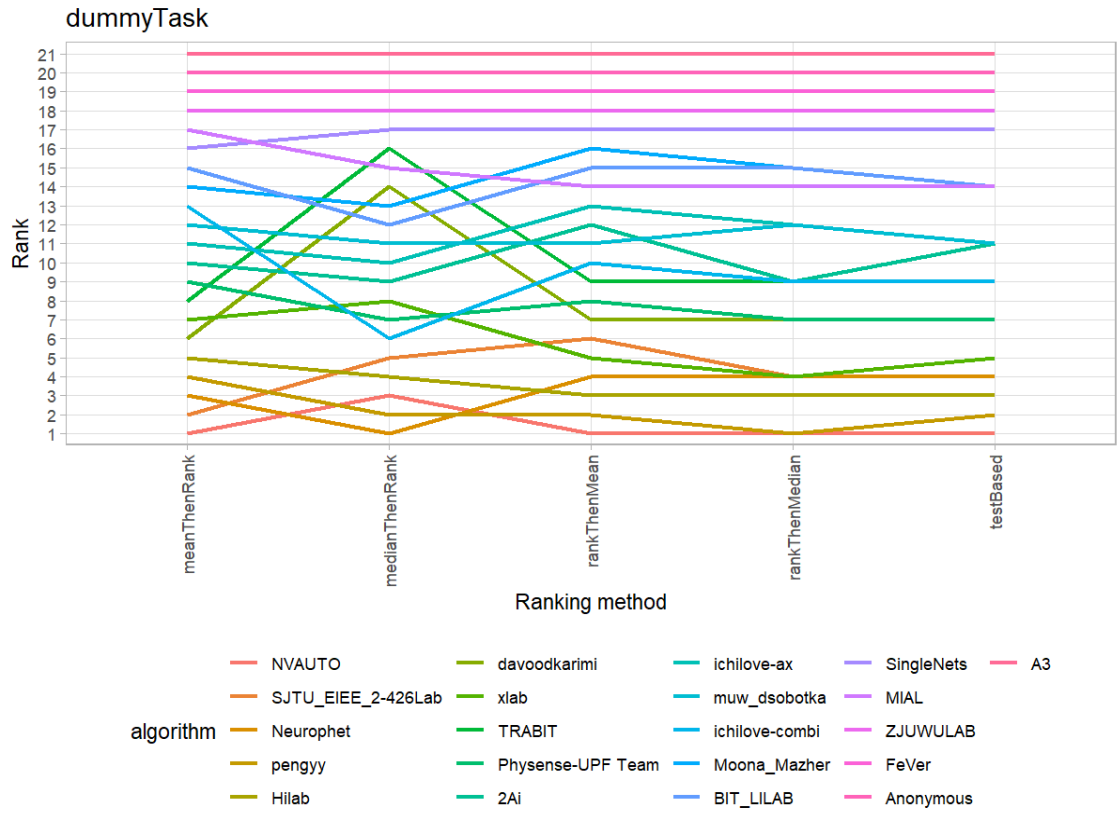

## 2.4 References

Wiesenfarth, M., Reinke, A., Landman, B.A., Eisenmann, M., Aguilera Saiz, L., Cardoso, M.J., Maier-Hein, L. and Kopp-Schneider, A. Methods and open-source toolkit for analyzing and visualizing challenge results. *Sci Rep* **11**, 2369 (2021). https://doi.org/10.1038/s41598-021-82017-6

M. J. A. Eugster, T. Hothorn, and F. Leisch, "Exploratory and inferential analysis of benchmark experiments," Institut fuer Statistik, Ludwig-Maximilians-Universitaet Muenchen, Germany, Technical Report 30, 2008. [Online]. Available: http://epub.ub.uni-muenchen.de/4134/.

### 3. Benchmarking report for multiTaskChallengeHD_combined

created by challengeR v1.0.2

12 April, 2022

This document presents a systematic report on the benchmark study "multiTaskChallengeHD_combined". Input data comprises raw metric values for all algorithms and cases. Generated plots are:

- Visualization of assessment data: Dot- and boxplot, podium plot and ranking heatmap
- Visualization of ranking stability: Blob plot, violin plot and significance map, line plot

Details can be found in Wiesenfarth et al. (2021).

#### 3.1 Ranking

Algorithms within a task are ranked according to the following ranking scheme:

*aggregate using function ("mean") then rank*

The analysis is based on 21 algorithms and 280 cases. 0 missing cases have been found in the data set.

Ranking:

|  | Hausdorff_mean | rank |
|---|---|---|
| NVAUTO | 14.01218 | 1 |
| Hilab | 14.56878 | 2 |
| 2Ai | 14.62533 | 3 |
| SJTU_EIEE_2-426Lab | 14.67063 | 4 |
| pengyy | 14.69852 | 5 |
| TRABIT | 14.90064 | 6 |
| Physense-UPF Team | 15.01775 | 7 |
| xlab | 15.26159 | 8 |
| Neurophet | 15.37497 | 9 |
| ichilove-combi | 16.03872 | 10 |
| davoodkarimi | 16.75480 | 11 |
| muw_dsobotka | 17.15931 | 12 |
| BIT_LILAB | 18.16190 | 13 |
| Moona_Mazher | 18.54792 | 14 |
| ichilove-ax | 21.32919 | 15 |
| MIAL | 25.10674 | 16 |
| SingleNets | 26.12090 | 17 |
| ZJUWULAB | 27.94799 | 18 |
| FeVer | 34.41890 | 19 |
| Anonymous | 37.38550 | 20 |
| A3 | 39.60763 | 21 |

## 3.2 Visualization of raw assessment data

### 3.2.1 Dot- and boxplot

*Dot- and boxplots* for visualizing raw assessment data separately for each algorithm. Boxplots representing descriptive statistics over all cases (median, quartiles and outliers) are combined with horizontally jittered dots representing individual cases.

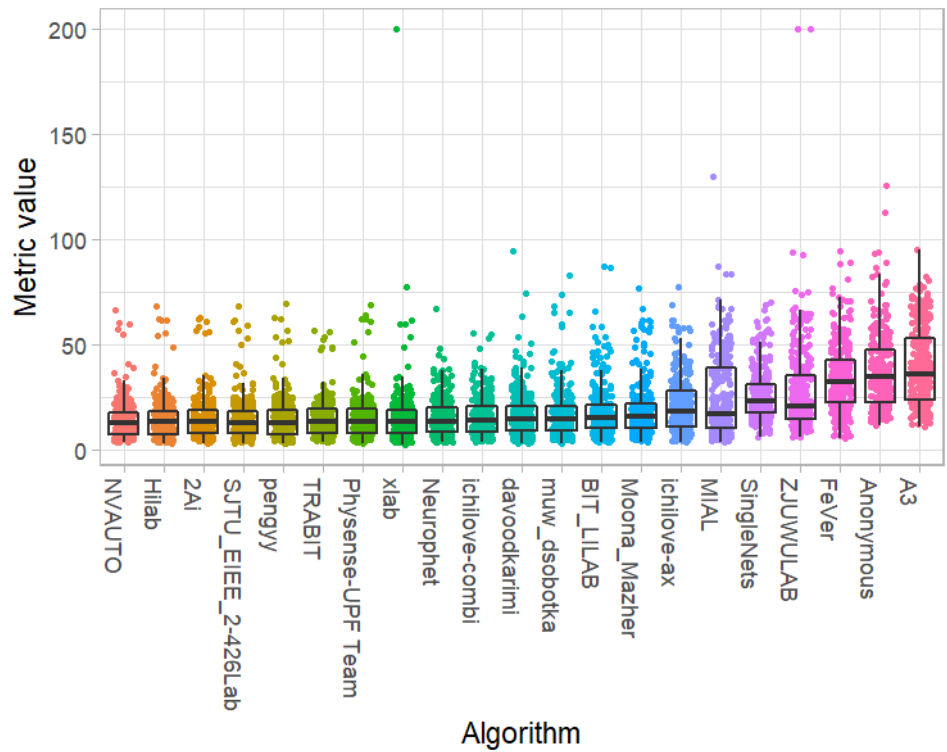

### 3.2.2 Podium plot

*Podium plots* (see also Eugster et al., 2008) for visualizing raw assessment data. Upper part (spaghetti plot): Participating algorithms are color-coded, and each colored dot in the plot represents a metric value achieved with the respective algorithm. The actual metric value is encoded by the y-axis. Each podium (here: $p$=21) represents one possible rank, ordered from best (1) to last (here: 21). The assignment of metric values (i.e. colored dots) to one of the podiums is based on the rank that the respective algorithm achieved on the corresponding case. Note that the plot part above each podium place is further subdivided into $p$ "columns", where each column represents one participating algorithm (here: $p = 21$). Dots corresponding to identical cases are connected by a line, leading to the shown spaghetti structure. Lower part: Bar charts represent the relative frequency for each algorithm to achieve the rank encoded by the podium place.

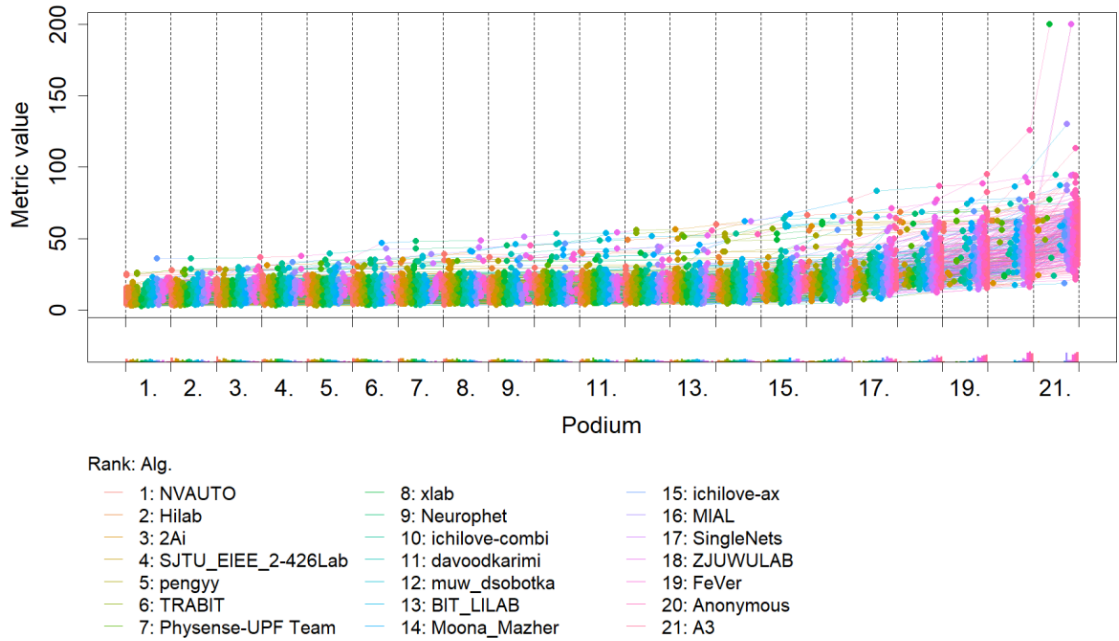

### 3.2.3 Ranking heatmap

*Ranking heatmaps* for visualizing raw assessment data. Each cell $(i, A_j)$ shows the absolute frequency of cases in which algorithm $A_j$ achieved rank $i$.

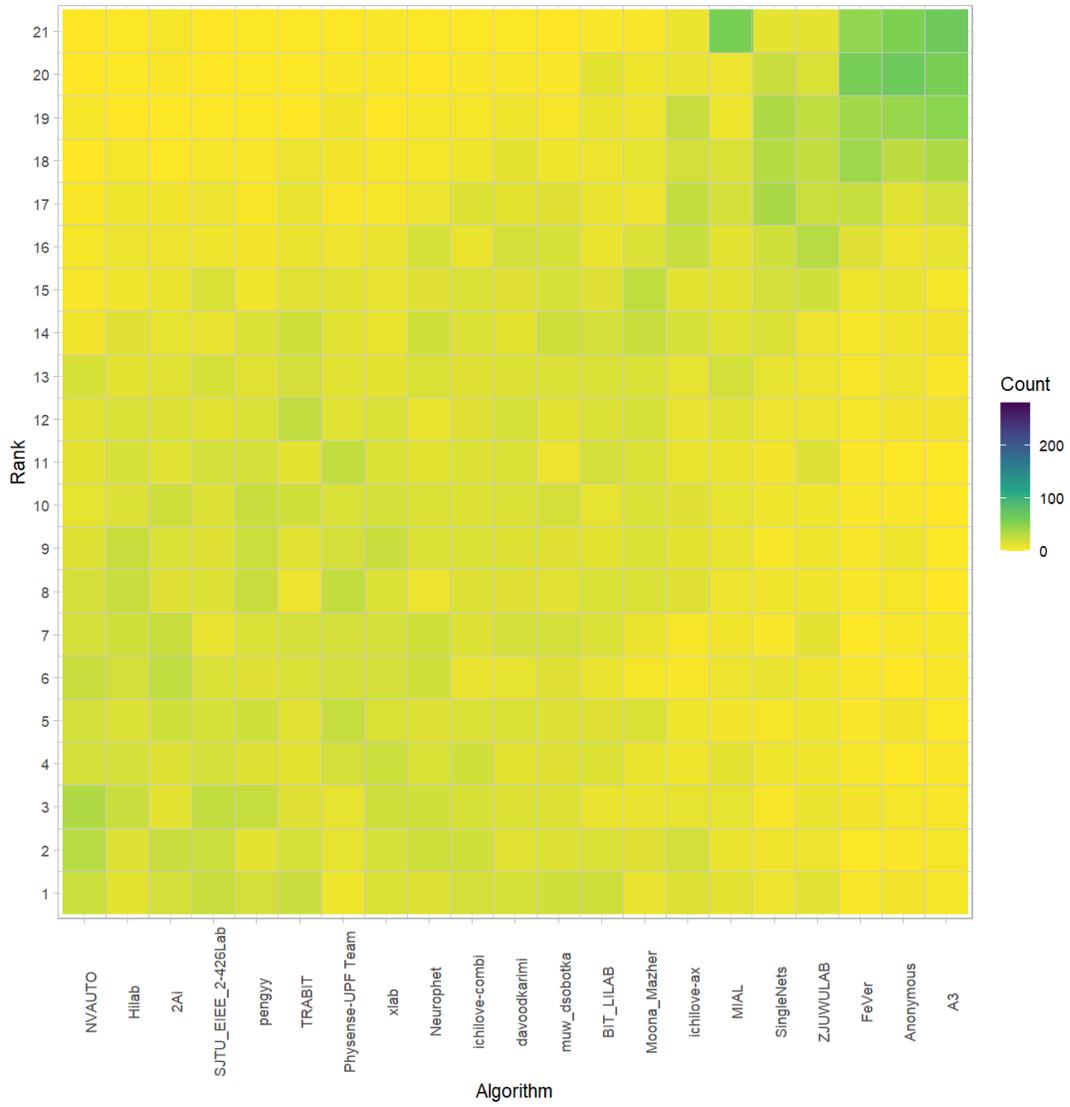

### 3.3 Visualization of ranking stability

#### 3.3.1 *Blob plot* for visualizing ranking stability based on bootstrap sampling

Algorithms are color-coded, and the area of each blob at position $(A_i, \text{rank } j)$ is proportional to the relative frequency $A_i$ achieved rank $j$ across $b = 1000$ bootstrap samples. The median rank for each algorithm is indicated by a black cross. 95% bootstrap intervals across bootstrap samples are indicated by black lines.

```
## Warning: `guides(<scale> = FALSE)` is deprecated. Please use `guides(<scale> =
## "none")` instead.
```

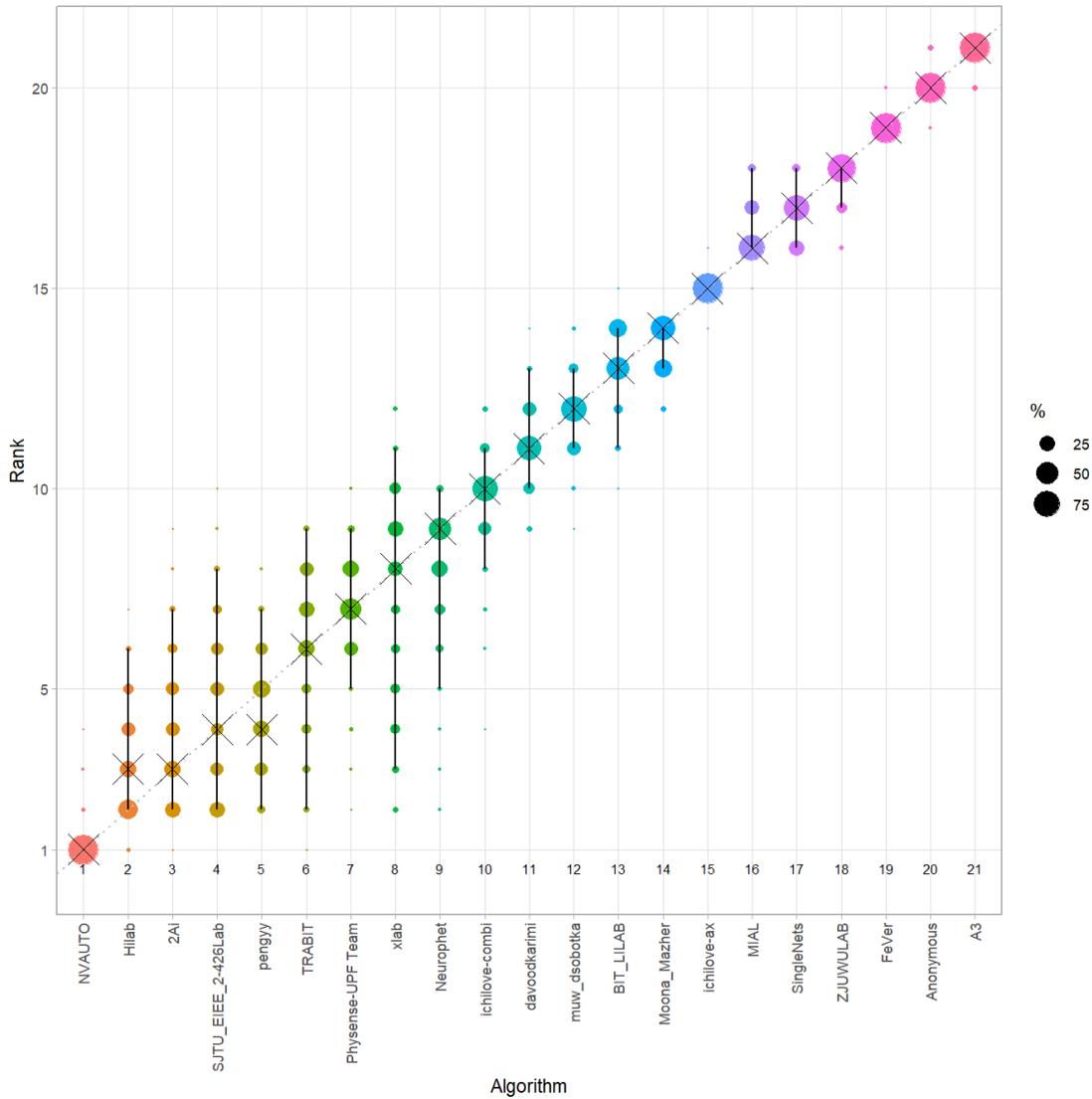

### 3.3.2 *Violin plot* for visualizing ranking stability based on bootstrapping

The ranking list based on the full assessment data is pairwise compared with the ranking lists based on the individual bootstrap samples (here $b = 1000$ samples). For each pair of rankings, Kendall's $\tau$ correlation is computed. Kendall's $\tau$ is a scaled index determining the correlation between the lists. It is computed by evaluating the number of pairwise concordances and discordances between ranking lists and produces values between $-1$ (for inverted order) and $1$ (for identical order). A violin plot, which simultaneously depicts a boxplot and a density plot, is generated from the results.

Summary Kendall's tau:

| Task | mean | median | q25 | q75 |
|---|---|---|---|---|
| dummyTask | 0.9334476 | 0.9333333 | 0.9142857 | 0.952381 |

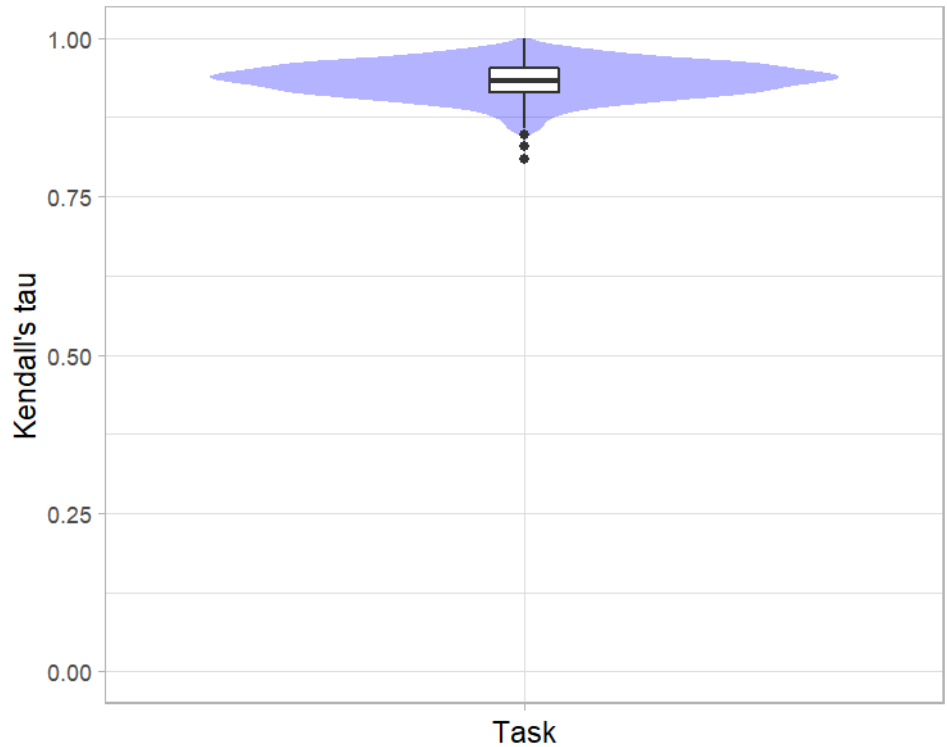

3.3.3 *Significance maps* for visualizing ranking stability based on statistical significance

*Significance maps* depict incidence matrices of pairwise significant test results for the one-sided Wilcoxon signed rank test at a 5% significance level with adjustment for multiple testing according to Holm. Yellow shading indicates that metric values from the algorithm on the x-axis were significantly superior to those from the algorithm on the y-axis, blue color indicates no significant difference.

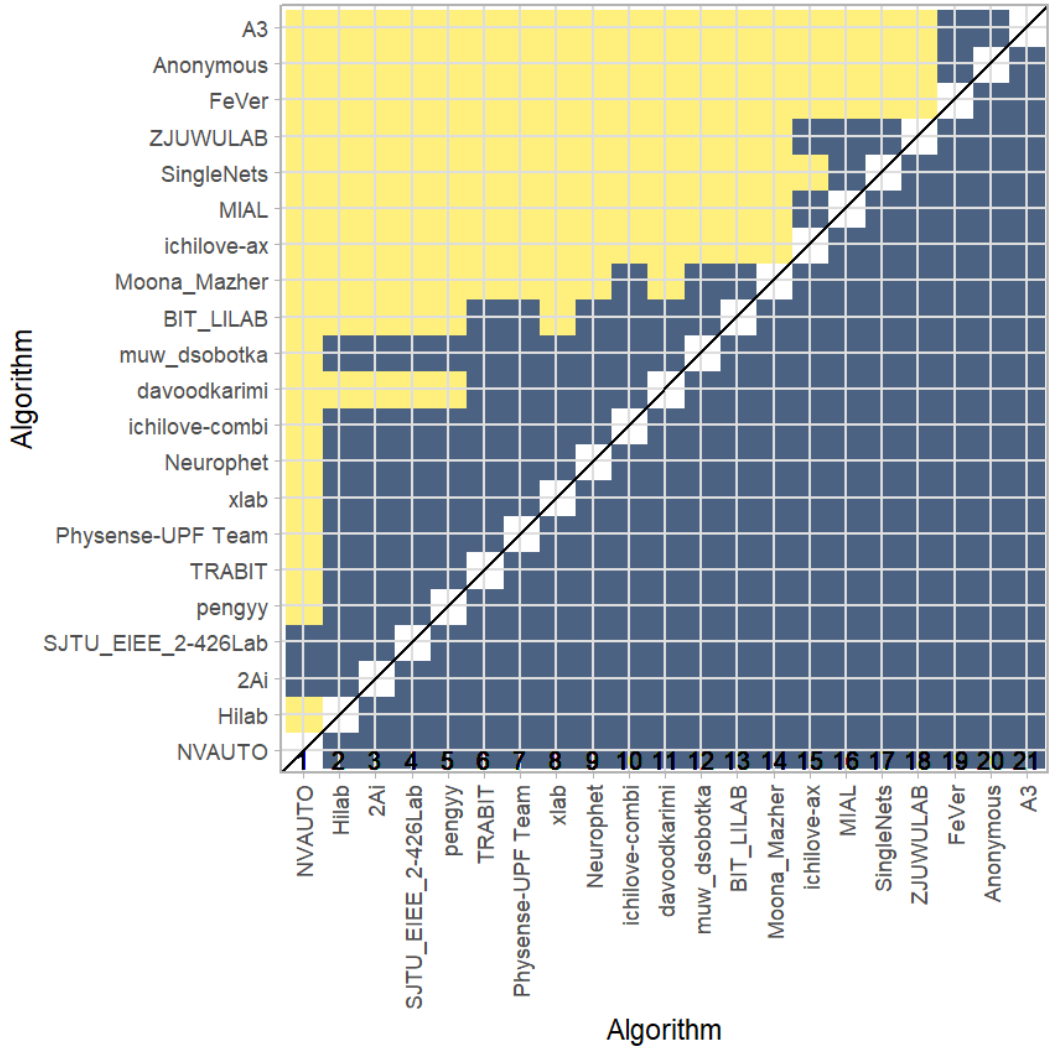

### 3.3.4 Ranking robustness to ranking methods

*Line plots* for visualizing ranking robustness across different ranking methods. Each algorithm is represented by one colored line. For each ranking method encoded on the x-axis, the height of the line represents the corresponding rank. Horizontal lines indicate identical ranks for all methods.

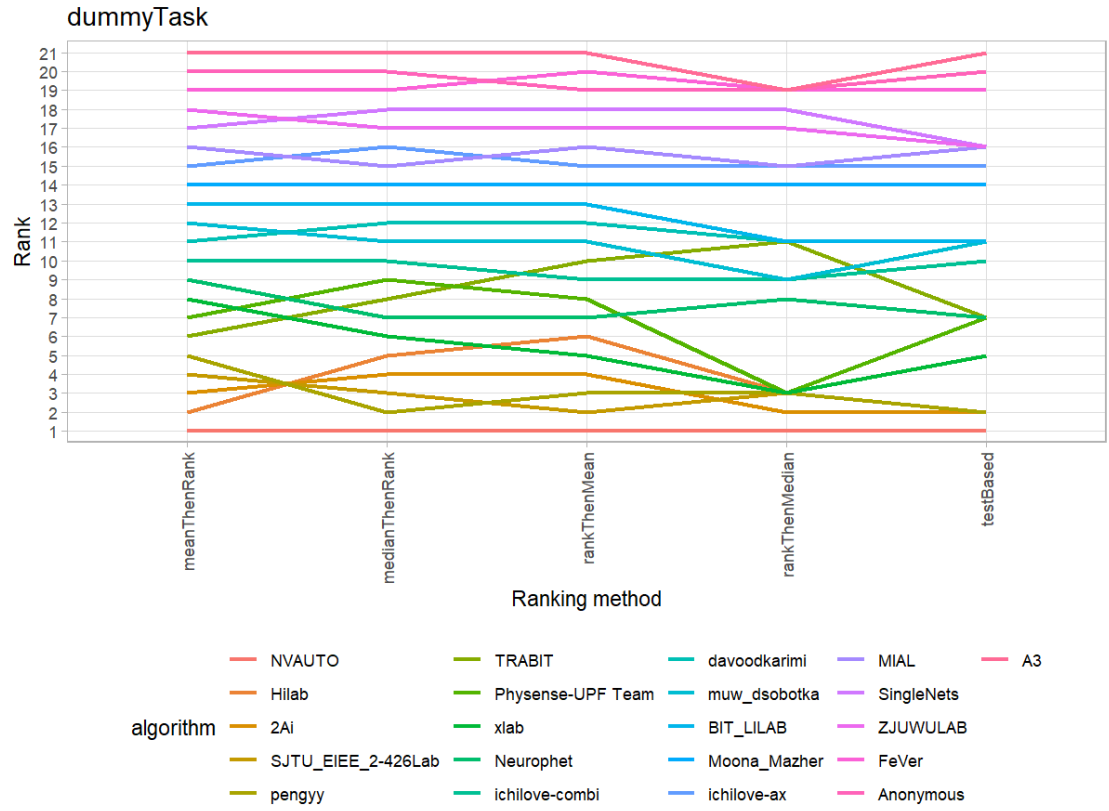

## 3.4 References

Wiesenfarth, M., Reinke, A., Landman, B.A., Eisenmann, M., Aguilera Saiz, L., Cardoso, M.J., Maier-Hein, L. and Kopp-Schneider, A. Methods and open-source toolkit for analyzing and visualizing challenge results. *Sci Rep* **11**, 2369 (2021). https://doi.org/10.1038/s41598-021-82017-6

M. J. A. Eugster, T. Hothorn, and F. Leisch, "Exploratory and inferential analysis of benchmark experiments," Institut fuer Statistik, Ludwig-Maximilians-Universitaet Muenchen, Germany, Technical Report 30, 2008. [Online]. Available: http://epub.ub.uni-muenchen.de/4134/.

## 4. Benchmarking report for multiTaskChallengeVolSim_combined

created by challengeR v1.0.2

12 April, 2022

This document presents a systematic report on the benchmark study "multiTaskChallengeVolSim_combined". Input data comprises raw metric values for all algorithms and cases. Generated plots are:

- Visualization of assessment data: Dot- and boxplot, podium plot and ranking heatmap
- Visualization of ranking stability: Blob plot, violin plot and significance map, line plot

Details can be found in Wiesenfarth et al. (2021).

### 4.1 Ranking

Algorithms within a task are ranked according to the following ranking scheme:

*aggregate using function ("mean") then rank*

The analysis is based on 21 algorithms and 280 cases. 0 missing cases have been found in the data set.

Ranking:

|  | Volume_Similarity_mean | rank |
|---|---|---|
| ichilove-ax | 0.8879231 | 1 |
| NVAUTO | 0.8849992 | 2 |
| SJTU_EIEE_2-426Lab | 0.8829732 | 3 |
| davoodkarimi | 0.8817364 | 4 |
| Neurophet | 0.8767972 | 5 |
| SingleNets | 0.8758778 | 6 |
| pengyy | 0.8746371 | 7 |
| muw_dsobotka | 0.8736648 | 8 |
| ichilove-combi | 0.8732283 | 9 |
| Hilab | 0.8732049 | 10 |
| xlab | 0.8731572 | 11 |
| BIT_LILAB | 0.8675333 | 12 |
| 2Ai | 0.8671930 | 13 |
| TRABIT | 0.8657770 | 14 |
| Moona_Mazher | 0.8657734 | 15 |
| Physense-UPF Team | 0.8633648 | 16 |
| MIAL | 0.8447757 | 17 |
| ZJUWULAB | 0.8353674 | 18 |
| FeVer | 0.8278280 | 19 |
| Anonymous | 0.8012755 | 20 |
| A3 | 0.7909992 | 21 |

## 4.2 Visualization of raw assessment data

### 4.2.1 Dot- and boxplot

*Dot- and boxplots* for visualizing raw assessment data separately for each algorithm. Boxplots representing descriptive statistics over all cases (median, quartiles and outliers) are combined with horizontally jittered dots representing individual cases.

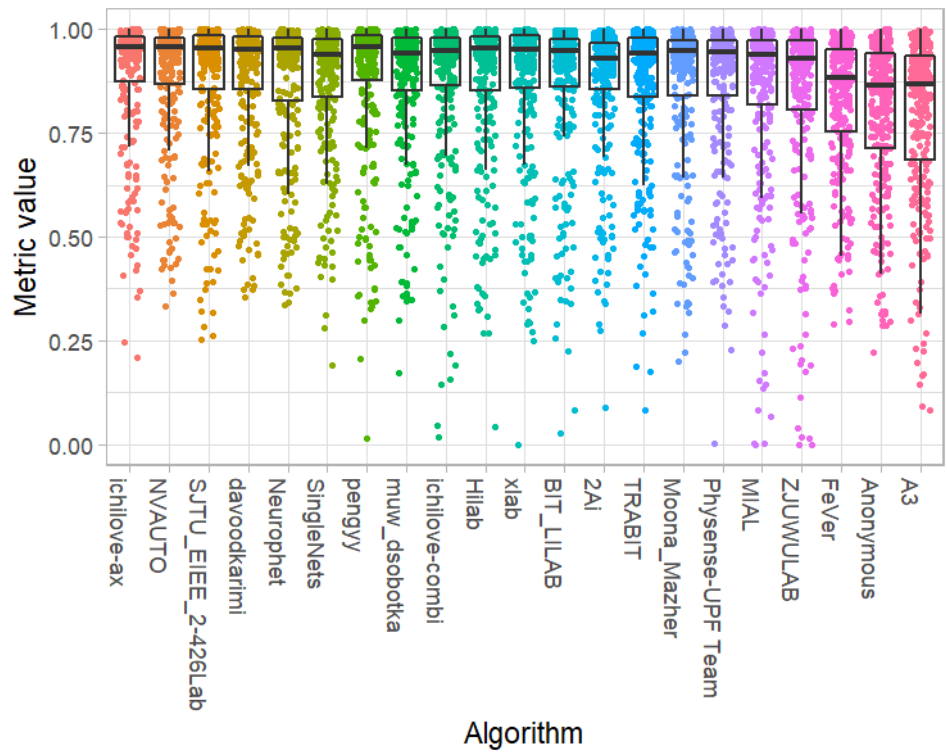

### 4.2.2 Podium plot

*Podium plots* (see also Eugster et al., 2008) for visualizing raw assessment data. Upper part (spaghetti plot): Participating algorithms are color-coded, and each colored dot in the plot represents a metric value achieved with the respective algorithm. The actual metric value is encoded by the y-axis. Each podium (here: $p$=21) represents one possible rank, ordered from best (1) to last (here: 21). The assignment of metric values (i.e. colored dots) to one of the podiums is based on the rank that the respective algorithm achieved on the corresponding case. Note that the plot part above each podium place is further subdivided into $p$ "columns", where each column represents one participating algorithm (here: $p = 21$). Dots corresponding to identical cases are connected by a line, leading to the shown spaghetti structure. Lower part: Bar charts represent the relative frequency for each algorithm to achieve the rank encoded by the podium place.

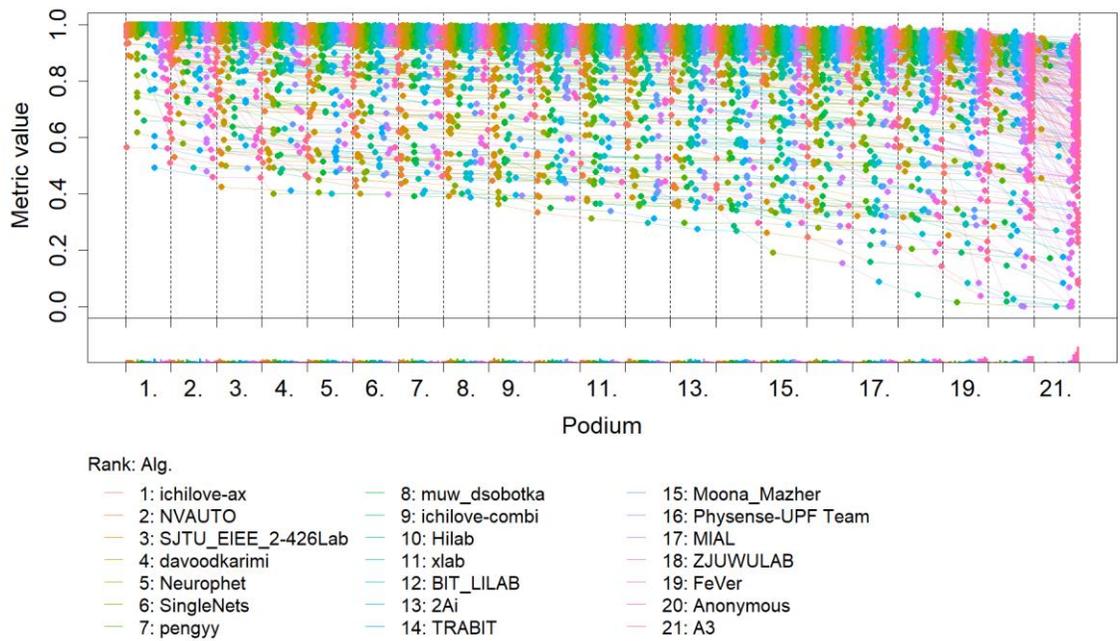

### 4.2.3 Ranking heatmap

*Ranking heatmaps* for visualizing raw assessment data. Each cell $(i, A_j)$ shows the absolute frequency of cases in which algorithm $A_j$ achieved rank $i$.

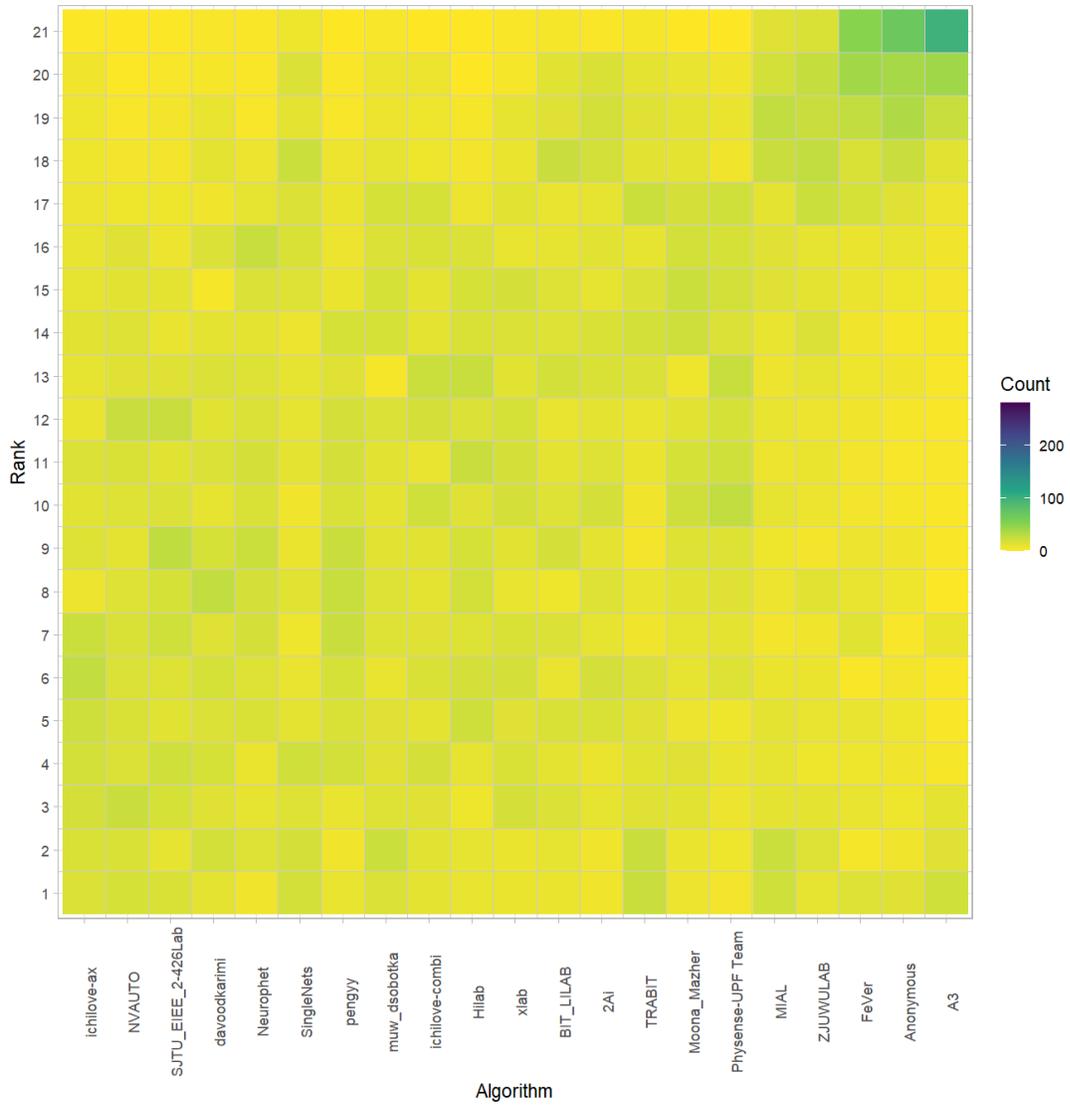

### 4.3 Visualization of ranking stability

#### 4.3.1 *Blob plot* for visualizing ranking stability based on bootstrap sampling

Algorithms are color-coded, and the area of each blob at position $(A_i, \text{rank } j)$ is proportional to the relative frequency $A_i$ achieved rank $j$ across $b = 1000$ bootstrap samples. The median rank for each algorithm is indicated by a black cross. 95% bootstrap intervals across bootstrap samples are indicated by black lines.

```
## Warning: `guides(<scale> = FALSE)` is deprecated. Please use `guides(<scale> =
## "none")` instead.
```

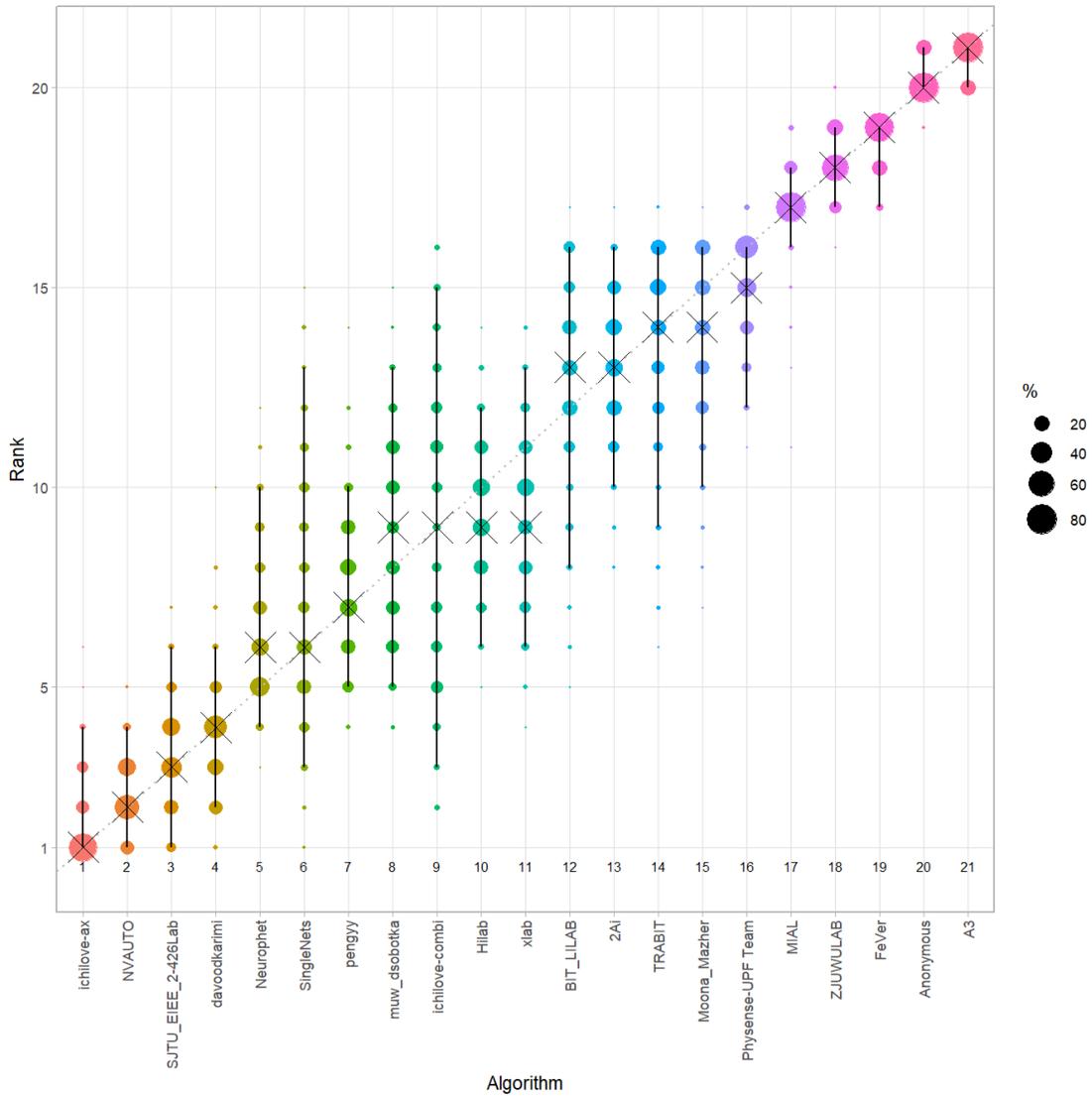

4.3.2 *Violin plot* for visualizing ranking stability based on bootstrapping

The ranking list based on the full assessment data is pairwise compared with the ranking lists based on the individual bootstrap samples (here $b = 1000$ samples). For each pair of rankings, Kendall's $\tau$ correlation is computed. Kendall's $\tau$ is a scaled index determining the correlation between the lists. It is computed by evaluating the number of pairwise concordances and discordances between ranking lists and produces values between $-1$ (for inverted order) and $1$ (for identical order). A violin plot, which simultaneously depicts a boxplot and a density plot, is generated from the results.

Summary Kendall's tau:

| Task | mean | median | q25 | q75 |
|---|---|---|---|---|
| dummyTask | 0.8657238 | 0.8666667 | 0.8380952 | 0.8952381 |

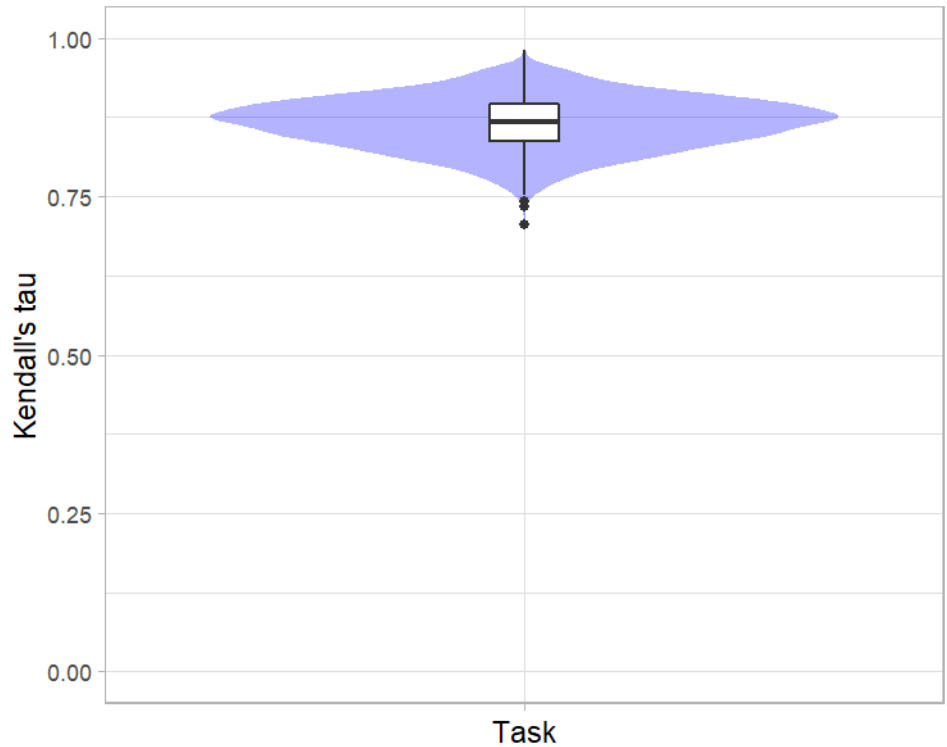

4.3.3 *Significance maps* for visualizing ranking stability based on statistical significance

*Significance maps* depict incidence matrices of pairwise significant test results for the one-sided Wilcoxon signed rank test at a 5% significance level with adjustment for multiple testing according to Holm. Yellow shading indicates that metric values from the algorithm on the x-axis were significantly superior to those from the algorithm on the y-axis, blue color indicates no significant difference.

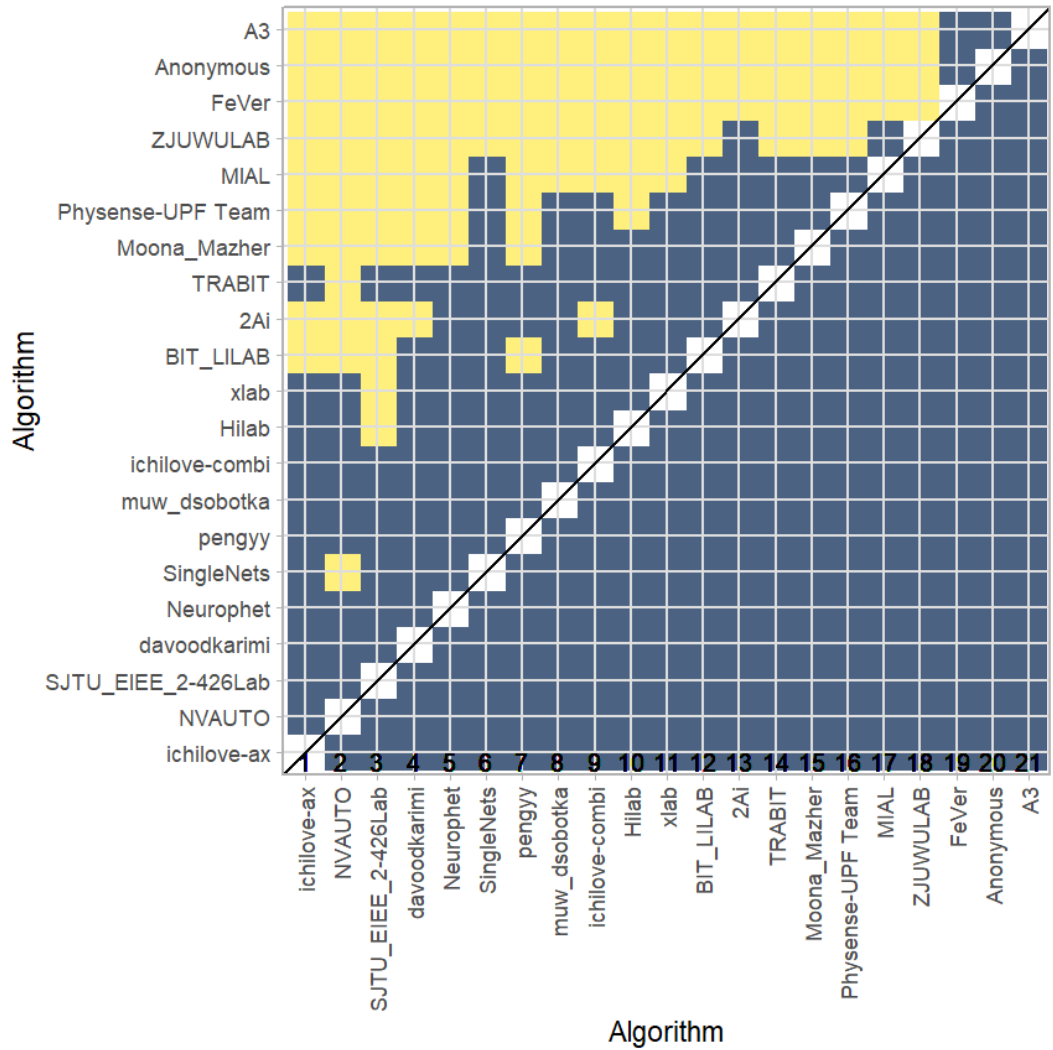

### 4.3.4 Ranking robustness to ranking methods

*Line plots* for visualizing ranking robustness across different ranking methods. Each algorithm is represented by one colored line. For each ranking method encoded on the x-axis, the height of the line represents the corresponding rank. Horizontal lines indicate identical ranks for all methods.

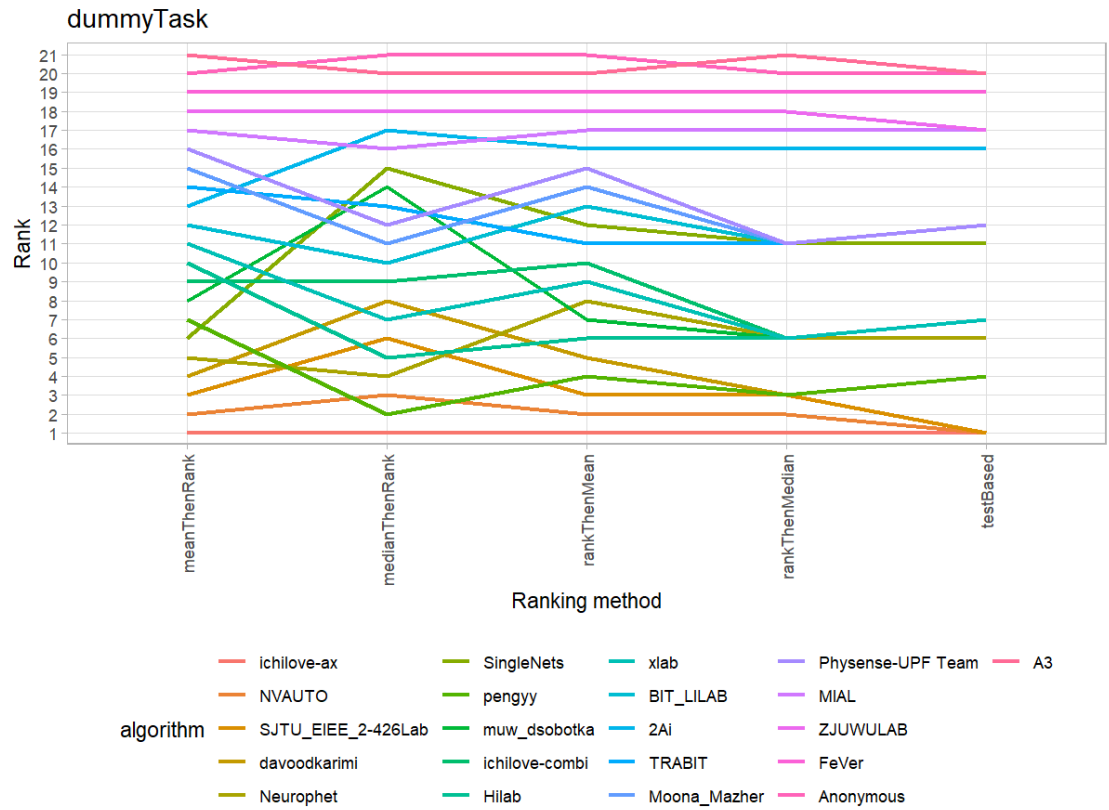

## 4.4 References

Wiesenfarth, M., Reinke, A., Landman, B.A., Eisenmann, M., Aguilera Saiz, L., Cardoso, M.J., Maier-Hein, L. and Kopp-Schneider, A. Methods and open-source toolkit for analyzing and visualizing challenge results. *Sci Rep* **11**, 2369 (2021). https://doi.org/10.1038/s41598-021-82017-6

M. J. A. Eugster, T. Hothorn, and F. Leisch, "Exploratory and inferential analysis of benchmark experiments," Institut fuer Statistik, Ludwig-Maximilians-Universitaet Muenchen, Germany, Technical Report 30, 2008. [Online]. Available: http://epub.ub.uni-muenchen.de/4134/.